\documentclass[12pt]{article}
\usepackage{color,amsmath}
\usepackage{verbatim}
\usepackage{multirow} 

\newif\ifhyprf
\hyprftrue
    
\RequirePackage{ifpdf}

\ifpdf
    \pdfoutput=1        
    \pdftrue
\else
    \pdffalse           
\fi

\ifpdf
  \usepackage[pdftex]{graphicx}
  \definecolor{rltred}{rgb}{0.75,0,0}
  \definecolor{rltgreen}{rgb}{0,0.3,0}
  \definecolor{rltblue}{rgb}{0,0,0.75}
  \definecolor{rltdarkgreen}{rgb}{0.1,0.6,0.1}
  \ifhyprf
     \usepackage[pdftex,
         colorlinks=true,
         urlcolor=rltblue,       
         filecolor=rltgreen,     
         linkcolor=rltred,       
         citecolor=rltdarkgreen, 
         pagebackref,
         pdfpagemode=None,
         pdftitle={Dark matter search in a Beam-Dump eXperiment (BDX) at Jefferson Lab},
         pdfauthor={many},
         pdfsubject={Beam dump experiment JLab Hall A},
         pdfkeywords={Hall A, Beam dump, dark photon}]{hyperref}
  \fi
  \usepackage{pdfcolmk}
  \DeclareGraphicsExtensions{.pdf,.png,.jpg}
\else
  \usepackage{graphicx}
  \DeclareGraphicsExtensions{.eps,.epsi,.ps,.eps.gz,.epsi.gz,.ps.gz}
\fi

\usepackage{graphicx,lscape,rotating}
\usepackage{amsmath}
\usepackage{amssymb,amsfonts,textcomp}

\usepackage{lineno}

\newcommand{\be}{\begin{eqnarray}}
\newcommand{\ee}{\end{eqnarray}}
\def\lsim{\mathrel{\rlap{\lower4pt\hbox{\hskip 0.5 pt$\sim$}}
\raise1pt\hbox{$<$}}}                
\def\gsim{\mathrel{\rlap{\lower4pt\hbox{\hskip1pt$\sim$}}
\raise1pt\hbox{$>$}}}

\newcommand{\s}{{\rm s}}
\newcommand{\m}{{\rm m}}
\newcommand{\apr}{{A^\prime}}
\newcommand{\MeV}{{\rm MeV}}
\newcommand{\GeV}{{\rm GeV}}

\makeatletter
\newcommand\arraybslash{\let\\\@arraycr}
\makeatother
\newcommand{\thedate}{\today}

\def\gevc2{(GeV/c)$^2$}
\newcommand*{\jlab}{Jefferson Lab, Newport News, VA 23606, USA}
\newcommand*{\nhs}{University of New Hampshire, Durham NH 03824, USA}
\newcommand*{\perimeter}{Perimeter Institute for Theoretical Physics, Waterloo, Ontario, Canada, N2L 2Y5}

\newcommand*{\frascati}{Istituto Nazionale di Fisica Nucleare, Laboratori Nazionali di Frascati, P.O. 13, 00044 Frascati, Italy}
\newcommand*{\genova}{Istituto Nazionale di Fisica Nucleare, Sezione di Genova\\ e Dipartimento di Fisica dell'Universit\`a, 16146 Genova, Italy}
\newcommand*{\sanita}{Istituto Nazionale di Fisica Nucleare, Sezione di Roma e Gruppo Collegato Sanit\`a,
 e  Universit\`a La Sapienza, Italy}

\newcommand*{\infnba}{Istituto Nazionale di Fisica Nucleare, Sezione di Bari e Dipartimento di Fisica dell'Universit\`a, Bari, Italy}
\newcommand*{\infnfe}{Istituto Nazionale di Fisica Nucleare, Sezione di Ferrara e Dipartimento di Fisica dell'Universit\`a, Ferrara, Italy}

\newcommand*{\ipn}{Institut de Physique Nucleaire d'Orsay, IN2P3, BP 1, 91406 Orsay, France}

\newcommand*{\ohio}{Ohio University, Department of Physics, Athens, OH 45701, USA}
\newcommand*{\gwu} {The George Washington University, Washington, D.C., 20052}

\newcommand*{\odu}{Old Dominion University, Department of Physics,
Norfolk VA 23529, USA}

\newcommand*{\edinb}{Edinburgh University, Edinburgh EH9 3JZ, United Kingdom}
\newcommand*{\glasgow}{University of Glasgow, Glasgow G12 8QQ, United Kingdom}

\newcommand{\sassari}{Universit\`a di Sassari e Istituto Nazionale di Fisica Nucleare, 07100 Sassari, Italy}
\newcommand{\torvergata} {Istituto Nazionale di Fisica Nucleare, Sezione di Roma-TorVergata e Dipartimento di Fisica dell'Universit\`a, Roma, Italy}
\newcommand{\catania} {Istituto Nazionale di Fisica Nucleare, Sezione di Catania,  Catania, Italy}
\newcommand{\torino} {Istituto Nazionale di Fisica Nucleare, Sezione di Torino,  Torino, Italy}
\newcommand{\padova} {Istituto Nazionale di Fisica Nucleare, Sezione di Padova,  Padova, Italy}
\newcommand*{\hu}{Department of Physics, Hampton University, Hampton VA 23668, USA }
\newcommand*{\canisius}{Canisius College, Buffalo NY 14208, USA}
\newcommand*{\occidental}{Occidental College, Los Angeles, California 90041, USA}
\newcommand*{\unm}{University of New Mexico, Albuquerque, New Mexico, NM}
\newcommand*{\mainz}{Institut fur Kernphysik, Johannes Gutenberg-Universitat Mainz, 55128 Mainz, Germany}
\newcommand*{\slac}{Stanford Linear Accelerator Center (SLAC), Menlo Park, CA 94025, US}
\newcommand*{\fnal}{Center for Particle Astrophysics, Fermi National Accelerator Laboratory, Batavia, IL 60510}
\newcommand*{\msu}{Mississippi State University, Mississippi State, MS 39762, USA}
\newcommand*{\stony}{C.N. Yang Inst. for Theoretical Physics, Stony Brook University, NY}
\newcommand*{\idaho}{Dept. of Physics, Idaho State University, Pocatello, ID 83201  USA}
\newcommand*{\spaolo}{Instituto de Fisica, Universidade de S\~ao Paulo, Brasil}
\newcommand*{\northw}{Northwestern University, Evanston, IL 60208, USA}
\newcommand*{\julich}{Nuclear Physics Institute and Juelich Center for Hadron Physics, Forschungszentrum Juelich, Germany}
\newcommand*{\mito}{Massachusetts Institute of Technology, Cambridge, MA 02139, USA}

\usepackage{alphalph}
\makeatletter
\newcommand*{\fnsymbolsingle}[1]{%
\ensuremath{%
\ifcase#1%
\or *%
\or \dagger
\or \ddagger
\or \mathsection
\or \mathparagraph
\else
\@ctrerr \fi
}%
}
\makeatother
\newalphalph{\fnsymbolmult}[mult]{\fnsymbolsingle}{}

\begin{document}

\begin{center}
{\tiny \leftline{V2.0}}
{\tiny\leftline{\thedate}}
\date{\today}
\vskip 1.0cm
{\bf\huge Dark matter search in a \\*[0.2cm]
Beam-Dump eXperiment (BDX) \\*[0.2cm]
at Jefferson Lab}
\vskip 1.cm
{  \large \it The BDX Collaboration }
\vskip 0.5cm

{M.~Battaglieri\footnote{Contact Person, email: Marco.Battaglieri@ge.infn.it}\footnote{Spokesperson}, A.~Bersani, B.~Caiffi, A.~Celentano$^\dag$, R.~De~Vita$^\dag$,  E.~Fanchini,  L.~Marsicano, P.~Musico, M.~Osipenko, F.~Panza, M.~Ripani, E.~Santopinto, M.~Taiuti\\}
{\small\it\genova}
\bigskip

{V.~Bellini, M.~Bond\'i, M.~De Napoli$^\dag$,  F.~Mammoliti, E.~Leonora, N.~Randazzo, G.~Russo, M.~Sperduto, C.~Sutera, F.~Tortorici\\}
{\small\it\catania\\}
\bigskip 

{N.Baltzell, M. Dalton, A. Freyberger, F.-X.~ Girod, V. Kubarovsky, E.~Pasyuk,
 E.S.~Smith$^\dag$, S.~Stepanyan, M. Ungaro, T.~Whitlatch\\}
{\it\small\jlab}
\bigskip

{E. Izaguirre$^\dag$\\}
{\small\it\perimeter}
\bigskip

{G. Krnjaic$^\dag$\\}
{\small\it\fnal}
\bigskip

{D.~Snowden-Ifft\\}
{\it\small\occidental}
\bigskip
\newpage
{D.~Loomba\\}
{\it\small\unm}
\bigskip

{M.~Carpinelli, V.~Sipala\\}
{\small\it\sassari}
\bigskip

{P. Schuster, N. Toro\\}  
{\small\it\slac}
\bigskip

{R.~Essig\\}
{\it\small\stony}
\bigskip

{M.H.~Wood\\}
{\it\small\canisius}
\bigskip

{M.Holtrop, R.~Paremuzyan\\}
{\it\small\nhs}
\bigskip

{G.~De~Cataldo, R.~De~Leo, D.~Di~Bari, L.~Lagamba, E.~Nappi, R.~Perrino\\}
{\small\it \infnba}
\bigskip

{I.~Balossino, L.~Barion, G.~Ciullo, M.~Contalbrigo, P.~Lenisa, A.~Movsisyan, F.~Spizzo, M.~Turisini\\}
{\small\it \infnfe}
\bigskip

{F.~De~Persio, E.~Cisbani, F.~Garibaldi, F.~Meddi, G.~M.~Urciuoli\\}
{\small\it \sanita}
\bigskip

{D.~Hasch, V.~ Lucherini, M.~Mirazita, S.~Pisano\\}
{\small\it \frascati}
\bigskip

{G.~Simi\\}
{\small\it\padova\\}
\bigskip

{ A.~D'Angelo, L.~Lanza, A.~Rizzo, C.~Schaerf, I.~Zonta \\}
{\small\it \torvergata}
\bigskip
\newpage

{A.~Filippi\\}
{\small\it\torino}
\bigskip

{S.~Fegan\\}
{\it\small\mainz}
\bigskip

{M.~Kunkel\\}
{\it\small\julich}
\bigskip

{M.~Bashkanov, P.~Beltrame, A.~Murphy, G.~Smith, D. Watts, N.~Zachariou, L.~Zana\\}
{\it\small\edinb}
\bigskip

{D. Glazier, D.~Ireland, B.~McKinnon, D. Sokhan\\}
{\it\small\glasgow}
\bigskip

{L.~Colaneri\\}
{\it\small\ipn}
\bigskip

{S.~Anefalos Pereira\\}
{\small\it \spaolo}
\bigskip

{A.~Afanasev, B.~Briscoe, I.~Strakovsky\\}
{\it\small\gwu}
\bigskip

{N.~Kalantarians\\}
{\it\small\hu}
\bigskip

{L.~Weinstein\\}
{\it\small\odu}
\bigskip

{K. P. Adhikari, J. A. Dunne, D. Dutta, L. El Fassi, L. Ye\\}
{\it\small\msu}
\bigskip 

{K.~Hicks\\}
{\it\small\ohio}
\bigskip

{P.~Cole\\}
{\it\small\idaho}
\bigskip

{S.~Dobbs\\}
{\it\small\northw}
\bigskip

{ C.~Fanelli\\}
{\it\small\mito}
\bigskip

\newpage
\begin{abstract}
\textcolor{red} {}
MeV-GeV dark matter (DM) is theoretically well motivated but remarkably unexplored.  
This proposal presents the MeV-GeV DM discovery potential for a $\sim$1 m$^3$ segmented CsI(Tl) scintillator detector placed downstream of the Hall A beam-dump at
Jefferson Lab, receiving up to 10$^{22}$ electrons-on-target (EOT) in 285 days. This experiment (Beam-Dump eXperiment or BDX)
would be sensitive to elastic DM-electron and to inelastic DM scattering at the level of 10 counts per year, reaching the limit of the neutrino irreducible
background. The distinct signature of a DM interaction will be an electromagnetic shower of few hundreds of MeV, together with a reduced activity in the
surrounding active veto counters.
A detailed description of the DM particle $\chi$ production in the dump and subsequent interaction in the detector has been performed by means of Monte Carlo simulations.
Different approaches have been used to evaluate the expected backgrounds:
the cosmogenic background has been extrapolated from the results obtained with a prototype detector running at INFN-LNS (Italy),  while the beam-related background
has been evaluated by GEANT4  Monte Carlo simulations.
The proposed experiment will be sensitive to large regions of DM parameter space, exceeding the discovery potential of existing and planned experiments in the MeV-GeV
DM mass range by up to two orders of magnitude.

\end{abstract}

\vskip 1.0cm
 
\end{center} 

\newpage
\tableofcontents\newpage


\section{Introduction}

We propose a beam-dump experiment to search for light (MeV-GeV) Dark Matter (DM). DM in this mass range is motivated by both experimental and theoretical considerations. On the theory side, simple extensions to the Standard Model (SM) can accommodate DM-SM interactions that yield the observed DM cosmological abundance. On the experimental side, such models also generically feature particles 
 that explain the currently discrepant value of the muon's anomalous magnetic moment and resolve anomalies in astrophysical observations, while simultaneously evading cosmological and direct-production constraints. 
  
This experiment could be performed by placing a detector downstream of one of the JLab experimental Halls to detect DM particles that could be
produced by the electron beam in the dump, pass through surrounding shielding material, and deposit visible energy inside the detector by 
scattering off various target particles or  --- if unstable --- by decaying inside the detector volume. 
A new underground facility placed $\sim 20m$ downstream of the beam dump of the experimental Hall-A
will host the detector, serving as a general-purpose facility for any future beam-dump experiments.
The run would be completely parasitic without affecting the normal operations and the physics program of the Hall. The most striking signal that this experiment would look for consists of events with $\sim$ GeV electromagnetic energy deposition. With the detector and the experimental set-up we are proposing, this signal will be easily detected over a negligible background.
This striking signature can arise in two classes of models: in those where DM scatters elastically off atomic electrons in the detector, and in those where the DM can scatter inelastically in the detector and subsequently de-excite in the active detector material into GeV-scale electron pairs, leading to the electromagnetic energy deposition. It will also be possible to detect the small signal produced by a light DM particle scattering off a nucleon. However the detection thresholds need to be fixed at values as low as possible ($\sim$ MeV), where 
 spurious signals from beam-related (neutrinos) and cosmogenic (muons, 
neutrons and neutrinos) backgrounds limit the measurement sensitivity of the DM-nucleon scattering channel. Nevertheless sensitivity to a broad range of possible DM interactions  could provide a tool for systematic and consistency checks.

The physics motivation for light DM is presented in Sec.~\ref{sec:theory} and, for completeness, more details are reported in Appendix~\ref{appx:phenomenology}.  Also in Sec.~\ref{sec:theory}, we describe the uniqueness of BDX to test a wide class of DM models and how its sensitivity can exceed that of other experiments proposed at FNAL, CERN, and LNF.  
Section~\ref{sec:proposed-measurement} describes the proposed experimental set-up: the Hall-A beam-dump, the detector, the data acquisition, the off-line analysis and the proposed new underground facility downstream of the dump.
Section~\ref{sec:setup} describes the model for DM production in the dump and DM interaction in the detector as well as detailed estimates of beam unrelated (cosmogenic) and beam-related backgrounds.
Cosmogenic backgrounds have been evaluated extrapolating results obtained in a dedicated measurement performed at INFN-LNS with a prototype of the BDX detector. All details are reported in Appendix~\ref{Section:BDX-protoype}. Beam-related background estimates were based 
 on GEANT4 Monte Carlo simulations of the beam interaction in the  Hall-A beam-dump and the downstream shielding.
Projections, counting rates for signal and background, and the expected reach of BDX are reported in Sec.~\ref{sec:fullexp}. 
Finally, an alternative detection technology based on a gaseous TPC (the DRIFT-BDX detector) is described in Appendix~\ref{sec:drift}.   We illustrate how a prototype that fits in the new  proposed facility would complement the main BDX detector, elaborate on the detection concept and show how it could provide powerful cross-checks on backgrounds in the beam dump lab.

\clearpage
\section{LDM search in  beam-dump experiments}
\label{sec:theory}
In this section we motivate the search for sub-GeV light Dark Matter (LDM) using an electron beam at Jefferson Lab. The theoretical viability of LDM
as well as a more thorough description of the simple models that can accommodate all existing data are presented in Appendix\,\ref{appx:phenomenology}.
In this proposal, we focus on models of LDM where LDM has non-gravitation interactions with the Standard Model (SM). These additional interactions are responsible for generating the correct DM abundance through the well known mechanism of thermal freeze-out \cite{PhysRevLett.39.165}.

The requirement of a thermal origin is a strict requirement on LDM models, as it sets a {\it minimum} interaction strength between DM and the SM. Combined with the requirement to only consider extensions of the SM that respect the known SM symmetries, the model parameter space in the paradigm of thermal-origin DM greatly simplifies. By and large the most viable model of LDM for exploration is that where LDM interacts with the SM through a ``dark photon'',  $\apr$. The dark photon, $\apr$,  kinetically-mixes with the Standard Model (SM) hypercharge. LDM particles, denoted by Dirac-fermions $\chi$ and $\bar \chi$ for concreteness, are produced via the real or virtual decay of the $\apr$. Note that the phenomenology discussed below is equally applicable to spin 0 LDM as well.

We use this scenario as the basis for studying the sensitivity of the experiment. The Lagrangian for this setup is
\cite{Holdom:1985ag}
 \be
 \label{eq:lagrangian}
{\cal L}_{\apr} &\supset & 
-\frac{1}{4}F^\prime_{\mu\nu} F^{\prime\,\mu\nu} + \frac{\epsilon_Y}{2} F^\prime_{\mu\nu} B_{\mu \nu} + \frac{m^2_{A^\prime}}{2} A^{\prime}_\mu A^{\prime\, \mu} + g_D \apr_\mu J^\mu_\chi  +  g_Y B_\mu J^\mu_Y   ,
\ee
where  $F^\prime_{\mu\nu} \equiv \partial_\mu A^\prime_\nu -  \partial_\nu A^\prime_\mu$ is the dark photon field strength,
$B_{\mu\nu} \equiv \partial_\mu B_\nu -  \partial_\nu B_\mu$ is the hypercharge field strength,
  $g_D \equiv \sqrt{4\pi \alpha_D}$ is the dark gauge coupling, and $J^\mu_\chi$ and $J^\mu_Y$ are the DM and SM hypercharge matter currents, respectively.
 After electroweak symmetry breaking, the kinetic mixing term, proportional to $\epsilon_Y$, induces mixing with the photon  and $Z$ boson 
 \be
  \frac{\epsilon_Y}{2} F^\prime_{\mu\nu} B_{\mu \nu}  ~~ \longrightarrow~~  \frac{\epsilon}{2} F^\prime_{\mu\nu} F_{\mu \nu} +  \frac{\epsilon_Z}{2} F^\prime_{\mu\nu} Z_{\mu \nu}  ~,~~
\ee
where  $\epsilon \equiv \epsilon_Y \cos\theta_W$, $\epsilon_Z \equiv \epsilon_Y \sin\theta_W$, and $\theta_W$ is the weak mixing angle.  As a result, we get dark photon interactions with dark and visible matter  
\be 
g_D \apr_\mu J^\mu_\chi  +  g_Y B_\mu J^\mu_Y ~~\longrightarrow~~  A^\prime_\mu ( g_D J^\mu_\chi + \epsilon e J^\mu_{\rm EM})~,
\label{eq:darkcurrent1}
\ee
where  $J^\mu_{\rm EM}$ is the usual SM electromagnetic current and we have omitted terms higher order in $\epsilon$. The phenomenological features of this model are:

\begin{itemize}
\item The SM fermions acquire an effective ``milli-charge'' under the short-range force carried by $\apr$, namely $\epsilon e$.
\item The phenomenology in the LDM sector will depend on how DM couples to $\apr$, {\it i.e.,} what is the dark current $J^\mu_\chi$. Two broad categories could ensue: DM couples to $\apr$ {\it diagonally}, or {\it off-diagonally}. For instance, if DM is a spin 1/2 fermion, as written in Eq.~\ref{eq:lagrangian}, these two categories correspond to whether DM is Dirac or Majorana, respectively. For Dirac-like DM, DM, which we denote also by $\chi$, can be produced along with its antiparticle, $\bar\chi$, through on-shell or off-shell production of $\apr$. In the Majorana scenario, the DM, now represented by $\chi_1$, is produced along with an excited state $\chi_2$ also through the $\apr$. We will focus on these two categories separately, as they lead to distinct signatures. We note that the phenomenological signatures we focus on are quite generic, as they also can be realized in the case where DM is a spin 0 boson as well. We discuss these scenarios in more detail in Appendix\,\ref{appx:phenomenology}.

\end{itemize}

\begin{figure}[t!]
\center
\includegraphics[width=10cm]{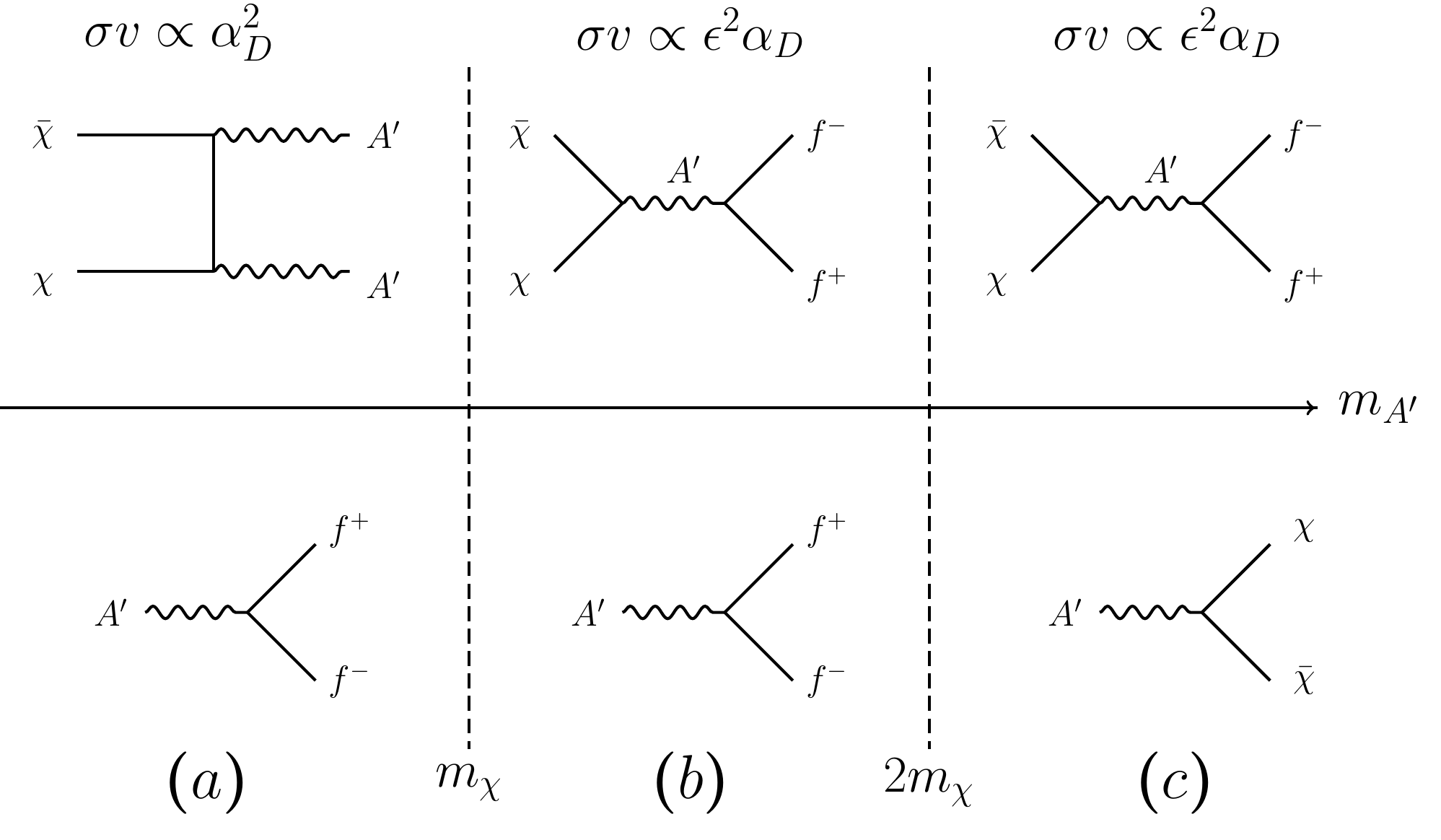}
\caption{ Classification of dominant DM annihilation and mediator decay channels in the benchmark dark photon ($\apr$) mediated scenario
 for different $m_{\apr}/m_\chi$ ratios were $f$ is a charged SM fermion -- similar categorizations exist for other mediators. Also, the same classification holds for Majorana-DM, with the substitution $(\chi,\bar\chi)\rightarrow (\chi_1,\chi_2)$. {\bf (a)} In the left column, the mediator is lighter than the DM, so for $\epsilon e \ll g_D$ the dominant annihilation is in the ``secluded" channel, which is independent of the mediator coupling to the SM. This scenario has no direct thermal target; every arbitrarily small values of  $\epsilon$ are compatible with a thermal annihilation rate. {\bf (b)} The middle column represents the $m_\chi < m_{\apr} < 2 m_\chi$ window in which the annihilation rate is sensitive to $\epsilon$ but the mediator decays visibly. This regime has a predictive thermal relic target, which can be tested by probing sufficiently small values of $\epsilon$ in searches for visibly decaying dark photos (e.g. HPS, APEX, Belle II). {\bf (c)} The right column where $m_{\apr} > 2 m_\chi$ offers ample parameter space with a predictive thermal target and features mediators that decay {\it invisibly} to DM states. Since $\sigma v \propto \epsilon^2 \alpha_D$ this scenario has a thermal target which can be probed by testing sufficiently small values of this combination at BDX, whose
 signal yield scales as the same combination of input parameters.   }
\label{fig:breakdown}
\end{figure}

\begin{figure}[t!]
\center
\hspace{-1cm}\includegraphics[width=7.7cm]{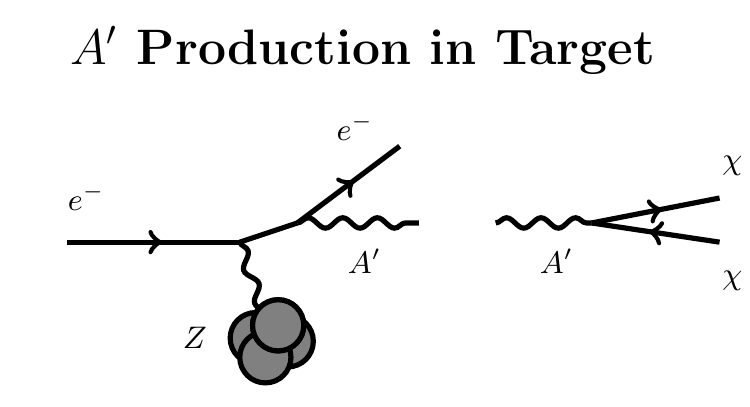} \hspace{1.5cm}
\includegraphics[width=6.3cm]{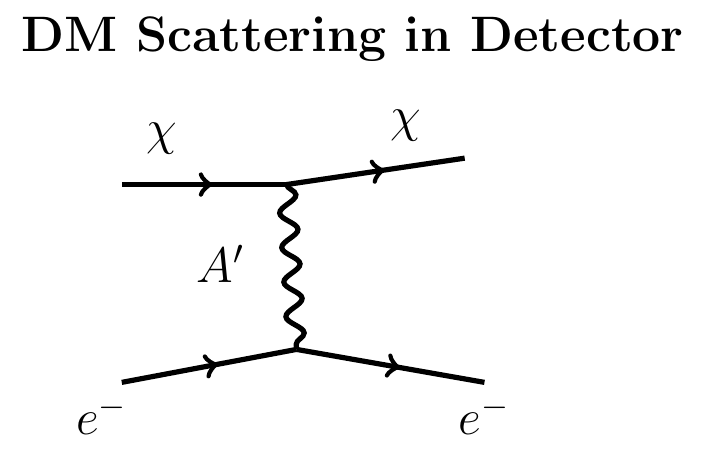} 
\caption{   a) $\chi \bar \chi$ pair production in electron-nucleus collisions via the Cabibbo-Parisi radiative process (with $A'$ on- or off-shell) and b)
$\chi$  scattering off an electron in the detector. 
\label{fig:prod} }
\end{figure}

In the paradigm of a thermal origin for DM, DM would have acquired its current abundance through annihilation directly/indirectly into the SM. Here, we focus on the direct annihilation regime, in which $m_{\chi} < m_{\rm MED.}$, where $m_{\rm MED.}$ would correspond to $m_\apr$ in the model we are focusing on (see Appendix\,\ref{appx:phenomenology} for more details).  The annihilation rate scales as  (see Fig.~\ref{fig:breakdown})
\be
\hspace{1cm}({\rm ``direct" ~annihilation})~~~~\langle \sigma v \rangle   \sim       \frac{      g_{D}^2 \,  g_{\rm SM}^2  \,  m_{\chi}^2 }{   m_{\rm MED}^4  \!\!}~~ ,~~~~~~~~~  ~~~~
\ee
and offers a clear, predictive target for discovery or falsifiability since the dark coupling $g_{D}$  and mass ratio $m_{{\chi}}/m_{\rm MED}$ are 
at most ${\cal O}(1)$ in this $m_{\rm MED} > m_{\chi}$ regime, so there is a {\it minimum} SM-mediator coupling compatible with a thermal history; larger values of $g_D$ require non-perturbative dynamics in the mediator-SM coupling or intricate model building.

In the direct annihilation regime, up to order-one factors, the minimum annihilation rate requirement translates into a minimum value of the dimensionless combination 
\be\label{eq:generic-thermal-target1}
\boxed{
y \equiv  \frac{    g_D^2 \, g_{\rm SM}^2  }{4\pi} \, \left(\frac{m_\chi}{m_{\rm MED}}\right)^4 \gtrsim  \langle \sigma v \rangle_{\rm relic}   \, m_\chi^2~,~~
}
\ee
which, up to order one factors, is valid for every DM/mediator variation provided that $m_{\rm DM} < m_{\rm MED.}$.
We will use this target throughout this document to assess experimental sensitivity to various LDM scenarios;
 reaching at least this benchmark sensitivity is necessary to discover or falsify a large class of simple direct annihilation models.

\subsection{Important Variations}

\subsubsection{Inelastic Dark Matter (iDM)}

If the $A^\prime$ couples to a DM fermion with both Dirac and Majorana masses, the leading interaction is generically off-diagonal and 
\be
A_\mu^\prime  J^\mu_{DM} ~ \to ~ A_\mu^\prime  \bar \chi_1 \gamma^\mu \chi_2~,~
\ee
where the usual Dirac fermion $\chi$ decomposes into two Majorana (``pseudo-Dirac") states $\chi_{1,2}$ with masses $m_{1,2}$ split
by an amount $\Delta$. This kind of scenario is well motivated for LDM which is safe from CMB constraints \cite{Izaguirre:2015yja}.
and has striking implications for possible signatures at BDX.

\subsubsection{Leptophilic $A^\prime$ and Dark Matter}
A similar scenario involving a vector mediator arises from gauging the difference  between electron and muon numbers under the abelian $U(1)_{e-\mu}$ group. 
Instead of kinetic mixing, the light vector particle here has direct couplings to SM leptonic currents 
\be
A_\beta^\prime J^\beta_{SM} ~\to~   g_V   A_\mu^\prime \left(  \bar e \gamma^\beta e  +  \bar \nu_e \gamma^\beta \nu_e -
\bar \mu \gamma^\beta \mu  +  \bar \nu_\mu \gamma^\beta \nu_\mu \right)  ~,~~
\ee
where $g_V$  is the gauge coupling of this model, which we normalize to the electric charge,  $g_V \equiv \epsilon e$ and consider
parameter space in terms of $\epsilon$, like in the case of kinetic mixing. Note that here, the $A^\prime$ does not couple
to SM quarks at tree level, but it does couple to neutrinos, which carry electron or muon numbers.
Note also that this scenario is one of the few combinations of SM quantum numbers that can be gauged without requiring additional field content.
Assigning the DM $e-\mu$ number yields the familiar $g_D A^\prime_\beta J^{\beta}_{\rm DM}$ interaction as in Eq.~\ref{eq:lagrangian}.
 Both of these variations can give rise to thermal LDM as discussed above.

 \subsection{Muon Anomalous Magnetic Moment}
 It is well known that a light, sub-GeV scale gauge boson (either a kinetically mixed dark photon, or a leptophilic gauge boson that couples to muons)
  can ameliorate the $\sim 3.5 \sigma$ discrepancy between the
theoretical prediction and experimental observation of the muon's anomalous magnetic moment \cite{Pospelov:2006sc}. 
Although there are many active efforts to search for dark photons independently of their connection to dark matter, the
success of these efforts relies on the assumption
that the $A^\prime$ is the lightest particle in its sector and that its primary decay channel only depends on $\epsilon$. Furthermore, if the 
$A^\prime$ decays predominantly to SM particles, this explanation of the $(g-2)_\mu$ anomaly has been ruled out (see discussion in Sec.~\ref{sec:fullexp}).

If, however, the $A^\prime$  couples to a light DM particle $\chi$  ($m_{A^\prime} > m_\chi$), then the parameter space for reconciling 
theory and experiment with regard to $(g-2)_\mu$ remains viable. For large values of $\alpha_D$, this explanation of the 
anomaly is  under significant tension with existing constraints, but for $\alpha_D \ll \alpha_{\rm EM}$ this explanation is still viable and 
most of the remaining territory can be tested with BDX@JLab (see discussion in Sec.~\ref{sec:fullexp}).
 
In the remainder of this section, we review  the salient
features of  LDM production at an electron fixed-target facility. Secondly, we give an overview of the status of LDM models parameter space, and the capabilities of present, and near future proposals to make progress in the field. Finally, we highlight how BDX uniquely fits in this developing field.
 \begin{figure}[t]
\center
\includegraphics[width=11.0cm]{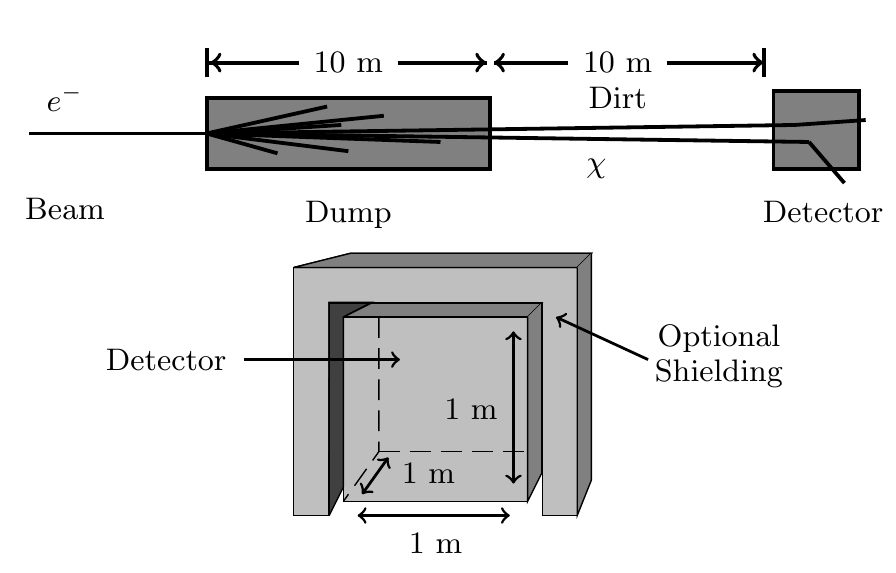}
\caption{Schematic of the experimental setup. A high-intensity multi-GeV electron beam impinging on a beam-dump produces a secondary beam of dark sector states.
In the basic setup, a small detector is placed downstream with respect to the beam-dump so that muons and energetic neutrons are entirely ranged out.
\label{fig:setup} }
\end{figure}

\subsection{Production and Detection \label{sec:elastic}}
Whether the dark sector is quite simple or has a rich structure of light particles, the fixed-target phenomenology of stable $\chi$s (or unstable $\chi$s with lab-frame
lifetimes $\gtrsim \mu\s$) is well-described by the simplest case --- the Lagrangian from Eq.~\ref{eq:lagrangian}. Here the label $\chi$ could refer to scalar or fermion LDM, with diagonal or off-diagonal couplings to $\apr$.

In this theory, $\chi$s can therefore be pair-produced radiatively in electron-nucleus collisions in the dump (see Fig.~\ref{fig:prod}a).  A fraction of these relativistic
particles then scatter off nucleons, nuclei, or electrons in the detector volume (see Fig.~\ref{fig:prod}b), positioned downstream from the dump or target.

If $m_{A^\prime} < 2 m_\chi$,  the dominant $\chi$ production mechanism in
an electron fixed-target experiment is the radiative process illustrated in 
Fig. \ref{fig:prod}a) with off-shell $A'$. In this regime, the $\chi$ production yield scales as $\sim \alpha_D \epsilon^2/m^2_\chi$ ($\alpha_D\equiv{g_D}^2/4\pi$),
while $\chi$-nucleon scattering in the detector via $A^{\prime}$ exchange (see Fig. \ref{fig:prod}b))
 occurs with a rate proportional to $ \alpha_D  \epsilon^2/m_{A'}^2$ over most of the mass range. 
 Thus, the total signal yield scales as
\be
N_{\chi} \sim   \frac{  \alpha_D^2 \epsilon^4 }{m_{\chi}^2m_{A'}^2}.  ~~
\ee

If $m_{A^\prime} > 2 m_\chi$, the secondary $\chi$-beam arises from
 radiative on-shell $A^{\prime}$ production followed by $A^\prime \to \bar \chi \chi$ decay.  In this regime, the $\chi$ production and
  the detector scattering rates are respectively proportional to $\epsilon^2/m_{A'}^2$  and $  \alpha_D  \epsilon^2/m_{A'}^2$ and the signal yield scales as  
\be
N_{\chi} \sim  \frac{ \alpha_D \epsilon^4}{m_{A'}^4} ~~.
\ee
Thus, for each $\alpha_D$ and $m_{A'}$, we can extract an $\epsilon$-sensitivity corresponding to a given scattering yield.

\begin{figure}[htb!]
\center
\hspace{-1.cm}\includegraphics[width=7.8cm]{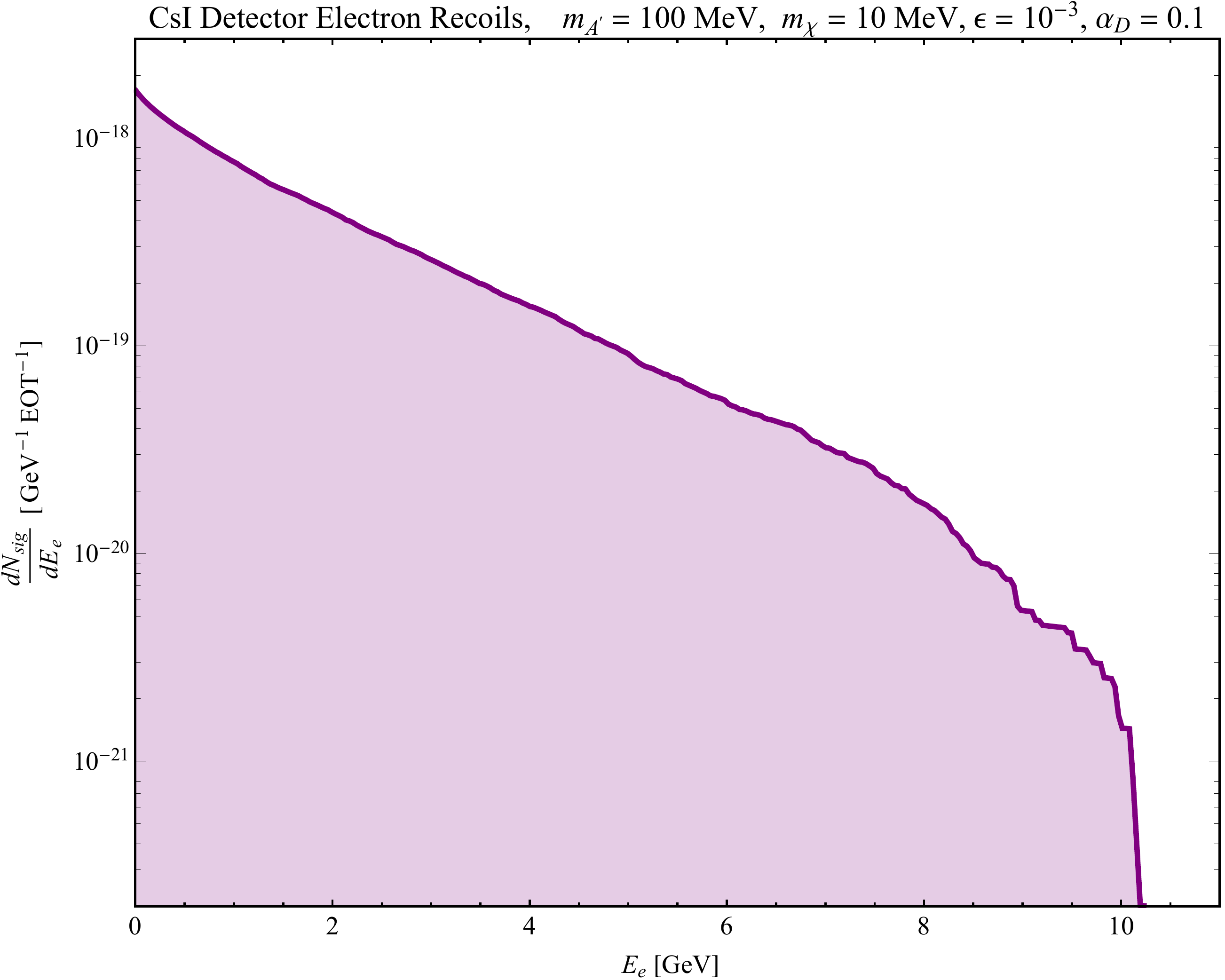}
\hspace{0.4cm}\includegraphics[width=7.8cm]{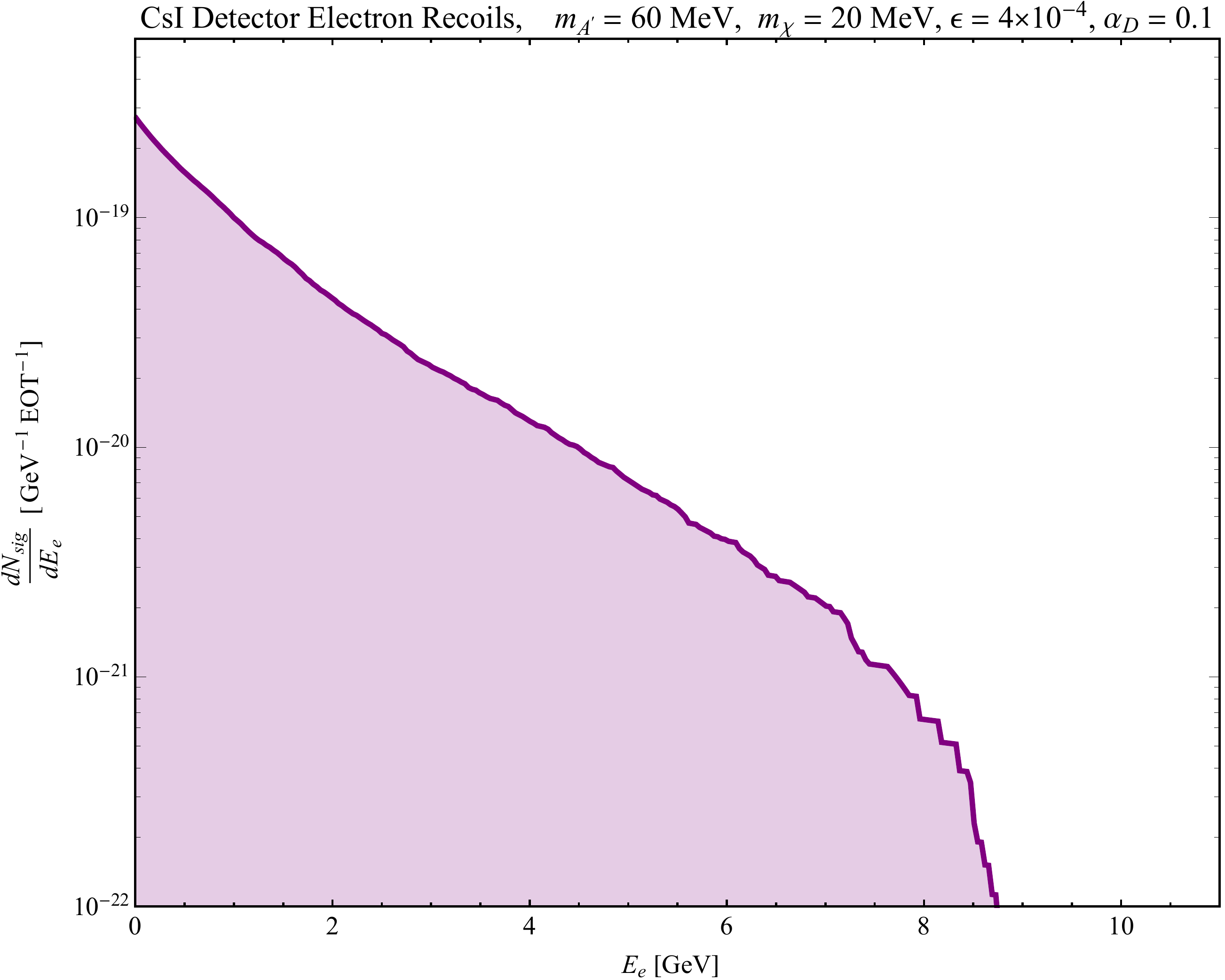}
\caption{Signal energy distributions of scattered electrons for the two choices of M$_\chi$ and M$_{A'}$. The
distributions are based on a simulated population of $\chi$ particles after applying geometric acceptance.
\label{fig:energyE} }
\end{figure}

A generic sketch of a beam dump experiment is shown in Fig.\,\ref{fig:setup}, where the DM particles are produced in the beam dump and
traverse unimpeded through sufficient material that eliminates all SM particles aside from neutrinos. They then scatter in the shielded detector downstream.
The experimental signal is an electromagnetic shower induced by $\chi$-electron scattering.  Because the electron is light and the $\chi$ are energetic,
scattered electrons typically carry GeV-scale
energy and are therefore subject to much lower backgrounds than nucleon scattering.
Figure~\ref{fig:energyE} shows the lepton recoil energy for different choices of M$_\chi$ and M$_{A'}$.
Indeed, for models with kinetically mixed mediators which produce both electron- and nucleon-scattering signals, electron-recoil searches at BDX will have the greatest sensitivity. As such, the experimental detection of the low-energy scattered nucleons is a secondary goal of the experiment because it provides an alternative probe of LDM production and is sensitive to some models that have distinctive signals primarily in nucleon-scattering. However, backgrounds
from cosmic-ray neutrons are expected to limit the sensitivity in this channel.

\begin{figure}[t!]
\center
\hspace{-1.4cm} \includegraphics[width=7.7cm]{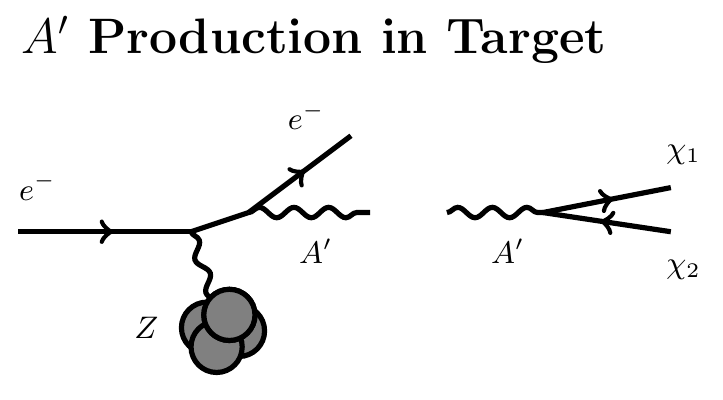}\\ 
\vspace{1cm}
\includegraphics[width=9.5cm]{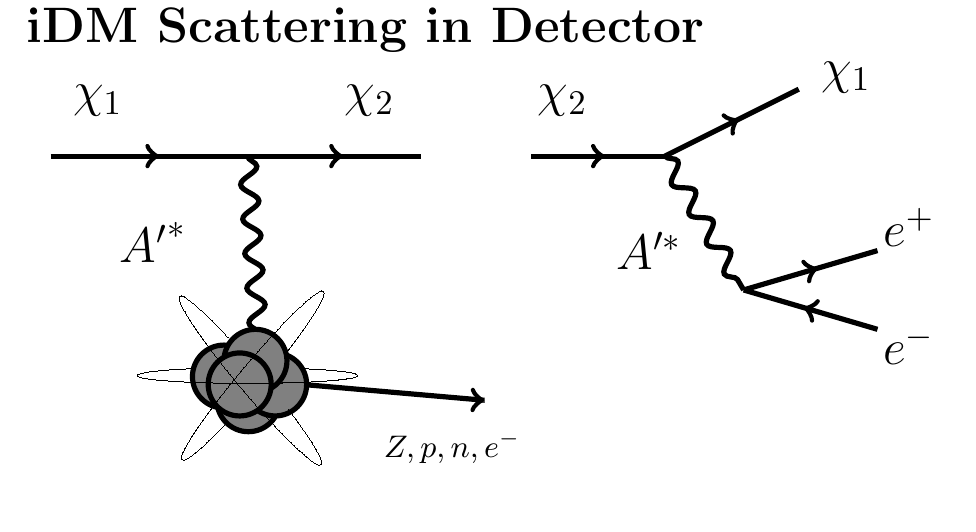} 
\caption{   
 Top: Same as Fig.~\ref{fig:prod}, but for an {\it inelastic} Majorana DM scenario in which the $A^\prime$ decays to a pair of
 different mass eigenstates. The unstable $\chi_2$ decays in flight, so the flux at the detector is dominated by $\chi_1$ states
 which upscatter off electron, nucleon, and nuclear targets (bottom) to regenerate the $\chi_2$ state. Subsequently, the $\chi_2$ promptly de-excites in a 3-body
 $\chi_2\to \chi_1 e^+e^-$ process, depositing significant $\sim$ GeV scale electromagnetic signal inside the BDX detector. 
\label{fig:prodIDM} }
\end{figure}

To close this discussion of dark matter models and their signals, we comment on the simple and well-motivated case of majorana LDM with off-diagonal couplings. In this scenario, (a) the electron beam produces $\chi_1 \chi_2$ pairs, (b) for sufficiently large mass splittings $\Delta\equiv m_2 - m_1$, the $\chi_2$ decays to $\chi_1 e^+ e^-$ inside the detector, and (c) the  $\chi$-scattering processes in the detector are inelastic (e.g. $\chi_1 p \rightarrow \chi_2 p$), with a total deposited energy that is often dominated by the energetic $e^+e^-$ pair from the subsequent $\chi_2$ decay. 
Like the electron scattering process, this inelastic scattering signal can be searched for with very low background rates. Fig.~\ref{fig:prodIDM} illustrates the production and detection signature of models of majorana LDM.

\subsection{Overview of experimental searches}
\begin{figure}[t!]
\center
\hspace{-1cm}\includegraphics[width=7.7cm]{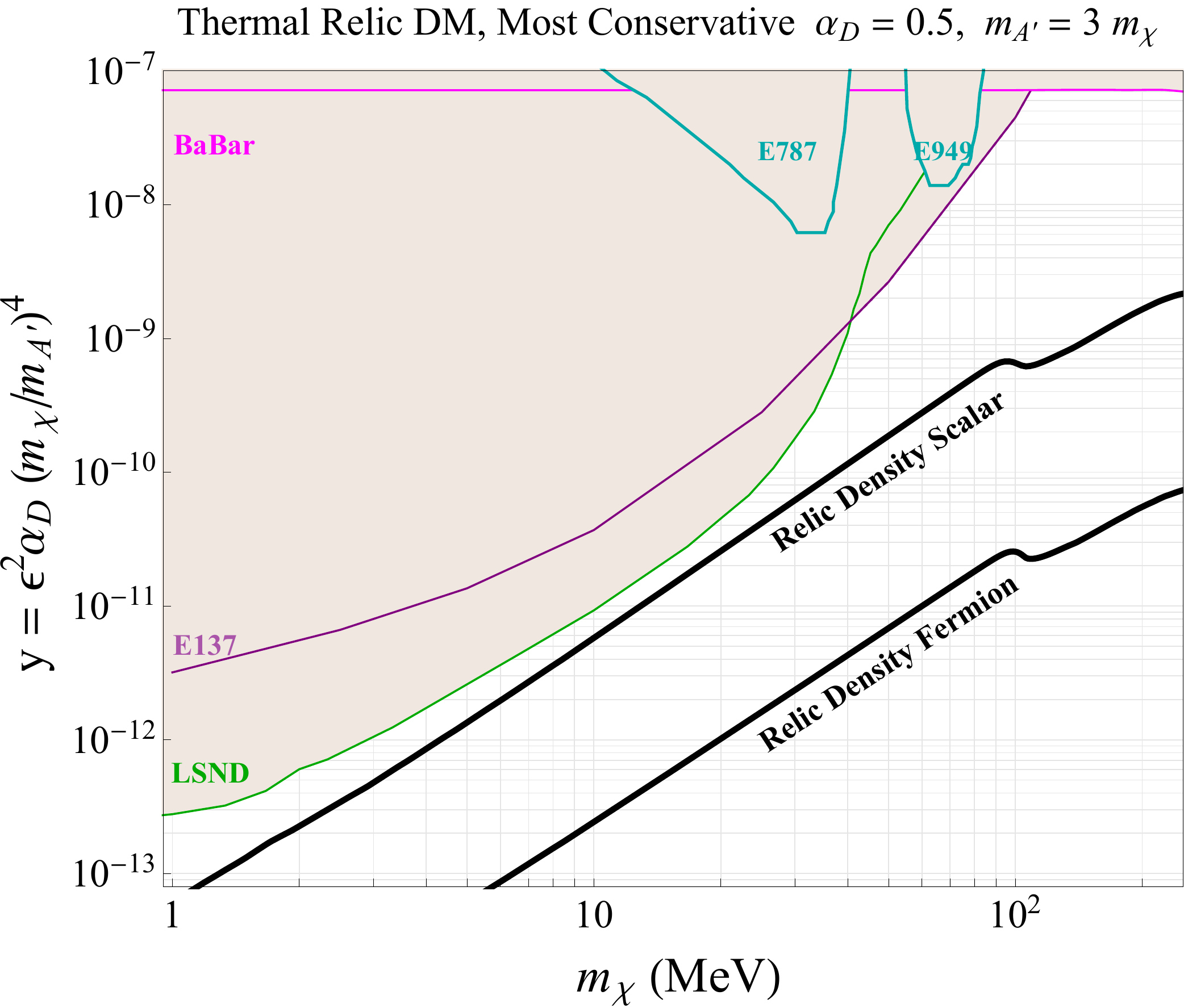}
\includegraphics[width=7.7cm]{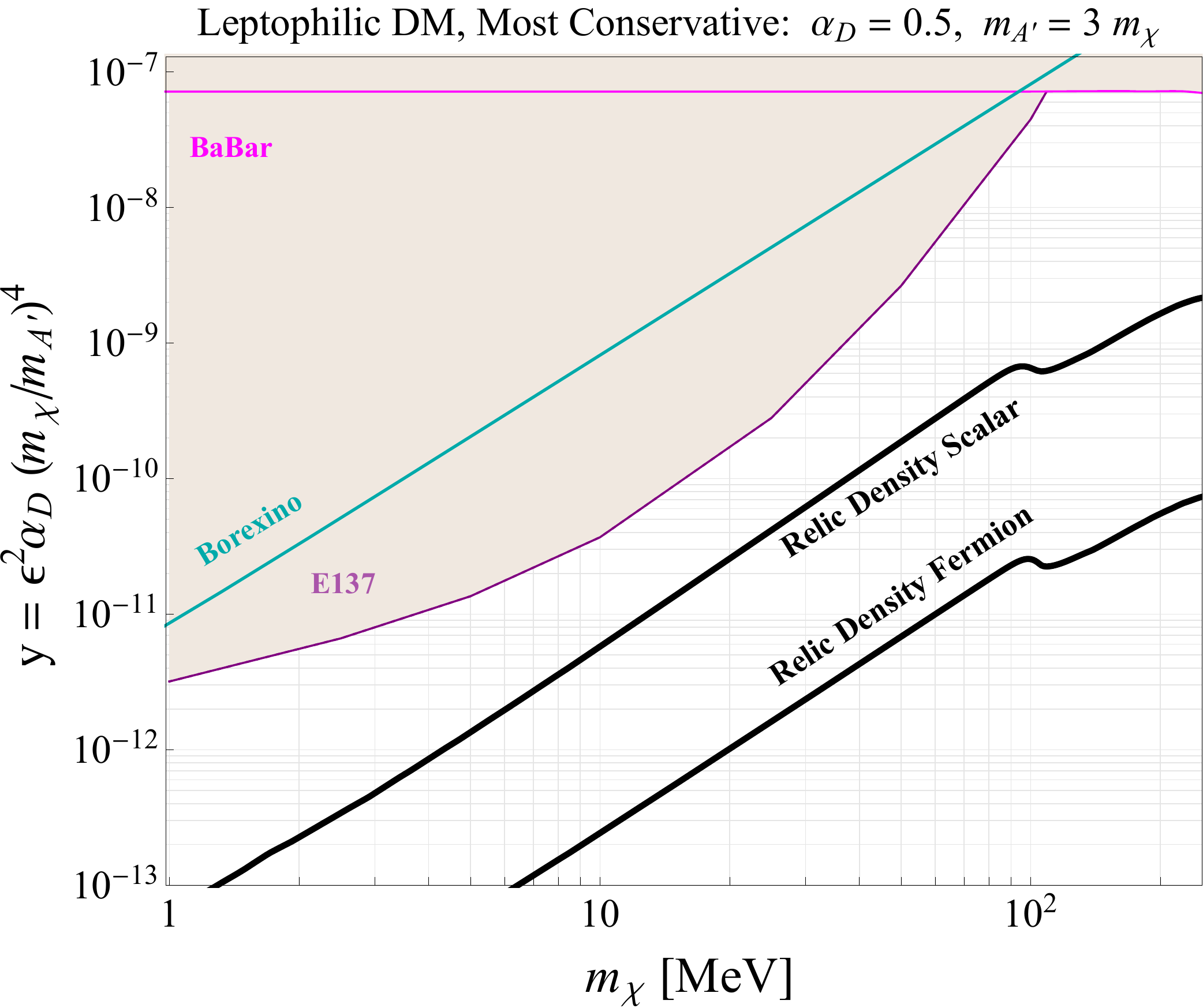}
\caption{Example of the viable parameter space for light dark matter in the representative kinetically-mixed and leptophilic scenarios alongside appropriate
constraints. The parameter space is
characterized by two masses ($m_\apr$ and $m_\chi$), the coupling of the $\apr$ to the LDM particle $\chi$, $\alpha_D$, and the kinetic mixing
represented by $\epsilon$. The ``$y$" variable on the vertical axis is chosen because it is proportional to the annihilation
 rate, so the thermal target (solid black) is fixed for a given choice of $m_\chi$.
 As we will see, for low background rates, BDX becomes sensitive to unexplored regions of the parameter space.
\label{fig:main-yplot}   }
\end{figure}

\begin{figure}[t!]
\center
\hspace{-1cm}
\includegraphics[width=8cm]{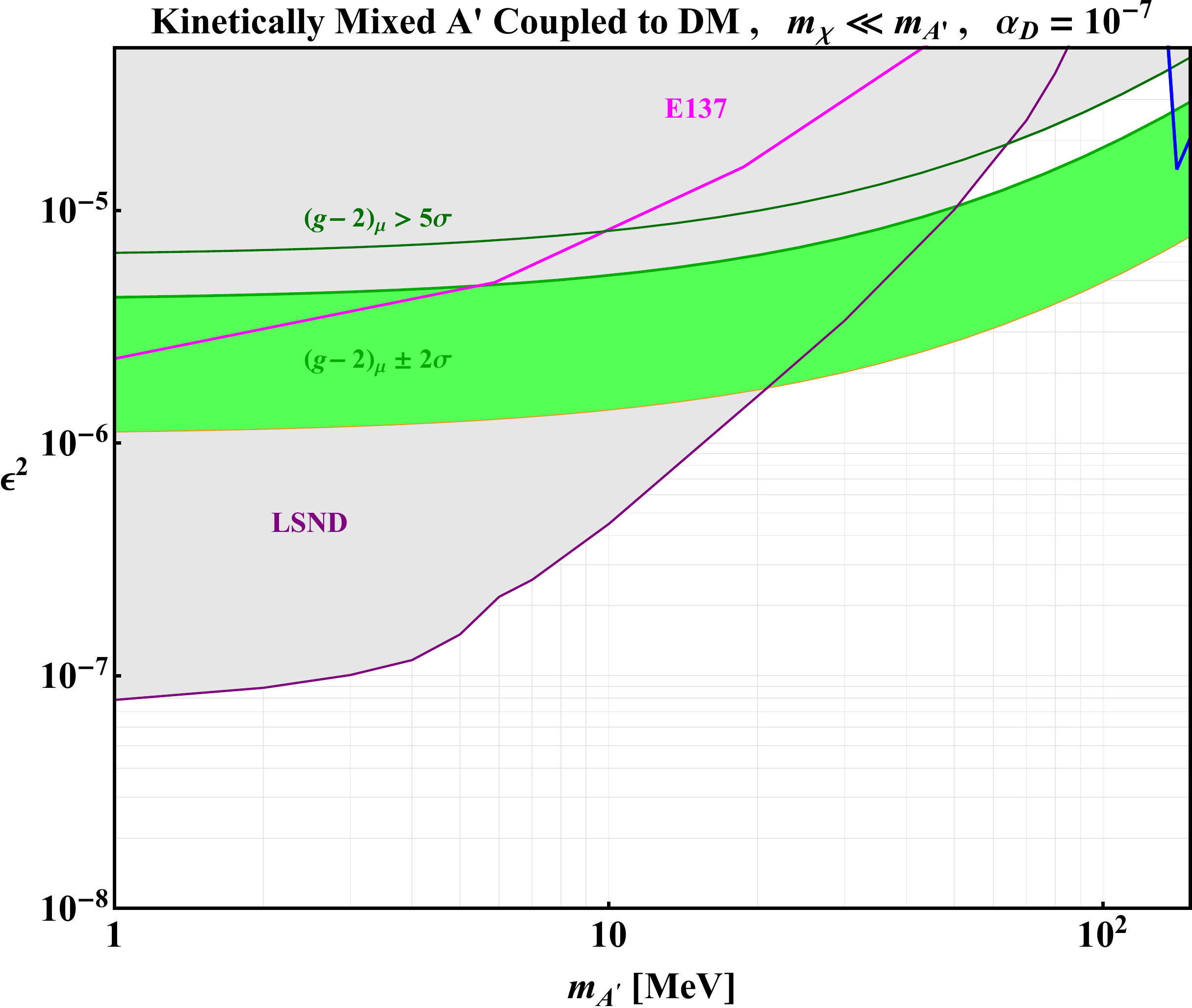}
\caption{Viable parameter space for sub-GeV DM coupled to a kinetically mixed $A^\prime$ for $\alpha_D \ll \epsilon$. Here,  $\alpha_D \equiv g_D^2/4\pi$, where $g_D$ is the DM coupling to the mediator. The green band represents the parameter space that can explain $g-2$ of the muon.
   \label{fig:traditional}   }
\end{figure}

In this section, we discuss current and near future experimental programs and highlight the uniqueness of BDX at JLab. The search for LDM covers the space
represented by two masses ($m_\apr$ and $m_\chi$) and two couplings ($\alpha_D$ and $\epsilon$), an example of which is shown in Fig.\,\ref{fig:main-yplot}. The colored
areas have been ruled out by various experiments, but leave open regions which can be probed by BDX. Moreover, Fig.~\ref{fig:traditional} illustrates some of the parameter space in LDM models that can still explain the discrepant value of $(g-2)$ of the muon, in particular the $m_{\apr} \gg m_\chi$ and $\alpha_D \gg \epsilon$ regime.

In the following we describe the various searches and comment on their
sensitivity. The paradigm of DM interactions with the SM offers three broad possibilities to search for it: accelerators, direct, and indirect detection. The first relies on production of DM,
either directly, or through the production and decay of a mediator such as the $\apr$. The second approach seeks to directly detect the interaction of DM particles from the halo,
as they pass through the earth. In the third, DM annihilation in the early Universe could affect cosmological observations; or alternatively, in the present day, DM could annihilate
in dense regions such as the center of our galaxy --- giving rise to final state SM particles that one can look for. We briefly discuss previous, current, and near-future efforts in
the search for LDM. For more details, see Ref.~\cite{Izaguirre:2015yja}.

\subsubsection{CMB}
While DM annihilation freezes out before the era of recombination, 
residual annihilations can re-ionize hydrogen and distort the high-$\ell$ CMB power spectrum \cite{Finkbeiner:2011dx, Lin:2011gj, Galli:2011rz, Madhavacheril:2013cna, Ade:2013zuv}. These data can be used to constrain the total power injected by DM annihilations \cite{Ade:2013zuv}, which scales as the DM annihilation cross-section (hence proportional to $y$) and can be invariantly compared with the relic density target. Dirac fermion DM annihilating through an $s$-channel $\apr$ is ruled out by Planck 2015 data \cite{Ade:2015xua}, but the other scenarios remain viable. In particular, the Majorana LDM scenario is viable because the DM annihilation rate during the CMB epoch is sharply suppressed relative to its  value at freeze out, and the scalar DM scenario is allowed because annihilation in this case is suppressed by the DM velocity $v^2$, {\it e.g.,} p-wave annihilation.

\subsubsection{Light Degrees of Freedom}
There is an indirect bound on light DM $\lsim 10$ MeV that remains in thermal equilibrium with SM radiation (but not neutrinos) during BBN \cite{Nollett:2013pwa}. This 
bound is more model dependent than accelerator probes because it can be evaded with additional sources
of dark sector radiation (e.g. sterile neutrinos). 

\subsubsection{B-Factories}
Mono-photon and missing-energy production at B-factories sets a limit on models of LDM.
The BaBar search for an (untagged) 
$\Upsilon(3S)\rightarrow \gamma+\rm{invisible}$ \cite{Aubert:2008as}
 constrains the process $e^+e^-\rightarrow \gamma+A'^{(*)}\rightarrow  \gamma \chi\bar\chi$ \cite{Izaguirre:2013uxa,Essig:2013vha}.
 Since the $\apr$ production rate only depends  on $\epsilon$ and the beam energy,  one must make a choice for     
  $m_\varphi / m_{A'}$ and $\alpha_D$ to build the $y$ variable using these data; smaller choices of either quantity would overstate the BaBar constraint. 

\subsubsection{High Energy Colliders} 
 Electroweak precision tests at LEP can constrain the existence of a new massive photon. In particular, kinetic mixing induces a shift in the mass of the $Z^0$ boson, and the constraint depends on $\epsilon$ and only mildly on $m_{A'}$ \cite{Hook:2010tw, Curtin:2014cca}.
At the LHC, light DM can be produced in association with a QCD jet.  Recasting a CMS DM search \cite{Khachatryan:2014rra}  in the monojet and missing energy channel places a constraint
on the $y$ vs. $m_\chi$ plane. These constraints do not scale with $y$ so one must choose a specific  value of $m_\chi / m_{A'}$ and $\alpha_D$ in constructing $y$ for colliders.

\subsubsection{Visible Decays}
Direct searches for the new mediator, {\it resonance searches} of $\rm{Mediator} \rightarrow$ SM SM, can also target models of LDM. This realm can arise in a model where the mass hierarchies are $m_{\rm{DM}} > m_{A'} > m_{e^{\pm}}$ or whenever $\epsilon \gg g_D$. Many experiments have been performed searching for an $A^{\prime}$ boson with mass in the range $1 - 1000$ MeV and coupling, $\varepsilon$, in the range $10^{-5} - 10^{-2}$.
Several different and complementary approaches were proposed (for a summary, see \cite{Bjorken:2009mm} and references therein), and indeed JLab has a strong presence with APEX \cite{Essig:2010xa}, HPS \cite{Battaglieri:2014hga}, and DarkLight \cite{Balewski:2014pxa}.

\subsubsection{Solar Neutrino Bounds}
If the $A^\prime$ is a leptophilic gauge boson that  also couples to neutrinos (e.g. as the mediator of a $U(1)_{e-\mu}$ gauge group, or similar variations including $U(1)_{\mu -\tau}$ or $U(1)_{B-L}$ which
are in the same class of models), it can affect the rate of solar neutrino scattering in the Borexino detector \cite{Izaguirre:2014dua}. This  constraint 
is shown in Fig.~\ref{fig:traditional} (bottom) in the context of the $U(1)_{e-\mu}$ model. 

\subsubsection{Missing Energy Experiments}

A recently proposed experiment at CERN SPS \cite{Andreas:2013lya,Gninenko:2013rka}, now known as NA64, would also search for invisible $\apr$ decay. The experiment employs an innovative technique, by having the primarily $e^{-}$ beam from the SPS, with energy between 10 and 300 GeV, impinging on an \textit{active} beam-dump, made by a calorimeter based on scintillating fibers and tungsten, ECAL1. An almost-hermetic detector would be  located behind the active beam-dump. The detector is made by a charged particle veto counter, a decay volume, two scintillating fiber counters, a second electromagnetic calorimeter ECAL2, and an hadronic calorimeter. The primary goal of the experiment is to search for the $\apr$ production in the active dump trough a Brehmstrahlung-like process, followed by the decay to $e^{+} e^{-}$. The signature for these events is a signal in ECAL1 and two clusters in ECAL2, from the $\apr$ decay products. 
The same experiment, could also search for $\apr$ invisible decays by exploiting the detector hermeticity, and requiring a single hit in ECAL1 from the  $e^{-}$ radiating  the $\apr$. The projected sensitivity for $3\cdot 10^{12}$ electrons makes this experiment one of BDX's direct competitors for search of LDM with diagonal couplings to $\apr$. 

Note, however, that the case of Majorana LDM (off-diagonal couplings) can be problematic for missing energy experiments, because the DM signal in this scenario,  $e^- + \rm{target} \rightarrow e^- + \rm{target}+ \chi_1 \chi_2$ with $\chi_2$ later decaying to $\chi_1 e^+ e^-$, is mimicked precisely by their most problematic background from the reaction $e^- + \rm{target} \rightarrow e^- + \rm{target}+ \gamma$ with $\gamma$ converting to $e^+ e^-$ in their detector.

A class of experiments that looks for missing mass from the reaction $e^+ + e^- \rightarrow \gamma + (A'\rightarrow \bar\chi\chi)$ originating from a positron beam have been proposed at both Frascati and at Cornell \cite{Raggi:2014zpa}. While this signature is a rather clean one, namely a bump search, these experiments are limited to energies and rates that may limit their ability to constraint parameter space consistent with DM's current abundance \cite{Raggi:2015gza}.

\subsubsection{Direct Detection Experiments}

Elastic DM-nuclear interactions are constrained by recent results from CRESST  \cite{Angloher:2015ewa}, whose low threshold allows for sensitivity down to a few 100s of MeV in DM mass. New ideas for direct detection of DM off of electrons have been proposed in recent years however, and while these searches are currently background-limited \cite{Essig:2012yx,Essig:2015cda}, new techniques have the potential to in principle also target the thermally-motivated parameter space of light DM that BDX can access for elastically-scattering DM. However, for majorana LDM, the sensitivity of direct detection experiments is quite limited. This is because tree-level scattering is inelastic and kinematically forbidden for mass splittings of order $\Delta \gsim $KeV; elastic scattering arises from a one-loop box diagram which scales as $y^2$ and is also invariant on the $y$ vs. $m_\chi$ plane.

\subsubsection{Beam dump experiments}

We now discuss beam-dump experiments. First we focus on results from the re-analysis of old data, and later on a current effort at a proton-beam-dump experiment.\\

 \noindent{\it \bf{Reanalysis of old data}}\\

The considerable sensitivity of beam-dump experiments to light dark matter is underscored by the reach of existing neutrino experiments \cite{Batell:2009di,deNiverville:2011it, deNiverville:2012ij,Dharmapalan:2012xp,Essig:2013lka}.  
For example, the LSND measurement of electron-neutrino scattering \cite{Auerbach:2001wg} can be used to derive the most stringent constraints to date on the parameter space for invisibly-decaying dark mediators that couple to both baryons and leptons \cite{deNiverville:2011it}. That experiment delivered $\sim 10^{23}$ 800 MeV protons to the LANSCE beam-dump.  
For very low mass $A'$s and dark matter sufficiently light ($100 \MeV \lesssim m_{A'}\lesssim 2 m_{\chi}$), the produced neutral pions have a small exotic decay into $A^\prime$s which 
then decay to $\chi$. The $\chi$ can then scatter off electrons in the LSND detector via $A'$-exchange.
However, the sensitivity of LSND vanishes if the mediator couples only to leptons or baryons and is weakened if its coupling to either is suppressed.   

Recently it was shown that electron-beam fixed target experiments could offer powerful sensitivity to a broad class of dark sector scenarios with particle dark matter in the $\MeV-\GeV$ mass range \cite{Izaguirre:2013uxa,Diamond:2013oda,Izaguirre:2014dua}. Electron beam-dump experiments are complementary to dedicated efforts at proton beam facilities, have comparable DM scattering yield, 
can run parasitically and on a smaller scale than proton-beam counterparts,
and benefit from negligible beam-related backgrounds. 
Such searches can dramatically improve sensitivity to MeV-to-GeV mass dark matter and other long-lived weakly coupled particles, extending well beyond the reach of proposed neutrino-factory experiments and Belle-II projections.
The power of electron beam dump experiments in this context is illustrated
by the existing sensitivity of the SLAC E137 experiment \cite{Bjorken:1988as}. That experiment was sensitive to invisibly decaying dark mediators produced in fixed target collisions involving 20 GeV electrons and the E137 beam-dump \cite{Batell:2014mga}.
Despite the rather high energy threshold ($\sim$ 3 GeV) required to see secondary scattering of dark matter particles off electrons, and the small geometric acceptance,
E137  has already probed mediator mixings beyond that probed by proton beam-dumps at intermediate masses. 
In a year of parasitic running, BDX will receive roughly 100 times the charge deposited on E137, with a comparable solid angle, higher-density detector, and lower energy threshold.\\ 

   \noindent{\it \bf{Current beam-dump experiments}}\\

The MiniBooNE experiment, originally designed to study neutrino oscillations, recently completed a test run to demonstrate the feasibility for MeV DM \cite{Dharmapalan:2012xp} search. In the experiment, the primary 8.9 GeV proton beam from the FNAL accelerator impinged on a 50-m long iron beam-dump. 
Dark matter particles are produced through neutral mesons decay ($\pi^{0}$, $\eta$), where one of the photons converts  to an $\apr$ that, in turns, decays to a $\chi \overline{\chi}$ pair. 
 These particles can then scatter on the electrons or nuclei in the MiniBooNE detector, placed 490 m downstream the beam-dump. The otherwise dominant neutrinos background,  generated by charged mesons decay in flight, was reduced by a factor of $\simeq 70$ directing  the proton beam straight on the dump, instead of  the original beryllium target. 
With the support of the FNAL PAC, MiniBoone is currently seeking to collect $\sim 10^{20}$ protons on target in this beam-dump mode and will continue taking data in this mode this year.\\

\subsection{The Unique Capabilities of BDX at JLab \label{sec:unique}}

While the field of light DM interacting with the SM is making rapid progress,  BDX at JLab offers a series of unique possibilities unmatched by proposed
competitors. First, thanks to the world-leading capability of CEBAF's intense beam,  electron beam-dump experiments have the potential to reach ever smaller couplings between the
mediator and the SM thanks to the luminosities offered by JLab's Hall A or C --- this is unmatched by any of the potential competitors that feature an electron beam with an
energy  in the few-GeV range. As to proton beam-dump experiments, electron beams offer comparable signal yields and do not suffer from the same level of neutrino backgrounds that proton beam-dump experiments do. In particular, BDX could improve the sensitivity over LSND for LDM masses above the $m_{\pi^0}/2$ threshold. Importantly however, in a real sense, electron beam-dump experiments target parameter space that's orthogonal to that probed by proton beam-dump
experiments, as the former are sensitive to models with $A'$ with leptophilic couplings. Similarly, proton beam-dumps have the ability to uniquely probe leptophobic models.

It is important to mount beam dump searches even when electron-scattering direct detection experiments may constrain overlapping parameter space. First, we re-iterate that direct detection experiments using noble-liquid detectors are currently background-limited, while semiconductor-based detectors still need to demonstrate sensitivity to detecting the expected single- or few-electron events from DM-electron scattering.  It therefore still needs to be demonstrated that direct detection can indeed cover thermal-origin motivated DM parameter space. In the event that these new techniques do achieve their potential, we actually view it as a strength of this program that a discovery in an accelerator-based experiment could also be observed by a direct detection counterpart. In fact, this multi-pronged approach is recognized as being essential in searching for Weak-scale WIMPs.  Moreover, there is a class of models that each technique is uniquely suited for. For example,
the Majorana DM scenario described above is a strength of BDX and a weakness of direct detection. In this class of models, the ground state $\chi_1$ is unable to upscatter into the excited state $\chi_2$ for mass splittings $\Delta$
above some KeV, thereby shutting off the leading interaction at direct detection experiments. Conversely, models of ultralight DM are a particular strength of direct detection. Therefore, the two approaches must be seen as complementary in nature.

Even within electron-beam accelerator experiments, a beam-dump setup can offer superior sensitivity to Majorana DM models than missing energy/mass experiments, despite the signal
yields for the latter scaling more favourably ($\epsilon^2$ vs $\epsilon^4 \alpha_D$ for missing energy/mass and beam-dump, respectively). For missing energy experiments, production
of DM in the active target would proceed via $e^- + N\rightarrow e^- (A'\rightarrow \chi_1 \chi_2)$. The de-excitation of the $\chi_2$ inside the active detector from the reaction
$\chi_2 \rightarrow \chi_1 \ell \ell$ would mimic the most problematic background for that class of experiments: bremsstrahlung events with a photon converting to $e^+ e^-$.
In contrast, at a beam-dump experiment, the $\chi_1$ produced in the dump could upscatter in the detector via the reaction $\chi_1 + \rm{target}\rightarrow \chi_2 +\rm{target}$.
If the excited state de-excites inside BDX, that would lead to an even more striking signal --- one recoil-target such as an electron, nucleus or nucleon, and the two electrons
from the de-excitation from $\chi_2$.

Moreover, the sensitivity of BDX for LDM with diagonal couplings is competitive with the projected reach of NA64's most advanced phase of running, proposed by NA64 to start running no earlier than in 2020 \cite{NA64Timeline} --- although not approved by CERN yet.

\clearpage
\section{Proposed measurement} \label{sec:proposed-measurement}
\begin{figure}[t!]
\center
\includegraphics[width=10cm]{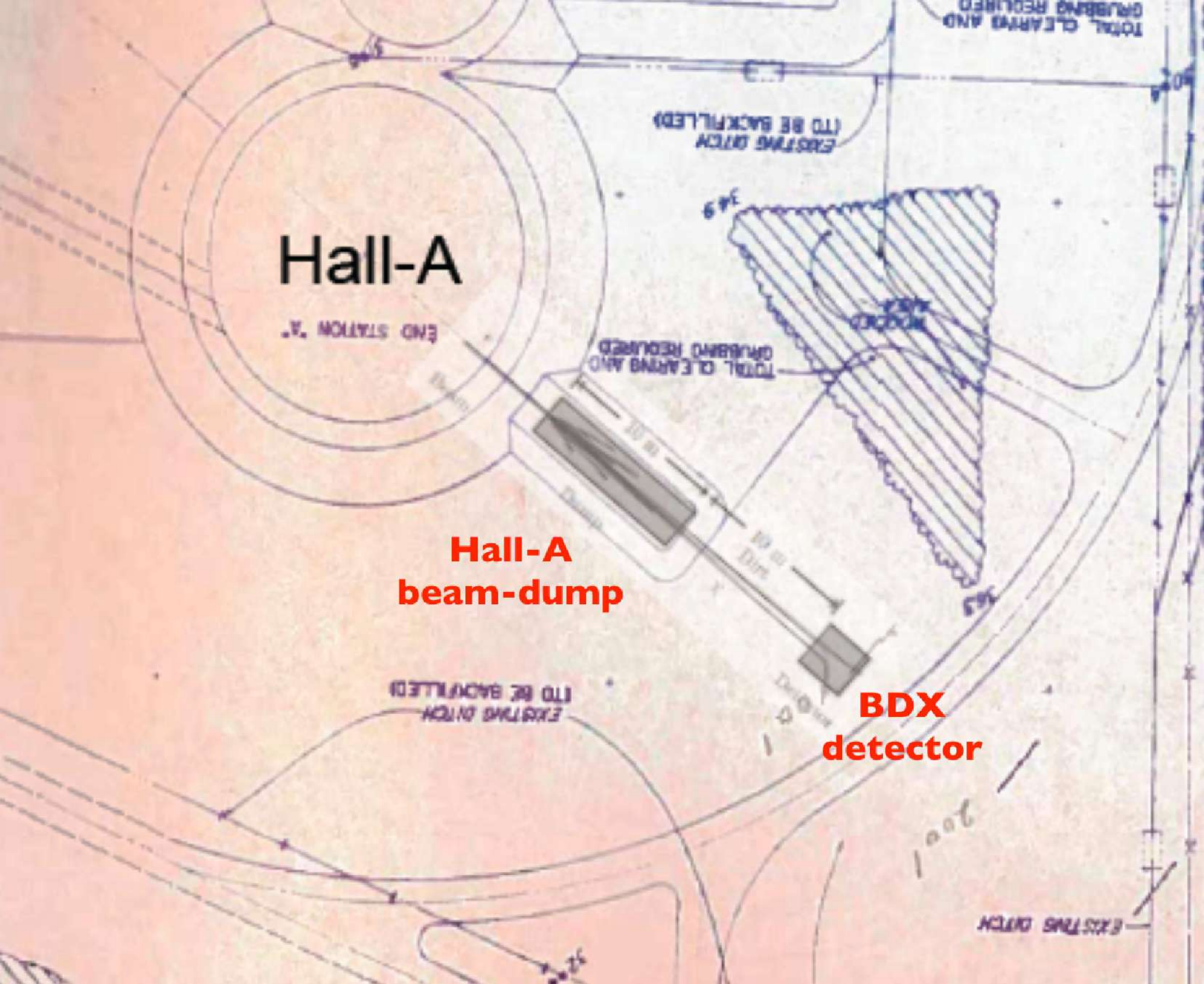}
\caption{The possible location of BDX detector at $\sim20$m from the Hall-A beam dump.}
\label{fig:setup_meas}
\end{figure}

The proposed experiment would require a $1\ \m^3$-scale detector volume, located $\sim 20$ meters downstream of the dump of a high-intensity
multi-GeV electron beam, and could run parasitically. We studied in detail the option of a new underground facility located downstream of the Hall-A beam dump.
See Fig.~\ref{fig:setup_meas} for a schematic representation of the experimental setup.
BDX will use the electron scattering of a DM $\chi$ particle  in a state-of-the-art electromagnetic calorimeter with excellent forward geometric acceptance, 
to greatly extend dark matter sensitivity beyond that available to the high threshold/low acceptance E137 setup or  to existing proton beam-dumps. 
Being also  sensitive to  low-energy nuclear recoil, BDX will use it  as  cross check of any possible findings.
The approach makes good use of Jefferson Lab upgrade to 11 GeV energies with  the new CEBAF scheduled to deliver up to about $65\mu$A currents.  
 
\subsection{The Hall-A beam dump}
The Hall-A at JLab is expected to receive from CEBAF a 11 GeV electron beam with a maximum current of about $65 \mu$A. The maximum available energy that focus the $\chi$ beam towards the detector together with a sizeable current that allows to collect the desired charge in the shortest amount of time, makes the Hall-A the optimal choice for a beam dump experiment at JLab.

\begin{figure}[t!]
\center
\includegraphics[width=12.5cm]{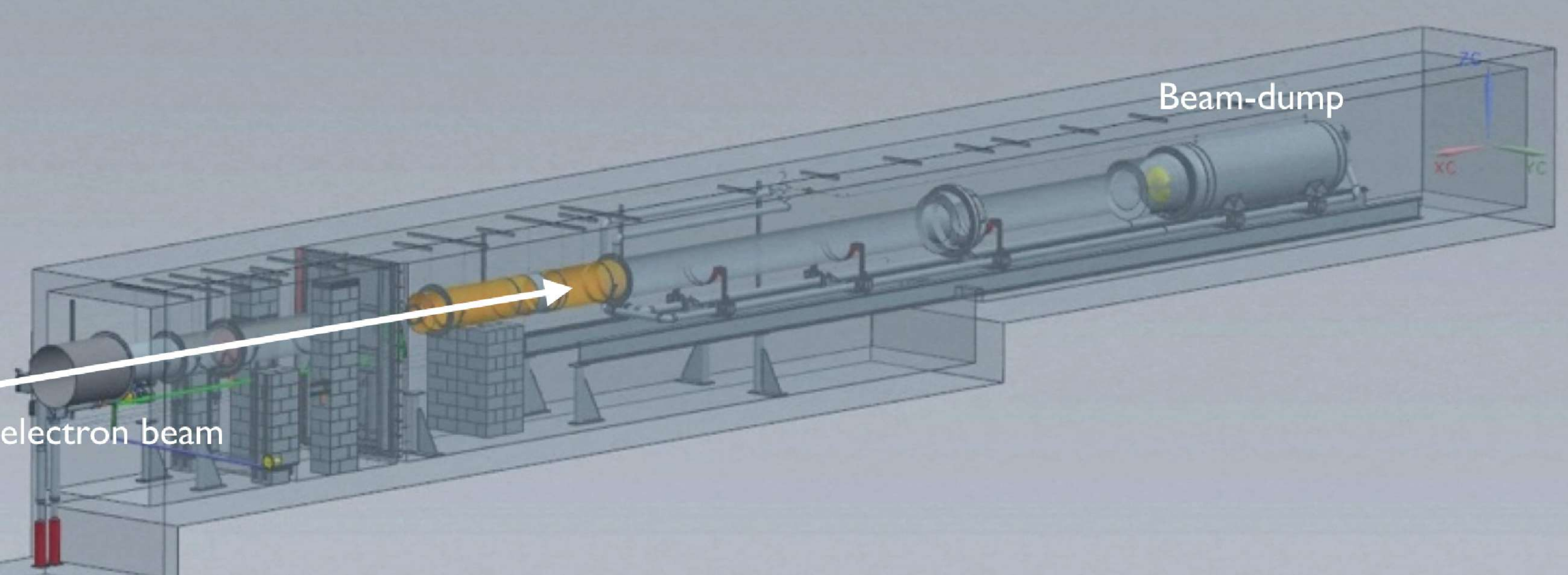}
\caption{The Hall-A beam-dump enclosure in the concreate tunnel.\label{fig:beam-dump-enclosure}}
\end{figure}

The Hall-A beam-dump is enclosed in a concrete tunnel at the end of the beam transport line. A rendering of the dump and the last fraction of the beam
line is shown in Fig.~\ref{fig:beam-dump-enclosure}.
\begin{figure}[t!]
\center
\includegraphics[width=12.5cm]{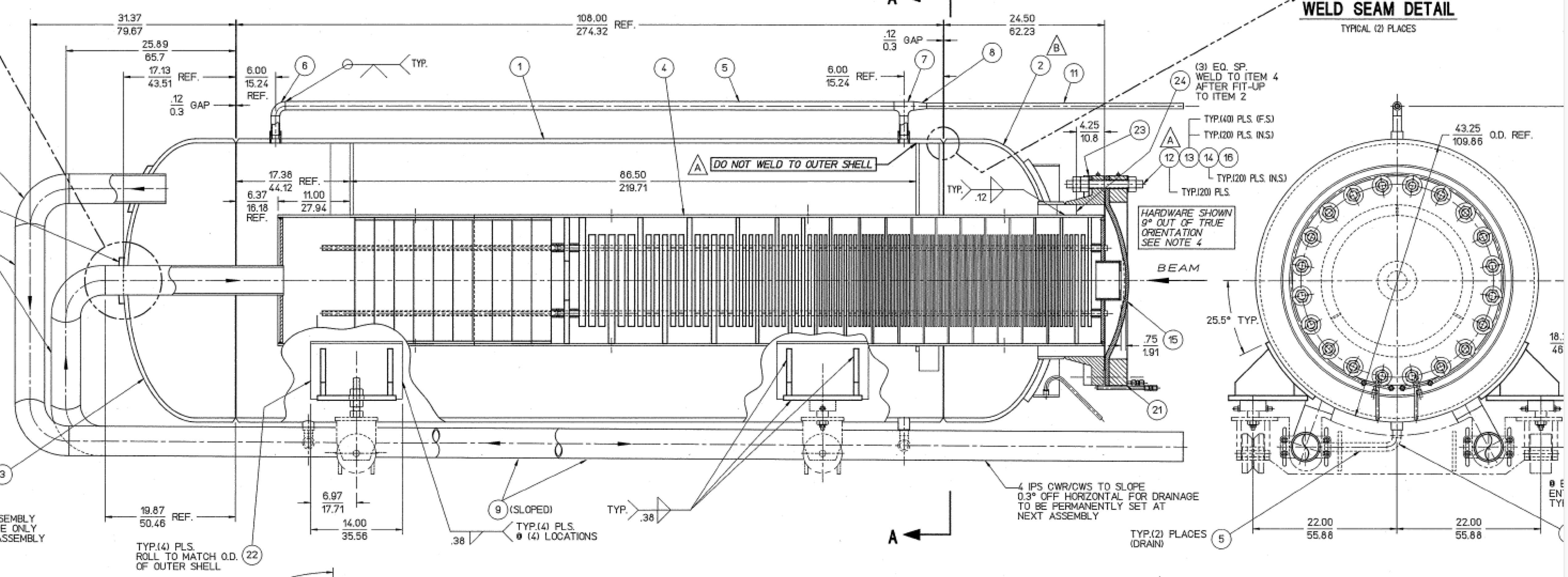}
\caption{ A detailed drawing of the Hall-A beam dump. \label{fig:beam-dump}}
\end{figure}
The dump is made by a set of about 80 aluminum disks, each approximately 40 cm in diameter of increasing thickness (from 1 to 2 cm), for a total length of approximately $200$ cm, followed by a solid Al cylinder 50cm in diameter and approximately $100$ cm long. They are both cooled by circulating water. The full drawing of the beam-dump is
shown in Fig.~\ref{fig:beam-dump}.
To increase the radiation shielding, the thickness of the concrete tunnel surrounding the Al dump is about 4-5 m thick.

\subsection{The BDX detector}
\begin{figure}[t!]
\center
\includegraphics[width=15cm]{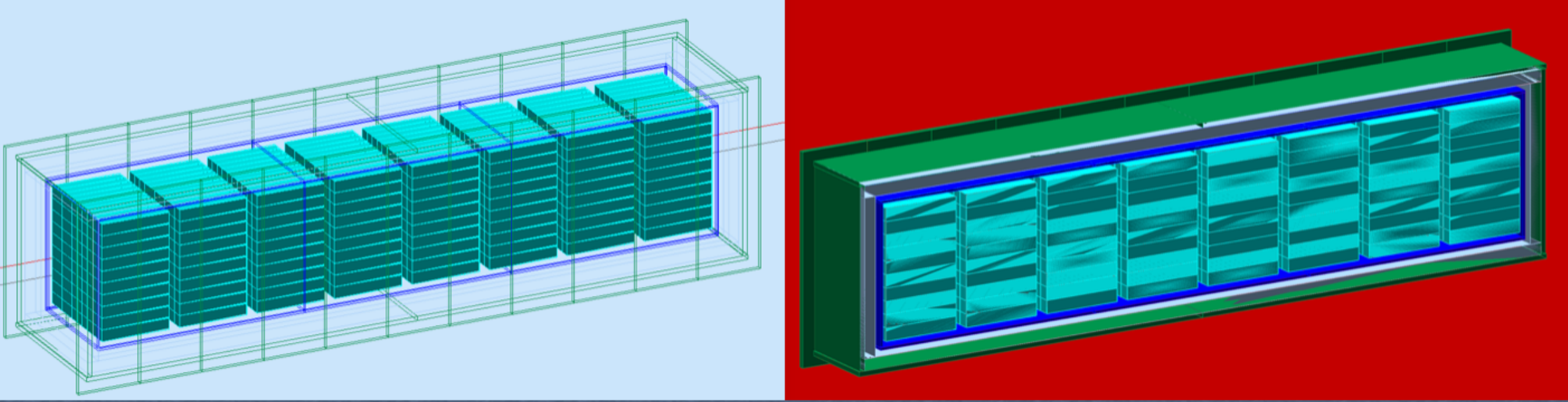}
\caption{A GEANT4 implementation of the BDX detector. On the right,  the Outer Veto is shown in green,  the Inner Veto in blue, the lead in gray and the crystals in cyan.}
\label{fig:full_detecor}
\end{figure}

The BDX detector is made by two main components: an electromagnetic calorimeter used to detect signals produced by the interacting dark matter, and a
veto detector used to reduce the cosmic background. The veto detector consists of a passive layer of lead sandwiched between two instrumented layers of scintillators.
The lead shielding reduces the sensitivity to low-energy environmental
background (mainly low energy photons).
A sketch of the BDX detector is shown in Fig.~\ref{fig:full_detecor}. The detector concept has been validated by a campaign of measurement at INFN - Sezione di Catania and Laboratori Nazionali del Sud (LNS) with a prototype,
extensively discussed in Appendix~\ref{Section:BDX-protoype}.

\subsubsection{The electromagnetic calorimeter} 
The core of the BDX detector is an electromagnetic calorimeter sensitive to both the $\chi$-electron and $\chi$-nucleon scatterings. The signal expected in the two cases
are quite different: a few GeV electromagnetic shower in the first and a low energy (few MeV) proton/ion recoil in the latter. Among the different options we chose a high-density,  inorganic crystal scintillator material to reduce the detector  footprint, fitting in the new proposed facility for beam-dump experiments at JLab (see Sec.~\ref{Sec:BD-facility}). The  combination of a low threshold (few MeV) sensitivity for high ionizing particles (light quenching not higher than few percents), a reasonable radiation length (few centimeters), together with a large light yield limits the choice to few options: BGO, BSO, CsI(Tl) and BaF$_2$ \footnote{We are not considering  some new very expensive crystals such as LYSO or LaBr.}. Considering that the request of about 1 cubic meter of active volume would  drive costs of any possible options in the range of few million dollars, and that the timeline for producing and testing thousands of crystals would be of the order of several years, we decided to reuse crystals from an existing calorimeter. Former experiments that still have  the desired amount of crystals available from decommissioned EM calorimeters include: BaBar at SLAC (CsI(Tl)), L3 at CERN (BGO)and CLEO at Cornell (CsI(Tl)). After consulting with the management of the different laboratories, we identified the BaBar option as the most
suitable for a BDX detector. In particular, the BaBar EM end-cap calorimeter, made by 820 CsI(Tl) crystals for an equivalent volume of about 1 cubic meter, matches perfectly the BDX requirement. The excellent performance of the BaBar calorimeter \cite{Aubert:2001tu}, together with the willingness of the SLAC management for an intra-DOE-Laboratories loan, makes this option technically suitable and practical, with minimal paperwork involved\footnote{An {\it Expression of Interest} for the BaBar  end cap calorimeter crystals has already been signed between the BDX Collaboration and SLAC management. As a consequence, 22 CsI(Tl) crystals have been shipped from SLAC to INFN in order to assemble a  4x5 ecal prototype.}. Details about the crystals dismounting procedure are not reported in this proposal, but from preliminary contacts with the SLAC personnel in charge of BaBar decommissioning \cite{Wisniewski}, we have been ensured to receive all the necessary support in term of tools (a manipulator to extract modules with crystals from the frame) and information for a safe and efficient procedure. Funds and labour to reassemble crystals in a suitable way will be provided by the BDX Collaboration.

\begin{figure}[t!]
\center
\includegraphics[width=8cm]{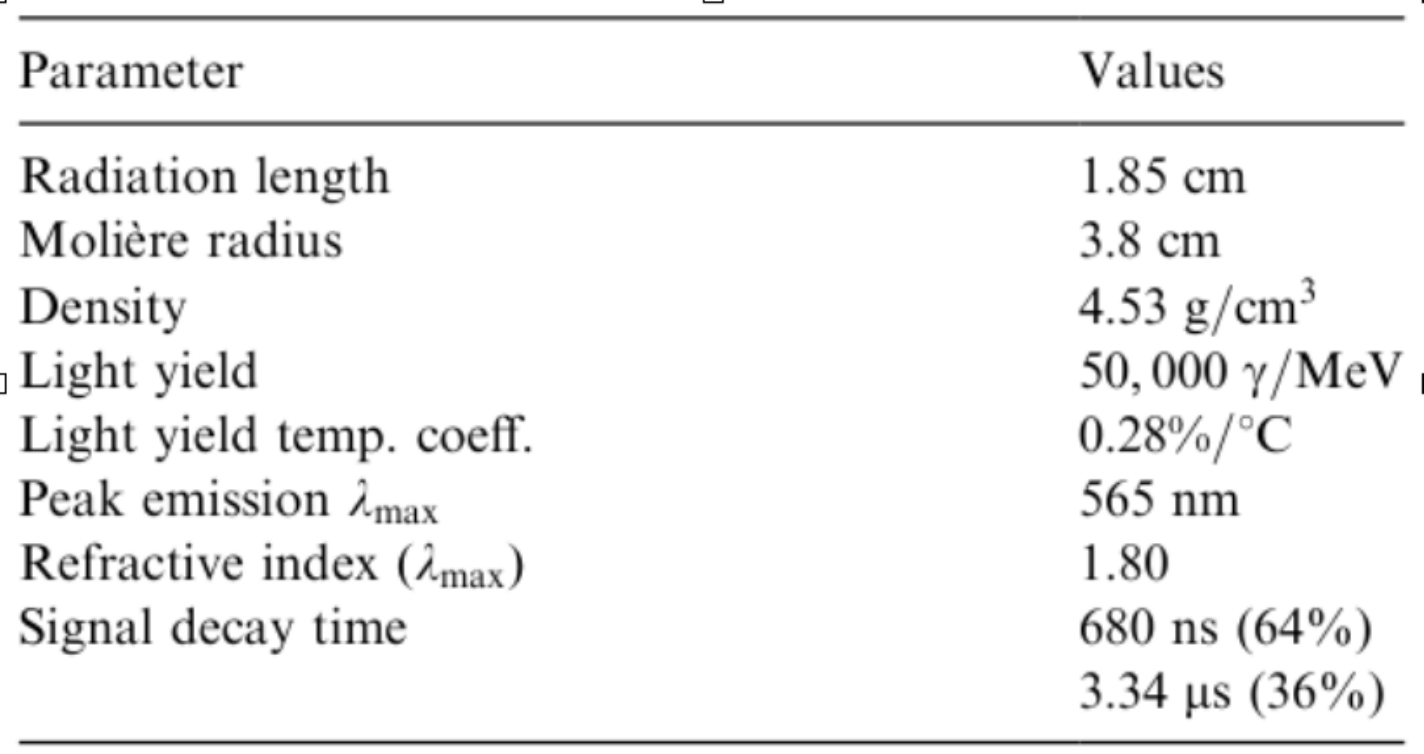}
\includegraphics[width=8cm]{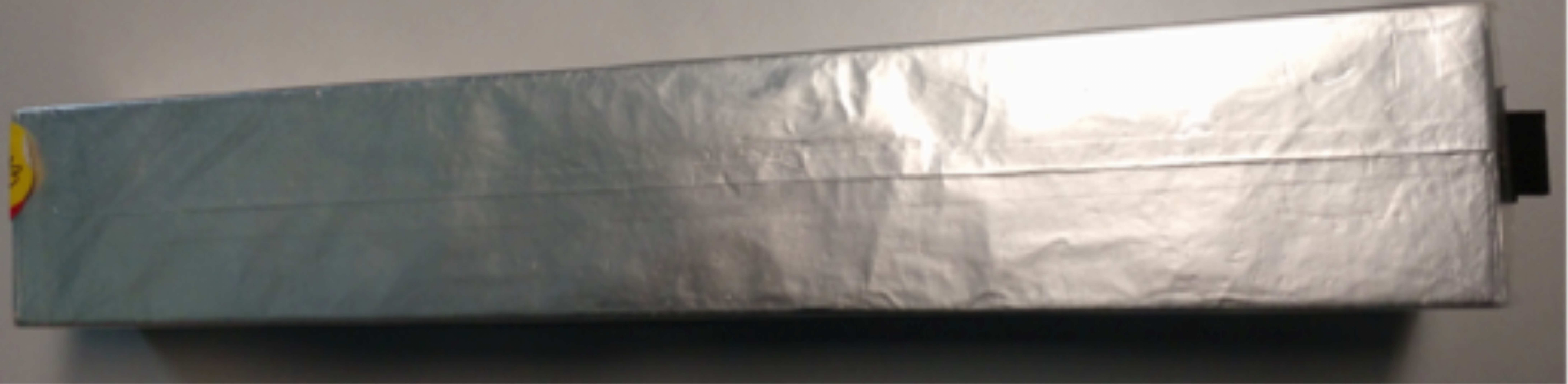}
\caption{Picture of a BaBar  CsI(Tl)  and table of properties.
\label{Tab:BaBar-CsI}}
\end{figure}

Crystals, that comes in different shapes and tapering, due to the projective geometry of the BaBar calorimeter, will be inserted in new regular-parallelepiped
aluminum alveolus in order to have regular elements easy to assemble in variable-size arrays.  The average size of each crystal is (4.7 x 5.4 x 32.5) cm$^3$ while the
alveolus size is (5 x 5.5 x 33) cm$^3$. Details about  individual  crystal properties are available in Refs.~\cite{Aubert:2001tu,Brose:1998hq}. The table in Fig.~\ref{Tab:BaBar-CsI} reports the main parameters of the BaBar CsI(Tl) crystals.

The reference setup used in this proposal foresees 8 modules of 10x10 crystals each (800 total), arranged with the long size along the beam direction. This arrangement has  a cross section of  50x55 cm$^2$ for a total length of 295 cm. Tacking advantage of the partially forward-focused $\chi$ beam, the parallelepiped shape of the detector is preferable to a cube-like arrangement to maximise the $\chi$-electron interaction length.
 
A former BaBar EM cal  crystals has been extensively tested in Genova to  assess  performances and to define the  most suitable readout. In fact, the BaBar readout scheme used a pair of silicon diodes (Hamamatsu S2744-08) with unitary gain, that required a sophisticated ASIC-based amplification (complemented with a custom CAMAC-based readout electronics) to provide the integrated signal  with no access to time information. The limited availability of the original FE electronic spares, as well as the absence of timing information, imposed a new readout scheme. We decided to bypass the pin diodes and place in the opposite crystal face a new readout sensor.
Considering the slow scintillation time of the CsI(Tl) ($\sim2-3\mu$s) we decided to use a fast readout sensor able to track the (faster) scintillation signal
rise time (10-20 ns).  Regular PMTs were excluded for the extra-length that would have been added to the crystal size. SiPMs represent a viable alternative providing an
excellent timing and single photo-electron sensitivity. Considering the sizeable CsI(Tl) light yield a small area sensor (e.g.  6x6 mm$^3$) satisfies the BDX requirements
providing  a cheap and high-performing solution. SiPM will be coupled to custom trans-impedence amplifiers already used in the prototype tested in Catania (see Appendix~\ref{Section:BDX-protoype}) 
\begin{figure}[t!]
\center
\includegraphics[width=14cm]{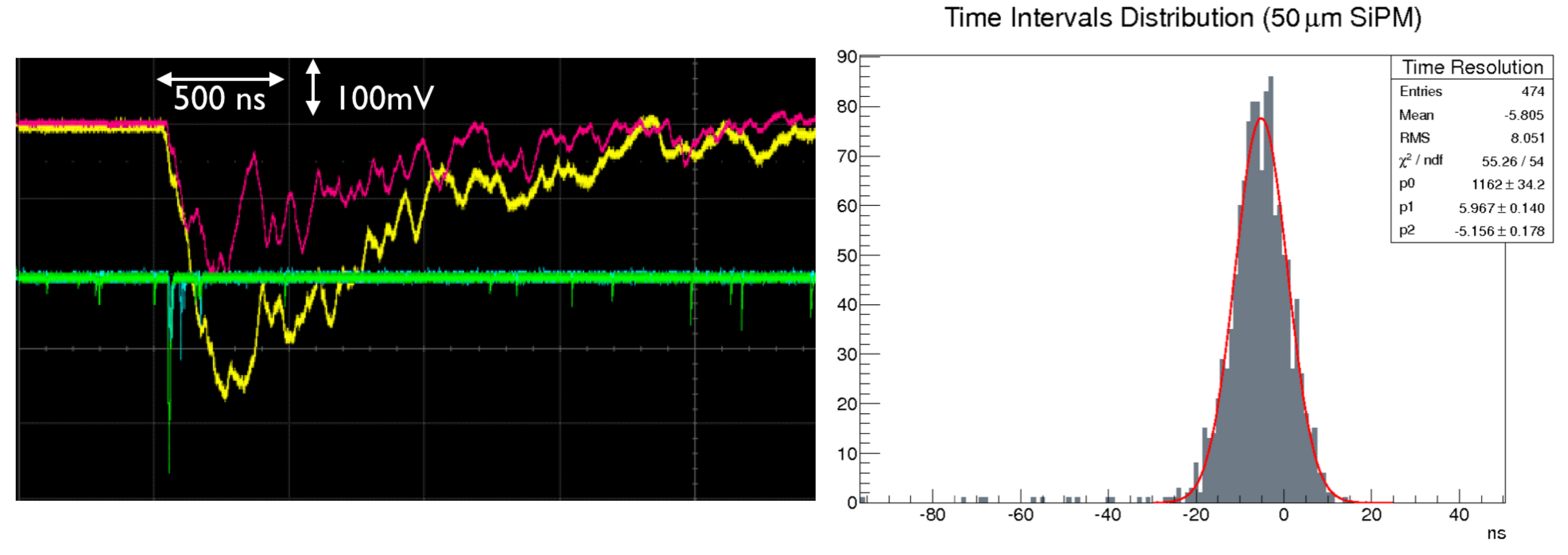}
\caption{Left: A typical signal released by a crossing muon. Right: time distribution of cosmic muons; the spread is all due to the crystal performance since the trigger
introduces negligible jitter.
\label{fig:babar-cry-performance}}
\end{figure}
Results of our tests performed  coupling a BaBar crystal to 3x3 mm$^3$ SIPMs with different pixel size (25 and 50 $\mu$m)  are reported in Fig.~\ref{fig:babar-cry-performance}. They show that a light yield of $\sim10$ phe/MeV and a time resolution of about 6-7 ns (for cosmic muons) is achievable. The use of 4-times larger sensors (6x6 mm$^3$), now commercially available, would make the SIPM option even more suitable for the BDX needs.
The limited timing achievable with CsI(Tl) crystals does not represent a limitation for BDX since a tight time coincidence between the detector and the beam (bunches separated by 4 ns) would require a time resolution of tens of ps, difficult to achieve with any organic crystals for small energy deposition.\footnote{Plastic scintillator would have been a good alternative in term of light yield, timing and costs but the reduced density would require a detector almost 5 time bigger
in  length  making this choice impractical.}.

All the results discussed in this proposal do not consider any change to the beam structure. However, from a preliminary discussion with JLab Accelerator Division \cite{Freyberger}, we concluded that it may be possible to operate the CEBAF accelerator in such a way that a reduction of a factor 5 in the beam-unrelated background would be feasible, even with the above timing resolution (see Sec.\ref{sec:fullexp}.)

\subsubsection{The active VETO system} 
\begin{figure}[t!]
\center
\includegraphics[width=7cm]{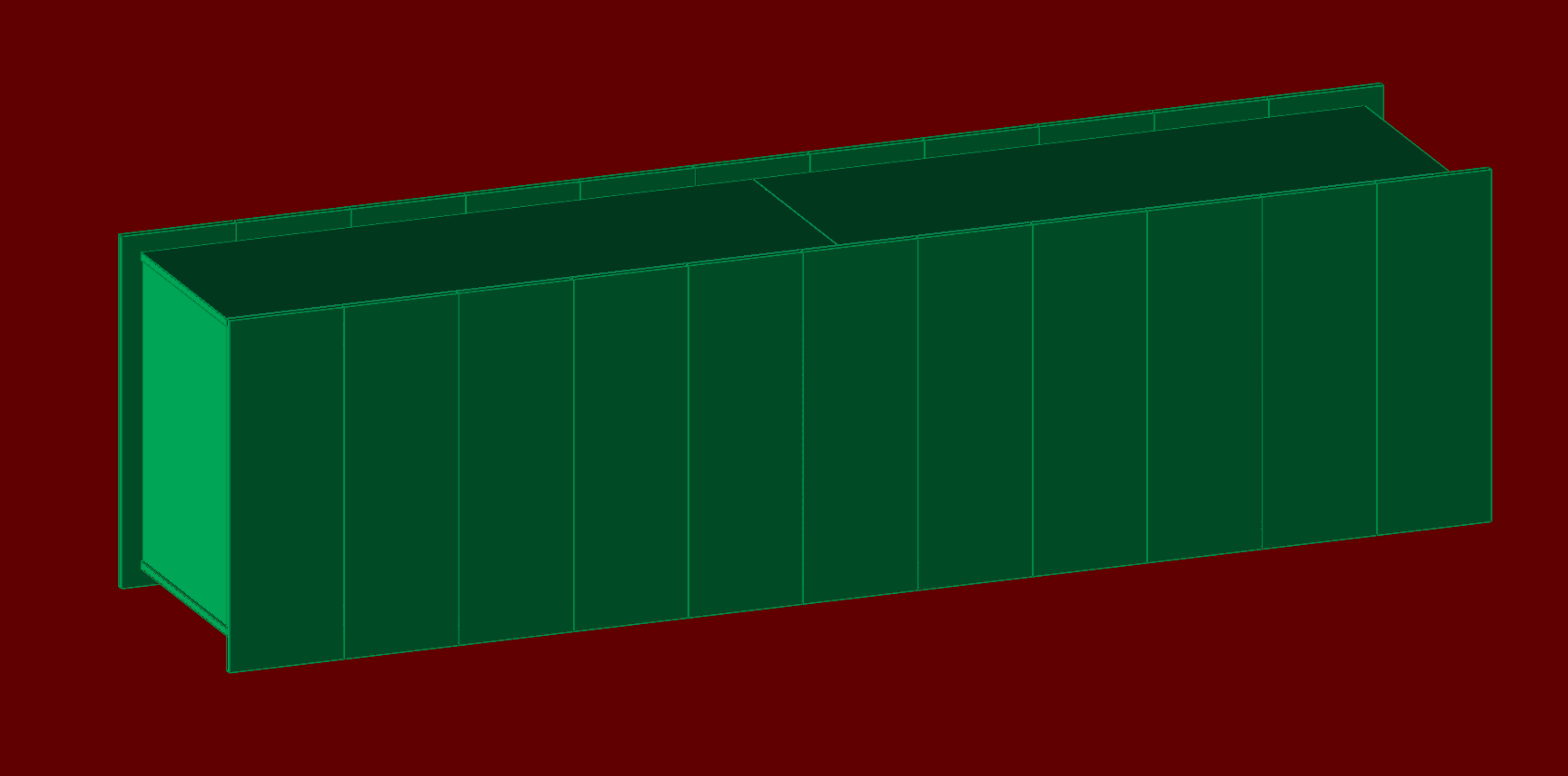}
\includegraphics[width=7cm]{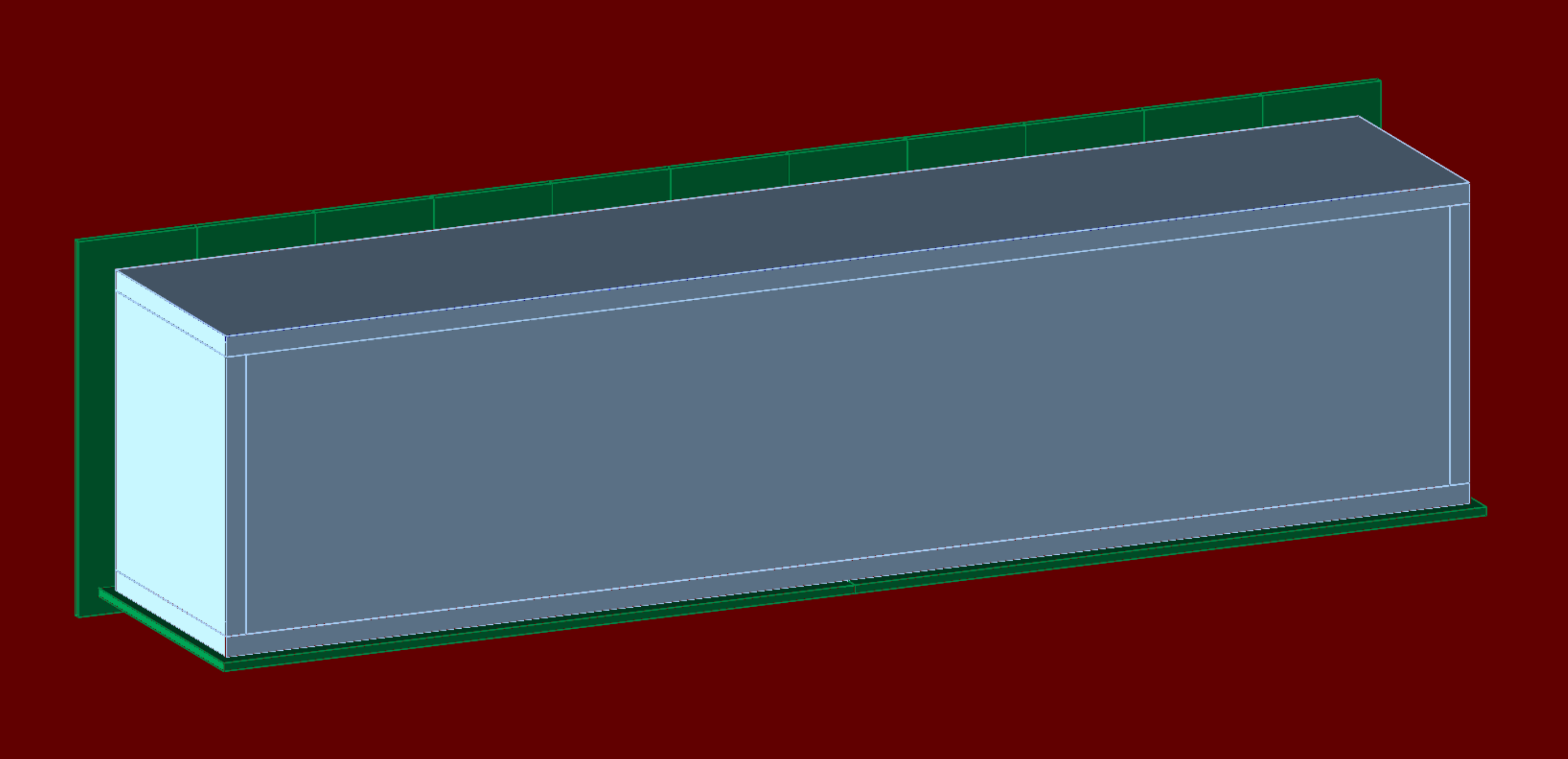}
\includegraphics[width=7cm]{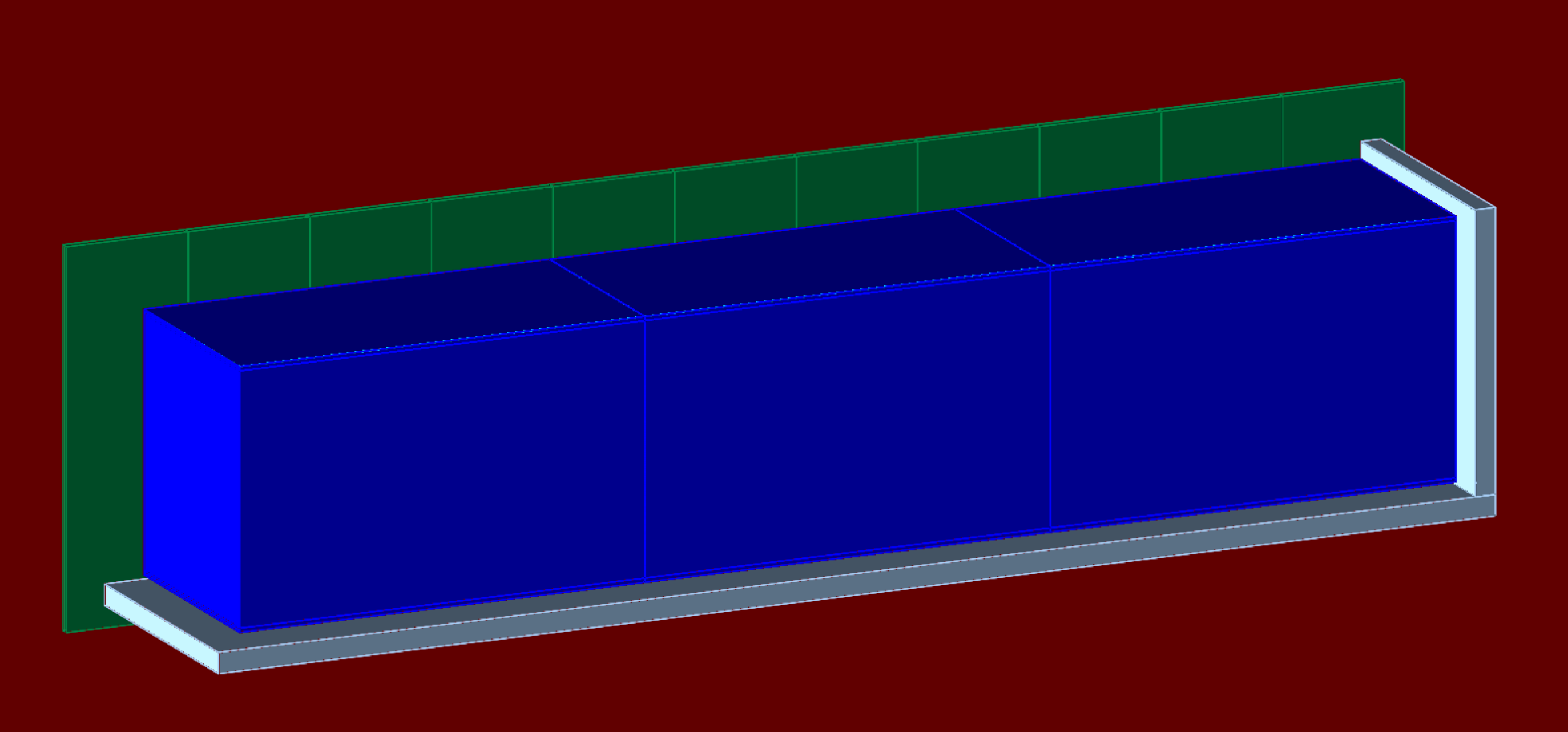}
\includegraphics[width=7cm]{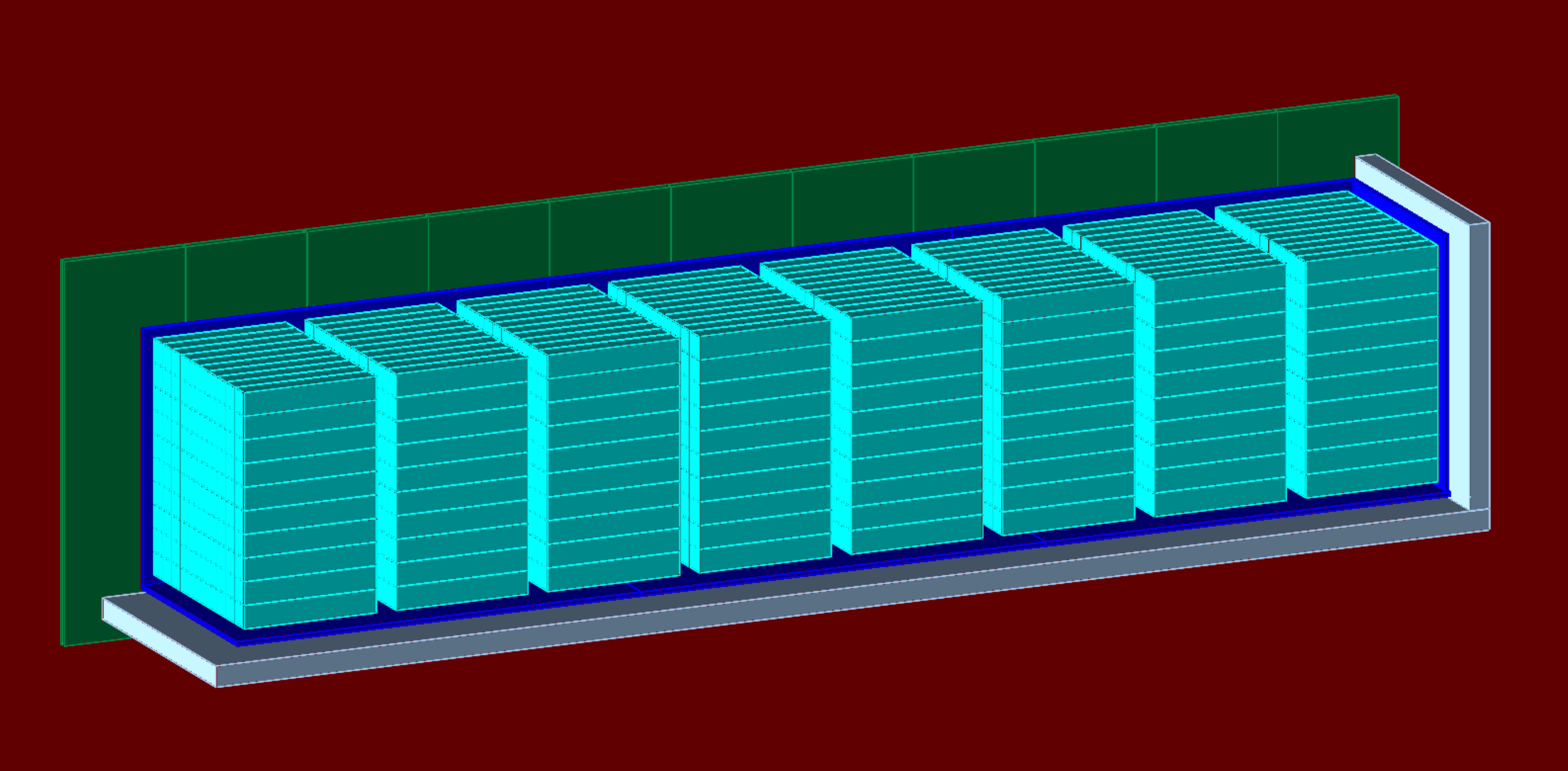}
\caption{The BDX detector. The Outer Veto is shown in green: side paddes are read from the top by a PMT coupled to a light guide, the top and the bottom paddles are read by
a PMT on the side coupled to a light guide and the upstream and downstream paddles are read by a PMT directly attached to the large face of the scintillator (PMT and guides
are not shown in the drawing). The Inner Veto is shown in blue: side, top and bottom scintillators are read by 4 SiPM placed on the sides,  coupled to WLS fibers glued to 4
parallel grooves; the upstream and downstream paddles have a spiral-like groove with two SiPM coupled to both sides (SiPM and grooves are not shown). The lead vault is shown
in gray. Crystals arranged in 8 blocks of 10x10, are shown in light blue. Each crystal is read by a SiPM directly attached on the front face (not shown).
\label{fig:vetos}}
\end{figure}

The EM calorimeter is operated inside two hermetic layers of plastic scintillator veto (see Fig.~\ref{fig:vetos}).
Between the Inner (IV) and Outer Veto's (OV) a layer of lead prevents low energy photons from hitting the crystal.
The OV consists of 2cm-thick plastic scintillator, coupled to a single-side PMT with a plexiglass light guide.
Due to the sizeable size requested to cover the whole calorimeter  and to preserve the possibility of changing the geometry, the OV is segmented in many different paddles.
In particular, the top and the bottom are divided in two parts while the lateral sides are made by 11 paddles per side. The upstream and downstream covers are made by the
same plastic scintillator thickness but read by a PMT located at the center, directly glued on the surface. The paddle's  geometry for the reference configuration has been
inspired by  size and arrangement of the prototype tested in Catania (see Appendix~\ref{Section:BDX-protoype}).

The IV consists of a hermetic box comprised of 6 1-cm thick plastic scintillator paddles. To avoid the PMT encumbrance and keep the size of the OV as small as possible, the lateral  paddles are read on one side by 4 SiPM. Grooves on the surface host two 1mm WLS fibers to convoy the scintillation light to the SiPM. This solution presents many advantages: the reduced attenuation length ($\sim$6 m) allow to limit  the number of paddles not requiring further segmentation; the redundancy resulting by the light transmission inside the clear plastic makes any single SiPM inefficiency negligible  (a hit on a paddle is acknowledged when at least one of the four  SiPMs fires). The upstream and downstream covers have a spiral-like  groove.  The WLS fibers are read by one SiPM per paddle located on the scintillator surface. The compact design of the IV results in a fully hermetic plastic scintillator box. As for the OV, all these solutions were tested on the prototype currently running in Catania (see Appendix~\ref{Section:BDX-protoype}).
\subsubsection{The lead vault}
In between the two active vetos a 5cm thick layer of lead prevents low energy photons to reach the crystal.
The ``sandwich'' configuration, with the lead between IV and OV, has been chosen   to track the cosmic muons in the OV that may produce spallation in the dead material. It also acts as radiator for high energy photons produced by cosmic muon spallation, whose charge component is then detected by the IV.  
The lead vault has only a tiny clear window on the bottom side to let the calorimeter and IV cables to exit. Some lead bricks placed in front of the aperture mitigates the direct exposure to the external background.
As discussed for the two veto's, also the lead vault has been implemented in the prototype and tested in Catania.

\subsection{Trigger and data acquisition}
The trigger and data acquisition (TDAQ) system must be compatible with the specificity of the experiment, where the traditional concept of ``event''
is less applicable. Basically, the main trigger condition will be defined as any crystal signal exceeding a certain threshold. When this
condition is met, the signal from crystals that crossed the threshold must be read, to determine hit energy and time, together with signals
from all the active-veto counters. For the latters, the system must be designed to acquire also potential signals that occurred up to
${\cal O}(10)\;\mu$s before the crystal hit time. In this way, it is possible to reconstruct a complete ``story'' of the event, thus allowing
identification and rejection of all the possible background topologies - including rare effects such as muon decays, delayed neutron hits, etc.

Given these requirements, we foresee a TDAQ system based on Flash to Analog Digital Converters (FADCs). These devices continuously
sample the input signal at a rate $R$, working as a multi-channel digital oscilloscope. In the basic operation mode, when a trigger
is delivered to a FADC board, all the samples within a programmable acquisition window - properly shifted with respect to the trigger
time - are reported. The use of a multi-buffer architecture allows the system to work with almost zero dead-time. Most FADC boards can also
run online signal-processing algorithms on an on-board FPGA, thus making these devices very versatile.

The TDAQ system design here discussed is based on the 16-channels, 250-MHz sampling rate VME FADC board developed by the JLab Fast Electronics group~\cite{FA250}. The main characteristics are reported in Table\,\ref{tab:fa250}. Each FADC in a crate can communicate with a central crate trigger board trough a fast (2.5Gb/s) serial line (VXS), both for trigger formation and for events readout.  
FADC and related ``ancillary'' trigger boards have been already extensively used to setup the TDAQ system for various experiments at Jefferson Laboratory: GlueX, CLAS12, HPS, \ldots. The latter, in particular, exploits the FADC online-processing capabilities in a sophisticated, FPGA based, trigger system.

\begin{table}[tbp]
\begin{center}
\begin{tabular}{|c|c|}
\hline
  Parameter & Value  \\   
\hline  
Sampling rate & 250 MHz \\
\hline
Voltage range & Selectable: 500 mV, 1 V, 2 V \\
\hline
Dynamics      &  12 bits \\
\hline
Acquisition buffer & 8 $\mu$s \\ 
\hline
Readout window & 2 $\mu$s (up to 8 $\mu$s with custom firmware\cite{CCuevas})\\
\hline
\end{tabular}
\caption{JLab FADC250 board main parameters\label{tab:fa250}}
\end{center}
\end{table}

The following parameters, obtained from the BDX prototype measurements in Catania, were considered while designing the system: 
\begin{itemize}
\item{1000 CsI(Tl) crystals, each read by a SiPM. Signal rate: 5 Hz / crystal}
\item{100 active veto channels, each read by a SiPM. Signal rate: 30 Hz / counter} 
\end{itemize}

The simplest TDAQ system for the BDX experiment, as discussed before, implements as main trigger condition the OR from the discriminated crystal signals above a certain threshold. This algorithm is already implemented in the standard FADC and trigger boards firmware. In order to maximize the information for each event, the largest readout window (8 $\mu$s) is employed for both crystal and active veto signals. The following readout modes are foreseen:
\begin{itemize}
\item{For the crystals, signals are read in so-called ``raw-mode'': for every trigger, all the 2048 samples in the $8\;\mu$s acquisition window are reported, with no online elaboration. This gives maximum flexibility for the off-line analysis.}
\item{For the active veto channels, signals are processed by an online pulse-integration algorithm implemented on the FPGA\footnote{This algorithm already exists in the FADC250 firmware, and has been extensively tested for short ($\cal{O}$(100 ns)-wide) pulses.}. The algorithm reports, for each pulse exceeding a programmed threshold, the pulse area, amplitude, and threshold-crossing time. Up to four distinct pulses per channel per event can be reported.}
\end{itemize}

The event and data rate are estimated with the very conservative assumption of having completely uncorrelated signals, i.e. adding the individual rates from all the crystals.
\begin{itemize}
\item{The overall trigger rate will be $R_{trg}=5$ $Hz$/crystal $\cdot 1000$ crystals = 5 kHz.}
\item{The data size of each crystal signal is: $D_{crs}=2048$ samples $\cdot 12$ bit/sample = 3 kB. The total data rate from crystals is: $DR_{crs}=D_{crs} \cdot R_{trg}=14$ MB/s.}
\item{The data size of a FADC-integrated pulse is $D_{veto}\simeq 12$B. Assuming (conservatively) that $N_{veto}/10$ veto counters report a pulse for each trigger, the total data rate from these is: $DR_{veto}=N_{veto}\cdot D_{veto} \cdot R_{trg}=1$ MB/s.}
\item{The total event rate is: $DR_{tot}\simeq 1.1\cdot(DR_{crs}+DR_{veto})=16$ MB/s. A 10$\%$ overhead has been assumed for event-related information (event time, indexes of channels, \ldots)}
\end{itemize} 
These values are completely compatible with the existing hardware, firmware and software, thus allowing the implementation of the TDAQ system with no major efforts. The list of the required equipment  is reported in Appendix~\ref{sec:costs}.

The proposed solution has the advantage of being robust and achievable with already-existing technologies. However, given the flexibility provided by FADCs and trigger boards - in particularly the newly developed VXS Trigger Processor (VTP) -, we also plan to investigate an alternative, trigger-less readout mode. The rough scheme of this evolved setup is the following: every time a crystal or a plastic scintillator counter signal exceeds a local threshold, a corresponding raw-waveform (within a proper acquisition window), is reported, via VXS bus, to the central VTP board, together with a fine time-stamp. The VTP sees, therefore, a ``continuous stream'' of raw-waveforms from all the channels in the VME crate, and will report it via a fast optical link (34 Gb/s) to one (or more) VTP boards in a master trigger crate, collecting data from all the detector channels. On this board, the continuous datastream is analyzed by different online algorithms, to identify and save ``events'' of interest. Preliminary estimates of the foreseen data rate show that this solution is, in principle, possible for the BDX detector. We plan to investigate it further, working in collaboration with the JLab fast electronic group.

\subsection{Computing resources}
The computing resources needed by the experiment were estimated as follows. Assuming a rate of 5 kHz, as outlined in the previous section, the total number of recorded triggers that we expect to accumulate in four years of data taking is of the order of $5\cdot 10^{11}$. Based on the reconstruction software performance achieved on the Catania prototype data and on Monte Carlo data, we estimated the time needed to reconstruct one event on a modern CPU to be of the order of 10 ms. The total number of CPU-hours needed to reconstruct the entire BDX data set is therefore of the order of $2\cdot10^6$ including 50\% contingency. Additional  $6\cdot10^6$ CPU hours will be needed for the Monte Carlo simulation of about $10^{11}$ EOT (single event simulation time of 200 ms).
Based on the estimated data size and rate, the overall volume of collected data for four years of running amounts to approximately 2 PB. We assume the data recorded in the first year, i.e. 20\% or 400 TB, would be permanently saved to tape. For the remaining 80\%, online software, that will be developed and optimized based on the first year of data,  will be used to filter the portion of the data where potentially interesting events are found, with a data reduction factor of one order of magnitude. The total permanent storage needed for real data will be of 600 TB, including 80 TB of reconstructed data and 20 TB of Monte Carlo data. In addition, approximately 100 TB of disk space would be needed to temporarily store raw and reconstructed data, and Monte Carlo events.

\subsection{The event reconstruction framework}
The main requirement of the BDX event reconstruction and framework is the possibility of developing a modular code, where different pieces, each related to a simple task, can be assembled together. This modular design allows to de-couple the problem of defining the global reconstruction scheme from the actual implementation of the single tasks, and, at the same time, permits to change the first without the necessity of rewriting the latter. Finally, a modular architecture is also well suited for a collaborative development effort.

The framework should also be designed to fully exploit modern computer-science technologies that can
significantly speed-up the reconstruction procedure (multi-threading, multi-processing, etc.). Finally,
the compatibility with other common software tools, such as ROOT, is highly desiderable.

After consulting with experts in software development, we identified the ``JLab Data Analysis Framework'' (JANA)
as a convenient software package to develop the reconstruction code \cite{JANA}. JANA is a software package
written in C++ that provides the mechanism by which various pieces of the reconstruction software are brought together to fully reconstruct the data. This is motivated in large part by the number of independent detector subsystems that must be processed in order to reconstruct an event. The choice of the Hall-D GlueX experiment to use this as the basis for the development of the corresponding reconstruction software clearly demonstrates the ``maturity'' of this code. Furthermore, the choice of using an already existing code - with proper modifications to tailor it for BDX  - has the clear advantage of avaibility of experts support and possibility to re-use already developed parts. A preliminary version of the reconstruction software, to handle data from the BDX prototype measurements (see Appendix \ref{Section:BDX-protoype}), has been already implemented.

The JANA framework is built upon the idea of data factories. The general idea is the following: when data is requested from a factory (i.e. an order is placed) the factory’s stock is first checked to see if the requested items already exist. In JANA, a factory only makes one type of object, so if the objects have already been made for this event, const pointers to them are passed back. Otherwise, it must manufacture (instantiate) the objects. The manufacturing procedure itself needs first to get the ``parts'' from which to build its own objects. These parts are objects produced by other factories. Eventually, one gets down to requesting objects that are not produced by a factory but rather, originate from the data source (event file or online TDAQ system).
In this scheme, therefore, the overall reconstruction scheme is defined by the specific chain of factories involved in the process, starting from the highest-level, i.e. the event builder, down to the lowest-level ones, i.e. those associated with each sub-detector. Multiple reconstruction schemes, possible sharing the same factories, can be implemented in parallel, each scheme being associated, for example, to analysis tasks.

\subsection{A new facility for beam-dump experiments at JLab}\label{Sec:BD-facility}
\subsubsection{Building and access to detector}
We present the reference layout of the infrastructure and civil construction needed for the BDX experiment behind Hall A. This concept, referred to as ``C1'', grew out of 
feedback and comments to early ideas for the civil design \cite{bdxnote001,bdxnote002} and the BDX Letter-of-Intent \cite{LOI_bdx}. This reference C1 is used to determine overall dimensions and shielding in order to estimate costs and locate the detector relative to the Hall A beamdump.  While dimensions are consistent with the proposed experiment, specific details fo the detector should be obtained from other sections. We expect that this concept will evolve as the results of simulation are feed back into the concept for civil construction. 

\begin{landscape}
\begin{figure}[tp]
\begin{center}
\includegraphics[height=12cm,clip=true]{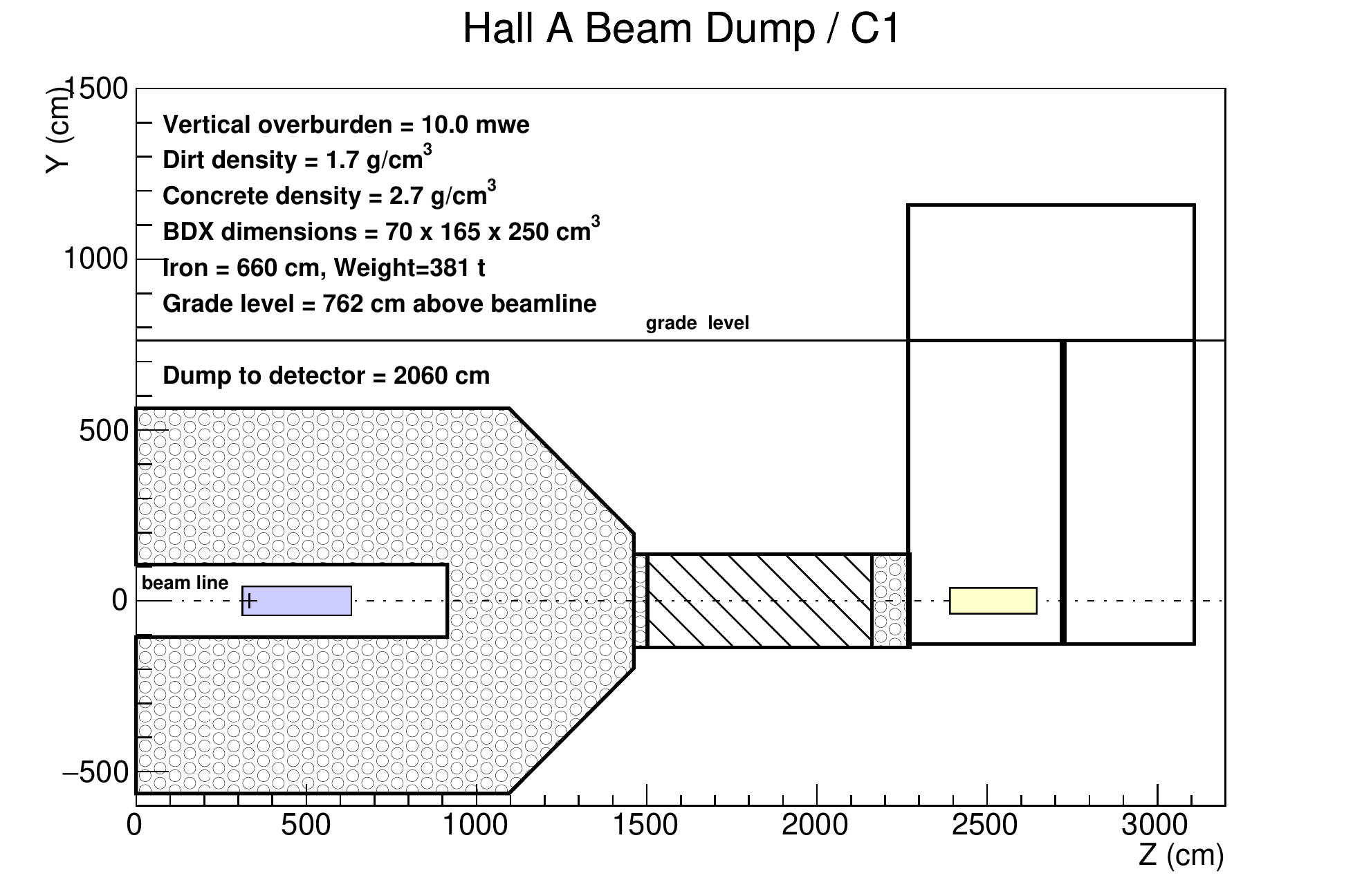}
\caption{Elevation sketch of civil concept C1. This view includes the Hall A beam dump proper and surrounding concrete, filling material between the beam dump and the BDX experiment,
the detector in the underground room and staircase area. General parameters and assumptions are specified on the drawing.
\label{fig:Draw_C1_HallA_BeamDump}}
\end{center}
\end{figure}
\end{landscape}

\begin{landscape}
\begin{figure}[tp]
\begin{center}
\includegraphics[height=12cm,clip=true]{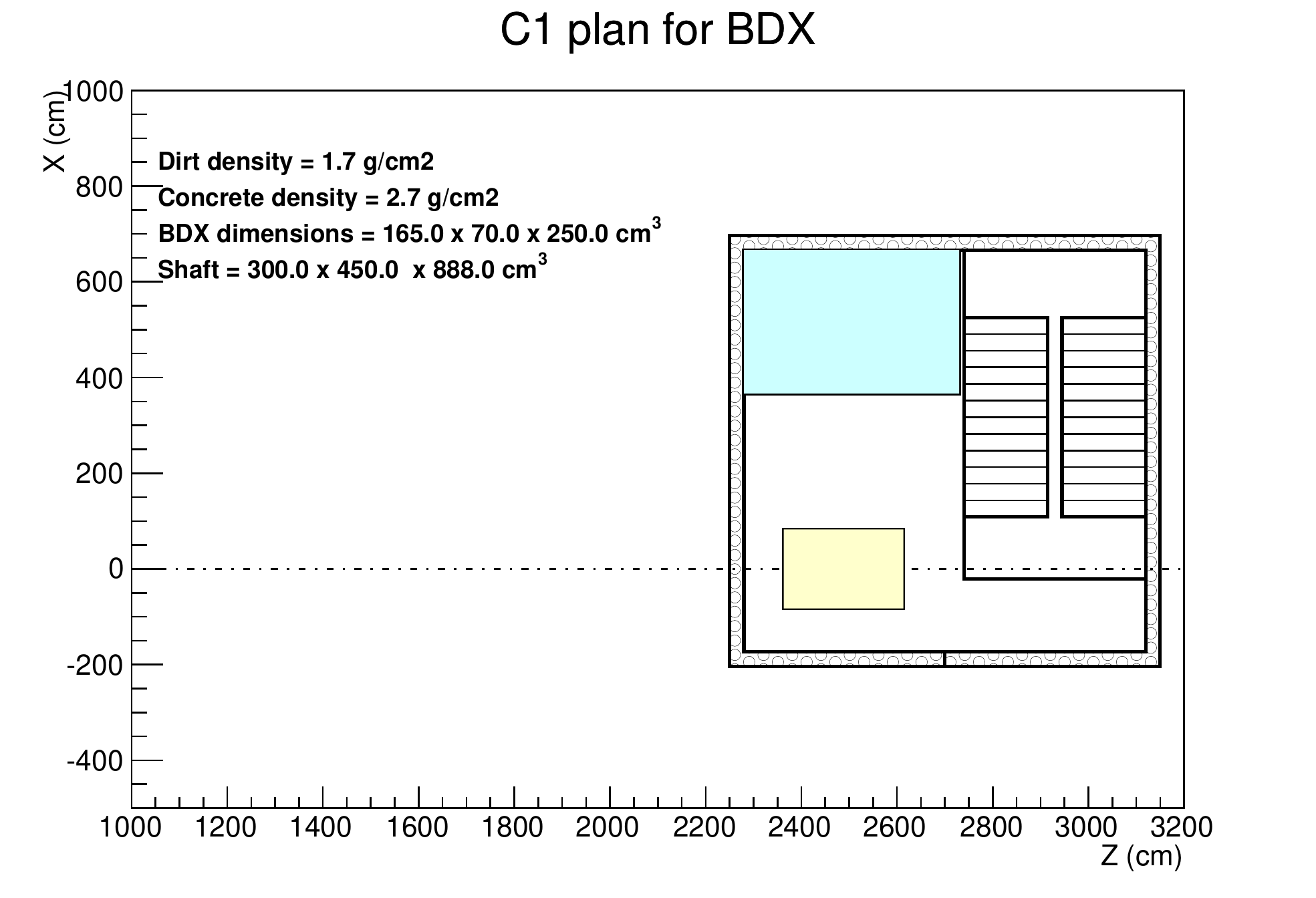}
\caption{Plan sketch of civil concept C1. The vertical shaft for access to the underground room is shown in light blue.
The detector outline, including cosmic-ray shield is shown in light yellow. Access stairs are included for access to the underground room.
General parameters and assumptions are specified on the drawing.
\label{fig:Draw_C1_plan}}
\end{center}
\end{figure}
\end{landscape}

\begin{landscape}
\begin{figure}[tp]
\begin{center}
\includegraphics[height=12cm,clip=true]{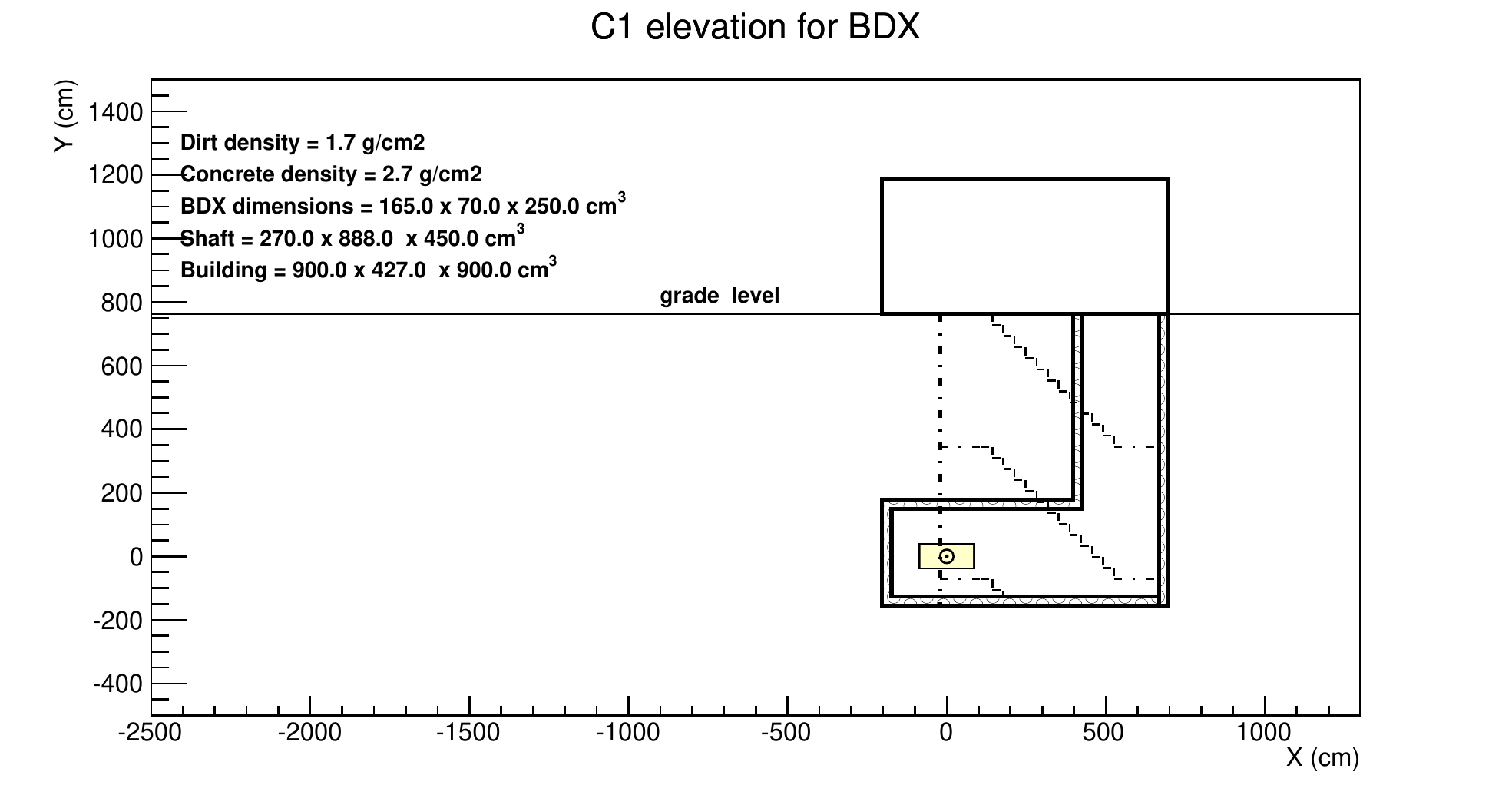}
\caption{Cross sectional sketch perpendicular to the beam of civil concept C1. The vertical shaft is shown to the north of the detector room. After lowering
the detector to the underground room, the detector would be rolled inside into the underground room that is shielded 
against cosmic-rays. The access stairs are shown schematically with dashed lines, which allow access to the underground 
room. General parameters and assumptions are specified on the drawing.
\label{fig:Draw_C1_elevation}}
\end{center}
\end{figure}
\end{landscape}

The present concept provides full personnel access to the detector at the basement level of the BDX building. The location of the detector is sufficiently deep, that it is unlikely that there would be any significant savings by installing the detector into a shaft without personnel access. There is only about 5 m of concrete shielding surrounding the Hall A dump (see Fig.\,\ref{fig:Draw_C1_HallA_BeamDump}) 
and therefore it is likely that additional shielding will be required if the detector is placed close to the beam dump source. Thus there may be considerable excavation required even if a very narrow shaft is used for the detector. This would also be true if the shaft is constructed at an angle (see Fig.\,4 of Ref. \cite{bdxnote001}) since it is probably easier to dig a large ditch and backfill than to dig a tunnel at an incline. The difficulties and complications that come from limited access and restoring cosmic-ray shielding every time the detector is serviced can be avoided by starting with a concept that has a shaft for moving the detector underground and a personnel access to service the detector once installed. A sketch of the plan view is given in Fig.\,\ref{fig:Draw_C1_plan} and a sketch of the elevation view is  
given in Fig.\,\ref{fig:Draw_C1_elevation}.

This concept includes an above-ground building with a crane to lower the detector underground and stair access to the underground room with the detector during operation (see Fig.\,\ref{fig:Draw_C1_plan}). The detector would be lowered down through an access shaft and then rolled into a room, which is shielded from cosmic rays by the overburden. This room is
accessible using the stair connection to the underground (see Fig.\,\ref{fig:Draw_C1_elevation}). Electronic racks would likely reside underground with only fiber connections to the outside world. The environment for the entire building would be controlled for proper operation of the detector and electronics.

\begin{table}[tbp]
\begin{center}
\begin{tabular}{|l|c|c|}
\hline
\hline 
  Component & Size & Number  \\   
\hline  
\hline 
CsI(Tl) crystals & $5\times5\times30$ cm$^3$ & 100/module \\ \hline 
Module & $15\times15\times30$ cm$^3$ & 10 \\ \hline
Calorimeter & $50\times50\times300$ cm$^3$ & 1 \\ \hline
Active veto & $100\times100\times350$ cm$^3$ & 1 \\
\hline
\end{tabular}
\caption{List of components of the reference detector and their dimensions. Note that the detector design has evolved considerably and this table should be taken only as a rough basis for infrastructure sizing. \label{tab:components}}
\end{center}
\end{table}

\subsubsection{Detector size}
The following detector design has been considered as the basis for infrastructure sizing and cost estimates. We underline that this is just a conservative approximation of the final detector design reported in the proposal, only used for the aforementioned purpose. We assumed that the detector will fit into a volume of 100 $\times$ 100 $\times$~350 cm$^3$. The detector was sized based on early concepts based on the promising and compact option to re-use the Thallium-doped CsI crystals from the BaBar end cap calorimeter at SLAC \cite{Aubert:2001tu}, with improved SiPM-based readout. 
The weight of the calorimeter is about 3.4 metric tons. The reference dimensions are shown in Table~\ref{tab:components}. 

For the active cosmic-ray shield we used reference dimensions from the Los Alamos beam dump experiment E645 \cite{Freedman:1993kz}. It consists of an outer layer of active scintillator and an inner passive layer of lead. 
The E645 experiment used about 15 cm for both the lead and scintillator layers. We took the thicknesses of each layer to be 5 cm, based on simulations for the proposal. The total weight of the lead shield is 5.5 metric tons.

The number of electronics channels for this detector is of order one thousand for the calorimeter and one hundred for the active veto (see Table\,\ref{tab:components}). The power consumption per calorimeter channel (SiPM preamp) is about 0.1 W. For the veto system, with the conservative assumption of using PMTs for both the inner and the outer veto, the power consumption is about 2 W/channel. This result in a total power consumption of about 300 W for the whole detector. The digitizing electronics power consumption (with about 70 16-channels digitizing boards) is about 4 kW\footnote{The power dissipated with an ``empty'' crate containing a CPU is about 80 W, and the nominal power consumption for the 16-channel JLab Flash 250 MHz 
is 58 W/board.}. These are only rough guesses, but more precise estimates must wait for a full electronics design.

\subsubsection{Shielding from the beam dump and overburden}
Various considerations for the shielding from the beam dump and overburden to shield from cosmic rays are discussed in Ref.~\cite{bdxnote001}. 
The detector is located sufficiently far from the beam dump 
and there is sufficient shielding in the form of concrete (150 cm) and iron (660 cm) to reduce any conventional particle source to acceptable levels. The amount of overburden is assumed to be a minimum of 10 meters of water equivalent (mwe). At the depth of the beam (762 cm), 10 mwe of overburden is achieved by  covering the underground room up to grade level.

\subsubsection{Cost estimate  \label{sec:civilcost}}
\begin{landscape}
\begin{figure}[p]
\begin{center}
\includegraphics[page=1,height=16cm,clip=true]{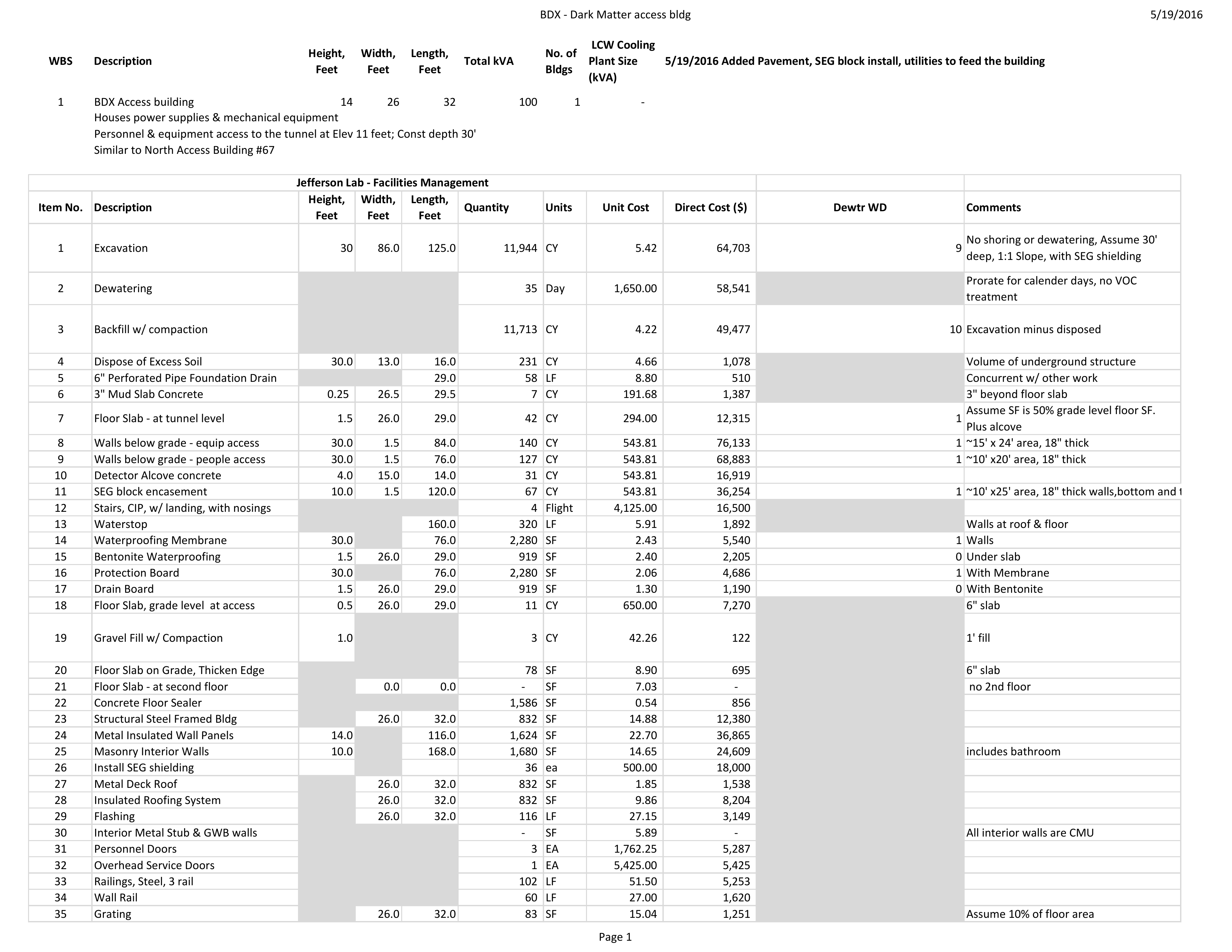}
\caption{Cost estimate for civil concept C1. Page 1.
\label{fig:BDX_Civil_Layou_Costv2_tw_1}}
\end{center}
\end{figure}
\end{landscape}

\begin{landscape}
\begin{figure}[p]
\begin{center}
\includegraphics[page=2,height=16cm,clip=true]{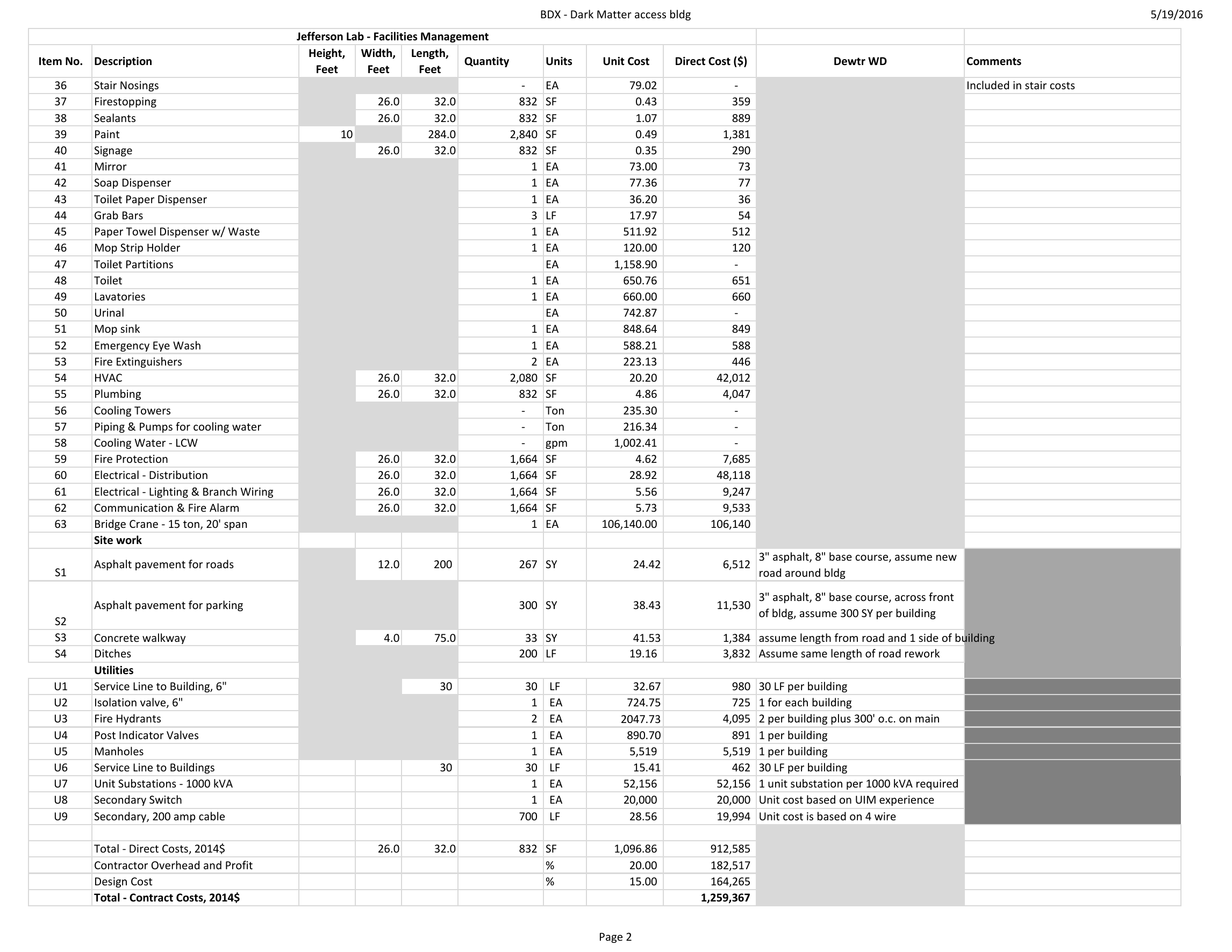}
\caption{Cost estimate for civil concept C1. Page 2.
\label{fig:BDX_Civil_Layou_Costv2_tw_2}}
\end{center}
\end{figure}
\end{landscape}

A preliminary cost estimate was conducted based on dimensions described above. The estimate is based on the cost of existing buildings,
estimates for the MEIC project and RSMEANS Facilities Construction Cost Data 2015. The cost estimate was made by T. Whitlatch. C. Whitlatch and
R. Yasky from Facilities Management kindly provided access to cost information
and helpful discussions on facility requirements. The estimate includes the following features
\begin{itemize}
\item excavation, dewatering, backfill and compaction of soil
\item walls below grade, underground room at beam level
\item personnel access and stairs
\item above-ground building with 15 t crane
\item HVAC, plumbing, piping, LCW, electrical distribution and lighting
\item Encasing and burying iron shielding blocks (assumes block availability)
\item Road, parking lot and power substations
\end{itemize}
The total estimated cost is \$1.3M. A breakdown of the cost estimate can be found in
Figs.\,\ref{fig:BDX_Civil_Layou_Costv2_tw_1} and \ref{fig:BDX_Civil_Layou_Costv2_tw_2}.

\clearpage
 \section{Signal and background rates}\label{sec:setup}

\subsection{Simulations of the experimental set-up}
\begin{figure}[t!] 
\center 
\includegraphics[width=12.5cm]{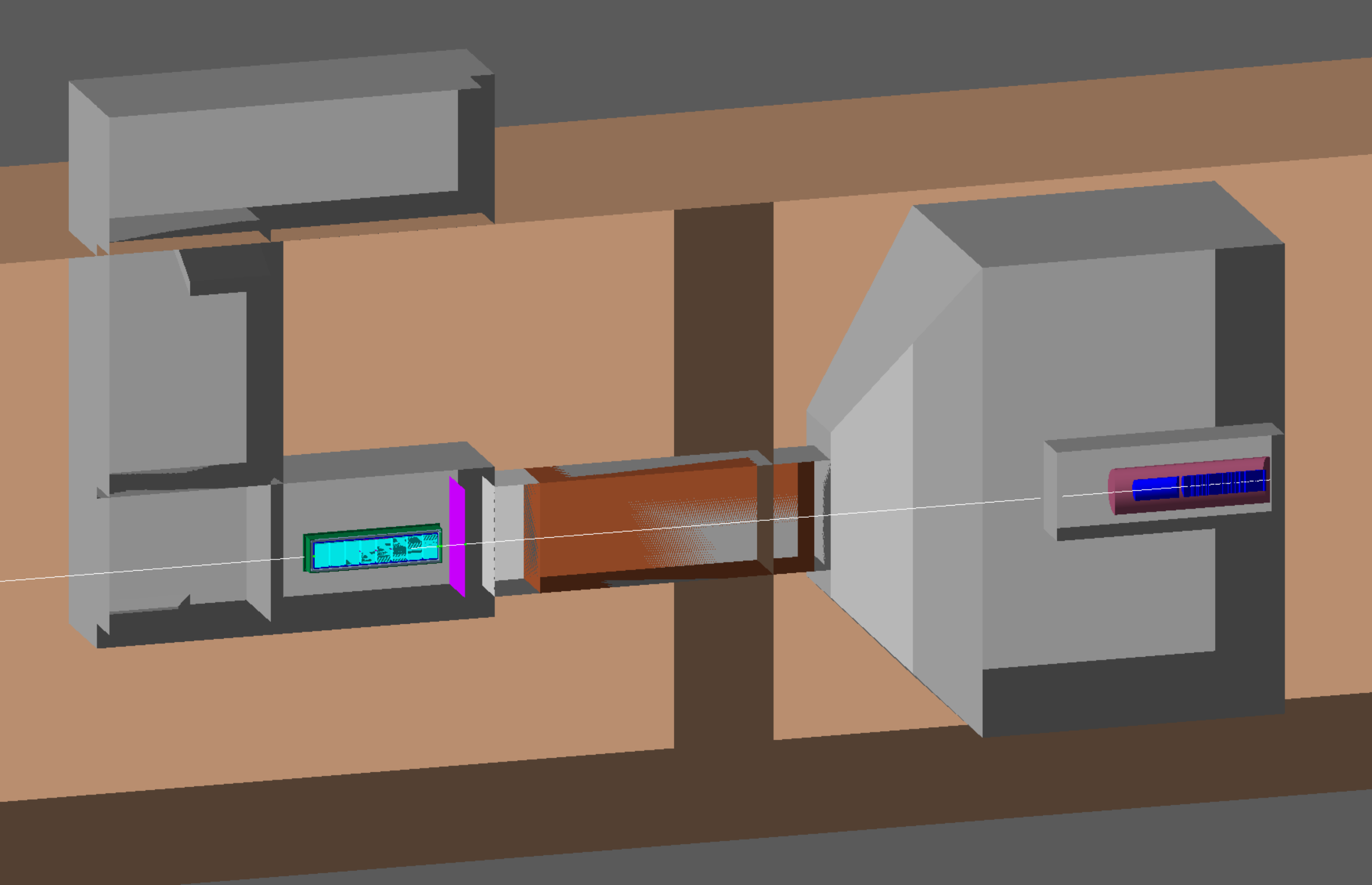} 
 \caption{Implementation of the BDX detector and Hall-A dump in
GEMC. The white line shows the beam centerline. \label{fig:bdx_gemc} }
\end{figure}

The proposed detector, the new underground facility and  the Hall-A beam-dump geometry have been
implemented in GEANT4 within GEMC simulation package~\cite{gemc}. In the following sections we 
present results concerning the expected rates from interaction of a $\chi$ particle, beam-related background and cosmogenic background. 
Figure~\ref{fig:bdx_gemc} shows the geometry as implemented in simulations.

\subsection{Signal}

The expected number of signal events measured in the detector was estimated trough a Monte Carlo calculation, according to the model for LDM production and detection described in Sec.~\ref{sec:elastic}. The calculation involves three steps. First, the evaluation of the number of $\chi$ particles electro-produced in the beam dump, trough  on-shell or off-shell $A^\prime$ mediatior. Then, the calculation of the interaction rate in the detector. Finally, the estimation of the actual detection efficiency for the scattered electrons and protons.
All these numbers -$\chi$ production yield, $\chi$ scattering rate, detector efficiency - depends on four parameters: the mass of the $\chi$ ($m_{\chi}$), the mass
of the exchanged $A^\prime$ ($m_{A^\prime}$), the coupling constant between the electron and the $A^{\prime}$ ($\varepsilon$) and the coupling constant
between the $\chi$ and the $A^\prime$ ($\alpha_D$). However, the kinematics only depends on the two masses; $\varepsilon$ and $\alpha_D$ are only related
to the absolute event yield. Therefore, calculations were performed with fixed values of these parameters, and final results were rescaled accordingly.

\subsubsection{$\chi$ production}

The $\chi$ production process in the beam-dump was simulated using a modified MadGraph4 version~\cite{madgraph}. The sofware was used to generate LDM events produced in electron-aluminum nucleus collisions, $e^- N \rightarrow e^-N A^\prime \rightarrow e^- N \chi \overline{\chi}$ (where $N$ is a nucleus with Z = 13, A = 27), and to calculate the total LDM production cross section. MadGraph was modified to include the aluminum nucleus form factor as found in \cite{nuc-ww}, which accounts for coherent scattering, as well as for nuclear and atomic screening.

The most common solution to account for the finite dump thickness adopted in similar calculations is to use the ``single-radiation length approximation'' (\cite{Izaguirre:2013uxa,Izaguirre:2014dua}, i.e. to consider an effective length equal to one radiation length, neglecting showering and energy loss effects. This strategy was, for example, adopted in the original E-137 re-analysis \cite{Batell:2014mga}. Instead, we performed a detailed study of the primary electron beam interactions in the beam dump, finding non-negligible corrections to the aforementioned approximation.

We proceeded as follows. First, we used the GEANT4 beam-dump simulation to sample the flux of electrons and positrons in the dump, as a function of energy, at
different depths (measured in radiation length units, $t$). To simplify the calculation, we adopted a simplified model of the beam-dump design, considering a
uniform Aluminum cilinder. However, since all the results are reported in radiation lenght units, and the showering process dependence on the material is almost
all contained in this quantity, we do not expect sizeable effects due to this choice. The result of the calculation was the differential energy spectrum of electrons
and positrons as a function of the depth in the dump, normalized to the number of primary particles, $\frac{dN}{dE}(t)$\footnote{Positrons were included in the
calculation since they can produce $\chi-\overline{\chi}$ pairs in the dump just like electrons.}.

From this quantity, the total number of $\chi-\overline{\chi}$ pairs produced per incident electron can be calculated as:
\begin{equation}
N_{\chi-\overline{\chi}} = \frac{N_{Av}}{A} \rho \cdot X_0 \int_0^{T_{dump}} dt \int^{E_0}_{E_{min}} dE \, \sigma(E) \frac{dN}{dE}(t)
\end{equation}
where $\rho \cdot X_0$ is the product of the $Al$ density and radiation length (24.01 g/cm$^2$), $T_{dump}$ is the dump length in $X_0$ units, $\sigma(E)$ is
the total cross section for the  $e^- N \rightarrow e^-N A^\prime \rightarrow e^- N \chi \overline{\chi}$ process, $E_0$ is the primary electron energy,
and $E_{min}$ is the threshold energy. Given the $t-$dependence of $\frac{dN}{dE}(t)$, it is safe to perform the calculation with $T_{dump} \rightarrow +\infty$.
This expression can be simplified by introducing the quantity $<\frac{dN}{dE}> \equiv \int_0^{T_{dump}} dt \frac{dN}{dE}(t)$, that does not depend on $\sigma(E)$.
The ``single-radiation length approximation'' is equivalent to $<\frac{dN}{dE}>=1\cdot\delta(E-E_0)$. In terms of this average flux the $\chi$ yield is:
\begin{equation}
N_{\chi-\overline{\chi}} = \frac{N_{Av}}{A} \rho \cdot X_0 \int^{E_0}_{E_{min}} dE \, \sigma(E) <\frac{dN}{dE}>
\end{equation}

\begin{figure}[t!] 
\center
\includegraphics[width=\textwidth]{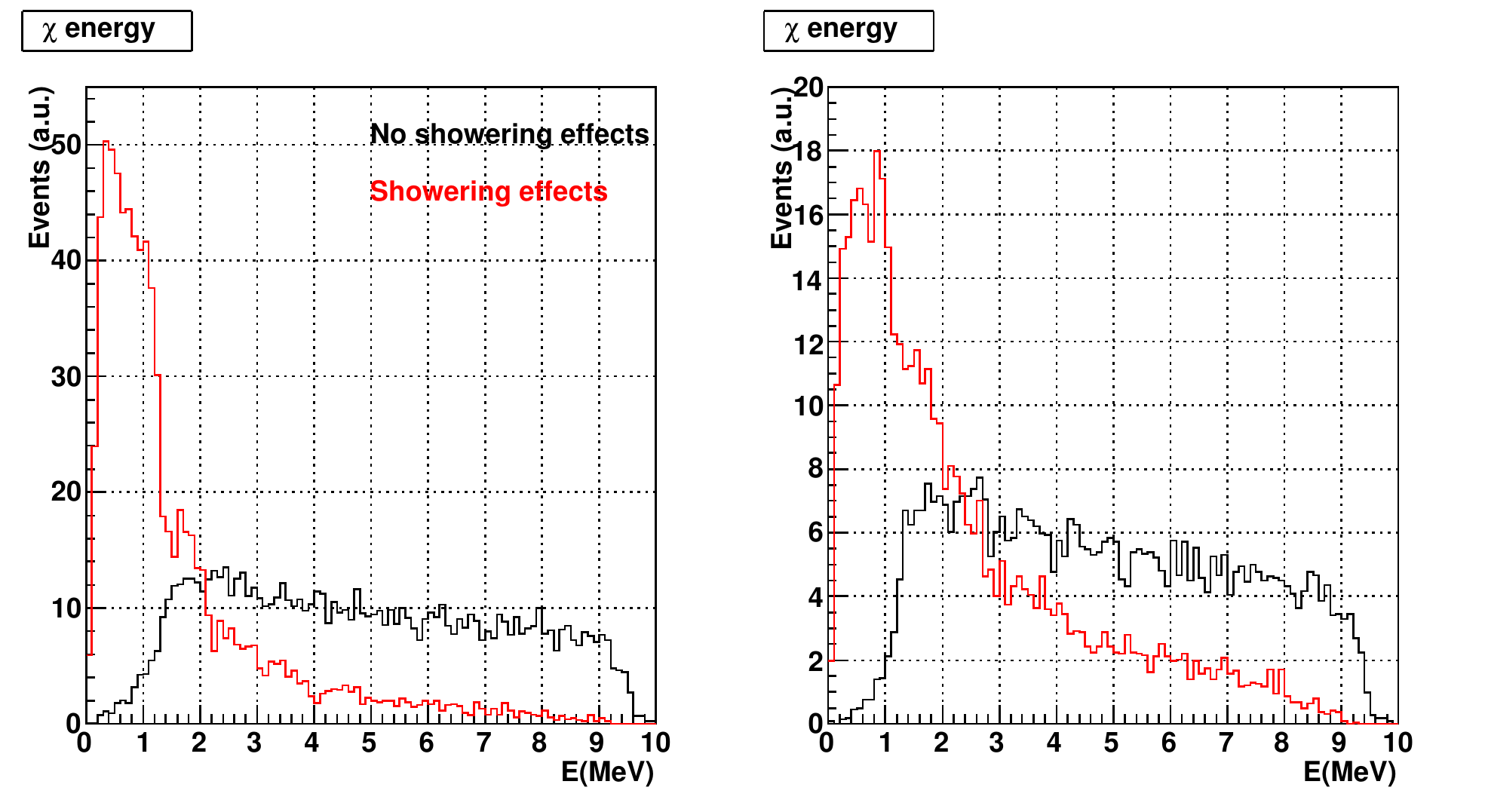}
\caption{Energy spectrum of $\chi$ particles produced in the beam dump, comparing the ``single-radiation length approximation'' (black) with the full beam-dump simulation (red). Left panel: $m_\chi=10$ MeV, $m_{A^\prime}=30$ MeV. Right panel: $m_\chi=100$ MeV, $m_{A^\prime}=300$ MeV. Vertical-axis units in the two panels are different.}\label{fig:chiSpectrumComparison} 
\end{figure}

For each $m_{A^\prime}-m_{\chi}$ combination, we estimated the total $\chi$ yield and the corresponding energy spectrum by numerically integrating the above expression. We performed multiple MadGraph4 simulations at discrete energies $E_i$, weighting each result by $<\frac{dN}{dE}>_{E_i}$, and then summing the different energy bins. The comparison of the $\chi$ energy spectra obtained with this procedure and with the ``single-radiation length approximation'' is shown in Fig.~\ref{fig:chiSpectrumComparison} for two choices of $m_\chi$ and $m_{A^\prime}$. The effect of including the showering mechanism in the calculation is clearly visible: the $\chi$ energy spectrum becomes more pronounced at lower energies and the absolute number of $\chi$ particles increases, due to emission from secondaries. The net effect of this in the foreseen event rate is discussed in the next session.

\subsubsection{$\chi$ interaction}
 
The $\chi$ interaction in the detector was evaluated trough a custom code, handling both the $\chi-e^-$ and the $\chi-p$ scattering processes. The program, given the incident $\chi$ flux for a given set of model parameters, computes the expected event rate within the detector for both topologies. It also provides a set of Monte Carlo events, containing scattered electrons and protons in the detector volume, generated according to the foreseen kinematics. These are then passed to the full detector simulation, implemented with GEANT4, in order to evaluate the experimental detection efficiency. 
 
\begin{figure}[t!] 
\center
\includegraphics[width=\textwidth]{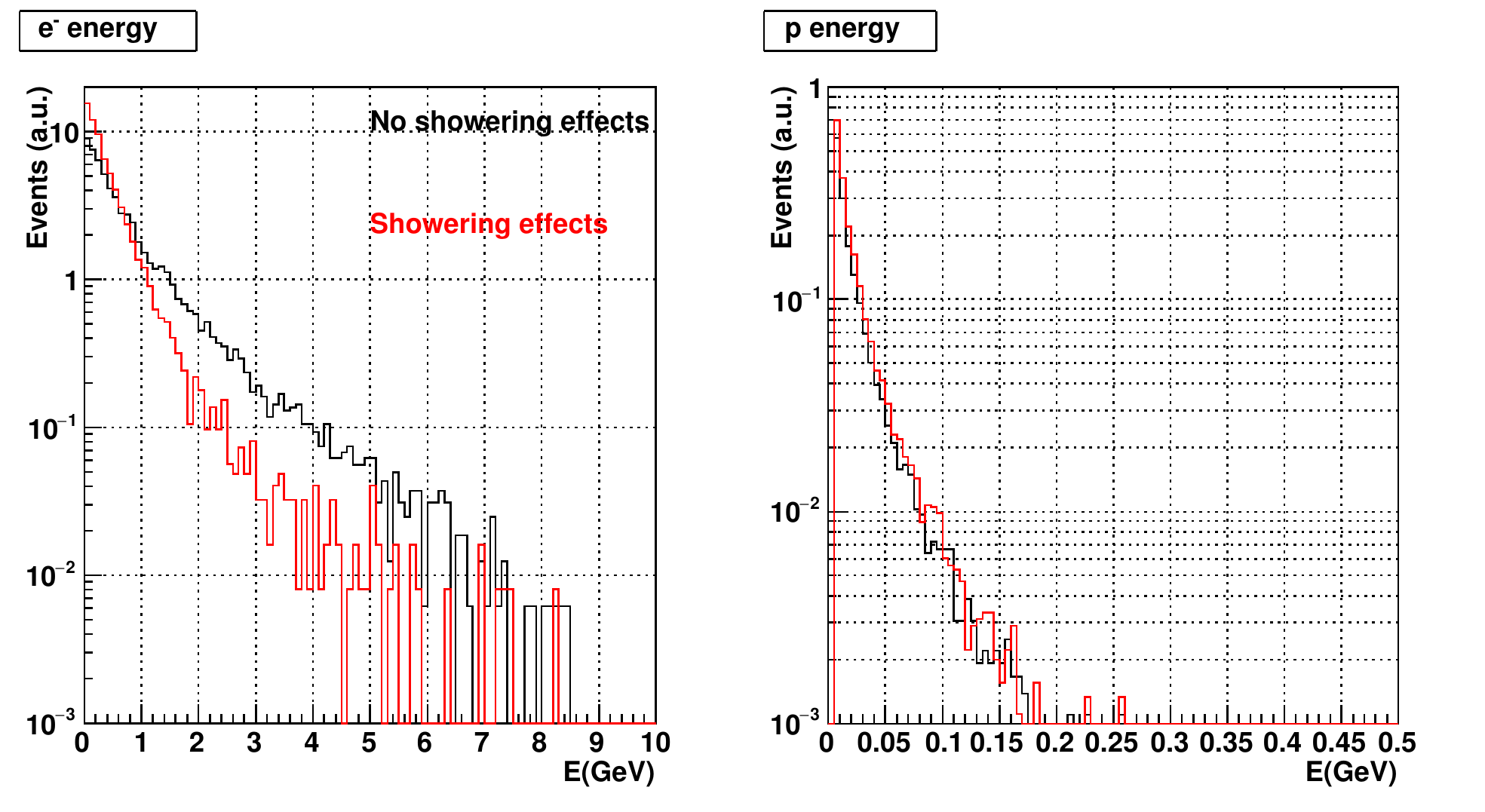}
\caption{Energy spectrum of scattered electrons (left) and protons (right) in the detector, comparing the ``single-radiation length approximation'' (black) with the full beam-dump simulation (red). Both plots correspond to $m_\chi=10$ MeV, $m_{A^\prime}=30$ MeV. Vertical-axis units in the two panels are differeent.}\label{fig:RecoilSpectrumComparison} 
\end{figure}

The corresponding cross-section were implemented in the code, according to the formulas described in \cite{Izaguirre:2013uxa}. For the $\chi-p$ quasi-elastic process, we parametrized the nuclear effects by introducing an ``effective'' nuclear form-factor and an average 8~MeV binding energy (computed from the corresponding values for Cs and I nuclei). The energy spectrum of recoiling electrons and protons, computed for $m_\chi=10$ MeV, $m_{A^\prime}=30$, is shown in Fig.~\ref{fig:RecoilSpectrumComparison}, comparing the ``single-radiation length approximation'' with the full beam-dump simulation. For the $\chi-e^-$ scattering process, the energy distribution is much more pronounced at lower values when the beam showering is considered. The total number of expected events is $30\%$ higher when the contribution of secondary electrons and positrons in the beam-dump is considered, with respect to the simple ``single-radiation length approximation''. However, the fraction of events with scattered electron energy greater than 300 MeV (500 MeV) is $40\%$ (70$\%$) less, resulting in a total event yield $10\%$  (20$\%$) lower (with the difference increasing if an even higher energy threshold is used). For the $\chi-p$ process the difference between the two cases is less pronounced.

\subsubsection{Detector response \label{sec:response}}

\begin{figure}[t!] 
\center
\includegraphics[width=10cm,clip=true]{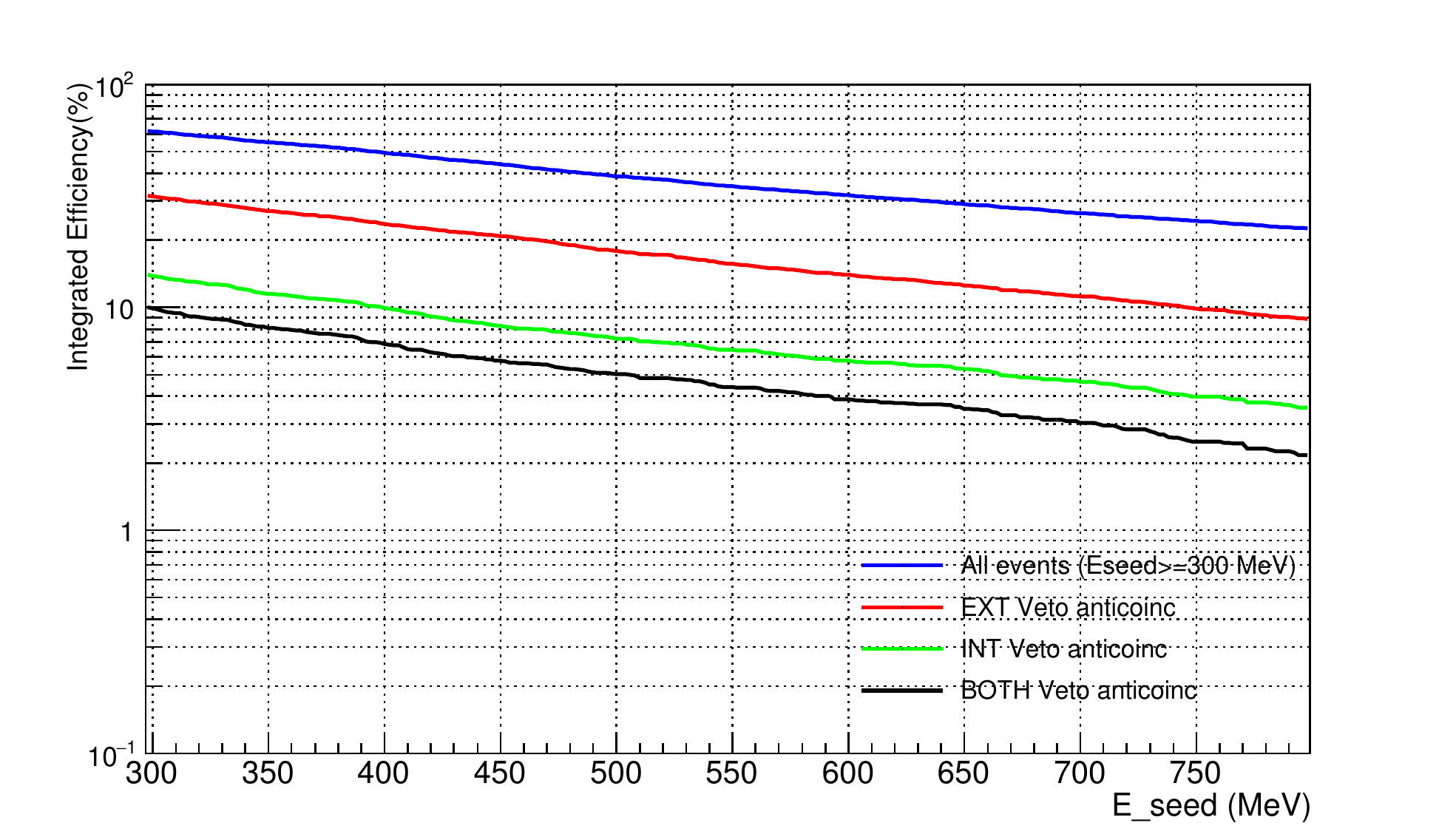}
\includegraphics[width=10cm,clip=true]{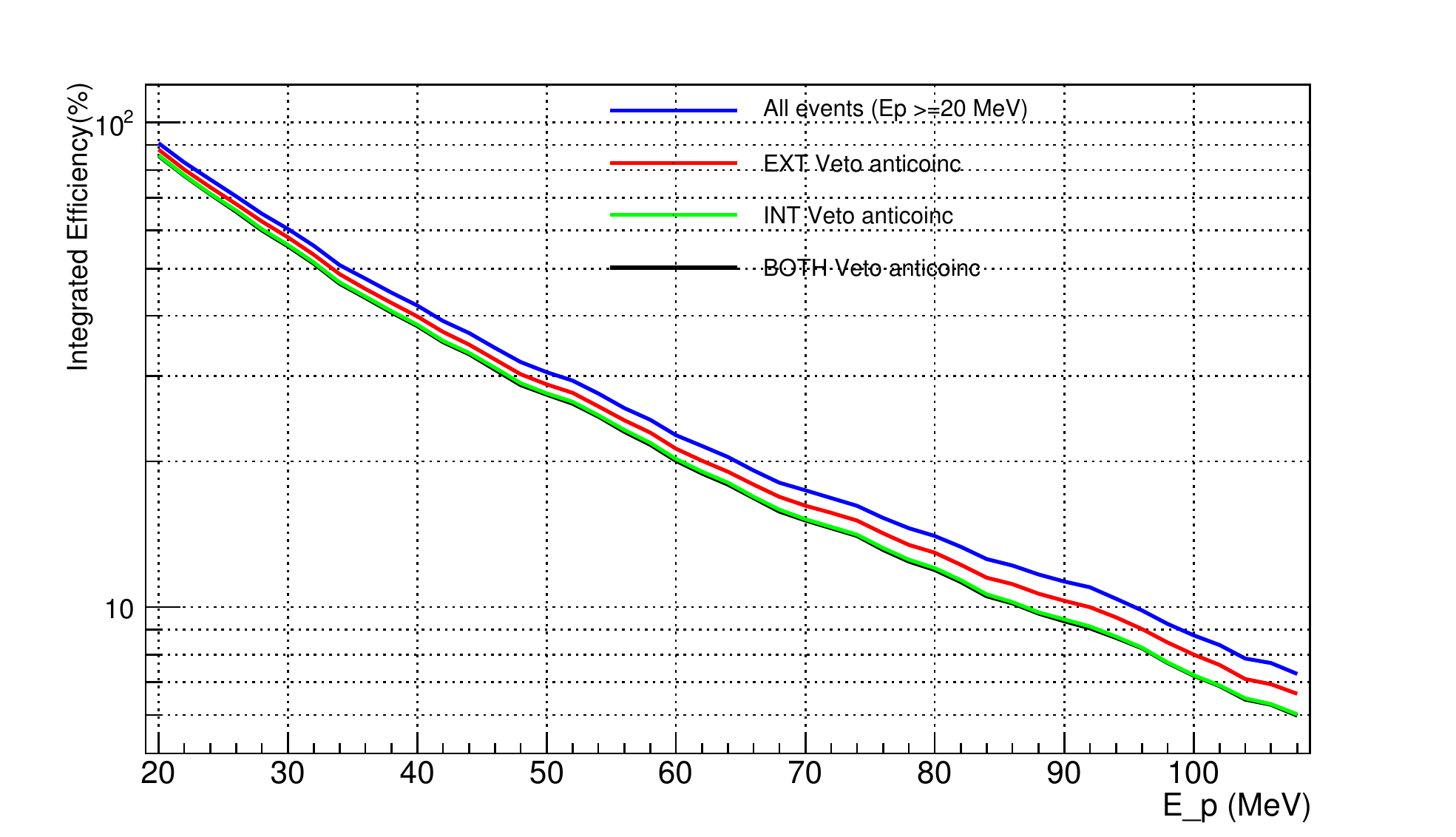} 

\caption{Top panel: Integrated detection efficiency for $\chi-e^-$ scattering events with $E_{e^{-}}>$300 MeV,  $m_\chi=30$ MeV and $m_{A^\prime}=90$ MeV,  as a function of $E_{seed}$. The blue curve indicates the efficiency obtained by requiring the presence of a seed with an energy deposited greater than 300 MeV. The red, green and black curves are the efficiencies when we also require the anti-coincidence of the outer, inner or both vetos, respectively. In the simulations we assumed a 1\% inefficiency for each veto detector. 
Bottom panel: Integrated detection efficiency for $\chi-p$ scattering events with $E_{p}>$20 MeV and for the same $m_\chi - m_{A^\prime}$ combination.}\label{fig:detection_efficiency} 
\end{figure}

\begin{figure}[t!] 
\center
\includegraphics[width=6.5cm,clip=true]{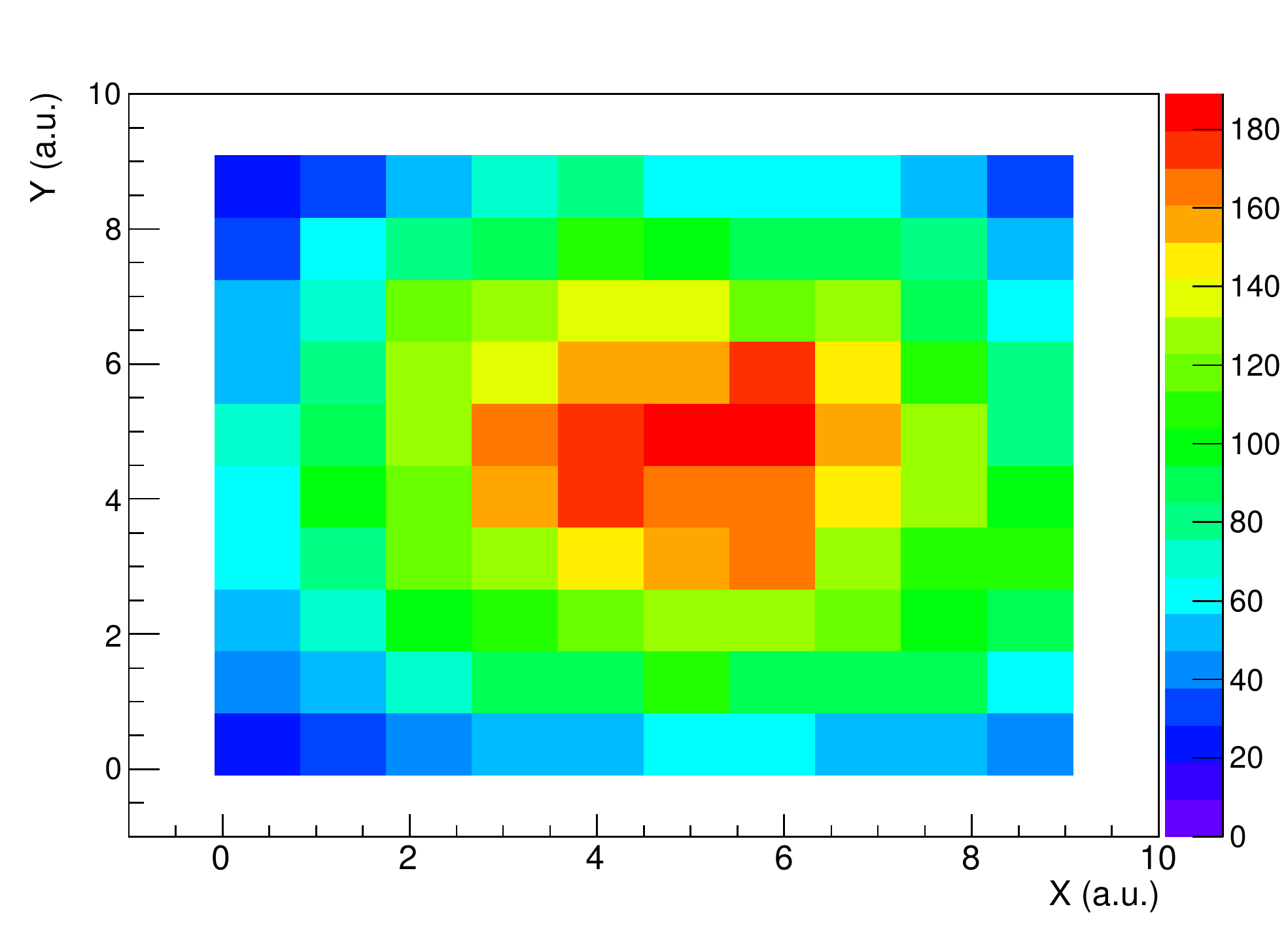}
\includegraphics[width=6.5cm,clip=true]{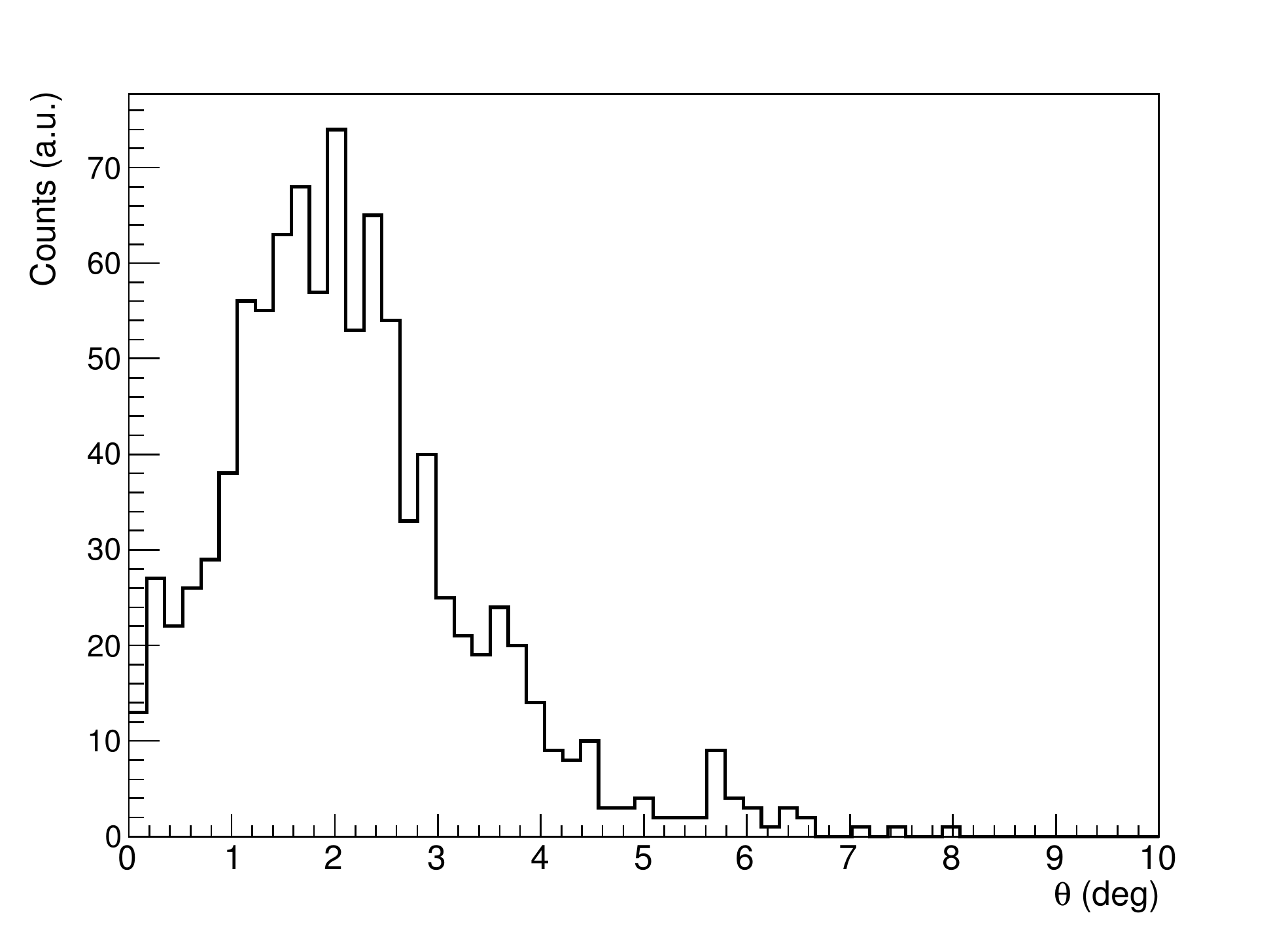} 

\caption{Left panel: an (x,y) view of the calorimeter where each square represents one crystals. The color scale indicates, in arbitrary units, the number of times a crystal is fired for $\chi-e^-$ scattering events. Right panel: Angular distribution of the e.m. showers  
with respect to the beam direction.}\label{fig:directionality} 
\end{figure}

The BDX detector response to scattered electrons and protons has been studied with the aforementioned GEANT4-based simulation code. For each combination $m_\chi - m_{A^\prime}$, the detection efficiency has been evaluated, for different selection cuts. 
For the  $\chi-e^-$ channel we considered the electrons scattered inside the detector volume with energy greater than 300 MeV. For this class of events, we evaluated the detection efficiency as a function of the energy deposited in a single crystal, which is the seed of the electromagnetic shower produced by the scattered electron inside the BDX calorimeter. Fig.~\ref{fig:detection_efficiency} (top panel) shows the integrated efficiency extracted for $m_\chi=30$ MeV and $m_{A^\prime}=90$ MeV. The blue curve represents the efficiency when we require the presence of a seed with an energy deposited greater than 300 MeV. The other curves indicate the efficiency when we also require the veto anticoincidences.  
In particular, the detection efficiency for $E_{seed}>$300 MeV with the Inner Veto ($\epsilon_{(E_{seed}>300MeV)}^{\overline{IV}}$) or the Outer Veto ($\epsilon_{(E_{seed}>300MeV)}^{\overline{OV}}$) in anti-coincidence is about 14\% and 31\%, respectively. In the $m_\chi$ - $m_{A^\prime}$ parameter space covered by BDX, the two efficiencies  vary in the following range: $\epsilon_{(E_{seed}>300MeV)}^{\overline{IV}}$ $\sim$ 10 $\div$ 20 \% and  $\epsilon_{(E_{seed}>300MeV)}^{\overline{OV}}$ $\sim$ 20 $\div$ 35 \%. 
It is worth noticing that the $\chi-e^-$ scattering generates events in the calorimeter with a clear topology. For example, Fig.~\ref{fig:directionality} shows two distinct features expected for these events.  Due to the natural collimation of the $\chi$ beam, most of the crystals involved in the e.m. showers are the central ones (left panel). Moreover, the e.m. showers are expected to be produced mostly along the beam direction. The right panel of Fig.~\ref{fig:directionality}  shows the angle between the beam axis and the direction of the e.m. shower, here defined as the direction formed by the energy weighted positions  $(x_{cl},y_{cl})$ of two clusters detected in two different modules of the calorimeter.
\\For the $\chi-p^-$ scattering events, the bottom panel of Fig.~\ref{fig:detection_efficiency}  shows the integrated detection efficiency as a function of the proton energy, for the same $m_\chi - m_{A^\prime}$ combination.  The efficiency is calculated for events where the proton recoil energy is greater than 20 MeV. In the $\chi-e^-$ events, where the e.m. showers produced inside the detector have a sizable probability to fire a veto detector, the anticoincidence conditions reduce the efficiency by a factors of 3/5. As expected, on the other side, the low recoil energy of the protons result in an efficiency which is only slightly reduced when we impose the absence of any signal in the veto detectors. 

\subsection{Beam related background} 
The beam-related background was studied with a multi-step approach, focusing first on the potential background for the electron recoil measurement and then on nuclear recoil. In the first case, we consider particles with energy of the order of 500 MeV or higher, while in the second case particles with energy greater than 10 MeV.
 
\begin{figure}[h!] 
\center
\includegraphics[width=0.95\textwidth]{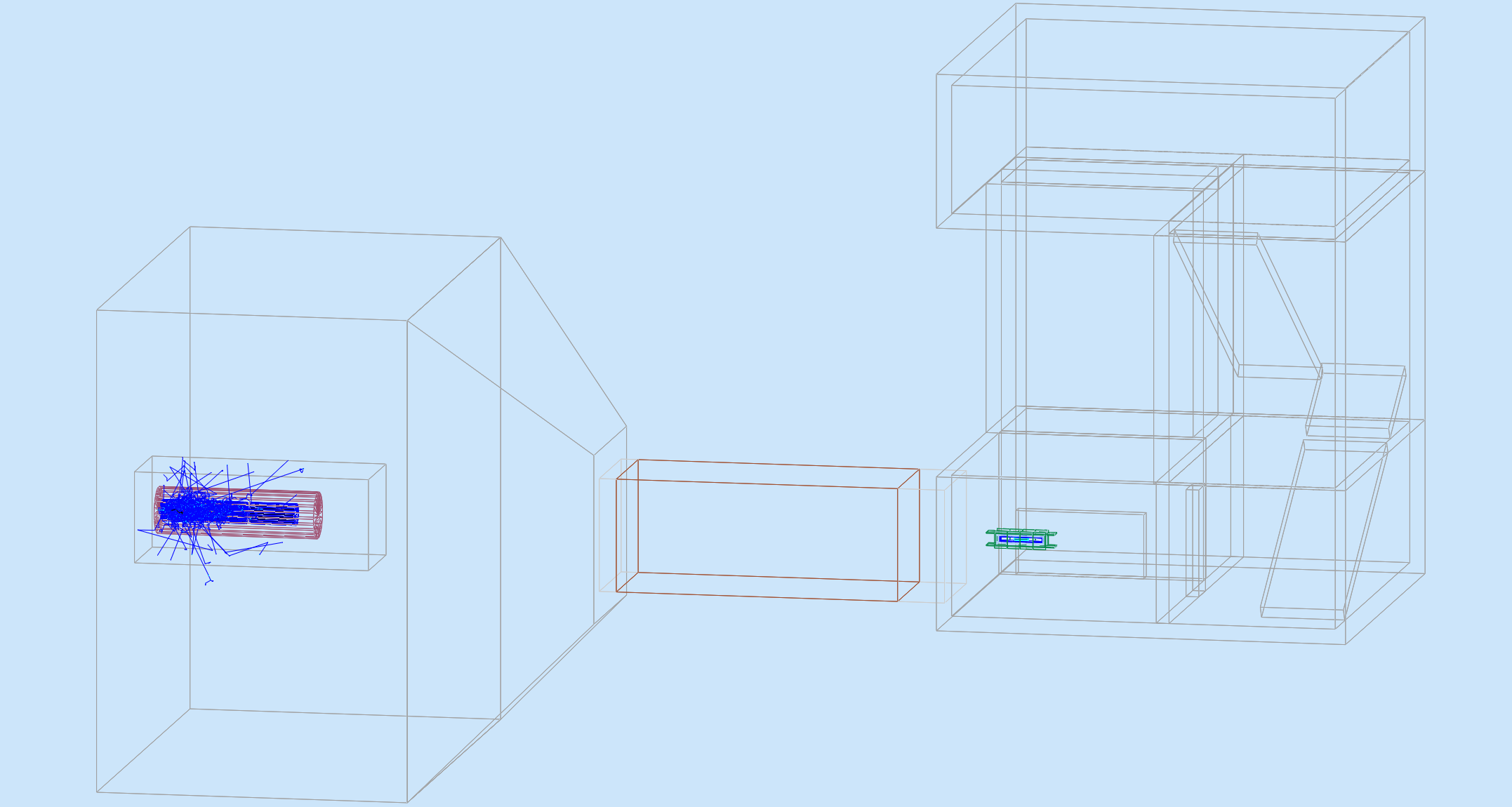} 
\vspace{0.1in}
\includegraphics[width=0.95\textwidth]{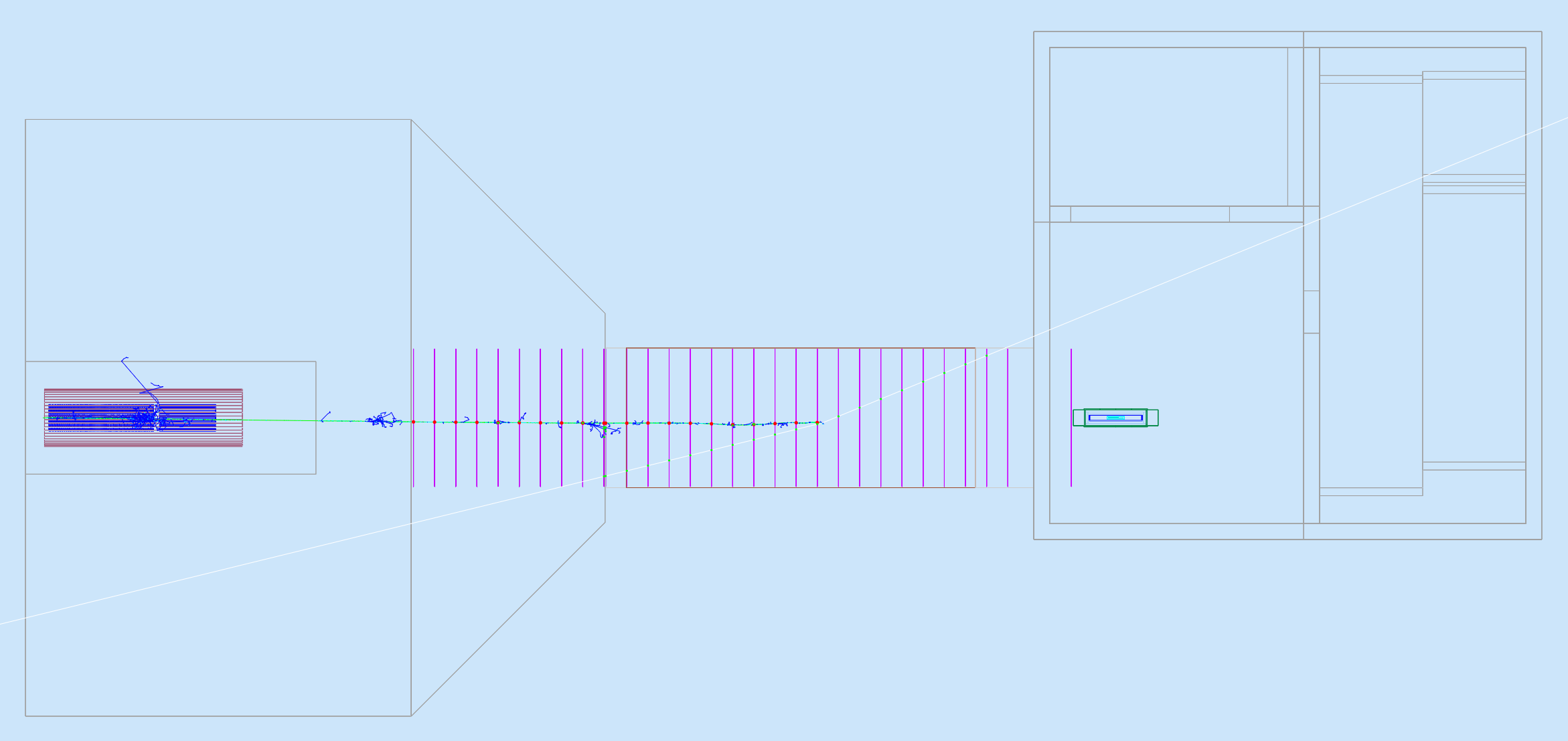} 
\caption{GEANT4 visualization of the electromagnetic shower
  produced by a 11 GeV electron in the dump (top) and of a 10 GeV
  muon (bottom). The vertical magenta lines indicate the position of the planes
where the secondary flux is sampled.}\label{fig:bdx_gemc_flux} 
\end{figure}
\begin{figure}[h!] 
\center
\includegraphics[width=0.48\textwidth]{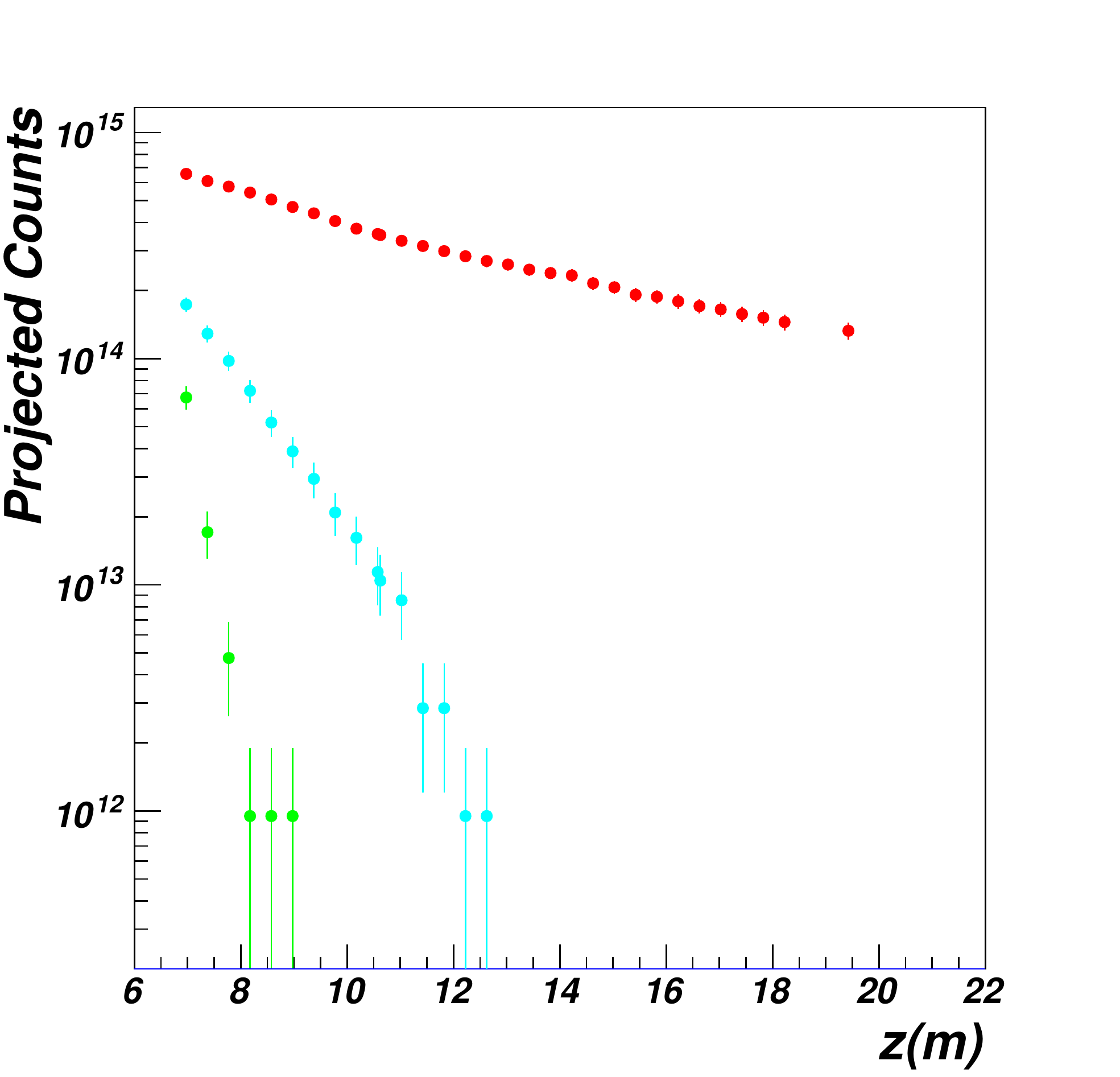} 
\includegraphics[width=0.48\textwidth]{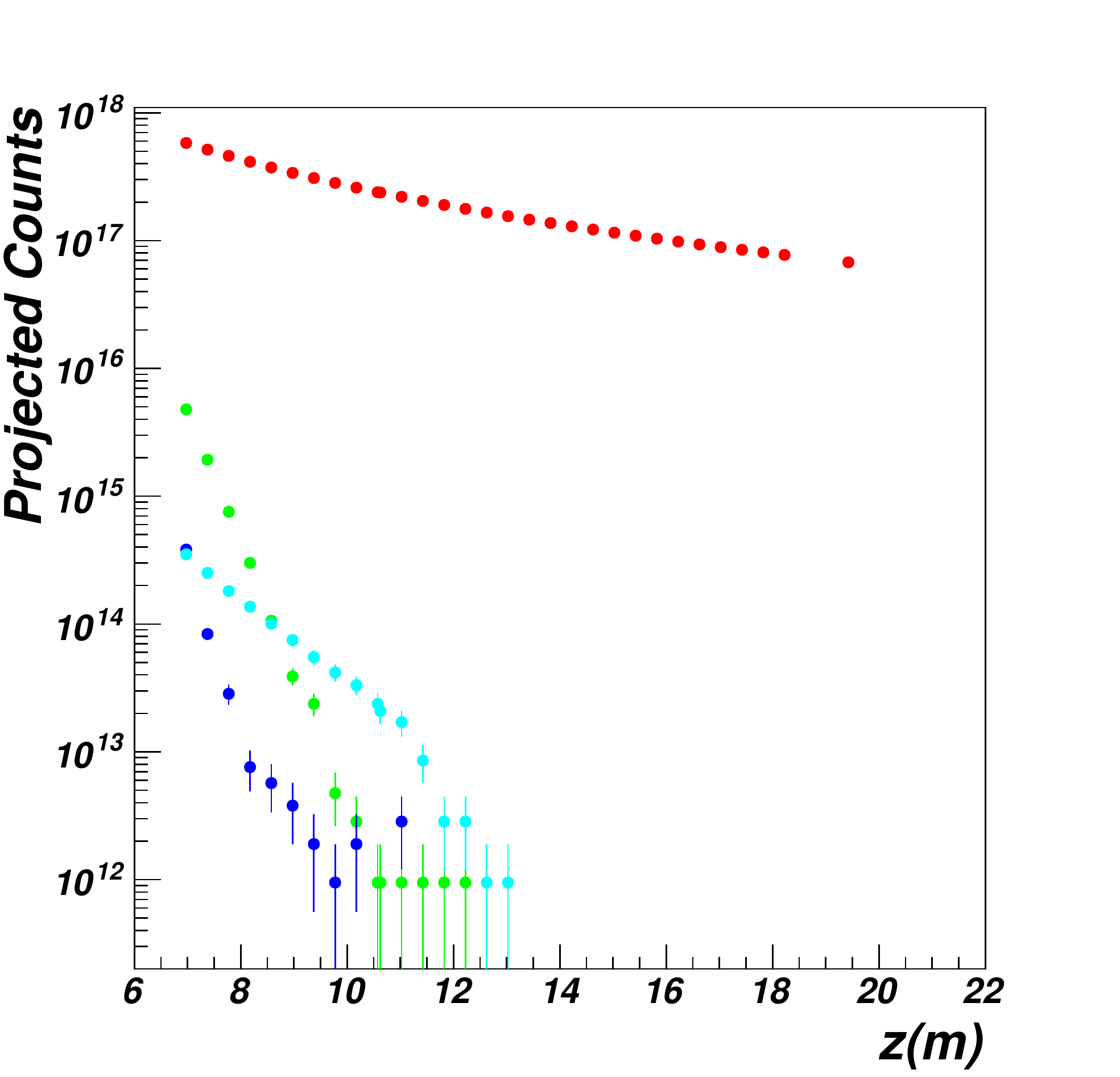} 
\caption{Number of secondary particles produced by the interaction of the 11 GeV electrons as a function of the distance from the upstream end of the dump.
The reported counts are normalized to $10^{22}$ EOT. The left and right plots show the results with a 500 MeV and 10 MeV threshold on the particle energy,
respectively. The different colors correspond to different particle types: neutrinos in red, muons in cyan, neutrons in green and photons in blue.
\label{fig:plot_gemc_profile}}
\end{figure}
\begin{figure}[h!] 
\center
\includegraphics[width=\textwidth]{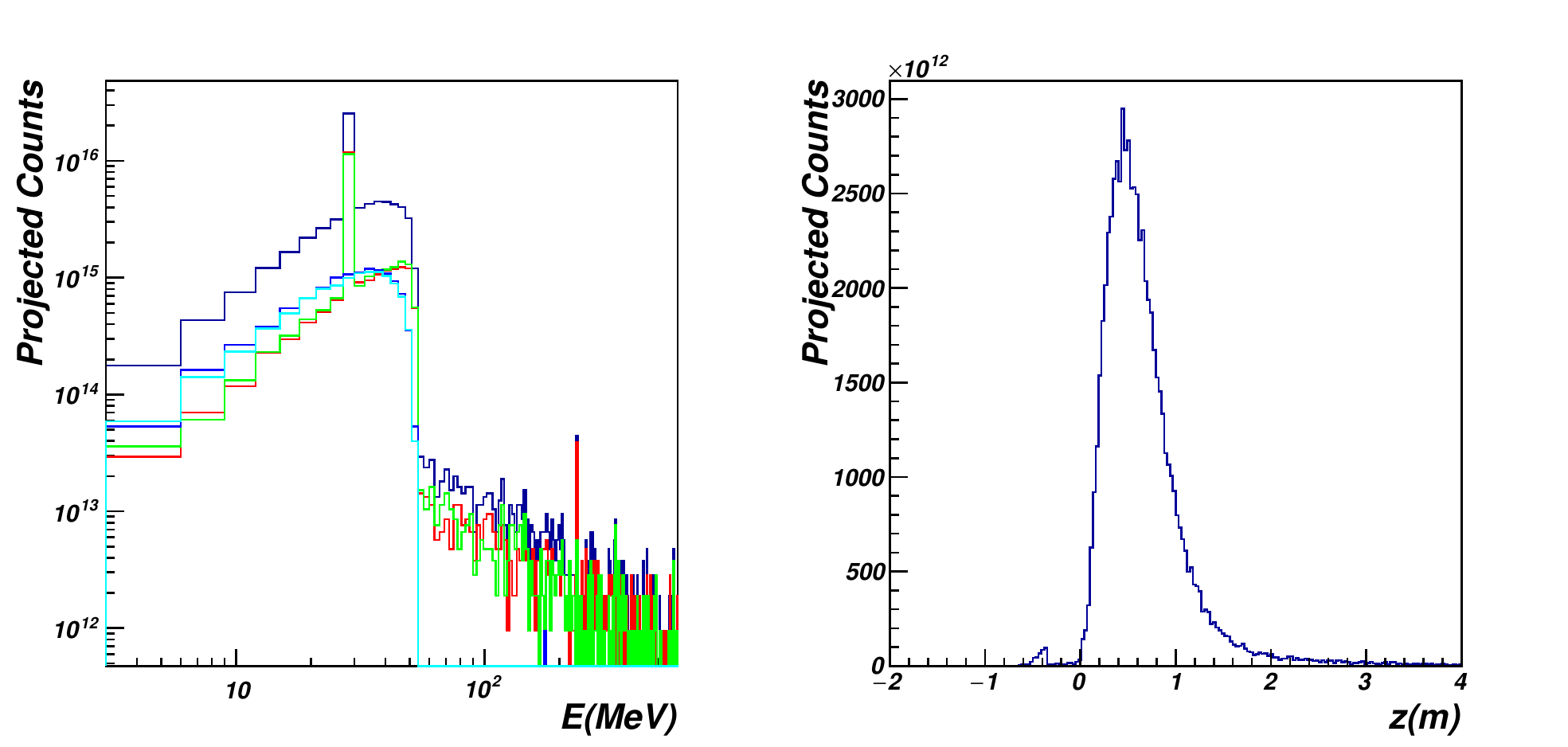} 
\caption{Left: energy spectrum of beam-related neutrinos reaching the
  detector; the dark-blue histogram shows the total spectrum while the
  red, green, blue and cyan histograms correspond to the $\nu_\mu$,
  $\bar\nu_\mu$, $\nu_e$ and $\bar\nu_e$ spectra,
  respectively. Right: the vertex of the decaying primary particles
  ($\mu$  and $\pi$) starting at the beam-dump location and extending for 1m
  downstream. The reported counts are normalized to $10^{22}$ EOT. }\label{fig:plot_gemc_neutrinos}  
\end{figure}

To evaluate these backgrounds, the interaction of the 11 GeV electron beam in the dump was simulated and the flux of secondaries was studied as a function of
the distance from the dump. For this purpose, the flux of particles was sampled over planes perpendicular to the beam direction as shown
in Fig.~\ref{fig:bdx_gemc_flux}. Since the simulation of the electromagnetic shower produced by the electron beam is very intensive
in terms of CPU usage, with this approach it is not possible to accumulate a statistics comparable with that expected for the present
experiment ($\sim 10^{22}$ EOT). A total of $\sim 0.5 \cdot 10^9$ 11 GeV EOT were simulated. The resulting particle fluxes for the 500 MeV and 10 MeV thresholds are shown in Fig.~\ref{fig:plot_gemc_profile}. In these plots, $Z=0$ corresponds to the upstream end of the aluminum dump, while the detector is located at $Z=20$ m.
Neutrinos are the dominant contribution to the overall particle flux. With the 500 MeV threshold, we also observe muons and neutrons that reach the
iron absorber, while no photons are found. On the other hand, with the lower threshold a non-zero photon flux is observed. With this statistics, the only
particles that are found reaching the detector area are $\nu$ and $\bar\nu$, predominantly from pion and muon decays at rest. This is shown by the energy
spectrum of the different neutrino species reported on the left panel of Fig.~\ref{fig:plot_gemc_neutrinos}. Pion decays at rest result in a monochromatic
set of $\nu_\mu$ and $\bar\nu_\mu$ of 30 MeV. Neutrinos from $\mu$ decay share a total energy of 105.1 MeV (M$_\mu$-M$_e$), which
results in a upper limit for the energy of the  $\nu_{\mu}$ and  $\nu_e$, and of their anti-particles.  The origin of these neutrinos inside the beam-dump are
shown in the right panel of Fig.~\ref{fig:plot_gemc_neutrinos}.

\subsubsection{Beam-related background for electron recoil}
To evaluate the expected background for the electron recoil measurement, we first evaluated the contribution from neutrinos. As shown by the left panel of
Fig.~\ref{fig:plot_gemc_profile}, the expected number of neutrinos is of the order of $\sim 1.7\cdot 10^{14}$. These are mostly $\nu_\mu$ and $\bar\nu_\mu$,
with only about 3-4\%  $\nu_e$ and $\bar\nu_e$. In this energy range, neutrinos interact mostly via charged current, with the dominant processes being quasi-elastic
scattering and resonance production on the nucleon, and cross sections of the order of $\sigma_{\nu N} \sim 10^{-38}$ cm$^{2}$~\cite{formaggio:2012}. These
neutrinos can therefore scatter in the BDX detector, producing a muon or an electron of similar energy that can be detected. Muons with energy of the order of
hundreds of MeV will lose energy via ionization leaving a ``track'' in the detector that can be distinguished from the electron recoil signal. Electrons will instead
induce an electromagnetic shower, resulting in the same signature of the $\chi$ interaction. The expected rate of such electrons produced by neutrino interaction
can be estimated as:
\be
N_{BG}^{\nu}(E_\nu\ge E_{min}) &=& N(\nu_e, E_\nu\ge E_{min}) \;\sigma_{\nu N}\; {\cal{N}}_{A} \; \rho_{CsI} \; L\; \epsilon_{e}
\ee
where $N(\nu_e,E_\nu\ge E_{min})$ is the number of $\nu_e$ and $\bar\nu_e$, with energy above a minimum value $E_{min}$ corresponding to the chosen electron-recoil threshold, that reach the detector, $\rho_{CsI}=4.51$ g/cm$^3$ is the CsI density, $L=260$ cm is the detector length and $\epsilon_e$ is the efficiency for the detection of the electromagnetic shower that we have assumed to be of the order of 20\%, similarly to the efficiency estimated for the $\chi$-induced electron recoil showers. The results for different values of $E_{min}$ are reported in Table~\ref{tab:nu-ng-counts} and indicate that the neutrino background for $10^{22}$ EOT is $\cal{O}$(10).

\begin{table}[tpb]
\centering
\begin{tabular}{|c|c|c|}
\hline
$E_{min}$ & $N(\nu_e, E_\nu\ge E_{min}) $ & $N_{BG}^{\nu}(E_\nu\ge E_{min})$\\
\hline
200 MeV & $7.7\cdot10^{12}$ & 11\\
\hline
250 MeV & $7.4\cdot10^{12}$ & 11\\
\hline
300 MeV & $7.0\cdot10^{12}$ & 10\\
\hline
350 MeV & $6.6\cdot10^{12}$ & 9\\
\hline
\end{tabular}
\caption{\label{tab:nu-bg-counts} Beam-related background for the electron-recoil measurement due to neutrinos. The reported counts are for $10^{22}$ EOT. $N(\nu_e, E_\nu\ge E_{min}) $ and $N_{BG}^{\nu}(E_\nu\ge E_{min})$ correspond to the number of $\nu_e$ and $\bar\nu_e$ that reach and interact in the detector, respectively. In estimating the number of neutrino interactions, we conservatively used a cross section of $10^{-38}$ cm$^2$, independently of the minimum energy.}\label{tab:nu-ng-counts} 
\end{table}

While this is an irreducible background for the experimental configuration we are considering, additional simulation studies we performed indicate that the neutrino production rate would be strongly suppressed if the dump was made by high Z material instead of aluminum: with the same dump layout but replacing aluminum with copper or tungsten for instance, the estimated neutrino background would be a factor 5 or 20 lower, respectively, i.e. of $\cal{O}$(2) or $\cal{O}$(1). The replacement of the dump material may be therefore considered for possible upgrades of the experimental setup.

\begin{figure}[h!] 
\center
\includegraphics[width=0.48\textwidth]{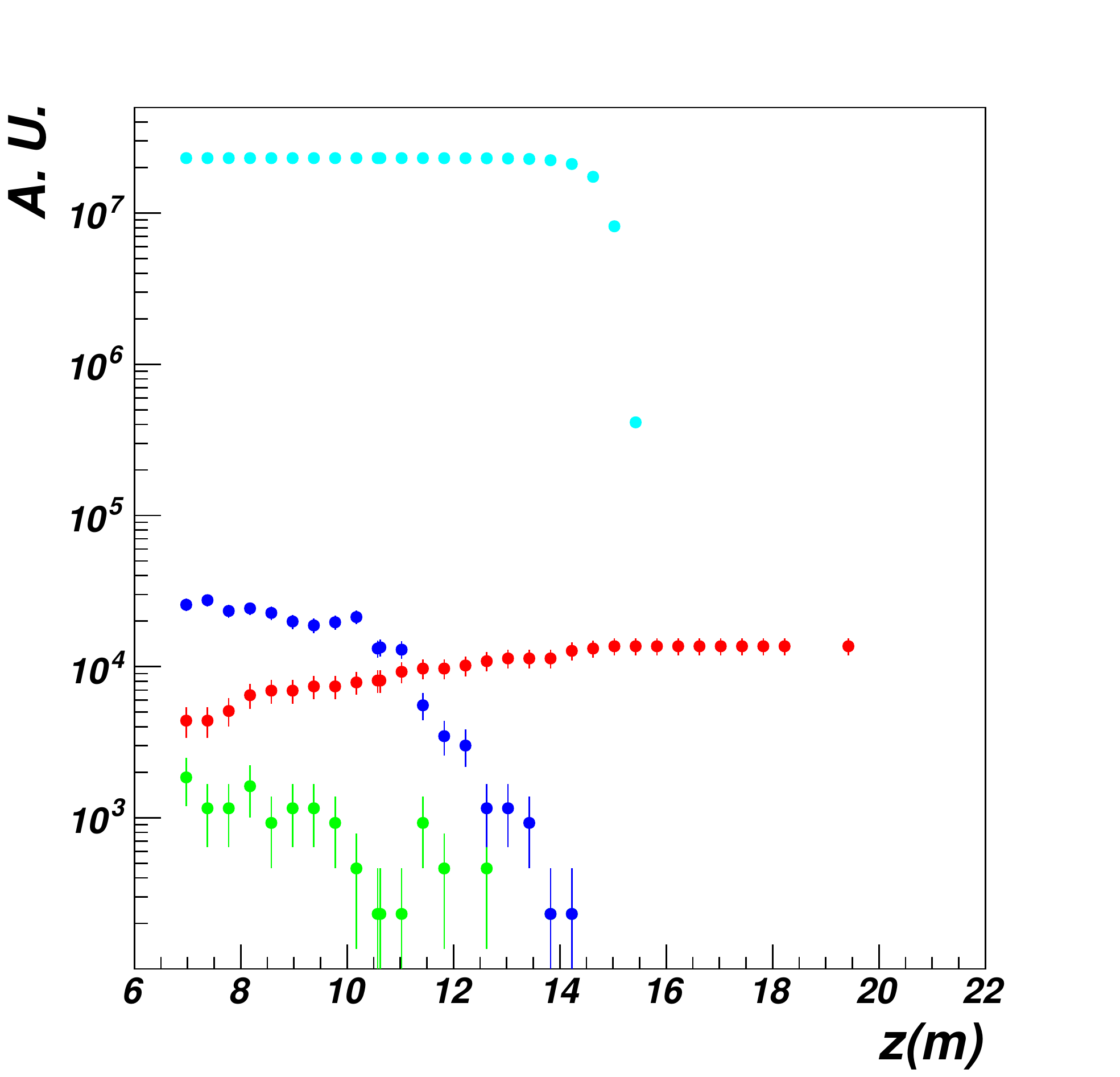} 
\includegraphics[width=0.48\textwidth]{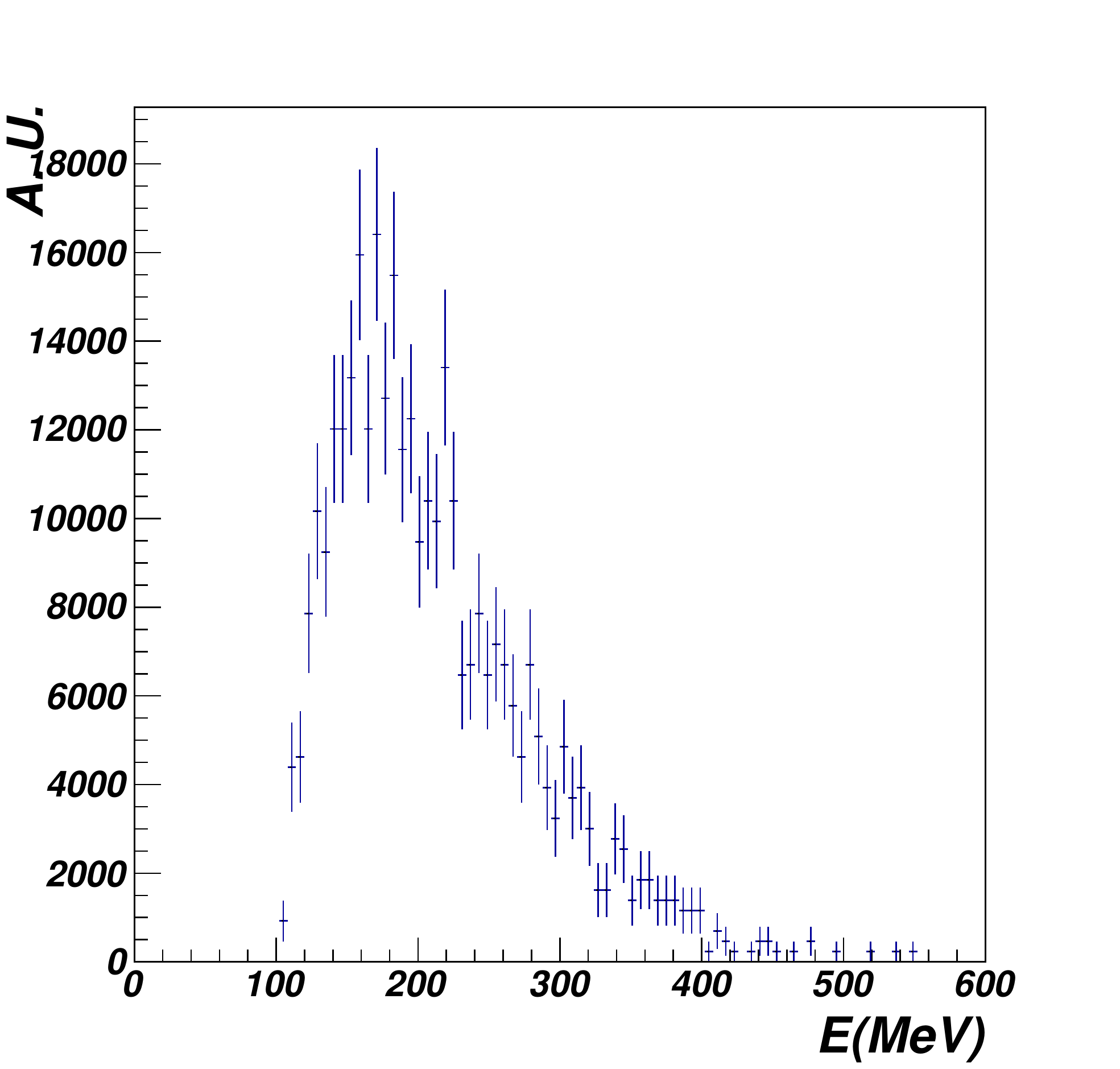} 
\caption{Left: neutrino (red), neutron (green), gamma (blue) and muon (cyan)
fluxes as a function of the distance from the dump resulting from the
simulations of 10 GeV muons. Here only particles with kinetic energy greater than 500 MeV are considered. Note
that the BDX detector would be located at 20 m from the dump. Right: muon energy spectrum at $Z\sim16$ m}\label{fig:plot_gemc_muons}  
\end{figure}
Additional background that may affect the electron-recoil measurement, may come from energetic muons or neutrons. As shown by the left panel of Fig.~\ref{fig:plot_gemc_profile}, the number of neutrons decreases very rapidly as the distance from the dump increases and, performing an exponential extrapolation of the projected counts, we estimated that no neutrons with energy of the order of hundreds of MeV would reach the detector. Muon counts decrease more slowly, consistently with the high penetetring power of these particles, and range out at $Z\sim12$ m.  To further check whether high energy muons or their secondaries can reach the detector area, 10 GeV muons, that could be produced via pair production by high energy photons, were generated at the center of the dump and the flux of particles produced was sampled as a function of the distance from the dump as in the original simulations. The resulting particle profiles shown in
Fig.~\ref{fig:plot_gemc_muons} indicate that the muon range is of the order of 16 m, i.e. significantly smaller than the distance between the detector and the dump. Close to the maximum range, the energy of the muons is strongly degraded being less than 300 MeV in average at $Z\sim16$ m. The flux of secondaries produced by the 10 GeV muons, with the exception of neutrinos, dies within a similar distance. 

\subsubsection{Beam-related background for nuclear recoil}
Potential contributions to the nuclear recoil measurement are due to particles with kinetic energy greater than the measurement threshold of 10 MeV. As shown by the right panel of Fig.~\ref{fig:plot_gemc_profile}, the most abundant particles are also in this case neutrinos, with total projected counts of $\sim 7\; 10^{16}$ for $10^{22}$ EOT and approximately equal counts for the different neutrino species. Neutrinos with energy in the range 10-100 MeV mainly interact by CC interaction  ($\bar\nu A \to e^+ A' $) with a cross section of about $\sigma_{\bar\nu A}\sim10^{-41}$ cm$^2$. Assuming a detection efficiency of $\epsilon_{e^+}=5\%$ with 10 MeV threshold, the corresponding background rate can be estimated as:
\be
N_{BG}^{\nu}(E_\nu = 10-100 MeV) = N(\nu_e, E_\nu = 10-100 MeV)\;\sigma_{\bar\nu A}\; {\cal{N}}_{A} 1/A \; \rho_{CsI} \; L\; \epsilon_{e^+}\sim 0.25,
\ee
which is negligible with respect to the background counts expected from higher energy neutrinos estimated in the previous section.

\begin{figure}[h!] 
\center
\includegraphics[width=0.5\textwidth]{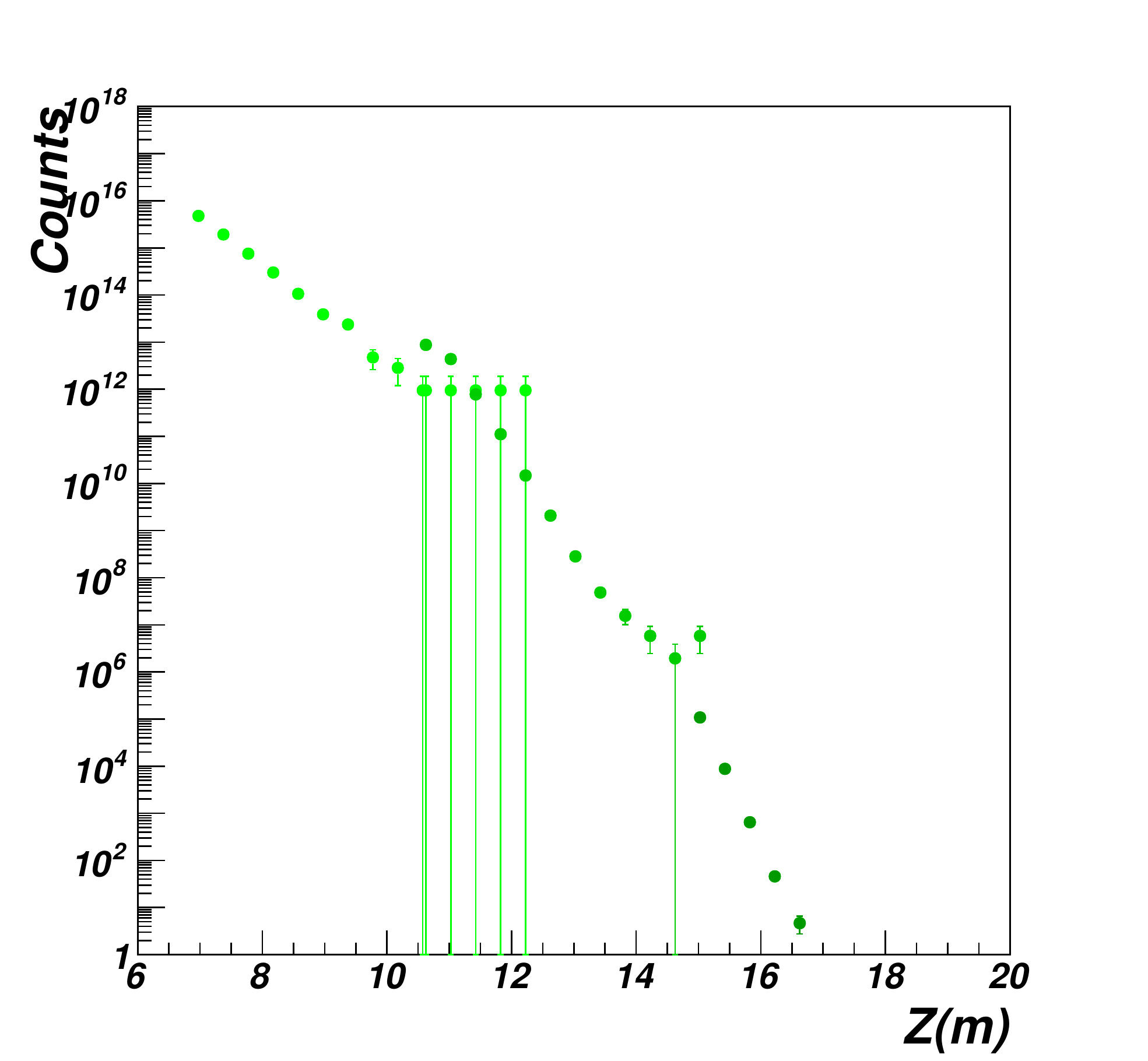} 
\caption{Projected neutron counts as a function of the distance from the dump upstream end. See the text for details on the simulation procedure. The estimated counts in the detector area, i.e. $Z=20$ m, are negligible with respect to other contributions.}\label{fig:plot_gemc_neutrons}  
\end{figure}

Other contributions could arise from photons and neutrons that may propagate through the iron absorber. Based on the simulations of the 11 GeV electron simulation, we found that a significant fraction of the photons and neutrons that are found at the largest Z values are muon secondaries. To verify whether these could reach the detector area, a dedicated simulation study was performed generating muons at $Z=16$ m with energy corresponding to the average of the spectrum shown in the right panel of Fig.~\ref{fig:plot_gemc_muons}: for 1 M muons generated, we observed no photons and neutrons exiting the iron absorber. Another estimate can be performed extrapolating the particle counts shown in the right panel of Fig.~\ref{fig:plot_gemc_profile}. For photons, the exponential extrapolation indicate that the expected counts at the detector position are less than $10^{-5}$ for $10^{22}$ EOT, i.e. negligible with respect to other contributions. We should note that, performing this extrapolation, we are not considering the degradation of particle energies as they propagate in the absorber. To take this into account for neutrons, we peformed additional simulation studies with the following procedure. We selected the largest $Z$ position where significant neutron statistics was found and we simulated mono-chromatic neutrons, with energy corresponding to the maximum kinetic energy previosuly observed, originating at the same position and traveling along the $Z$ axis. This procedure was repeated twice, allowing us to study the neutron propagation probability to the detector area. The result, shown in Fig.~\ref{fig:plot_gemc_neutrons}, indicate that the expected neutron counts in the detector area are negligible.

\subsection{Beam unrelated background}\label{sec:beam_unrelated_bck}
Beam-unrelated background is mainly due to cosmic neutrons, cosmic muons and their decay products, including  rare decays of muons producing gamma's between the passive shield and the active veto. Both direct cosmic flow (muon and neutron) and secondaries particles (muon, neutron and gamma) contribute to the beam-unrelated background count rate in the detector.

\subsubsection{Cosmic background estimate}
The cosmogenic  background rates expected in the BDX experiment have been evaluated by extrapolating the results obtained with the BDX prototype in Catania LNS measurements. 
The similar experimental set-up (including overburden) as well as a full prototype that incorporates all the elements of the BDX detector (active veto's, lead shielding, BaBar CsI(Tl) crystals) provide a solid base for a realistic, although conservative, estimate of the expected rates.
Details of the experimental conditions and data analysis are reported in Appendix~\ref{Section:BDX-protoype}. 
The extrapolation has been performed by scaling the experimental rates of a single crystal to the 800 crystals comprising the full detector.
This is certainly an upper limit on the expected rates since this assumes  crystal-to-crystal fully uncorrelated counts, which overestimates the case for
$\chi$-e scattering.  Tables~\ref{tab:bg-rates} and \ref{tab:bg-rates-lowE} report some of  the rates at different threshold energies for high
energy and low energy, which are relevant for $\chi$-e scattering and  $\chi$-nucleon, respectively. In the tables there is also the projection of the counts integrated over the expected beam-on time. 
For energy thresholds higher enough, between 300-350 MeV, the number of cosmogenic background counts reduces to zero. Thus, by choosing the appropriate energy threshold for the $\chi$-e scattering channel we could consider zero background.
However, in order to be conservative, in the next Section we will determine the sensitivity and the reach of the proposed BDX experiment considering $N^{cosmic}$ = 3.
\begin{table}[tpb]
\centering
\begin{tabular}{|c|c|c|}
\hline
Energy threshold & Extrapolated rate  & Projected counts (285 days) \\
\hline
200 MeV & $(3.0\pm1.2)\cdot10^{-5}$ Hz & $(740 \pm 300)$\\
\hline
250 MeV & $(2.3\pm1.0)\cdot10^{-6}$ Hz & $(57 \pm 25)$ \\
\hline
300 MeV & $(1.9\pm0.9)\cdot10^{-7}$ Hz & $(4.7 \pm 2.2)$ \\
\hline
350 MeV & $(1.5\pm0.9)\cdot10^{-9}$ Hz & $(0.037 \pm 0.022)$ \\
\hline
\end{tabular}
\caption{\label{tab:bg-rates} The expected cosmogenic background rates on the BDX detector when the IV anti-coincidence is requested.}
\end{table}
\
\begin{table}[tpb]
\centering
\begin{tabular}{|c|c|c|}
\hline
Energy threshold & Extrapolated rate  & Projected counts (285 days) \\
\hline
5 MeV & $(37 \pm 1)\cdot10^{1}$ Hz  & $(9.1 \pm 0.2)\cdot10^{9}$ \\
\hline
10 MeV & $(1.5 \pm 0.1)$ Hz  & $(3.7 \pm 0.2)\cdot10^{7}$ \\
\hline
20 MeV & $(0.5 \pm 0.1)$ Hz  & $(1.2 \pm 0.2)\cdot10^{7}$ \\
\hline
50 MeV & $(0.3 \pm 0.1)$ Hz  & $(0.7 \pm 0.2)\cdot10^{7}$ \\
\hline
\hline
\end{tabular}
\caption{\label{tab:bg-rates-lowE} The expected cosmogenic background rates on the BDX detector when the  IV and the OV  anti-coincidence is requested.}
\end{table}

\subsubsection{Background reduction strategies}

\paragraph{Time-Correlation}
Beam-unrelated background can be, in principle,  rejected by requiring a time coincidence between the RF signal and the event recorded by the detector. 
The background reduction factor, $R$, can be expressed as the ratio between the time coincidence window width (3$\sigma_T$)  and $\Delta$T:
\be
R=\frac{3 \sigma_T}{\Delta T}
\ee

Considering the time response of CsI(Tl) crystals,
 the almost-CW structure of the CEBAF beam does not allow to take full  advantage of this technique. In fact the bunch separation expected with the 12 GeV operations is $\Delta$T=4.0 ns to be compared with the measured time  resolution of BDX detector of $\sigma_T\sim$6 ns for 30 MeV deposited energy. As already mentioned, a detector based on inorganic scintillators with such a good time resolution would be prohibitive in term of costs (at least for a first generation of beam dump experiments) while the use of faster organic scintillator would lead to a significant increase of the detector footprint (a factor of 5 in length) impacting on the size of the new facility.
Some dense crystals, such has the BaF$_2$ do have a fast scintillation component in the range of $\sim1$ ns although with lower light output ($\sim 10^3 \gamma/MeV$) and peaked in UV region ($\sim$220 nm) but further studies are requested to demonstrate the feasibility.
If we stick to the current detector design, a significant background reduction (a factor of 5-10) would only be achieved by running the CEBAF machine in  a different  mode with a beam macro-structure of 1$\mu$s and keeping the bunch micro-structure of 250 Mhz, preserving the average current of 40-50 $\mu$A.
While we are  investigating the technical feasibility of this option with the Accelerator Division, we are aware that this would probably require a dedicated beam time preventing  BDX to run  parasitically to the already approved Hall-A physics program. If technically possible this would represent a valuable alternative for a dedicated, cosmogenic background-free second-generation beam-dump experiment at Jefferson Lab.

In the following, we are not considering any background rejection due to the timing cuts.

\paragraph{Directionality} 
Beside time-correlation, directionality could help in reducing the cosmogenic background. In fact the angular distributions of  cosmic muons 
(and their decay products)  and cosmic neutrons are peaked around the vertical (the angular distributions are proportional to $\cos^2{\theta}$ and   $\cos^3{\theta}$
respectively) while
the $\chi$s are expected to follow the beam direction.  Furthermore, MC simulations indicate that an electromagnetic shower produced inside the detector, pointing
downstream, such as the one produced in
the $\chi$-e scattering, has a peculiar shape, direction and energy distribution easily distinguishable from a cosmic hit (see e.g. Fig.~\ref{fig:directionality}).
Thus, a thorough analysis of background events collected during the experiment will certainly allow to reject some of the cosmic background.
For example, this capability is heavily exploited by the DRIFT-BDX detector to enhance sensitivy, as described in Appendix\,\ref{sec:drift}.
However in the following, keeping a conservative attitude, we are not considering any background suppression associated with directionality using the calorimeter.


\clearpage
\section{Expected results} \label{sec:fullexp} 
In this Section we present the expected reach of the BDX experiment. Results are reported as upper limit on exclusion plot.
The region above the lines is excluded since the model, with the chosen parameters, predicts a number of counts larger than the observed. Each line corresponds to a model-predicted  yield comparable or exceeding   the expected background.
This is true in case of null results. If any excess is observed, a thorough statistical analysis will be necessary to claim  a positive result. Leaving this discussion to the future, in the last section we just report a list of systematic checks that the BDX experiment will be able to performe to  corroborate  any possible findings.
 
\subsection{BDX expected reach}
Here we  consider an experimental set-up that takes advantage of the maximum beam  current available at JLab ($\sim65  \mu$A) compatible with
the Hall-A beam-dump  power limit ($\sim$ MW), at the maximum available energy (11 GeV) for a full parasitic run that will collect 10$^{22}$ EOT.
This would  correspond  to a total time of 2.5 10$^7$ s (285 calendar days or 41 weeks).

\subsubsection{Expected signal and measured background}
In order to place limits, we must determine the uncertainties in the background yields. The cosmic-ray rate will be determined when the beam is off, both
during accelerator running periods (assuming a nominal 50\% delivery) and also between running periods when the accelerator is off. We assume that the experiment
will run over a period of about 4 years (208 weeks of calendar time), resulting in 167 weeks of data when the beam is off and 41 weeks with beam on. The
beam off data will be used to determine the average cosmic-ray background, $N_{BG}^{cosmic}$, and subtract it from the beam on data. The beam related backgrounds, primarily due to
$\nu_e$ interactions, will be calculated using MC and normalized to the measured rate of $\nu_\mu$ interactions, which are almost 30 times higher.
The uncertainty in the calculation of the beam background, $\sigma_{BG}^{beam}$, will dominate the limits that can be set. The estimated excess of events above background is given by
\be
N_{excess} & = & N^{beam} + N^{cosmic} - N_{BG}^{beam} - N_{BG}^{cosmic} \label{eq:excess} \\
\sigma_{excess} & \simeq & \sqrt{N^{beam} + N^{cosmic} + (\sigma_{BG}^{beam})^2 + \frac{1}{4}N^{cosmic} },
\ee
where the total number of registered counts, $N=N^{beam} + N^{cosmic}$, is separated out into its two components for ease of discussion. The estimated
beam and cosmic-ray backgrounds are denoted by consistent notation. The cosmic-ray background is estimated from beam off data assuming we have collected four times the
amount of beam on.

\begin{figure}[t!] 
\center
\includegraphics[width=12cm]{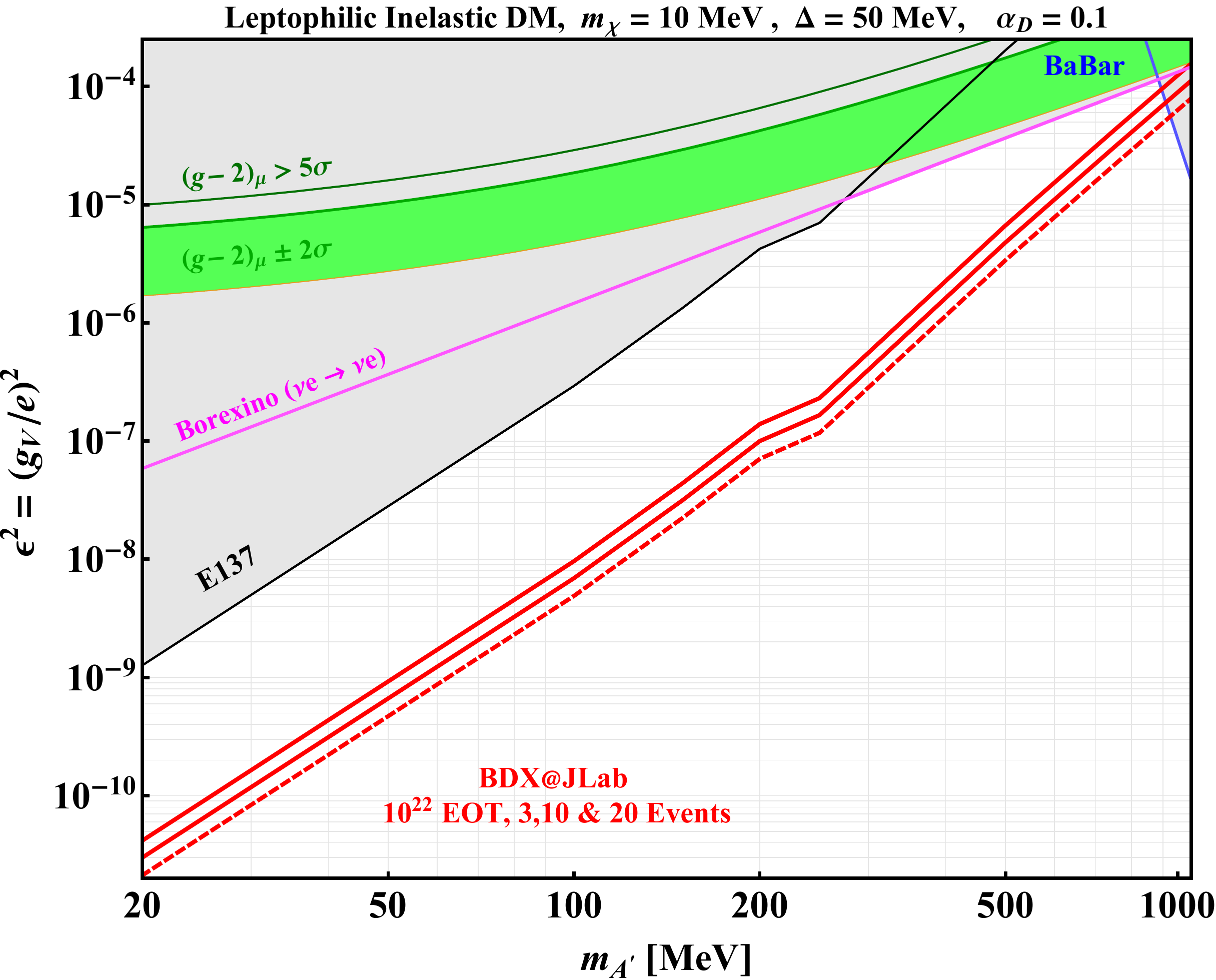} 
 \caption{Red curves show 3, 10, and 20 event BDX yield projections for leptophilic scenario.}
\label{fig:inelastic_leptophilic}
 \end{figure}
 
Curves on the exclusion plots report the predicted  counts by the models described in Sec.~\ref{sec:theory}. They all include the  efficiency  evaluated
in Sec.~\ref{sec:response} for the detection of an electromagnetic shower with energy deposited in the seed crystal greater than 300 MeV. The region on the plots
below the curves is excluded at the 2$\sigma$ level when
\be
N_{Model} & > & 2\, \sigma_{excess} \sim \,  11-17 {\rm \, counts},
\ee
where we have taken
\be
N^{beam} & = & 8 \\
N^{cosmic} & = & 3 \\
\sigma_{BG}^{beam} & \sim & (0.5-1)\times\;N^{beam}
\ee
The exclusion plots indicate levels of sensitivity between 3 and 20 counts, which span the expected range given above.
 
It is worth pointing out that the energy threshold for signal identification  (and beam-related background) as well as the number of counts related to the
cosmic background will be tuned during the experiment to the most favourable value that maximises the BG/SIGNAL ratio. In fact, with the proposed BDX
data acquisition scheme, the entire detector readout (full waveform of crystals and veto's)  will be triggered by any signals in any crystals corresponding
to an energy deposited of less than 1 MeV preserving  for the off-line analysis the most complete information about every collected event.

Thus the experiment will be detecting nucleon recoil events with thresholds below 10 MeV. However,
our present estimation of cosmic-ray backgrounds indicate that the sensitivity of this reaction is not competitive with the primary search for electron recoils.
Nevertheless, the data will be acquired and analysed to provide checks and complementary information to the main signal.
\begin{figure}[t!] 
\center
\hspace{-1.cm}\includegraphics[width=7.8cm]{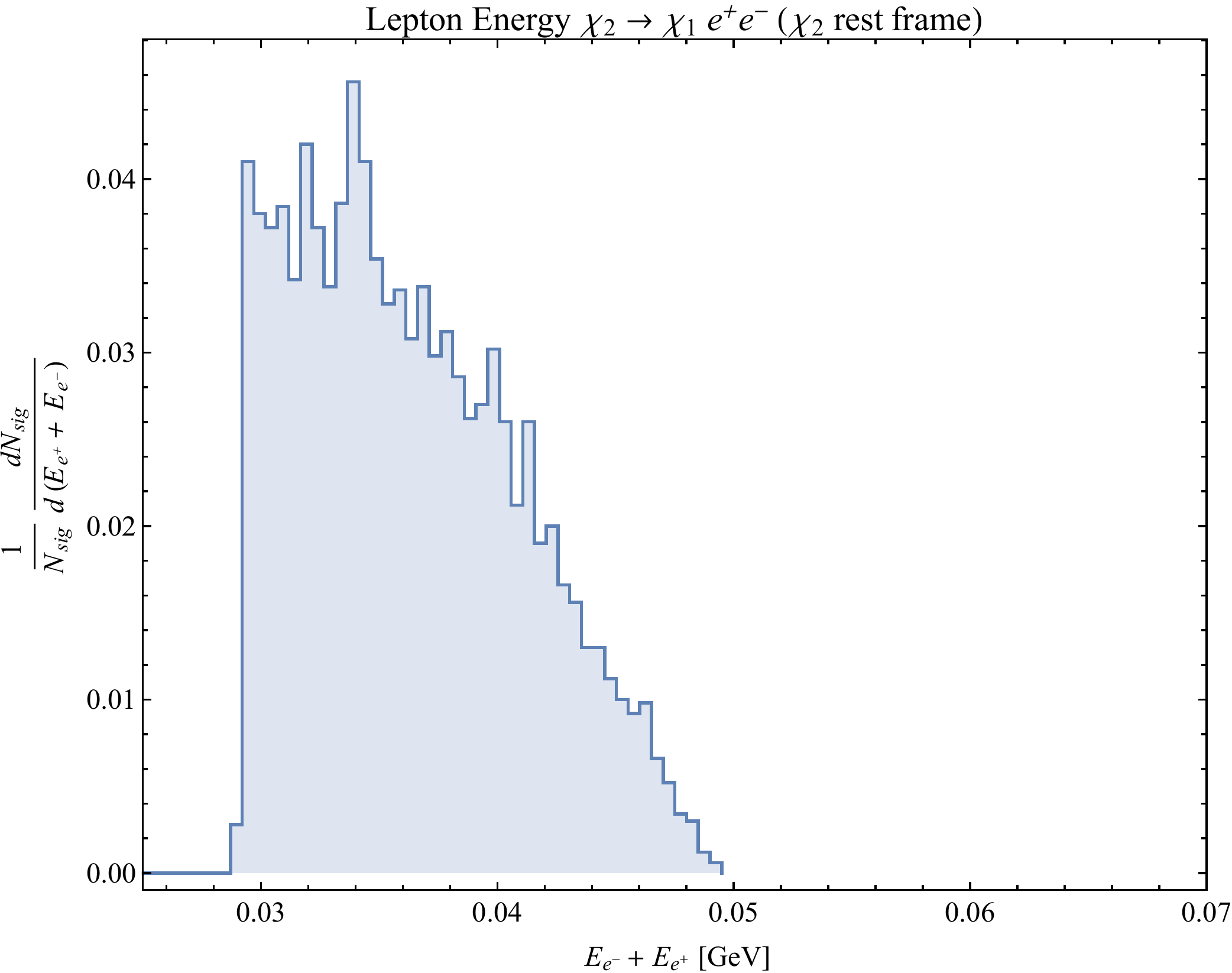} 
\hspace{-0.2cm}\includegraphics[width=7.8cm]{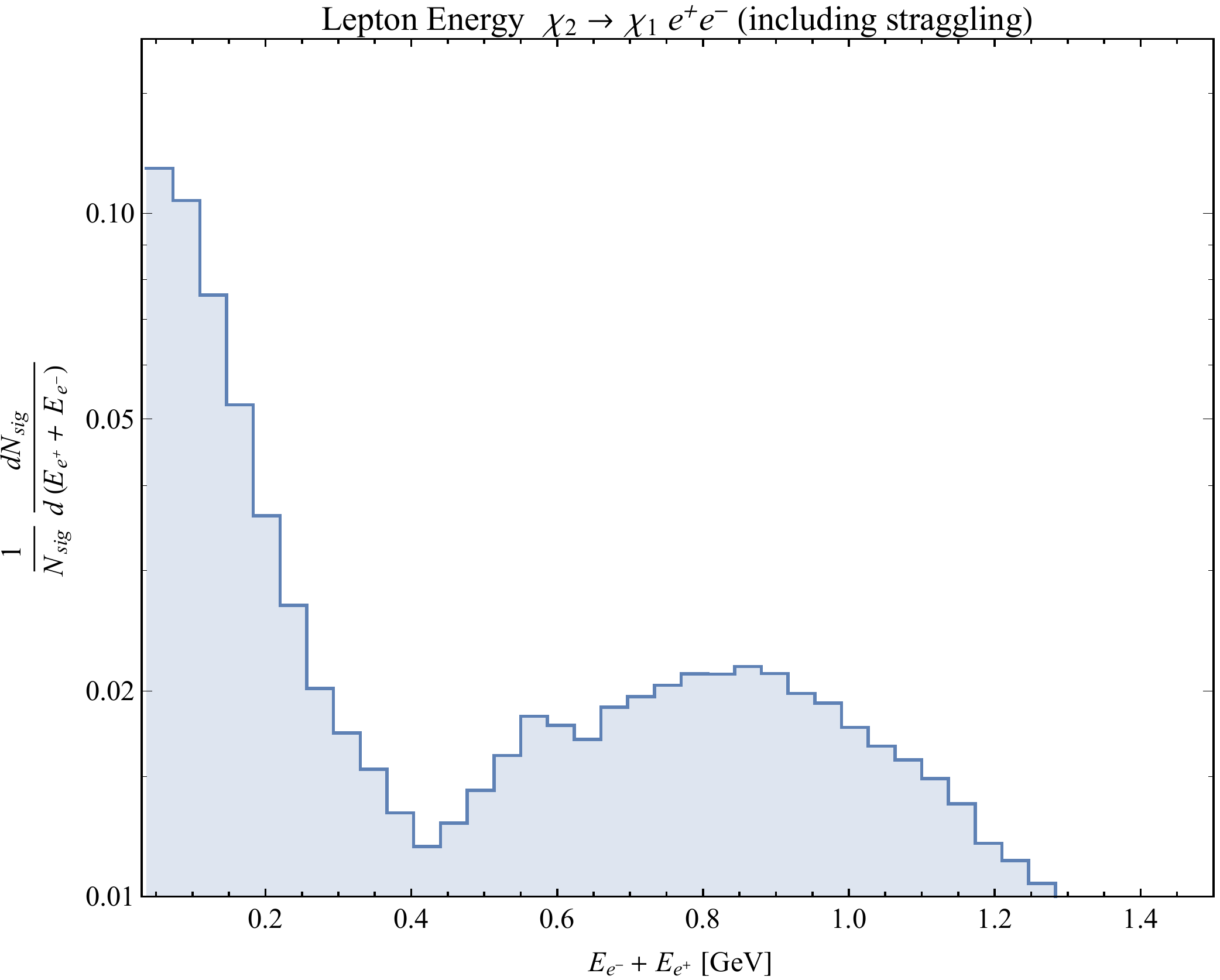} 
 \caption{Energy spectrum for $e^+e^-$ pair from the inelastic signal inside the BDX detector. For concreteness, here
 we assume $m_{\chi_1} = 10$ MeV, $m_{\chi_2} = 60$ MeV, $m_{A^\prime} = 100$  MeV, $\alpha_D = 0.1$ and $\epsilon = 8.3 \times 10^{-5}$,
 which corresponds to the BDX 10 event sensitivity for this parameter point in Fig.~\ref{fig:inelastic_leptophilic}.
}
\label{fig:IDMspectrum}
 \end{figure}

\subsubsection{The BDX reach}
In case of no positive observation, the accumulated data would provide very stringent limits on
the DM parameters space. 
\begin{figure}[t!] 
\center
\includegraphics[width=12cm]{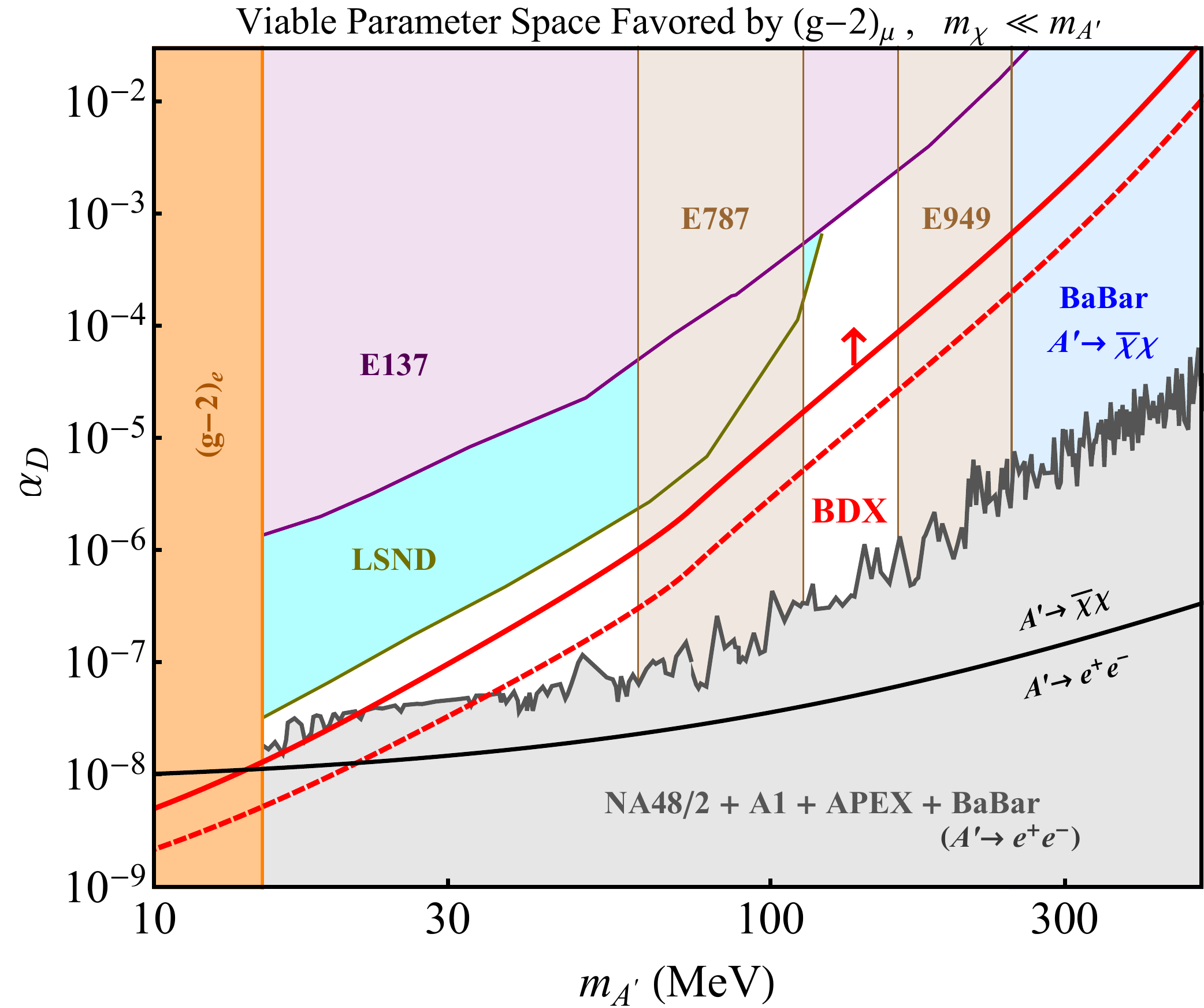} 
 \caption{ Red curves show 3 and 10 event BDX yield projections for a dark-photon ($A^\prime$) whose kinetic mixing parameter $\epsilon$ is  conservatively fixed to the {\it smallest} value that resolves the longstanding $(g-2)_\mu$ anomaly. 
 The black curve represents the ${\rm Br}(A^\prime \to e^+e^-) = {\rm Br}(A^\prime \to \chi \chi)$ contour. Testing 
 the remaining unshaded parameter space suffices to discover a dark photon responsible for the anomaly, or to decisively
 rule out such an explanation regardless of how the $A^\prime$ decays. The parameter space covered by BDX is the area above the red lines, as indicated by the arrow.
  \label{fig:gminus2}}
 \end{figure}
Figure~\ref{fig:inelastic_leptophilic} shows the BDX sensitivity to a leptophilic $U(1)_{e-\mu}$ gauge boson ($A^\prime$) coupled to a Majorana current of DM states charged under $e-\mu$ number with masses $\chi_1 =  10$ MeV and   $\chi_2 =  60$ MeV.  
In this scenario, the gauge boson is radiatively produced in electron-nucleus collisions in the
beam-dump and decays promptly to yield $\bar \chi_1 \chi_2$ pairs. The heavier $\chi_2$ state is unstable and short lived, so the flux of DM particles at the detector consists
entirely of $\chi_1$ states, which up-scatter off detector electrons, nucleons, and nuclei, thereby converting to $\chi_2$ states which decay  via $\chi_2 \to \chi_1 e^+e^-$ transitions and deposit significant ($\sim$ GeV) electromagnetic energy inside the detector. The energy for $e^+e^-$ pair from the inelastic signal inside the BDX detector is shown Fig.~\ref{fig:IDMspectrum}.
\begin{figure}[t!] 
\center
\includegraphics[width=10.cm]{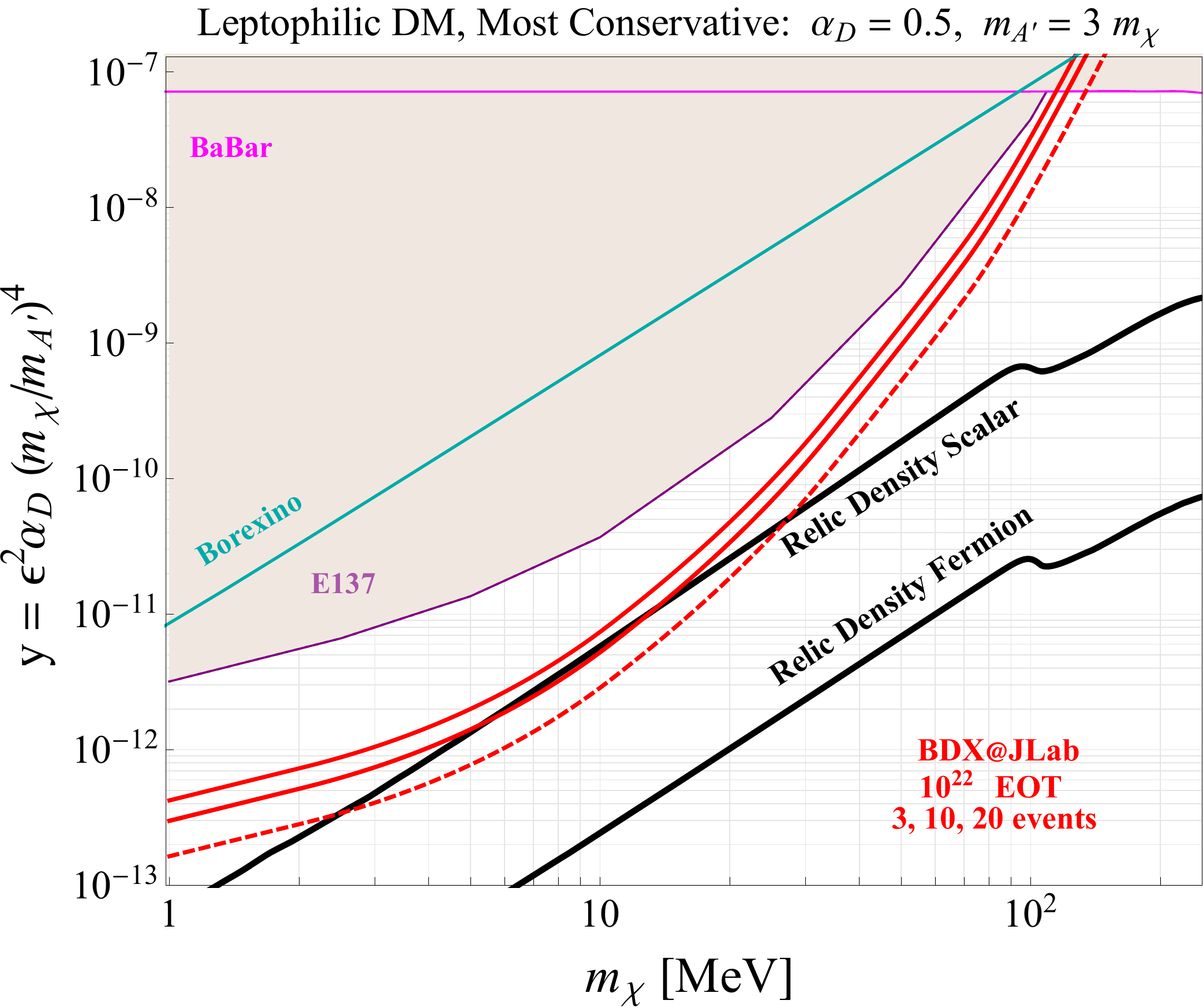} 
\includegraphics[width=10.cm]{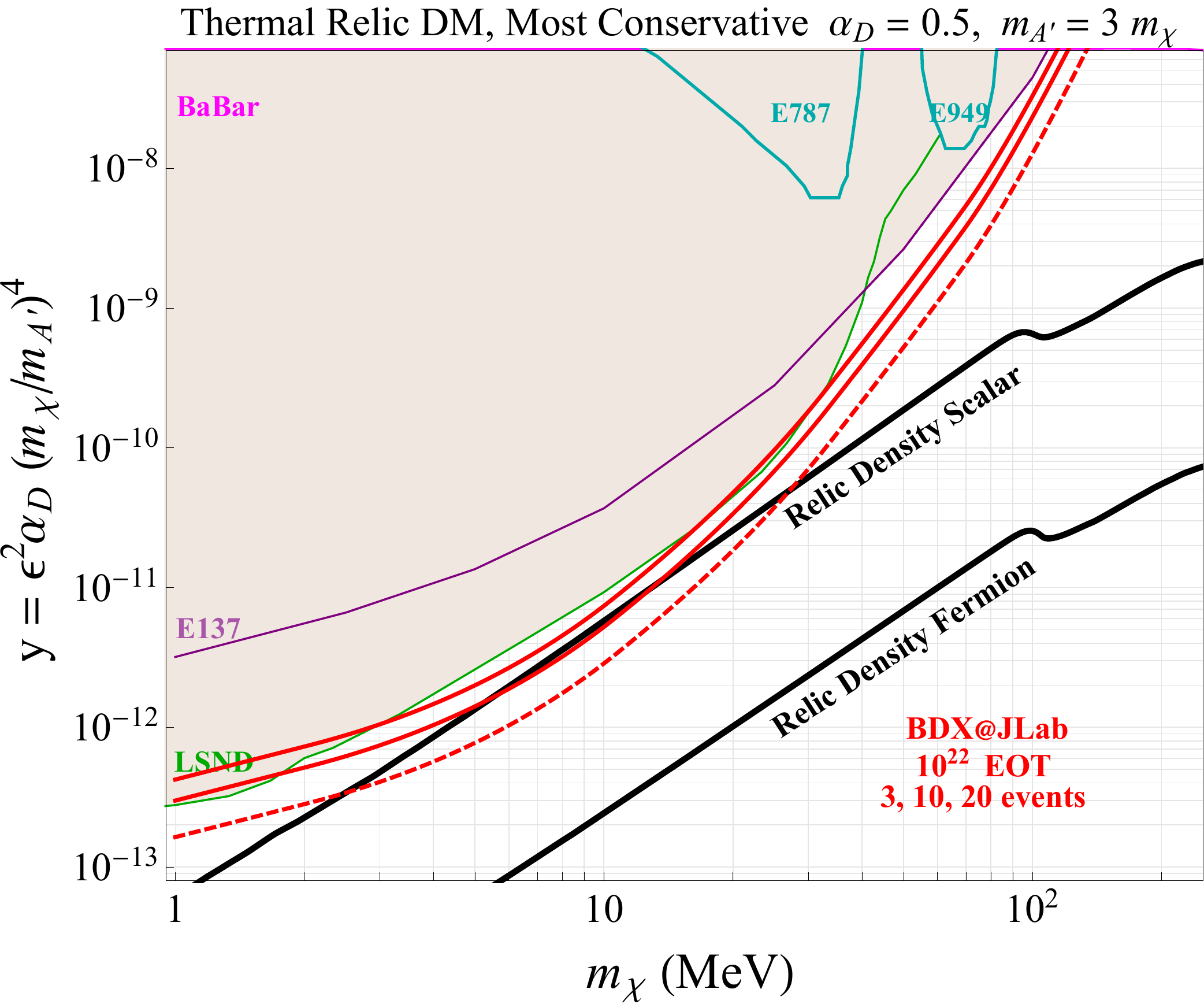} 
 \caption{ Red curves show 3, 10, and 20 event for BDX yield projections for electron scattering with a 300 MeV energy threshold for thermal relic DM in two representative scenarios. Top: thermal relic DM coupled to a  leptophilic $U(1)_{e-\mu}$ gauge boson ($A^\prime$). Bottom: here the  $A^\prime$ is a kinetically mixed dark photon coupled to the electromagnetic current. Here the thermal target --- where the model predicts the correct observed DM abundance --- is shown in solid black.}
\label{fig:ThermalTargets}
 \end{figure}

We now focus on the parameter space that can explain the discrepant $(g-2)$ of the muon. Figure~\ref{fig:gminus2} shows  the BDX projection for $10^{22}$ EOT in terms of  the $A^\prime$-DM coupling $\alpha_D$ as a function of  dark-photon mass in the $m_\chi \ll m_{A^\prime}$ limit. We also show the excluded parameter space from both visibly decaying $(A^\prime \to e^+e^-)$ and  invisibly decaying $(A^\prime \to \chi \chi)$ constraints;  for sufficiently small $\alpha_D$, the visible decays dominate so the constraint is independent of $\alpha_D$. Note that the 
parameter space for a purely visibly decaying dark photon is completely ruled out, so the only {\it viable} explanation is in the predominantly invisibly decaying region (white, unshaded);  everything below is excluded.

Figure~\ref{fig:ThermalTargets} shows the BDX yield projections for electron scattering with a 300 MeV energy threshold for thermal relic DM in two representative scenarios. 
For both plots, the relic target is proportional to the variable
on the $y$ axis, so it does not change as the assumption on $\alpha_D$ varies, but some experimental bounds do shift since they constrain
a different combination of couplings. Nonetheless, for $\alpha_D \sim \mathcal{O}(1)$, all the gaps in the parameter space are revealed; making $\alpha_D$
smaller only moves the projections down further, so the most conservative choice corresponds to a large, perturbative $\alpha_D = 0.5$ ; larger
values require UV completions with additional field content to avoid strong coupling near the GeV scale \cite{Davoudiasl:2015hxa}.
In the top plot, we show the projected sensitivity to thermal relic DM coupled to a  leptophilic $U(1)_{e-\mu}$ gauge boson ($A^\prime$).  Here we plot $y = \alpha_D \epsilon^2 (m_\chi/m_{A^\prime})^4$  vs. $m_{A^\prime}$ where $\epsilon \equiv g_V/e$, $g_V$ is the  $U(1)_{e-\mu}$ 
 gauge coupling, and $\alpha_D \equiv g_D^2/4\pi$ where $g_D$ is the $A^\prime$-DM coupling constant. Also plotted is the thermal relic target for the direct annihilation regime $m_{A^\prime} > m_\chi$.
 The bottom panel corresponds to the expected sensitivity to a dark photon, $A^\prime$. Here
 $\alpha_D$ is the dark photon's coupling to the DM and $e \epsilon$ is the effective coupling between $A^\prime$ and charged SM fermions. 
\begin{figure}[t!] 
\center
\includegraphics[width=10.5cm]{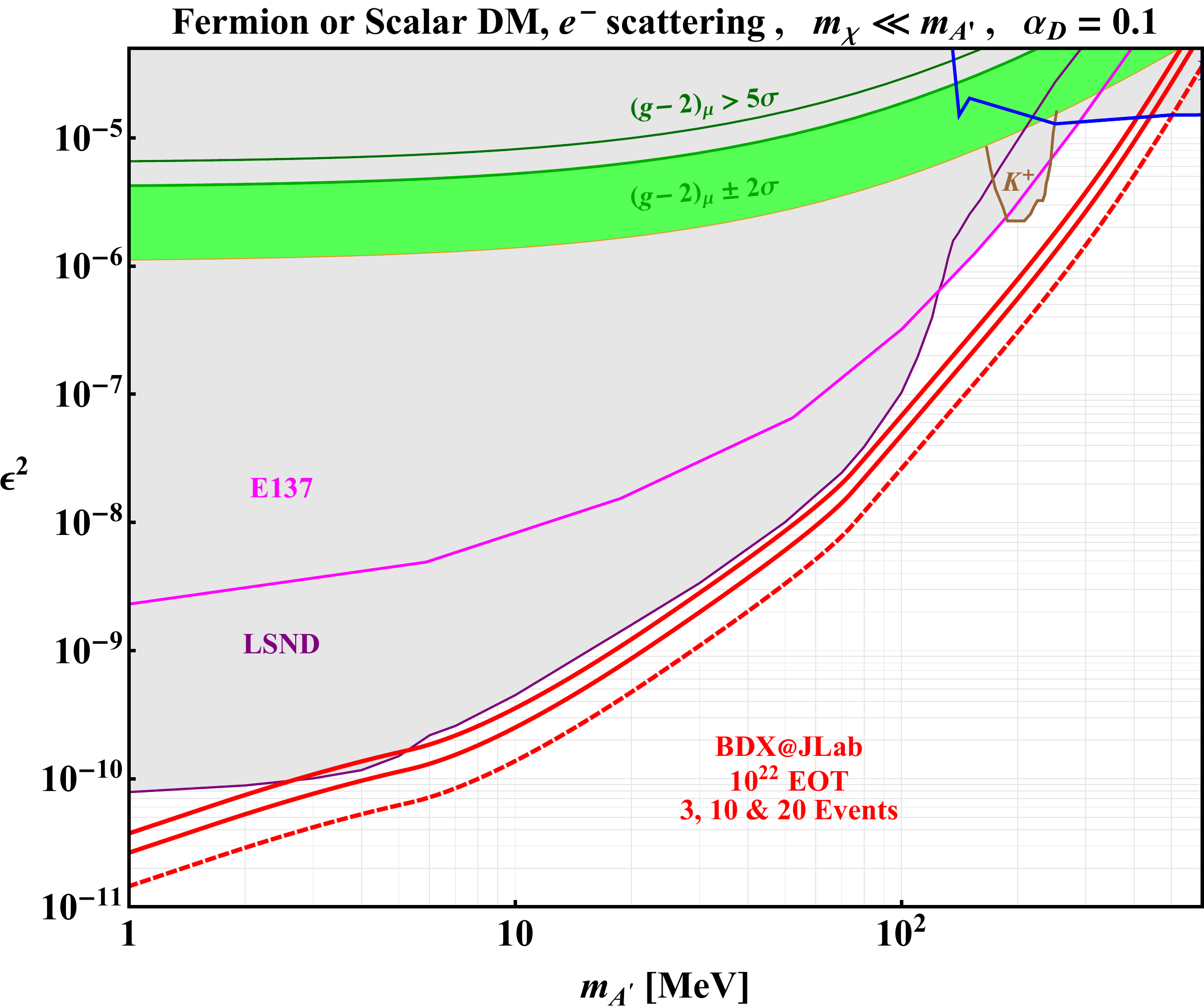} 
\includegraphics[width=10.5cm]{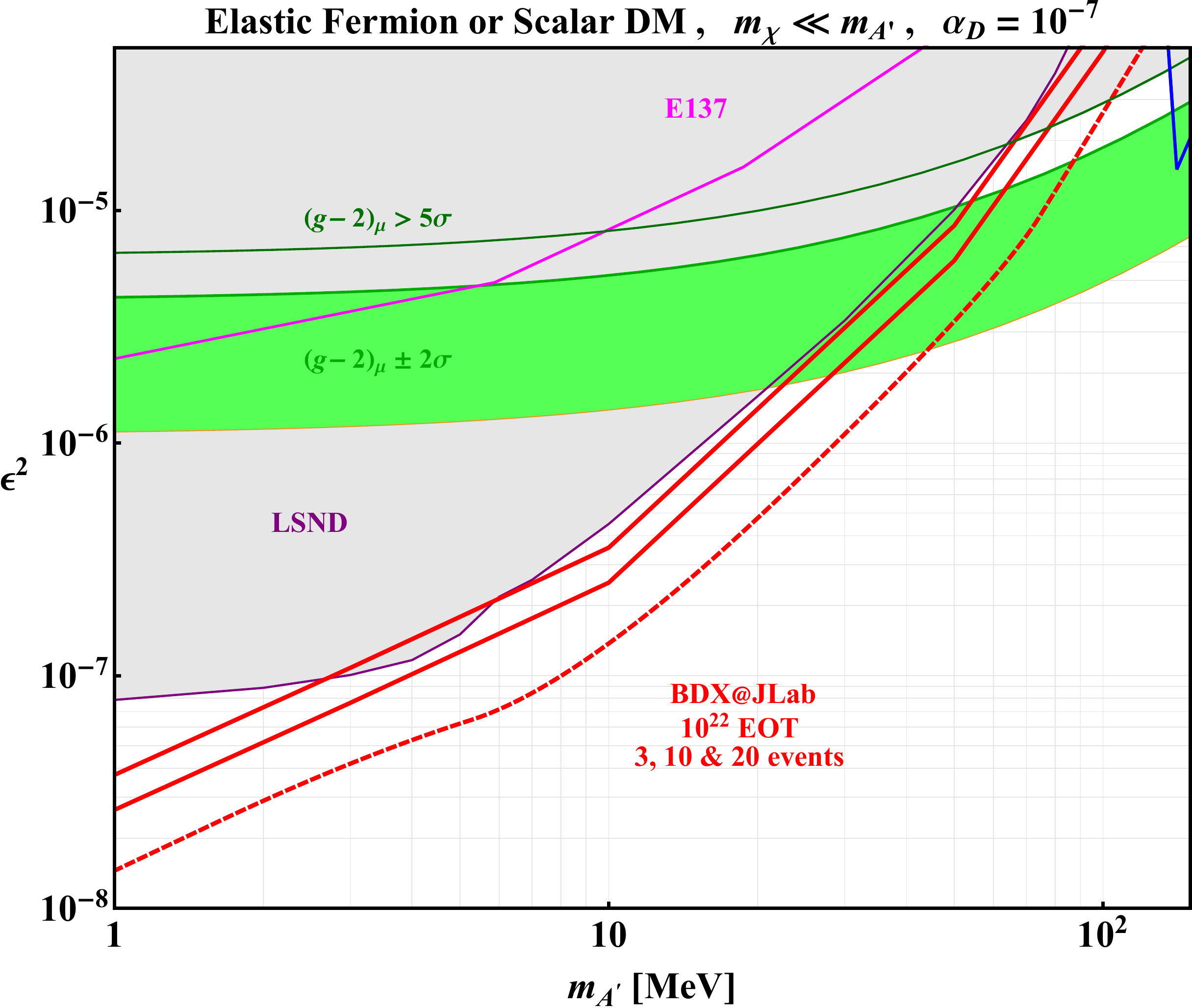} 
 \caption{
Red curves show 3, 10, and 20 event BDX yield projections for electron scattering with $10^{22}$ EOT and a 300 MeV recoil energy threshold. Here $A^\prime$ is  a
kinetically mixed dark photon coupled to DM with $\alpha_D = 0.1$ and $\alpha_D = 10^{-7}$.}
\label{fig:Traditional}
 \end{figure}

\begin{figure}[t!] 
\center
\includegraphics[width=10.cm]{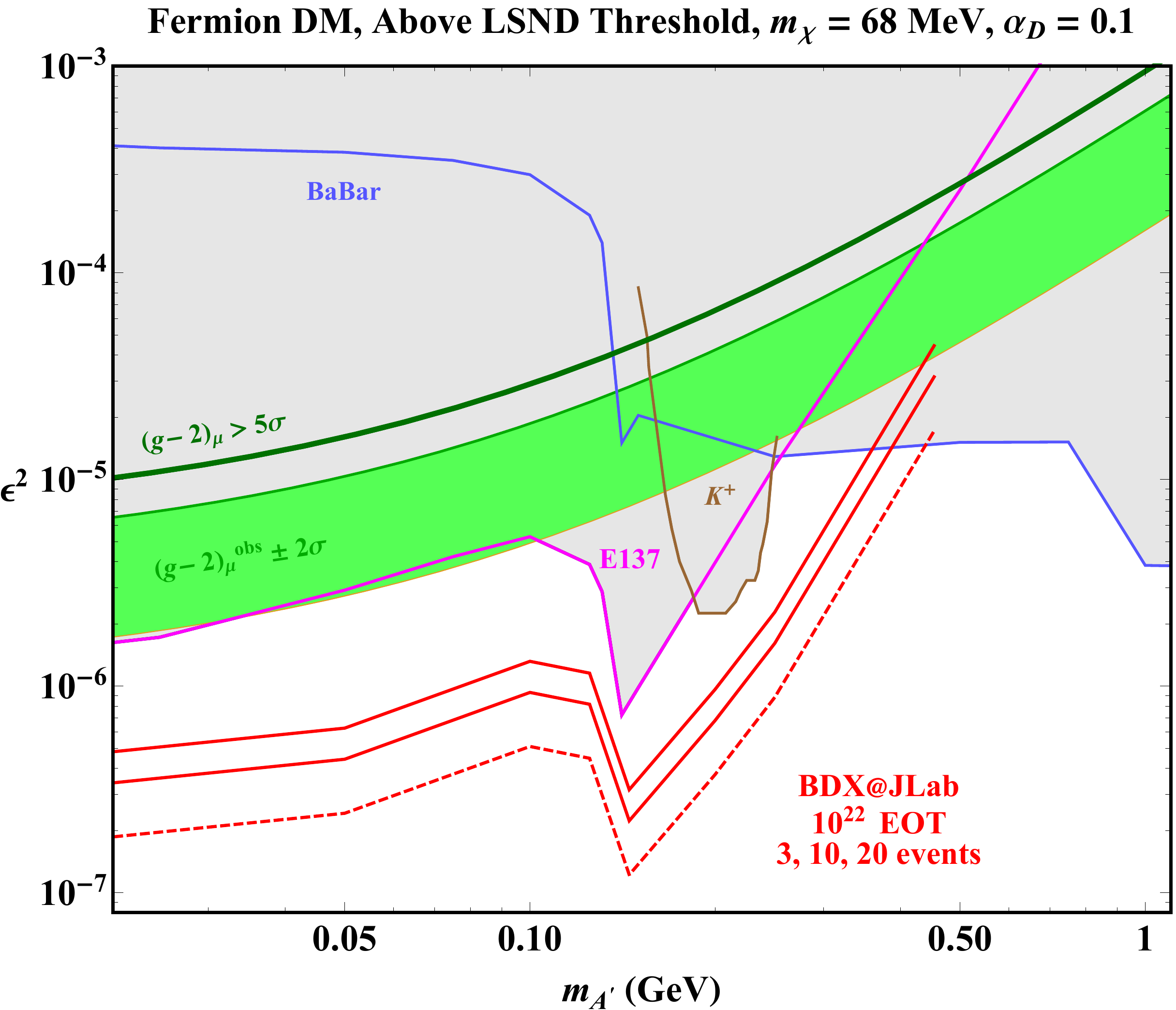} 
\includegraphics[width=10.cm]{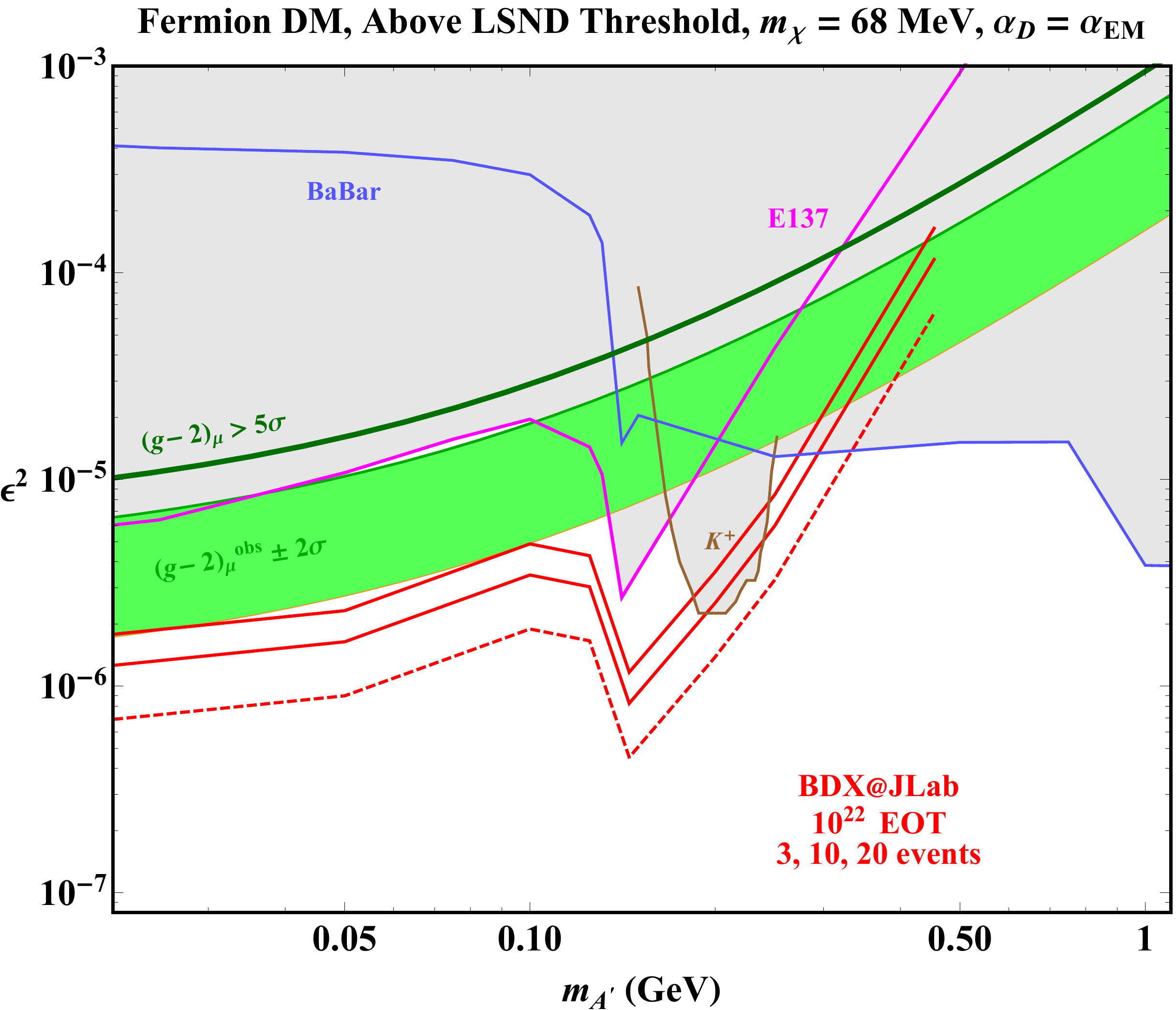} 
 \caption{
Same as Fig.~\ref{fig:Traditional} only here $m_\chi = 68$ MeV and we adopt $\alpha_D = 0.1$ and 
$\alpha_D = \alpha_{EM}$ for the two panels. This choice of $m_\chi$ represents the kinematic limit 
beyond which LSND can no longer produce pairs of $\chi$ via $\pi^0 \to \chi \chi$. Note that for $m_{A^\prime} < 2 m_\chi$ 
the dark photon will no longer decay to DM pairs and may be constrained by visible searches, but this is model
dependent.}
\label{fig:Traditional68}
 \end{figure}

In Fig.~\ref{fig:Traditional} we show the BDX reach in the parameter space $\epsilon^2$ - $m_{A^\prime}$ (for $m_\chi \ll m_{A^\prime}$). We 
show this parameter space for $\alpha_D = 0.1$ and $\alpha_D = 10^{-7}$ to illustrate how different constraints scale with different assumptions
about the $A^\prime$-DM coupling -- note that for sufficiently small values of $\alpha_D$, invisible decay bounds weaken whereas the $(g-2)_\mu$
favored region remains unaffected, since it is independent of the DM coupling. Thus, for the remaining portion of the viable parameter space, an $A^\prime$ explanation for this anomaly is revealed for $\alpha_D \sim 10^{-6}$ and smaller as displayed in Fig.~\ref{fig:gminus2}.   However, for even smaller values $\alpha_D\sim 10^{-10}$, the value of $\epsilon$ required to explain the anomaly is sufficiently large that the $A^\prime$ branching ratio is dominated by the 
visible channel $A^\prime \to e^+e^-$, for which the $(g-2)_\mu$ explanation has already been ruled out (as shown in lower gray shaded region in 
 Fig.~\ref{fig:gminus2}, so BDX has the potential to test nearly all of the remaining dark photon parameter space consistent with a dark-photon solution
 to this discrepancy. 

Finally, in Fig.~\ref{fig:Traditional68} we show the same $\epsilon^2$ vs. $m_{A^\prime}$ parameter space, but for 
a value of $m_\chi$ below the kinematic threshold of LSND, so that constraint does not bound this slice of the parameter space.

In all the aforementioned scenarios the region potentially covered by JLab would therefore significantly extend the parameter space already excluded by previous experiments.

\subsection{Systematic checks  \label{sec:systematics} }
In case of positive result, there is a list of possible checks that can be done to confirm that any observed excess of counts would be related to a real signal.
\begin{itemize}
\item{\bf Measurements during shielding installation:} Measurements of beam-related and cosmi-ray backgrounds can be made during construction as material is added to shield the detector both from
the beam stop and from cosmic-rays. These measurements can be used to validate the MC calculations.
\item{ \bf Beam-related backgrounds:}  due to the expected forward-peaked kinematics of the $\chi$, a measurement off-axis ($\sim 1$m)  will provide a check that the detected signal is really associated to the  electron beam interactions.
\item{\bf Cosmic background:} a precise measurement of the cosmic background in the detector will be possible by accumulating data
during the about 4 years of experiment time. In this way a more precise subtraction of the cosmic background will be possible.
\end{itemize}

The proposed BDX experiment, tacking advantage of the high intensity, high energy electron beam available at JLab has the unique capability of extending the possible reach by order of magnitude with respect to the previous (un-optimized) measurements getting close to the unreducible background due to the neutrinos produced in the beam-dump interaction.  The BDX experiment at Jlab may represent the ultimate beam-dump experiment with an intense electron beam proving a wide category of light DM models.

\clearpage
\section{Summary and Conclusions}
We propose to run an experiment to search for weakly interacting particles produced in the interaction of the electron beam in the dump.
The Beam Dump eXperiment (BDX)  will look for these particles using an electromagnetic  calorimeter surrounded by active and passive vetos in a new underground
facility located downstream of the Hall-A beam dump.
This experiment will have strong, unprecedented sensitivity to dark matter in the MeV -- GeV mass range.
The experimental setup proposed for BDX at JLab,
combining a state-of-the art electromagnetic calorimeter with the high energy and high intensity CEBAF electron beam, will hit the limit of
this class of experiment reaching the wall of the {\it irreducible} background produced by CC interactions of beam-related $\nu_e$ in the detector.

 Searches for particles in this mass range are motivated by models that  feature a dark matter particle $\chi$ whose  interactions with the Standard Model
(SM) through a new massive dark photon generically appear with strength $\epsilon$ near $10^{-4}-10^{-2}$ \cite{Holdom:1985ag}.
  Such models can also explain the persistent $3 - 4\sigma$ discrepancy between theoretical predictions and experimental observations of the muon's
anomalous magnetic moment.

The experiment would detect the elastic scattering of $\chi$s off atomic electrons 
in a detector situated about 20 m from the beam dump by measuring the electron recoil energies. The experiment would also be capable of accessing
complementary information from $\chi$-nucleon scattering. Additionally, the BDX experiment would be uniquely suited to look for DM models where the
DM scatters inelastically in the detector.

The sensitivity of BDX was evaluated by measuring the cosmic background under conditions similar to those proposed and estimating the beam-related
background using GEANT4  MonteCarlo simulations. Results were extrapolated (projected) to the requested accumulated charge of $10^{22}$ electrons on target.
In the absence of a signal, electromagnetic shower thresholds between 0.3 and 0.5 GeV can be used to set limits on the production of dark matter 
that exceed the expected sensitivity of previous, existing, and proposed experiments by up to two orders of magnitude.

\clearpage
\appendix
\section{Evidence and production of dark matter \label{appx:phenomenology}}

The overwhelming evidence for the existence of DM is based on multiple, independent astrophysical and cosmological observations. Stellar rotation curves in galaxies and dwarf-galaxies;  the power spectrum of temperature fluctuations in the Cosmic Microwave Background (CMB); the power spectrum of matter density fluctuations; the ratios of light
element yields from Big Bang Nucleosynthesis (BBN); the morphology of galaxy cluster collisions; and astrophysical mass measurements based on gravitational lensing, all consistently indicate that  85 \% of the matter and 25 \% of the total energy of our universe comprises an electrically neutral, non relativistic population of  ``dark matter" (for a comprehensive review of this evidence, see \cite{Agashe:2014kda}). 
 
 Although the Standard Model (SM) contains several neutral particles,
 none serves as a viable dark matter candidate. The Higgs boson and neutron are unstable with respective lifetimes of $\sim 10^{-22}$ and $\sim 10^{3}$ sec, so both decay too rapidly to accommodate a cosmologically metastable abundance. 
Neutrinos, whose masses satisfy $m_{\nu} \lesssim 0.1$ eV \cite{Ade:2015xua}, are relativistic throughout much of cosmological evolution and would have inhibited 
the formation of large scale structure had they constituted a significant fraction of DM. Thus,  the  existence of DM is  ``smoking gun" evidence of 
physics beyond the SM and uncovering its particle identity is top priority in fundamental physics.\footnote{ The accelerated expansion of the universe (``Dark Energy")  can be interpreted
as evidence of physics beyond the SM \cite{Agashe:2014kda}, but this phenomenon can be accommodated with a small cosmological constant within minimal General Relativity, so no new physics is strictly required. Similarly,  dynamically generating
the observed baryon asymmetry of the universe \cite{Morrissey:2012db} requires physics beyond the SM, however, in principle, this could be accommodated with fine-tuned initial conditions; a similar argument applies to the appeal for new physics motivated by cosmic inflation.  }

However, this task remains elusive because the entire body of evidence for dark matter is based on its
 gravitational influence on visible matter in astrophysical and cosmological contexts.
Given the weakness of the gravitational force, $G_N/G_F \sim 10^{-35}$, these data
 are unable to reveal DM's short distance properties, which 
remain completely unknown to date. Indeed, absent further assumptions about its non-gravitational interactions or cosmological history, the lower bound on 
the average DM particle mass is $m_{\rm DM } \gtrsim 10^{-22}$ eV; lower masses correspond to de Broglie wavelengths  larger than the smallest DM dominated dwarf galaxies \cite{Mateo:1998wg}.
The upper bound arises from the observed stability of large binary stellar systems, which requires $m_{\rm DM} < 100\, M_{\odot}$ \cite{Griest:1989wd}, so the viable mass window is dauntingly difficult to test without the existence of additional interactions between DM and the SM particles.    

There is a popular class of models that call for additional DM-SM interactions. In this paradigm, DM survives as a relic from an era of thermodynamic equilibrium with the SM in the early universe, and its abundance was set when its interaction rate with the SM became subdominant to the expansion rate of the universe --- a mechanism commonly known as ``freeze-out''.
A thermal origin imposes a requirement on the DM: its mass is restricted to be  $m_{\rm DM} > 10 $ keV. Otherwise
   it remains relativistic until late times and thereby erases the observed structure on small scales \cite{Peter:2012rz}.

\begin{figure}[t!]
\center
\includegraphics[width=15cm]{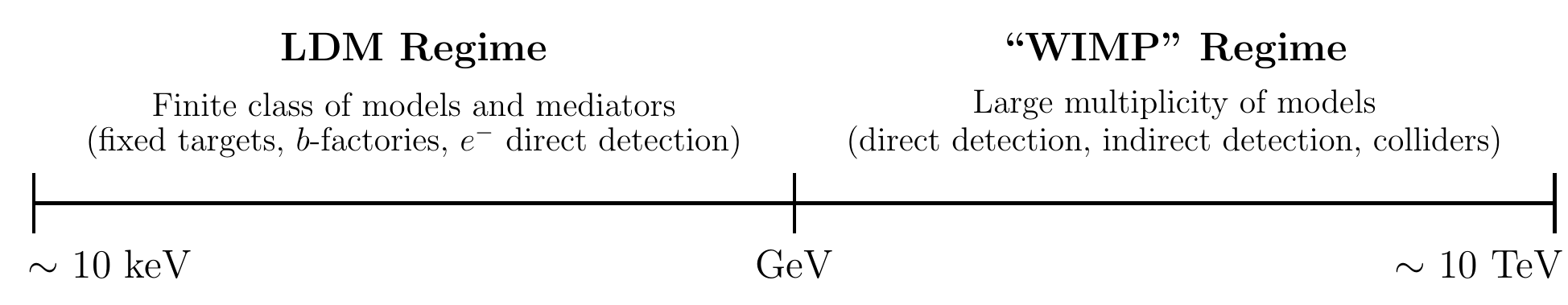}
\caption{The viable mass window for DM with a thermal cosmological history. The upper half of this parameter space covers traditional WIMP models 
in which DM carries SM charges under the weak force and is tested using direct detection (if it scatters elastically off nuclear targets), indirect detection (if its abundance is particle-antiparticle symmetric), and collider production.  The lighter
half of this window, the aim of the BDX collaboration, is comparatively less well studied, and can primarily be tested using intensity frontier methods including fixed target accelerator techniques. Below $\sim 10 $ keV, DM is too hot to accommodate viable large scale structure formation; above $\sim 10$ TeV, the DM annihilation rate compatible with a thermal origin violates perturbative unitarity and nontrivial model building is required.}
\label{fig:Schematic}
\end{figure}

In this paradigm of a thermal origin for DM, DM would have been in equilibrium with the SM in the early universe through DM annihilation into two or more SM final states. On general grounds, the annihilation cross section times velocity for such a process scales roughly as $\sigma v \sim g_{\rm DM}^4/m_{\rm DM}^2$, where $g_{\rm DM}$ is some DM coupling constant. 
 To achieve the observed DM abundance, $\Omega_{\rm DM} = 0.243 \pm 0.004 \propto 1/\langle \sigma v \rangle$ \cite{Ade:2015xua}, the thermally averaged annihilation cross section must be of order $\langle \sigma v \rangle \sim 3\times 10^{-26} \, {\rm cm}^3 {\rm s}^{-1}$ \cite{Jungman:1995df}. For  DM masses near $m_{\rm DM} \gtrsim 10$ TeV, an annihilation rate of this magnitude requires $g_{\rm DM}\gtrsim 4\pi$, corresponding to cross
 sections that violate perturbative unitarity in minimal models  \cite{Griest:1989wd}.\footnote{It is possible to circumvent such
 a hard requirement with composite or strongly interacting 
 dark sectors, but many additional degrees of freedom are required for a complete model \cite{Krnjaic:2014xza}. } Thus, the simple, physically motivated criterion of thermal equilibrium between dark and visible matter requires:
 \begin{itemize}
\item A minimum annihilation cross section $\langle \sigma v \rangle  \ge 3\times  10^{-26}  \, {\rm cm}^3 {\rm s}^{-1} \equiv \langle \sigma v \rangle_{\rm relic}$, where the equality corresponds to particle-antiparticle symmetric DM for which the annihilation is responsible for the full relic abundance and the inequality corresponds to asymmetric DM, for which annihilation merely eliminates thermal antiparticles. The asymmetric DM scenario that we refer to throughout this proposal refers to the case where there is an asymmetry between particles and anti-particles in the DM sector, in analogy to the baryon and lepton asymmetries in the SM.
\item  The DM mass must fall in the viable window between $\sim$ 10 keV $-$ 10 TeV, which dramatically focusses the scope of the experimental discovery effort regardless of other model-specific details.
\end{itemize}
The above considerations apply to {\it any} DM scenario that reaches thermal equilibrium with visible matter in the early universe.
  Figure \ref{fig:Schematic} shows a schematic diagram representing the viable mass range for thermal DM.
 
On the heavier half of this allowed mass range, $m_{\rm DM}  \in$ [few GeV, 10 TeV], the particle physics model landscape for DM and its interactions is often connected to the electroweak scale. For instance, heavy thermal DM can arise as the lightest neutral superpartner in supersymmetric extensions of the SM, motivated to address the electroweak hierarchy problem (e.g. the MSSM or similar model variants)  \cite{Dimopoulos:1981zb}. Other solutions to the electroweak hierarchy problem including composite Higgs models \cite{Kaplan:1983fs,Kaplan:1983sm,Georgi:1984af,Dugan:1984hq}, twin Higgs models \cite{Chacko:2005pe}, flat \cite{ArkaniHamed:1998rs} and warped \cite{Randall:1999ee} extra dimensions, also include weak scale DM candidates. In many such scenarios, the DM candidate is a Weakly Interacting Massive Particle (WIMP) which freezes-out via SM gauge interactions; this $\langle \sigma v \rangle_{\rm WIMP} \sim \langle \sigma v \rangle_{\rm relic}$ coincidence is the so-called ``WIMP miracle," for which no additional interactions are required to yield the observed DM abundance.

DM-SM interactions in this class of WIMP models can be tested with a rich variety of experimental techniques including direct detection, indirect detection, and 
 collider production searches. If the leading DM-SM non relativistic scattering process is elastic and spin independent, existing direct detection experiments like LUX \cite{Akerib:2013tjd}, XENON-100\cite{Aprile:2012nq}, and CDMS \cite{Ahmed:2009zw} and future experiments including XENON-1T \cite{Aprile:2012zx}, LZ \cite{Akerib:2015cja}, and Darwin\cite{Baudis:2012bc} can probe much of the remaining
 parameter space compatible with WIMP DM. If the DM abundance is particle/antiparticle symmetric with a thermal cosmological history, DM annihilation in 
 high density astrophysical regions (e.g. the galactic center) can be observed using ground and space based instruments, such as the {\it Fermi}-LAT \cite{Ackermann:2011wa}. Heavy DM 
can also be tested at the LHC in events with missing (transverse) energy in association with visible SM states \cite{Khachatryan:2014rra,ATLAS:2012ky}.
 
 However, for Light Dark Matter (LDM) on the lower half of the viable thermal window, $m_{\rm DM}  \in$ [10 keV, GeV], these well known experimental strategies become highly ineffective. Typical direct detection sensitivity thresholds require the incoming, elastically scattering DM candidate to deposit $E \gtrsim$ keV scale energies onto a nuclear target. However, the recoil energy of a 2-body scatter scales as $ E_{\rm rec.} \sim \mu v^2$, where $\mu = m_{N} m_{\rm DM}/(m_{N} + m_{\rm DM})$ is the nucleus-DM reduced mass and $v \sim 10^{-3} c$ is their relative
 velocity in the terrestrial lab frame.   Thus, for $m_{\rm DM} \lesssim$ GeV, this energy deposition is below typical sensitivity thresholds at direct detection experiments. For indirect detection, the energies of daughter particles resulting from DM annihilation are in the sub-GeV range where complicated astrophysical backgrounds
are poorly understood and difficult to distinguish from  a potential DM signal.  At collider experiments, LDM can appear as missing energy in association with
other visible objects; however, unlike heavier DM,  LDM does not significantly recoil against the visible object(s) and is, therefore, difficult to distinguish from SM backgrounds (e.g. $Z$ + jets or mis-measured missing energy in multijet events). 

\subsection{Generic features of the theory of Light Dark Matter}
We now focus our attention on the parameter space of LDM, in particular on the simple models that accommodate thermal-relic LDM.
If LDM were merely a WIMP with a smaller mass, its annihilation rate
via DM DM $\to Z^* \to$ SM SM would scale as 
\be
\langle \sigma v \rangle_{\rm LDM, WIMP} \sim G_F^2  m_{\rm DM}^2  =  1.3 \times 10^{-29} {\rm cm}^3 {\rm s}^{-1} \left(\frac{m_{\rm DM}}{100 \, {\rm MeV} } \right)^2 \ll \langle \sigma v \rangle_{\rm relic}~,~~
\ee
which is insufficient to accommodate efficient annihilation in the early universe and similar
arguments apply to other annihilation processes (e.g. virtual Higgs boson exchange). As a consequence, while one can still have the DM achieve thermal equilibrium, one will overproduce DM, so LDM WIMPs without additional interactions are not viable. 
Thus, thermal LDM is conceptually distinct from heavy WIMP DM in two key respects:

\begin{itemize}
\item LDM requires additional forces with correspondingly light, sub-GeV force carriers to achieve the observed DM abundance. 
\item Both the DM and the new force carriers (``mediators") must be neutral under the SM gauge group, otherwise they would have been discovered in direct searches at previous experiments ({\it e.g.,} LEP).
\end{itemize}
Given that there is no room for new SM charged matter at the GeV scale, the gauge and Lorentz symmetries of the SM greatly restrict the ways in which the mediator can couple to the SM. One expects the dominant interactions to be the so-called renormalizable portals: those interactions consisting of SM gauge singlet operators with mass dimension $< 4$:
\be
\hat {\cal O}_{\rm portal} ~ = ~~~H^\dagger H~~~,~~~ LH ~~~,~~~ B_{\mu \nu}~~~,
\ee
and a new SM-neutral degree of freedom, which can be a scalar $\phi$, a fermion $N$, or a vector $A'$.
Here $H$ is the SM Higgs doublet with charge assignment $(1, 2, +\frac{1}{2})$ under the SM gauge group $SU(3)_c \times SU(2)_L \times U(1)_Y$, $L$ is a lepton doublet of any generation
transforming as $(1,2, -\frac{1}{2})$, and $B_{\mu \nu} \equiv \partial_\mu B_\nu -\partial_\nu B_\mu$ is the hypercharge field strength tensor. Although there could also be higher dimension effective operators to connect to the mediators, direct searches for the states that resolve such operators  require suppression scales in excess of the electroweak scale, which generically would reintroduce the DM overproduction problem if these were the predominant interactions that set the DM relic abundance. 

If the mediator is a scalar particle $\phi$, the only allowed renormalizable interactions are through the Higgs portal via $\phi H^\dagger H$  and $\phi^2 H^\dagger H$ which induce mass mixing  between $\phi$ and the SM Higgs boson after electroweak symmetry breaking -- we consider this possibility in more detail in Sec.~\ref{sec:scalar-no-go}. 
 
 If the mediator is a fermion $N$, its interaction with the SM proceeds through the neutrino portal $\sim y_\nu LHN$ and it plays the role of a right handed neutrino with a Yukawa coupling $y_{\nu}$. If DM is {\it not}  thermal in origin, $N$ can itself be a viable, cosmologically metastable DM candidate in a narrow mass range \cite{Dodelson:1993je}. Since $N$ is stipulated to be sub-GeV, obtaining the observed neutrino masses (without additional field content) requires Yukawa couplings of order $y_\nu \lesssim 10^{-12}$, which are too small to allow thermalization to take place at early times \cite{Feng:2010gw}.
 
 If the mediator is a vector force carrier from an additional $U(1)_D$ gauge group under which  LDM is charged, the ``kinetic mixing" interaction $\epsilon_Y B^{\mu \nu} F^\prime_{\mu \nu}$ is gauge invariant under both $U(1)_D$ and $U(1)_Y$. Here $\epsilon_Y$ is, {\it a priori} a free parameter, though it often arises in UV complete models after heavy states charged under both groups are integrated out at a high scale, so it is generically expected to be small ($\epsilon_Y \sim 10^{-3} - 10^{-5}$ depending on the loop order at which it is generated). Such a radiative origin for $\epsilon_Y$ is required if either $U(1)$ group is a subset of a nonabelian group for which kinetic mixing is not a gauge invariant interaction; it can only be generated after a spontaneous symmetry breaking phase transition which preserves an unbroken $U(1)$ subgroup. 
 
\subsection{Defining thermal targets}

For all mediators and LDM candidates $\chi$, there is a basic distinction between ``secluded" annihilation to pairs of  mediators (via $\chi \chi$ $\to$ MED MED  for  $m_\chi > m_{\rm MED}$) followed by 
mediator decays to SM particles \cite{Pospelov:2007mp}, and ``direct" annihilation to SM final states (via virtual mediator exchange  in the $s$-channel, $ \chi \chi \to$ MED$^{*} \to$  SM SM for $m_{\chi} < m_{\rm MED}$) without an intermediate step. 

For the secluded process, the annihilation rate scales as 
\be 
\hspace{-1cm } ({\rm ``secluded" ~annihilation}) ~~~~~~ \langle \sigma v \rangle \sim \frac{  g_D^4 }{  m_\chi^4  } ~~ ~,~~
\ee
where $g_{\rm DM}$ is the coupling between the mediator and the LDM, and there is no dependence on the SM-mediator coupling $g_{\rm SM}$. Since arbitrarily small values of  $g_{\rm SM}$, the SM-mediator coupling, can be compatible with thermal LDM in this regime, the secluded scenario does not lend itself to decisive laboratory tests;
  
The situation is markedly different for  the direct annihilation regime in which $m_{\chi} < m_{\rm MED.}$ where the annihilation rate scales as  
\be
\hspace{1cm}({\rm ``direct" ~annihilation})~~~~\langle \sigma v \rangle   \sim       \frac{      g_{D}^2 \,  g_{\rm SM}^2  \,  m_{\chi}^2 }{   m_{\rm MED}^4  \!\!}~~ ,~~~~~~~~~  ~,~~
\ee
and offers a clear, predictive target for discovery or falsifiability since the dark coupling $g_{D}$  and mass ratio $m_{{\chi}}/m_{\rm MED}$ are 
at most ${\cal O}(1)$ in this $m_{\rm MED} > m_{\chi}$ regime, so there is a minimum SM-mediator coupling compatible with a thermal history; smaller values 
of $g_D$ require nonperturbative dynamics in the mediator-SM coupling or intricate model building.

In the direct annihilation regime, up to order-one factors, the minimum annihilation rate requirement translates into a minimum value of the dimensionless combination 
\be\label{eq:generic-thermal-target}
\boxed{
y \equiv  \frac{    g_D^2 \, g_{\rm SM}^2  }{4\pi} \, \left(\frac{m_\chi}{m_{\rm MED}}\right)^4 \gtrsim  \langle \sigma v \rangle_{\rm relic}   \, m_\chi^2~,~~
}
\ee
which, up to order one factors, is valid for every DM/mediator variation provided that $m_{\rm DM} < m_{\rm MED.}$.
We will use this target throughout this document to assess experimental sensitivity to various LDM scenarios;
 reaching this benchmark sensitivity suffices to decisively discover or falsify a large class of simple direct annihilation models.

\subsection{Excluding scalar mediated direct annihilation} \label{sec:scalar-no-go}
From the above listed class of portal mediators, the scalar Higgs portal scenario with a scalar mediator is compelling in its simplicity: a real singlet scalar $\phi$
couples to the SM by mixing with the SM Higgs boson and interacts with fermions in direct proportion to their masses. Assuming a fermionic DM particle for concreteness, the lagrangian for this theory is 
\be
{\cal L} \supset  g_D \phi \bar \chi \chi +\frac{m_\phi^2}{2} \phi^2 + A  H^\dagger H  \phi+  \frac{ m_f }{v} H \bar f_L f_R + {\it h.c.}~,~
\ee
where$f$ is a SM fermion with mass $m_f$, $A$ is a dimensional coupling constant, and $v$ is the SM Higgs vacuum expectation value (VEV). 
After electroweak symmetry breaking, the replacement $H\to v$ induces a mass mixing term between $\phi$ and the neutral component of the Higgs doublet. Diagonalizing this mixing yields a coupling between $\phi$ and SM fermions $g_{\rm SM} \equiv \sin\theta m_f/v$, where $\sin\theta$ is the higgs-$\phi$ mixing angle.

In the $m_\chi < m_\phi$, the relic density is achieved via  $\chi \chi \to \phi^* \to f f$ direct annihilation and the thermal target for a given final state $f$ 
is 
\be \label{eq:thermal-target-generic}
y \equiv  g_D^2 \sin^2\theta \left( \frac{m_f}{v} \right)^2  \left(  \frac{m_\chi}{m_\phi} \right)^4 > \langle \sigma v\rangle_{\rm relic}\, m_\chi^2~,~~
\ee
Since LDM requires annihilation to light SM fermions with $m_f \ll v$, the mixing angle must satisfy $\sin \theta  \sim {\cal O}(1)$ over
most of the $m_\chi  <$  GeV range \cite{Krnjaic:2015mbs}. However, such a large value predicts unacceptably large branching ratios for $B \to K \phi$ which 
contributes irreducibly  to $B^+ \to K^+ \nu \bar  \nu$ and  $ K^+ \to \pi^+ \nu \bar  \nu$ observables. This process, which is generated by loops of virtual tops whose coupling to $\phi$, is enhanced
by scales as $m_t/m_f$ relative to the corresponding coupling for $\chi \chi \to f f $ annihilation.  Thus, scalar mediated annihilation through the Higgs
portal is completely ruled out; ``secluded" annihilation for $m_\chi > \m_\phi$ via $\chi \chi \to \phi \phi$ is still viable, but offers no thermal target \cite{Krnjaic:2015mbs}.  

\subsection{Vector mediated models}

While scalar direct annihilation is ruled out and the neutrino portal is not easily compatible with thermal LDM, the vector portal, where a new vector $A'$
couples to the SM hypercharge, is largely unexplored for the parameter space of LDM.
Therefore, we take this scenario as a basis for determining the sensitivity of the experiment.
Consider a spin-1 mediator $\apr$, which kinetically mixes with the SM hypercharge via  $\epsilon_Y B^{\mu \nu} F^\prime_{\mu \nu}$. Here, $\apr$ can be
thought of as the gauge boson from a spontaneously broken $U(1)_D$ gauge group, and DM is charged under this gauge group \footnote{ Reinterpreting the
results from the kinetic mixing model into other scenarios -- for instance gauging both $\chi$ and the SM under one of the anomaly free combinations of
SM global quantum numbers: $U(1)_{B-L}$,  $U(1)_{e-\mu}$, $U(1)_{e-\mu}$,  $U(1)_{e-\tau}$,  $U(1)_{\mu-\tau}$ -- is trivially accomplished by appropriately
rescaling coupling constants by the kinetic mixing parameter $\epsilon$. The key phenomenological difference in these models is that the mediator couples
appreciably to neutrinos with equal strength, whereas the dark photon couples predominantly to the electromagnetic current}. The 1 for this setup is
\cite{Holdom:1985ag}
 \be
 \label{eq:lagrangian1}
{\cal L}_{\apr} &\supset & 
-\frac{1}{4}F^\prime_{\mu\nu} F^{\prime\,\mu\nu} + \frac{\epsilon_Y}{2} F^\prime_{\mu\nu} B_{\mu \nu} + \frac{m^2_{A^\prime}}{2} A^{\prime}_\mu A^{\prime\, \mu} + g_D \apr_\mu J^\mu_\chi  +  g_Y B_\mu J^\mu_Y   ,
\ee
where  $F^\prime_{\mu\nu} \equiv \partial_\mu A^\prime_\nu -  \partial_\nu A^\prime_\mu$ is the dark photon field strength,
$B_{\mu\nu} \equiv \partial_\mu B_\nu -  \partial_\nu B_\mu$ is the hypercharge field strength,
  $g_D \equiv \sqrt{4\pi \alpha_D}$ is the dark gauge coupling, and $J^\mu_\chi$ and $J^\mu_Y$ are the DM and SM hypercharge matter currents, respectively.
 After electroweak symmetry breaking, the hypercharge kinetic mixing $\epsilon_Y$ induces mixing with the photon  and $Z$ boson 
 \be
  \frac{\epsilon_Y}{2} F^\prime_{\mu\nu} B_{\mu \nu}  ~~ \longrightarrow~~  \frac{\epsilon}{2} F^\prime_{\mu\nu} F_{\mu \nu} +  \frac{\epsilon_Z}{2} F^\prime_{\mu\nu} Z_{\mu \nu}  ~,~~
\ee
where  $\epsilon \equiv \epsilon_Y/\cos\theta_W$, $\epsilon_Z \equiv \epsilon_Y/\sin\theta_W$, and $\theta_W$ is the weak mixing angle.  
Diagonalizing away this mixing yields dark photon interactions with dark and visible matter  
\be 
g_D \apr_\mu J^\mu_\chi  +  g_Y B_\mu J^\mu_Y ~~\longrightarrow~~  A^\prime_\mu ( g_D J^\mu_\chi + \epsilon e J^\mu_{\rm EM})~,
\label{eq:darkcurrent}
\ee
where  $J^\mu_{\rm EM}$ is the usual SM electromagnetic current and we have omitted terms higher order in $\epsilon$. 
   The dark photon $\apr$ couples to the LDM current $J_\chi^\mu$, which can represent
either a scalar or fermionic DM candidate. For the remainder of this document we will assume the predictive, direct annihilation regime $m_{\apr} > 2m_\chi$  (see right column of Fig.~\ref{fig:breakdown}). 

Regardless of the LDM candidate whose current is given by $J_\chi$, the thermal target from Eq.~(\ref{eq:thermal-target-generic}), corresponding to 
the direct annihilation in Fig. X (c), can be written as 
\be
\boxed{
({\rm vector~mediator~target})    ~~~~~  y \equiv \epsilon^2 \alpha_D \left(\frac{m_\chi}{m_{\apr}} \right)^4 ~,~
}~~~~~~~~~
\ee
where the precise value of this target depends on the choice of LDM candidate. 
This class of models is compatible with a finite set of LDM candidates which can be either fermions or scalars, cosmologically particle-antiparticle symmetric or asymmetric, and may couple elastically or inelastically to the $\apr$. We now consider these permutations in turn.

\subsection{ LDM candidates with vector mediator}

\bigskip
{\noindent \bf  Symmetric Fermion LDM} \\ \bigskip
If the LDM candidate is a fermion, the current in Eq.~(\ref{eq:lagrangian1}) is
\be
({\rm fermion~LDM ~current})    ~~~~~ J_{\chi}^\mu \equiv  \bar \chi \gamma^\mu \chi~,~~  
\ee
where $\chi$ is a four component Dirac fermion with mass $m_\chi$. The abundance in this scenario is symmetric with
respect to particles and antiparticles, so the annihilation rate must satisfy $\langle \sigma v\rangle \approx 3\times 10^{-26} {\rm cm}^2 {\rm s}^{-1}$
to achieve the observed relic abundance. 

However, for Dirac fermions, the annihilation cross section is $s$-wave 
\be
\langle \sigma v\rangle \propto \epsilon^2 \alpha_D \frac{m_\chi^2}{m_{\apr}^4} \sim \frac{ y }{m_\chi^2}~,~~
\ee
and therefore constant throughout cosmic evolution, including during the epoch of the CMB near $T \sim$ eV. Although
the abundance has frozen out by this point, out-of-equilibrium annihilations to SM particles can re-ionize hydrogen at
the surface of last scattering and leave an imprint in the SM ionized fraction \cite{Finkbeiner:2011dx}.
For a particle-antiparticle symmetric population of $\chi$, this scenario is comprehensively ruled out by measurements
of the CMB power spectrum so we will not consider it further \cite{Lin:2011gj}. 

{\noindent \bf Asymmetric Fermion LDM} \\ \bigskip
If the cosmic abundance of Dirac fermion $\chi$ is set by a primordial asymmetry, then the annihilation process depletes antiparticles during the CMB epoch so the effective abundance of antiparticles is suppressed by factors of 
 $\sim \exp({-\langle \sigma v \rangle})$. Thus, the CMB re-ionization bound does not rule out this scenario.

{\noindent \bf Majorana (Pseudo-Dirac) LDM}  \\\bigskip
The dark photon has a non-zero mass $m_{\apr}$, which means that the $U(1)_D$ in the dark sector is a broken gauge theory --- unlike the SM electromagnetism, which is unbroken, as evidenced by the photon being massless. A generic possibility in this broken gauge theory is that the fermion LDM candidate has both a  $U(1)_D$ preserving  Dirac mass and $U(1)_D$ breaking Majorana mass -- possibly
induced by the mechanism responsible for spontaneous symmetry breaking in the dark sector. This same mechanism could be responsible for giving the dark photon a mass $m_{\apr}$. As a result, the Weyl spinors in the 
four component $\chi$ will be split in mass and the dark photon couples predominantly to the {\it off-diagonal} current 
\be
({\rm off-diagonal~fermion~LDM ~current})    ~~~~~ J_{\chi}^\mu \equiv  \bar \chi_1 \gamma^\mu \chi_2 + h.c. ~,~~  
\ee
where $\chi_{1,2}$ are Majorana spinors split in mass by $m_2 - m_1 \equiv \Delta$; all interactions with the $\apr$ are off-diagonal
in this mass eigenbasis. Note that, like the Dirac scenario, this variant has the same degrees of freedom (2 Weyl spinors) 
but different global symmetries in the fermion mass sector, so this kind of coupling arises generically if those symmetries are broken.
Note that $\chi_2$ is unstable and decays via $\chi_2 \to \chi_1 f \bar f$ for $m_{\apr} > m_1 + m_2$, which is required for the direct 
annihilation scenario. 

Unlike the Dirac case, the direct annihilation for the pseudo-Dirac scenario requires {\it both} eigenstates to meet via $\chi_1 \chi_2 \to \apr \to f f$,
so at late times $T \ll \Delta$ this process shuts o
 For small $\Delta \ll m_\chi$ this annihilation rate has the same parametric scaling as the Dirac fermion scenario
\be
\langle \sigma v\rangle \propto \epsilon^2 \alpha_D \frac{m_\chi^2}{m_{\apr}^4} \sim \frac{ y }{m_\chi^2} + {\cal O}\left(\frac{\Delta}{m_\chi}\right)~,~~
\ee
so the same $y$ target applies. However, the CMB bound is now removed because the excited state $\chi_2$ is typically absent at late times, thus shutting off the tree-level annihilation $\chi_1 \chi_2 \to \apr \to f f$. It's worth noting that the setup of Majorana LDM with a dark photon that mixes with hypercharge falls within the popular class of models known as ``inelastic DM'' \cite{TuckerSmith:2001hy}.

{\noindent \bf Scalar LDM} \\ \bigskip
If the LDM candidate is a complex scalar coupled to the dark photon, the current in Eq.~(\ref{eq:darkcurrent}) is 
\be
({\rm scalar~LDM~current})    ~~~~~ J_{\chi}^\mu \equiv i ( \chi^* \partial^\mu \chi  -  \chi \partial^\mu \chi^*)~,~~  
\ee
where momentum dependence in the coupling to the $\apr$ leads to a $p$ wave annihilation rate $\chi \chi \to \apr \to f f$
\be
\langle \sigma v\rangle \propto \epsilon^2 \alpha_D \frac{ m_\chi^2 v^2}{m_{\apr}^4} \sim \frac{ y  v^2}{m_\chi^2}~,~~
\ee
where $v$ is the velocity. Note that compared to the fermion scenario, for fixed $m_\chi$,
the thermal target for $y$ is ${\cal O}(10)$ larger to compensate for the $v^2 \sim 0.1$ rate reduction due to the 
DM velocity at freeze out, $T\sim m_\chi/20$, when the relic abundance is set.

In summary, by far, a new spin-1 light degree of freedom, a $\apr$ that kinetically-mixes with the SM hypercharge, gives the most viable scenario
for models of LDM with a thermal origin.
This is the scenario targetted in the proposal and serves as a basis for defining the sensitivity for the experiment.


\clearpage
\section{The BDX prototype}\label{Section:BDX-protoype}
\begin{figure}[t!] 
\center
\includegraphics[width=10cm,clip=true]{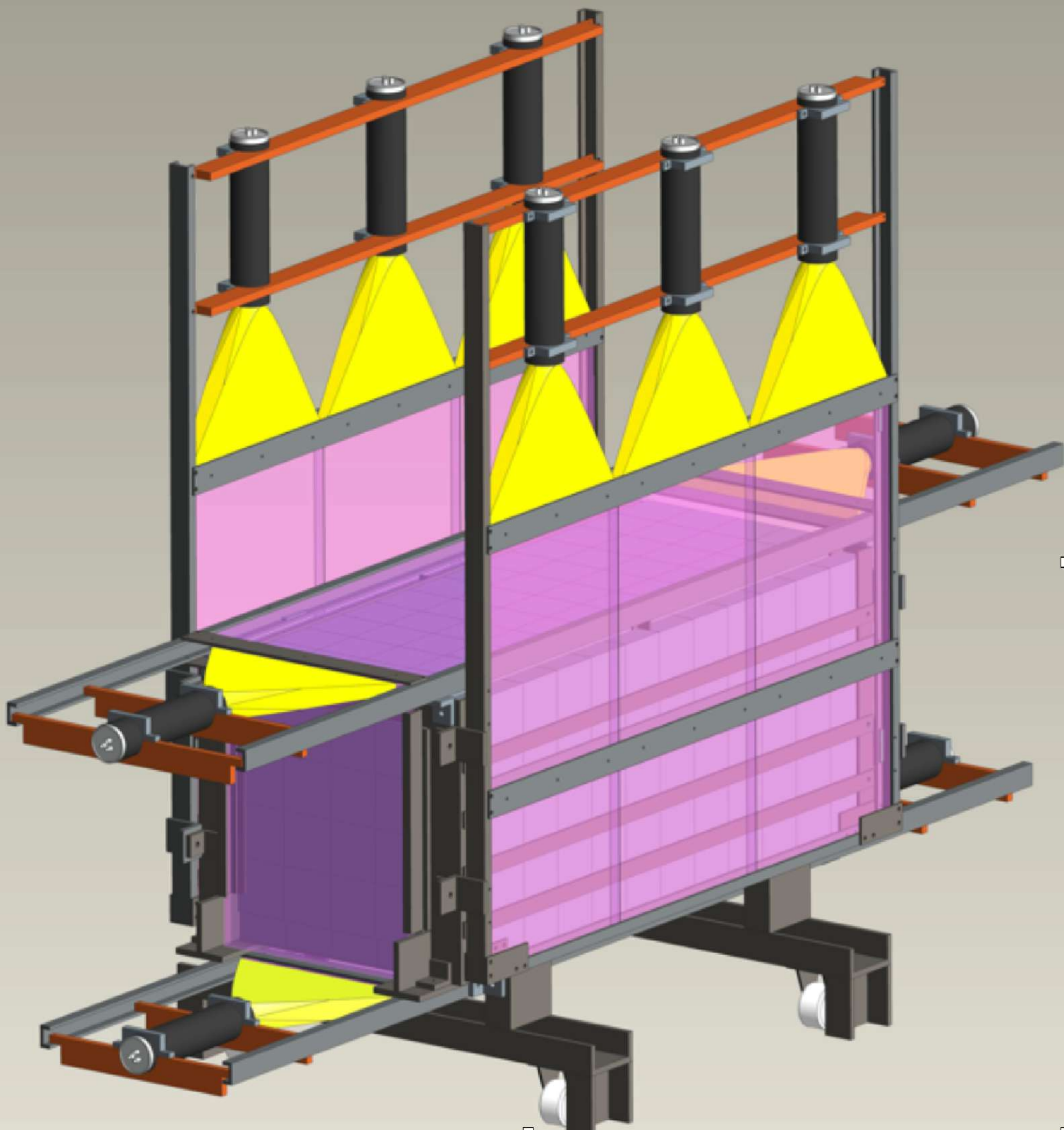}
 \caption{The CAD drawing of the BDX prototype detector: the Outer Veto (OV) with light guides and PMTs are visible together with the lead vault (in transparence).
 }\label{fig:proto-3d-cad}
\end{figure}
In this Appendix we report the results of a dedicated  campaign of measurements performed in Catania, Italy (INFN-CT) and Laboratori Nazionali del Sud (LNS) with a prototype of the  BDX detector. The measurement provided  information useful for the full detector design and expected performance in a realistic configuration. We tested  the proposed technology for the BDX detector: CsI(Tl) crystals read by SIPM, plastic scintillator read by PMT for the OV and plastic scintillator coupled to SIPM by WLS fibers; we validated the background model for cosmic muons and neutrons; we checked the effect of the lead shielding and the overburden, and,  eventually, we derived the single crystal rates as a function of the energy threshold in combination (anti-coincidence) with the veto system.  In Sec.~\ref{sec:beam_unrelated_bck} these measurements have been extrapolated to the full experiment (scaling to the detector size and the measurement  time) providing a reliable estimate of the expected cosmogenic background.
To validate Monte Carlo simulations in a standard and well-controlled configuration,  cosmic data were initially taken exposing the prototype to cosmic rays with  minimal shielding (15 cm of concrete roof of INFN-CT).  Then the prototype was moved into a bunker at LNS with  a surrounding  overburden of about 5 meters of concrete corresponding to an effective thickness  of 1080 g/cm$^2$ similar to 1165 g/cm$^2$ expected at JLab. 

In order to prove the detection capability of the BaBar crystals for low energy protons with the new improved readout sensor and electronics, we studied the response of  the CsI(Tl) crystal  to a low energy proton beam (T$_p$  from 2 MeV to 24 MeV)  provided by  the LNS-Tandem Van der Graaf. The results of these measurements are reported in the last paragraph.
\begin{figure}[t!] 
\center
\includegraphics[width=14cm,clip=true]{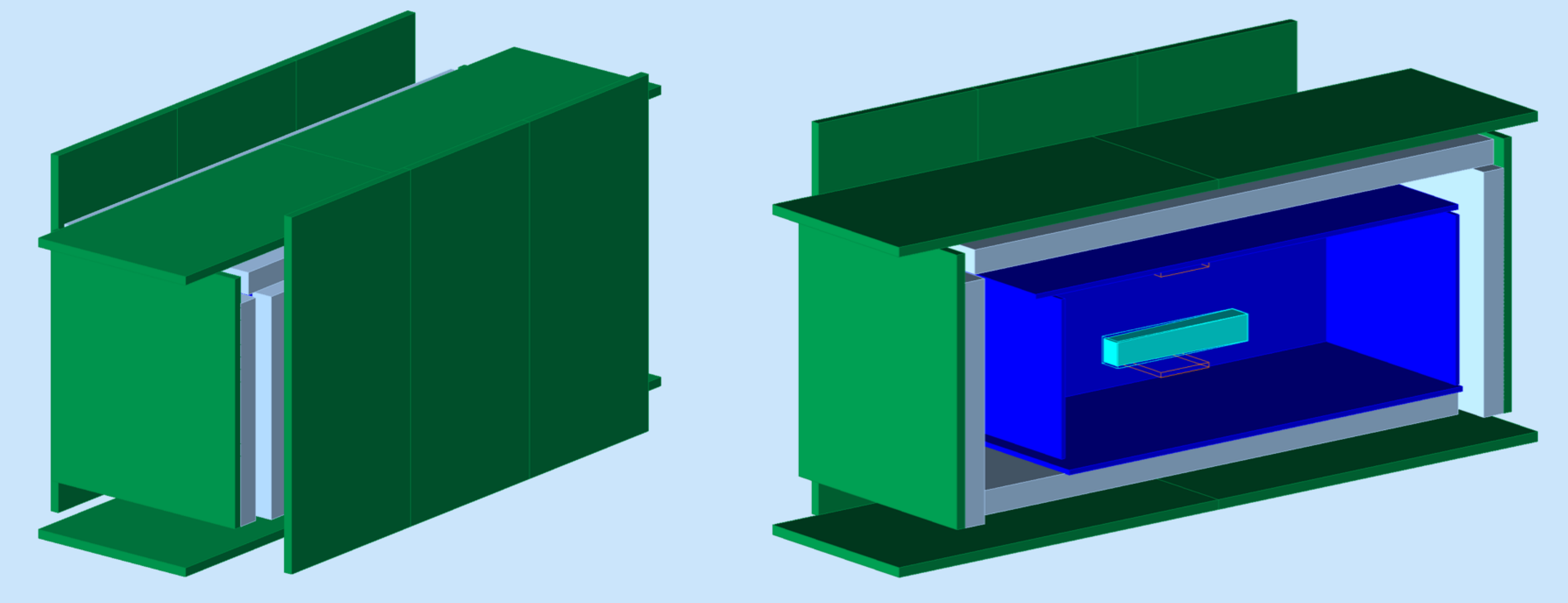}
 \caption{ The implementation of the BDX prototype in GEMC. The  OV (green), the lead vault (gray), IV (blue), and the crystal (cyan) are visible in the drowing on the right panel.
 }\label{fig:proto-3d}
\end{figure}

\subsection{The BDX protoype}
\begin{figure}[t!] 
\center
\includegraphics[width=12cm,clip=true]{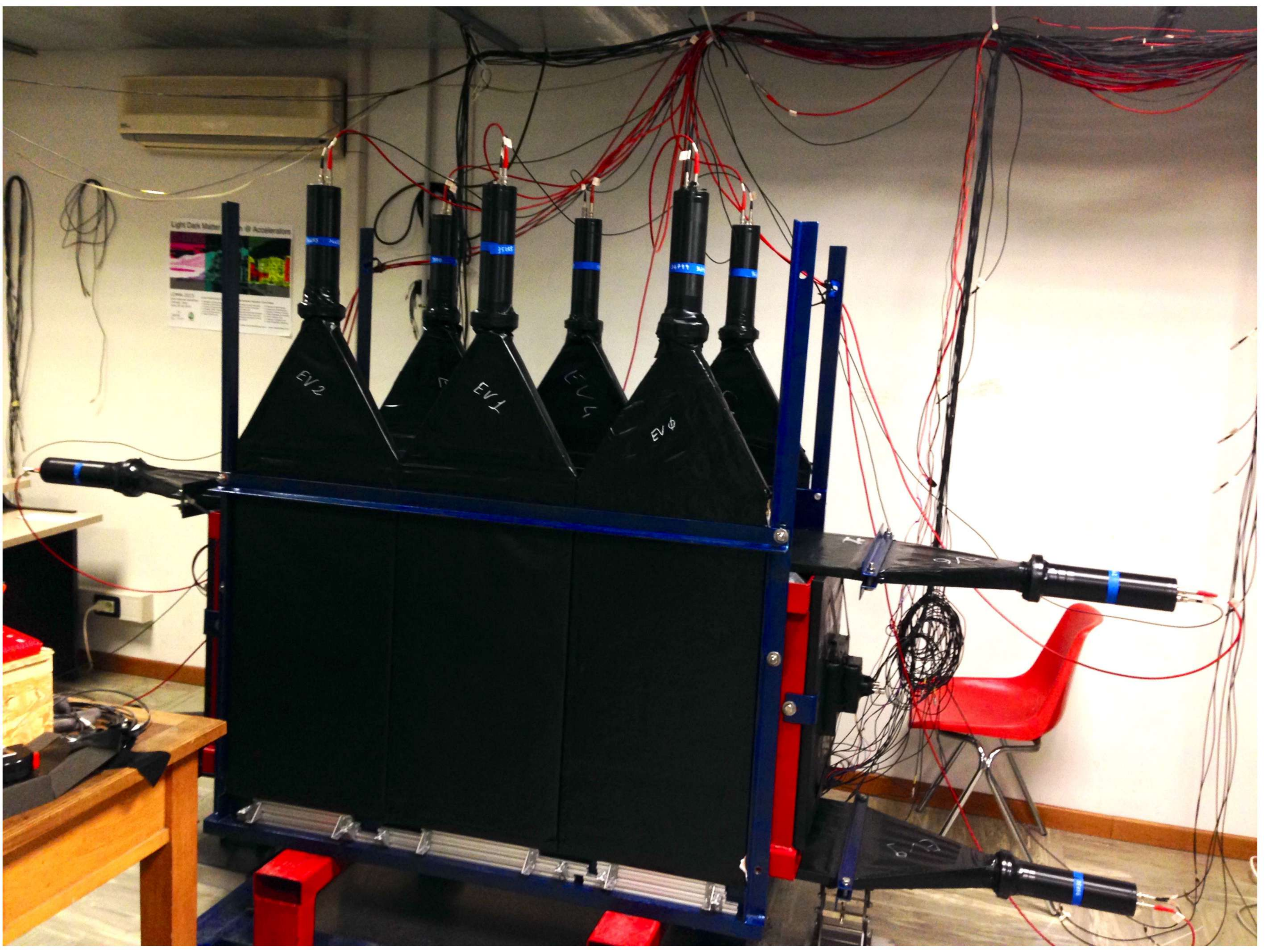} \hspace{0cm}
\includegraphics[width=6cm,clip=true]{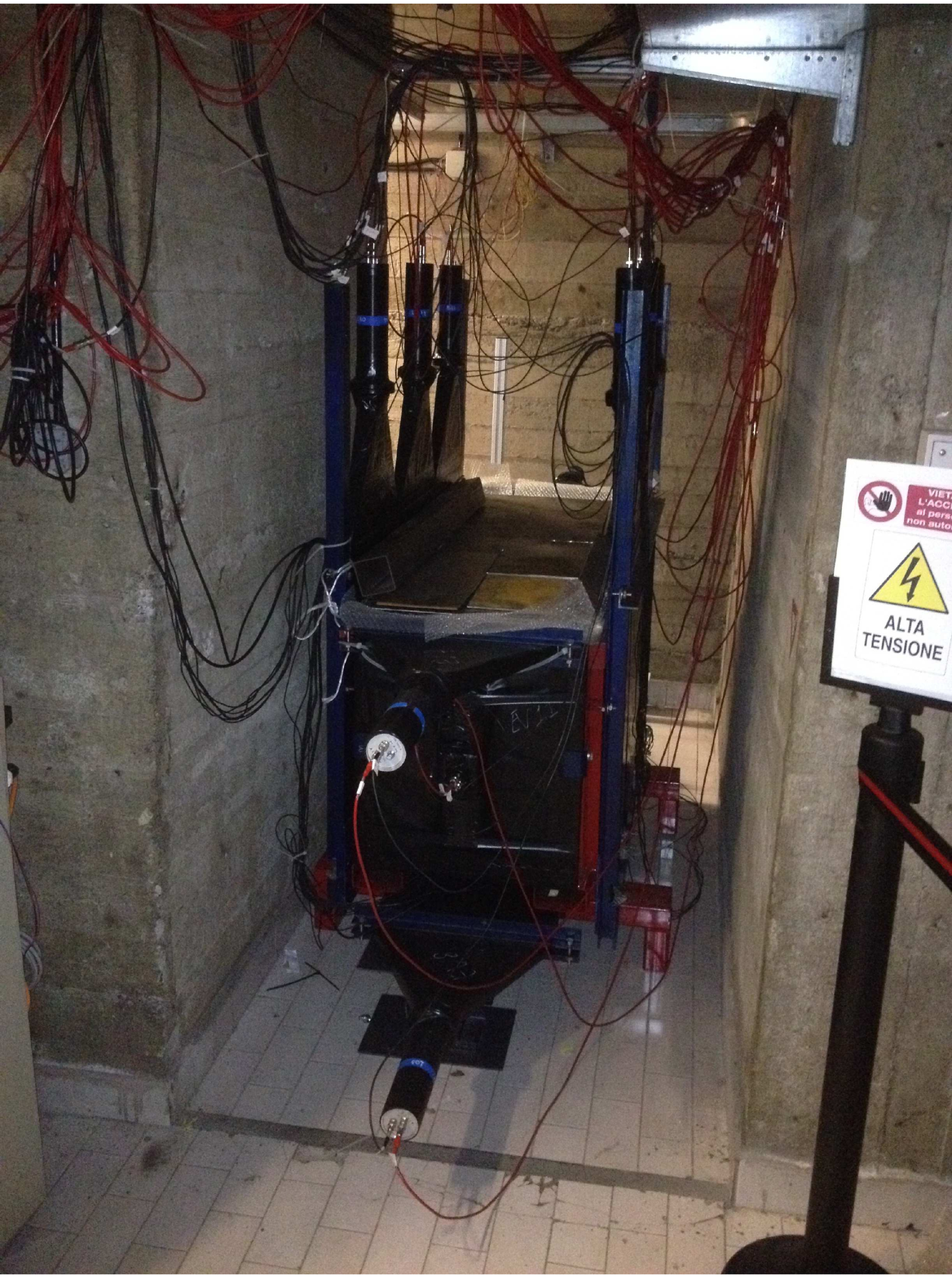} 
 \caption{The prototype mounted and cabled at the INFN- Sezione di Catania (top) and inside the LNS bunker (bottom).
 }\label{fig:prototype}
\end{figure}

The BDX prototype is made by a single BaBar CsI(Tl) crystal surrounded on all sides by a layer of veto detectors (Inner Veto or IV), a vault of lead bricks, and a second layer of veto detectors forming the Outer Veto  (OV, Fig.~\ref{fig:proto-3d-cad}). The combined use of two charged-particle veto-counter systems allows to compensate for their inefficiencies and better reject background. Between the inner and the outer veto, the 5cm thick lead vault shields the crystal from radiogenic low energy gammas. Two additional small-area plastic scintillator pads (12x12x1 cm$^{3}$) were placed inside the inner veto, one above and one below the CsI(Tl), to trigger on cosmic rays for energy calibration, timing and efficiency measurements. 
The full GEANT4 implementation of the BDX prototype is shown in Fig.~\ref{fig:proto-3d} while a picture of the experimental set-up mounted at INFN-CT and LNS is shown  in Fig.~\ref{fig:prototype}.

\begin{figure}[t!] 
\center
\includegraphics[width=4cm,clip=true]{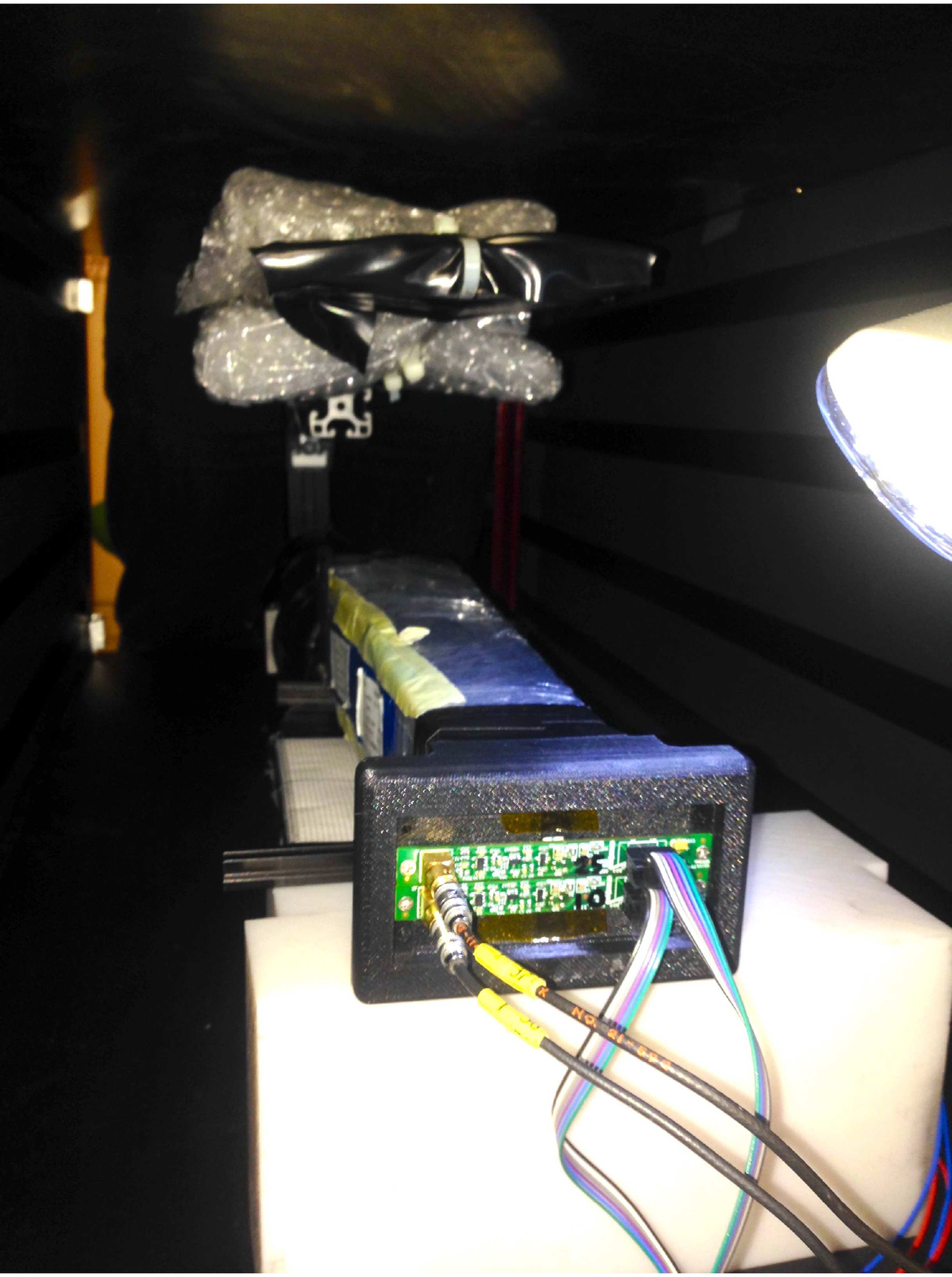}
 \caption{A picture of the CsI(Tl) crystal inside the prototype. The two charge-preamplifiers coupled to the SiPM are visible together with the two additional small plastic scintillators.
 }\label{fig:CsI}
\end{figure}

The CsI(Tl) crystal is one of those formerly used in the  BaBar Ecal end cap with a brand new SiPM-based readout. 
It is 31cm long and has a trapezoidal shape with a 4.7 x 4.7 cm$^{2}$ front face and a 6 x 6 cm$^{2}$ back face (Fig.~\ref{fig:CsI}).  Two 3x3 mm$^{2}$ SiPMs (Hamamatsu S13360-3025CS and S13360-1350CS), with pixel size of 25 and 50 $\mu$m, are glued to the  crystal front-face (leaving untouched the existing pin diode used by BaBar on the opposite side). The 50 $\mu$m cell-size has an higher PDE (35\%),  more suitable for low energy signals while the 25$\mu$m, having a larger number of pixels has a lower PDE (22\%)  but  results in a fairly linear response for higher energy signals. Both sensors are coupled to custom trans-inpedence amplifiers ~\cite{Orsay-sipmpreamp} with different gains: G$_{50\mu m}$=230 and G$_{25\mu m}$=40. The lower gain G$_{25\mu m}$ results in an extended dynamic range allowing the measurement of the high-energy part of the spectrum, up to about 500 MeV. Bias voltage for the two SiPMs was provided by a custom designed board, with an on-board tunable DC-DC converter, working with 5V input voltage.  

\begin{figure}[t!] 
\center
\includegraphics[width=13cm,clip=true]{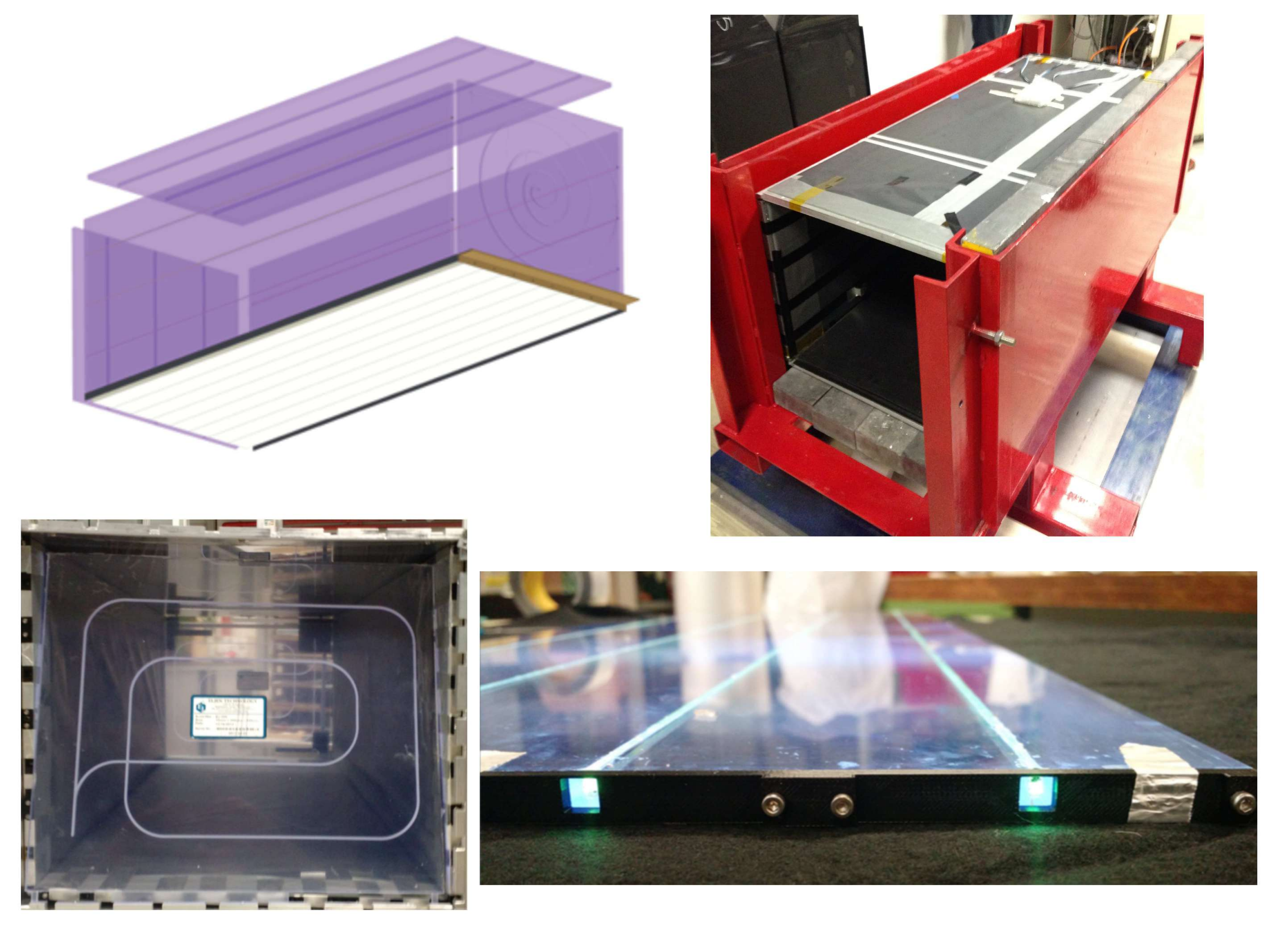}
 \caption{A sketch of the Inner Veto (top-left), the detector mounted inside the mechanical structure of the prototype (top-right), and two pictures of the upstream (bottom-left) and top (bottom-right) paddles.  
 }\label{fig:IV}
\end{figure}

The Inner Veto (IV)  is made by plastic scintillators, 1cm thick, forming a nearly hermetic parallelepiped (Fig.~\ref{fig:IV}). Two 35x42 cm$^{2}$ EJ200 scintillators are used for the downstream and upstream caps. On each of them a spiral groove hosts a WLS  fiber used to collect and transfer the light to a SiPM  (Fig. ~\ref{fig:IV} bottom-left). Three 35x140 cm$^{2}$ EJ200 scintillators form the top, left and right sides of the veto. In this case, the WLS fibers are inserted into four linear grooves running parallel to the long side of the plastic (Fig.~\ref{fig:IV} bottom-right). This solution results in an high detection efficiency ($>99.5\%$), almost independent  on the hit point, but still providing some  information also on the hit position, by correlating the quantity of light detected by each of the four independent SiPMs. Finally, in order to test another possible technology for the IV, the bottom side was made by four bars of extruded plastic scintillators, 8x140 cm$^{2}$, individually readout by a SiPM coupled to WLS fibers inserted in the middle of each bar. 

\begin{figure}[htbp]
   \centering
   \includegraphics[width=9.5cm,clip=true] {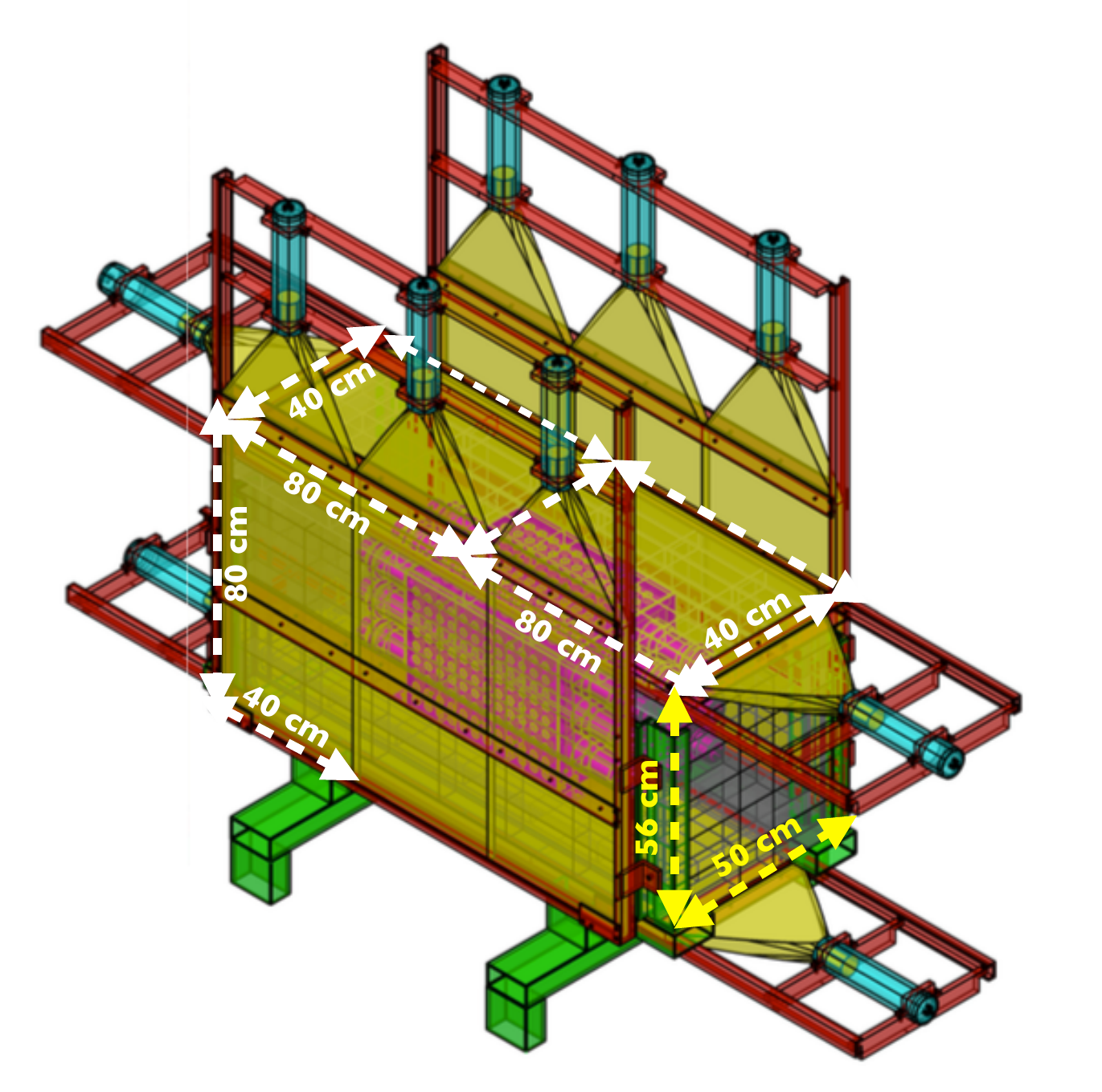}\hspace{-2cm}
   \includegraphics[width=7.5cm,clip=true] {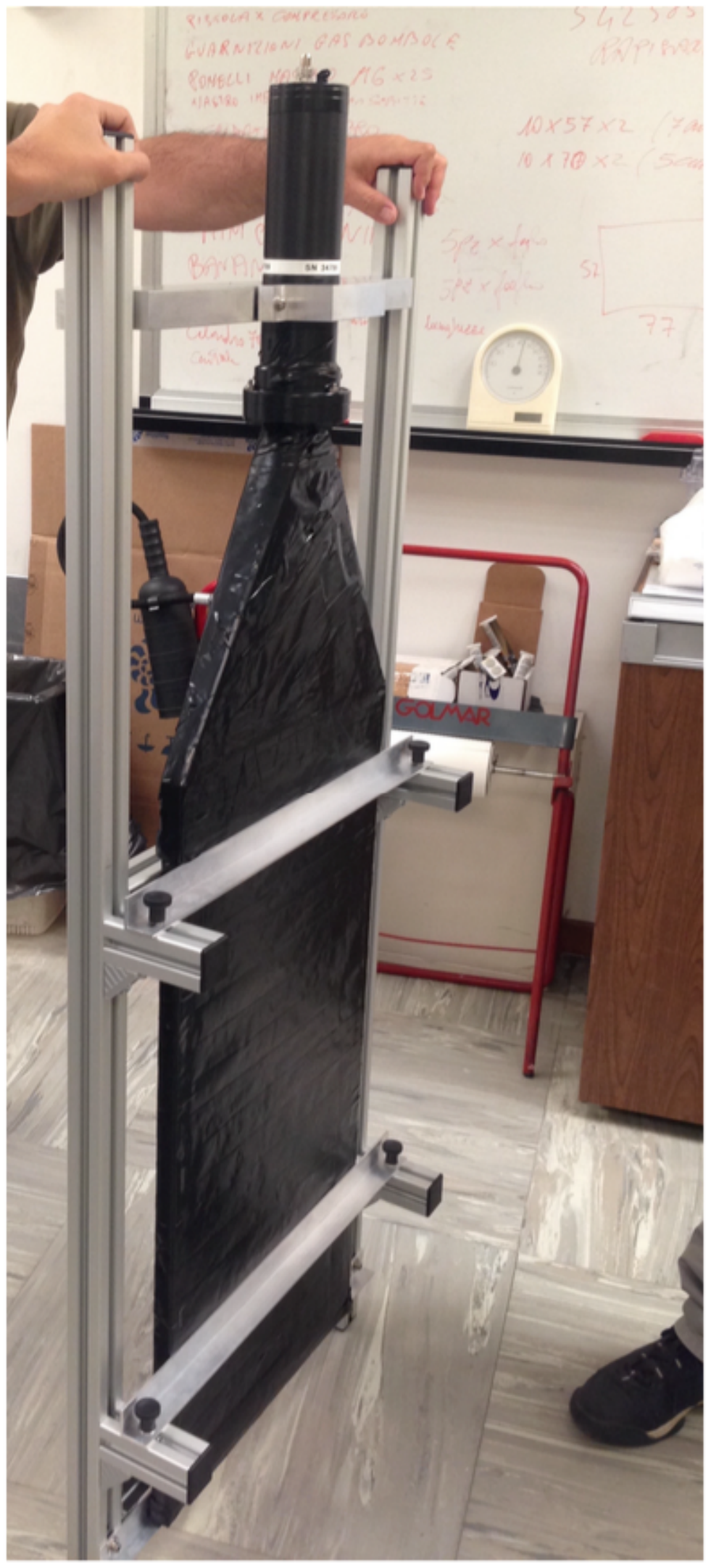}
   \caption{A CAD image of the prototype (left). The OV detectors and light guides are in yellow, the PMT in blue and their mechanical supports in red. On the right, an OV detector is shown inside the mechanical support used to glue the light-guide to the scintillator.}
   \label{fig:EV}
\end{figure}

The Outer Veto is made by 2cm thick NE110 plastic scintillators. The top (bottom) side is made by two 80x40 cm$^{2}$  paddles as shown in Fig. ~\ref{fig:EV}. Three scintillators of the same area are vertically arranged to cover each of the two lateral sides. A ``fish tail'' shaped PMMA light guide, glued on one side of the scintillator, directs the light to a 2" photomultiplier tube (Thorn EMI 9954A) optically matched with the light guide trough optical grease. A smaller paddle (56x50 cm$^{2}$) forms the upstream (downstream) cap. In this case, light is readout by a 1'' Photomultiplier tube (R1924A Hamamatsu) placed in the middle of the plastic surface and directly coupled to it through optical grease. A detection efficiency $>99.5\%$ was measured for each OV paddle for cosmic rays selected by triggering on the coincidence of two small paddles placed above and below the scintillator,  uniform over the whole scintillator surface.
\newpage
Data acquisition is based on VME-VXS JLab fa250 digitizers with 12 bit resolution, 250 MHz sampling rate and 2$\mu s$ readout window. The main trigger is generated by  a signal over threshold in the CsI(Tl) crystal, namely from the logic OR of the two SIPMs. The output signals are split by a 50 Ohm-50 Ohm divider: one copy is sent to the FADC and the other one to a Leading-Edge Discriminator with thresholds set to 15mV and 50mV for the 25 and 50 $\mu$m, respectively.  These thresholds correspond to about 5 p.e. for both SIPMs and, as derived  from the proton beam measurements (see Sec.~\ref{sec:onbeam}), they correspond to an energy threshold for protons of about 2 MeV. \\
Three other secondary triggers, conveniently pre-scaled, were also included for monitoring, calibration and efficiency studies: logic AND of the two small paddles, logic AND of two or more IV signals, logic AND of two or more OV detectors.   

\begin{figure}[t!] 
\center
\includegraphics[width=6.5cm]{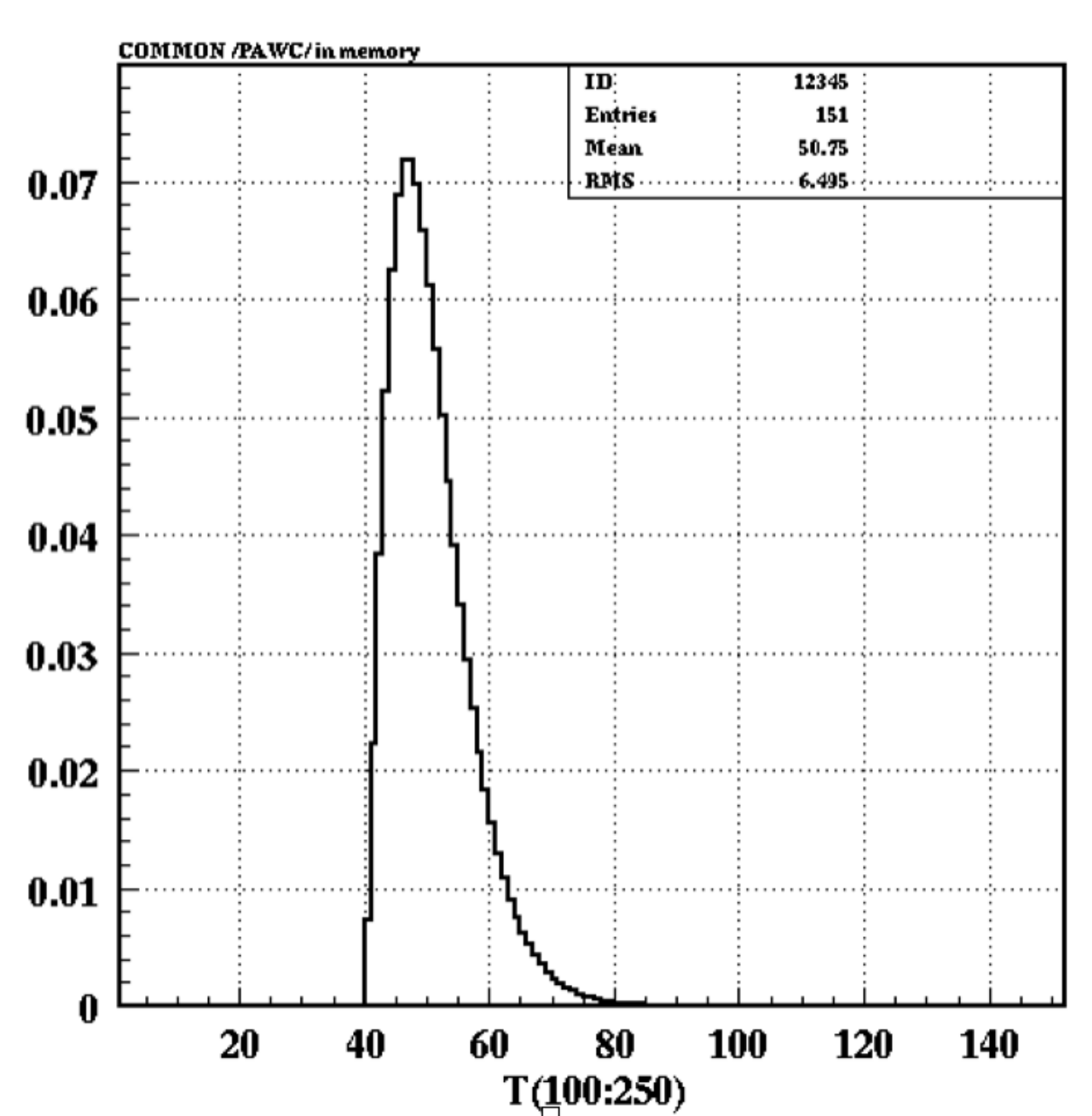}
\includegraphics[width=7cm]{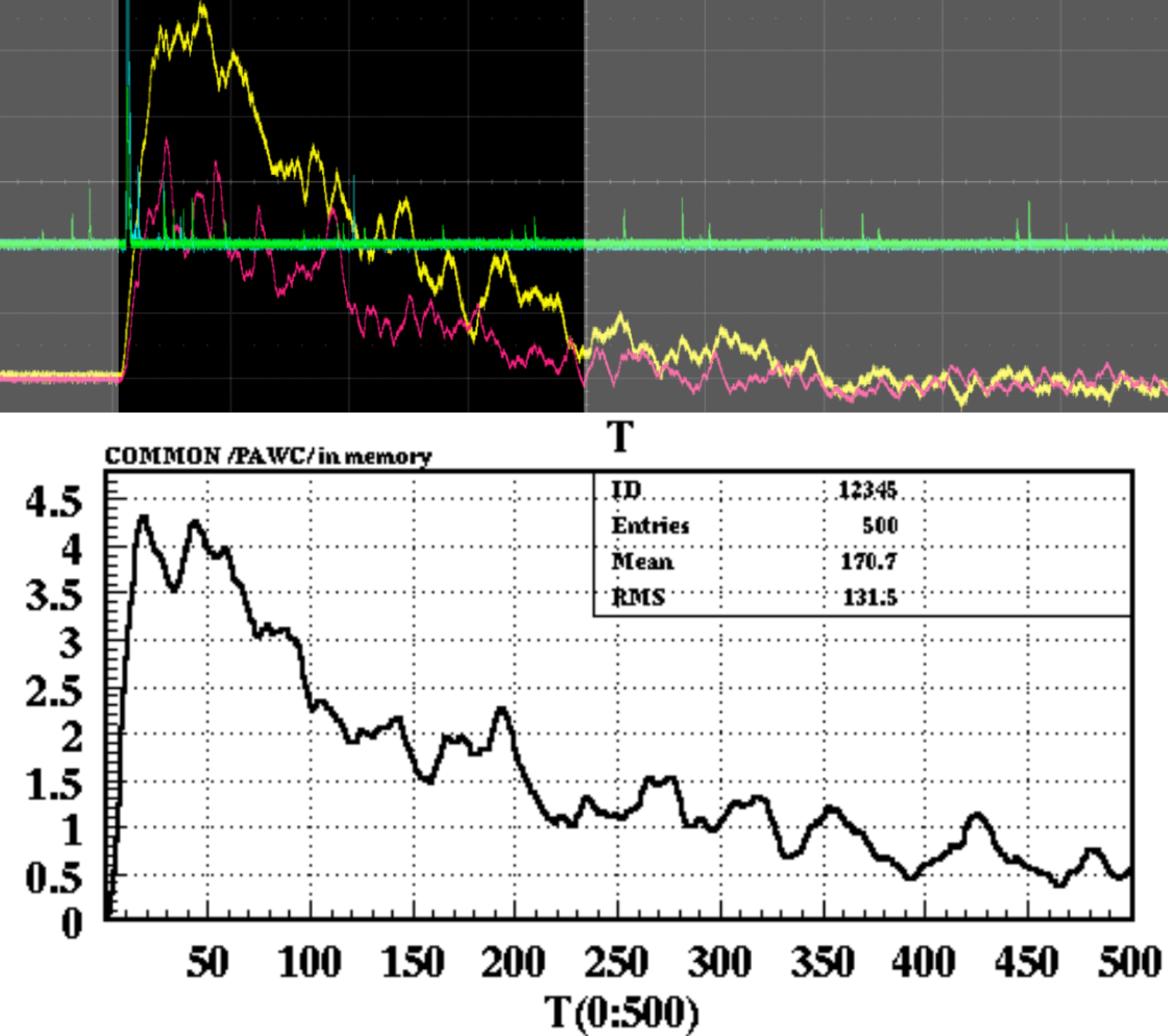}
 \caption{Left: Response of the preamplifier to a single p.e. (time is shown in 4 ns samples). Right: comparison of  the response to a crossing muon (top) and the result of the simulation (bottom). The simulation is limited to the highlighted  2$\mu s$ window (time is shown in 4 ns samples).}\label{fig:cry-res}
\end{figure}

\subsection{Simulation of the BDX prototype}
The realistic geometry as well as the material composition  of the BDX prototype have been implemented in GEMC (GEANT4) simulations. The response of individual components of the prototype (crystal, IV paddles and SIPM, OV paddles and lightguides plus PMTs)  have been measured by means of cosmic muons, parametrized, and included in simulations. The resulting  good agreement between data and MC for both cosmic muons and low energy protons will be  shown in the next Sections.
\begin{figure}[t!] 
\center
\includegraphics[width=15cm]{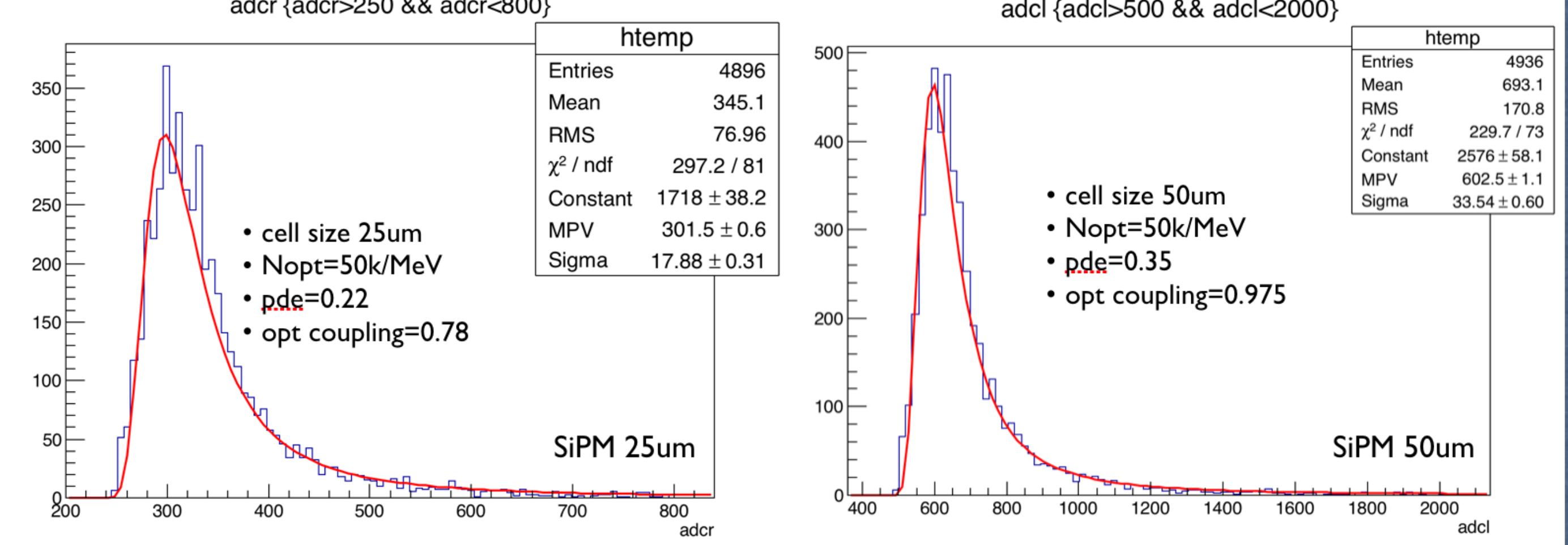}
 \caption{Simulated response of the crystal to crossing muons. Left and right panels correspond to the 25 and 50$\mu$m cell size respectively.
 }\label{fig:cry-res1}
\end{figure}
\subsubsection{The crystal response}\label{sec:crys-response}
The CsI(Tl) crystal is coupled to two 3x3 mm$^2$ SIPMs sensitive to a single p.e. The signal is then amplified by a trans-impedance preamplifier whose response to a single p.e. has been measured and parametrized  as reported in the left plot of Fig.~\ref{fig:cry-res}. Due to the sizable scintillation time of CsI(Tl), the response to  N$_{p.e.}$ can be described as a convolution of  the single p.e. response with the time distribution of the scintillation  signal.  A comparison of the measured signal for a cosmic muon and the results of the simulation is reported in the right plot of Fig.~\ref{fig:cry-res}. For highly ionizing particles, light quenching has been included in the simulation using a Birk constant of 3.2e-3 g/(MeV cm$^2$) (see Ref.~\cite{Tretyak:2009sr}). A light  emission yield of 50k $\gamma$/MeV, as reported by the BaBar Collaboration~\cite{Aubert:2001tu} and checked in Genova, and an attenuation length of $\sim60$cm, as measured exposing the crystal to a focused proton beam (see Sec.~\ref{sec:onbeam}), have been used in the simulations. The absolute energy calibration has ben obtained by matching the crossing muon Landau distribution to what obtained by the simulations. The resulting simulated signals measured by the two SIPMs are reported in Fig.~\ref{fig:cry-res1}

\subsubsection{The IV and  OV response}
\begin{figure}[t!] 
\center
\includegraphics[width=15cm]{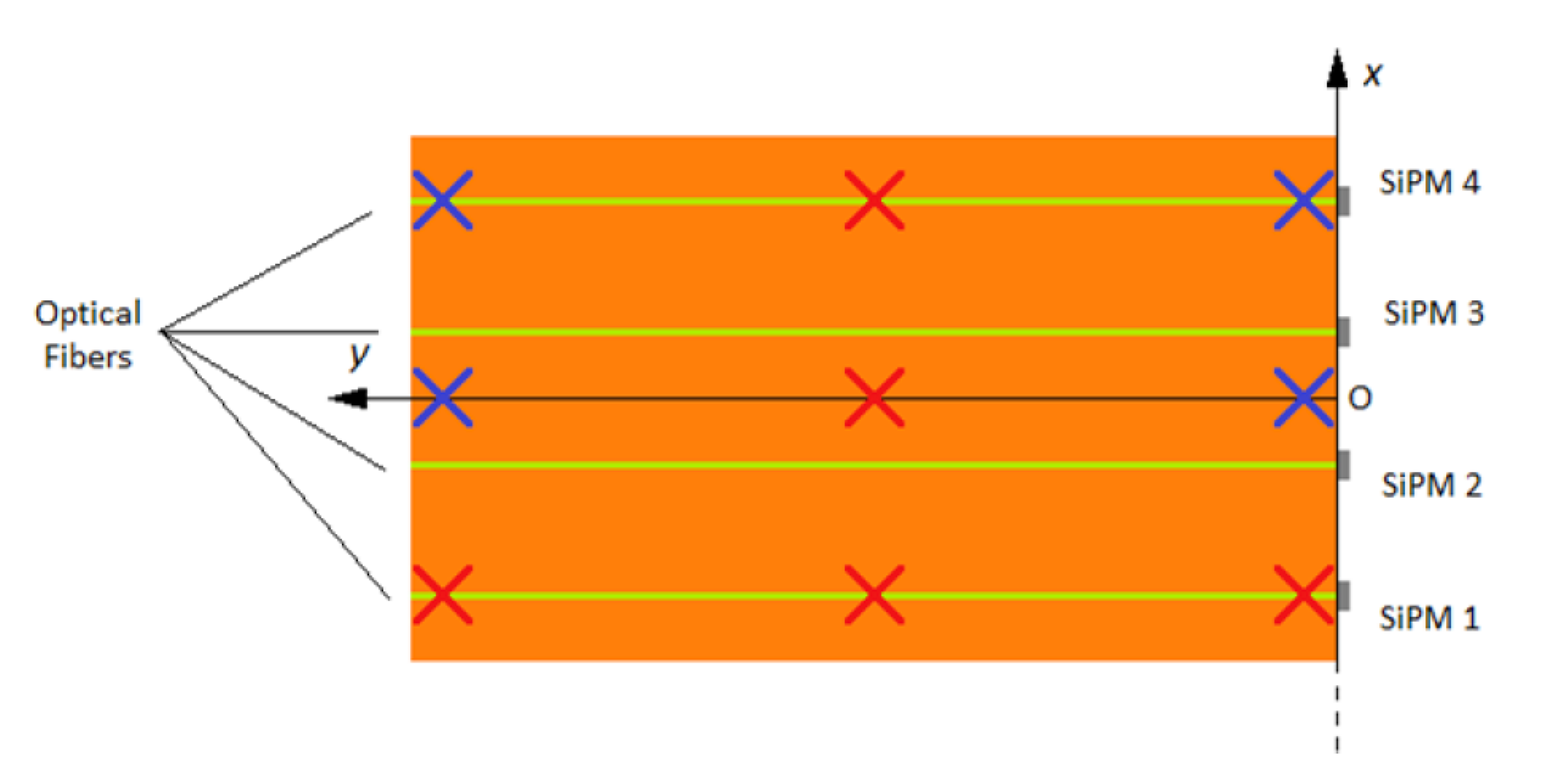}
 \caption{Position of the four fibers and SiPMs in the X-Y plane of an IV paddle.
 }\label{fig:IV-1}
\end{figure}
\begin{figure}[t!] 
\center
\includegraphics[width=7.5cm]{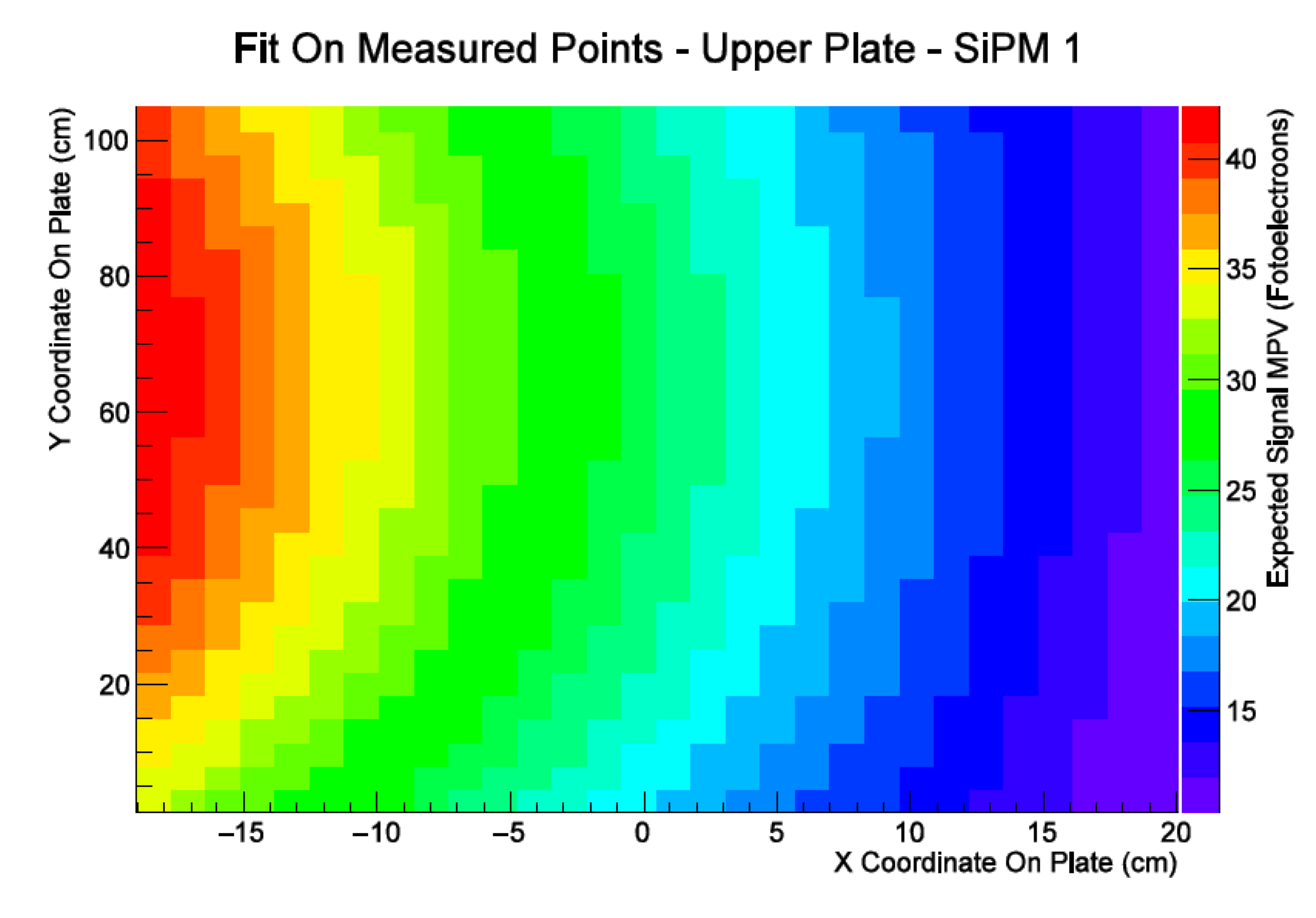}
\includegraphics[width=7.5cm]{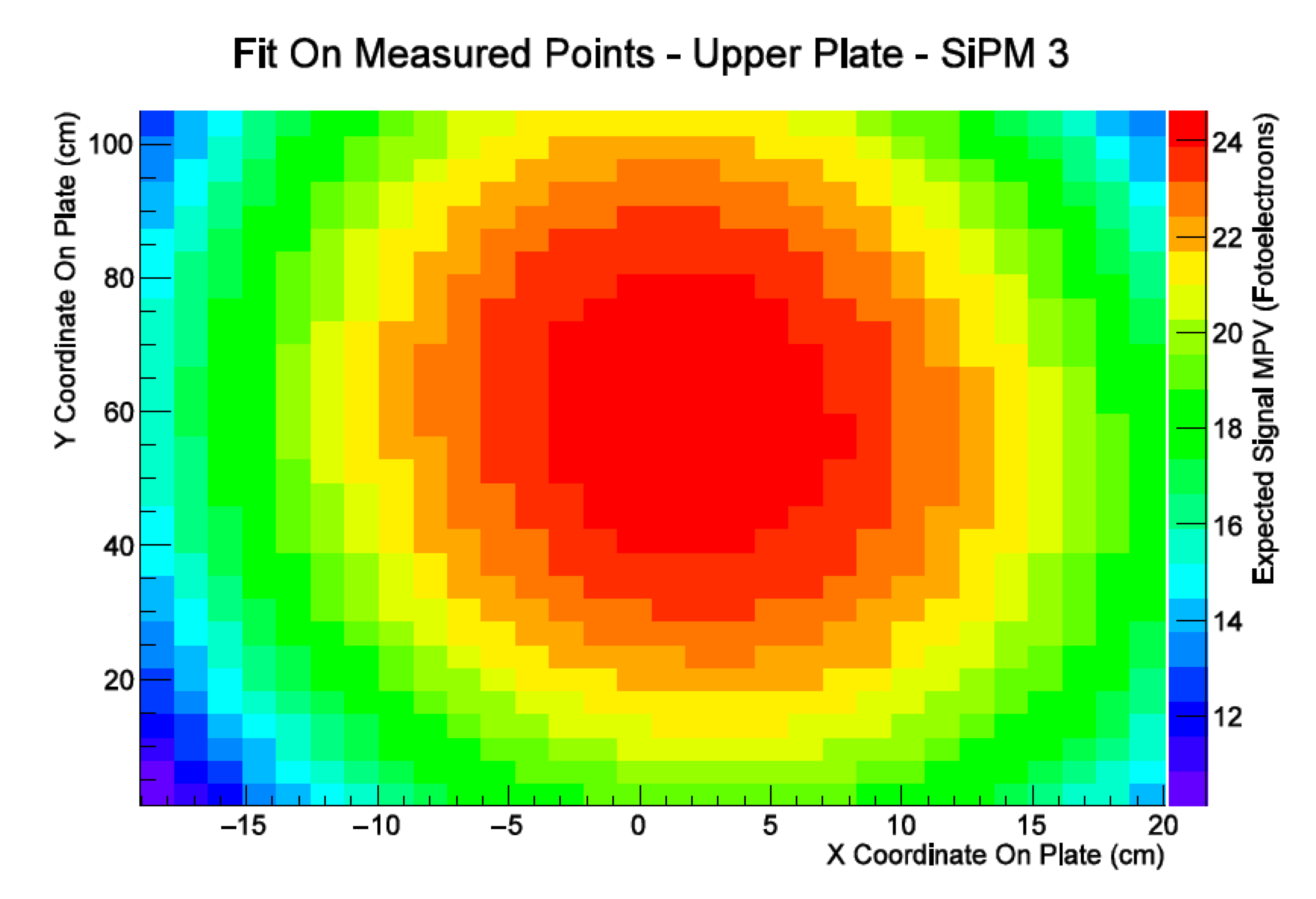}
 \caption{Response of SIPM 1 and 3 of the IV upper paddle. The SIPM is sensitive not only to the region in front of it but also to the all hit positions.
 }\label{fig:IV-3}
\end{figure}

Since a detailed description of the light collection in WLS fibers, transmission to the SiPM and coupling strongly depends on the manufacturing of the detector (groove polishing, fiber polishing, gluing ..), we decided to measure the response of the Inner Veto, parametrize it   and implement it  in the simulations. Figure~\ref{fig:IV-1} shows the position of the four fibers in the X-Y plane. Being the plastic transparent and the four grooves not optically separated, each SIPM  is mainly sensitive to the area around the fiber but can also detect  light produced farer. The combination (.OR.) of the four SiPM signals strongly reduce the paddle inefficiency to a crossing particle since the four independent photo sensor are always involved. The N$_{p.e.}$ response of each SiPM, for each paddle, as a function of the hit position has been measured, fitted to a polynomial function and included in the simulations. As an example, Fig.~\ref{fig:IV-3}  shows the response of the first and the third SIPM of the TOP paddle. Similar results were obtained for the other paddles. The simulated response include the N$_{p.e.}$ Poissonian statistics as well a Gaussian spread that has been derived by the shape of the measured Landau distributions for crossing cosmic muons.

Due to the more traditional  technology used in the OV, plastic scintillator couple to a PMT via a plexiglas light guide, the response of the Outer Veto was parametrized using some  standard techniques:  plastic light yield, attenuation length,  optical coupling, PMT quantum efficiency, ratio of the optical area. The results for crossing muons have been compared to measured distributions and the Landau peak positions adjusted to the data.

\subsection{The INFN-CT and LNS configurations}
\begin{figure}[t!] 
\center
\includegraphics[width=15cm]{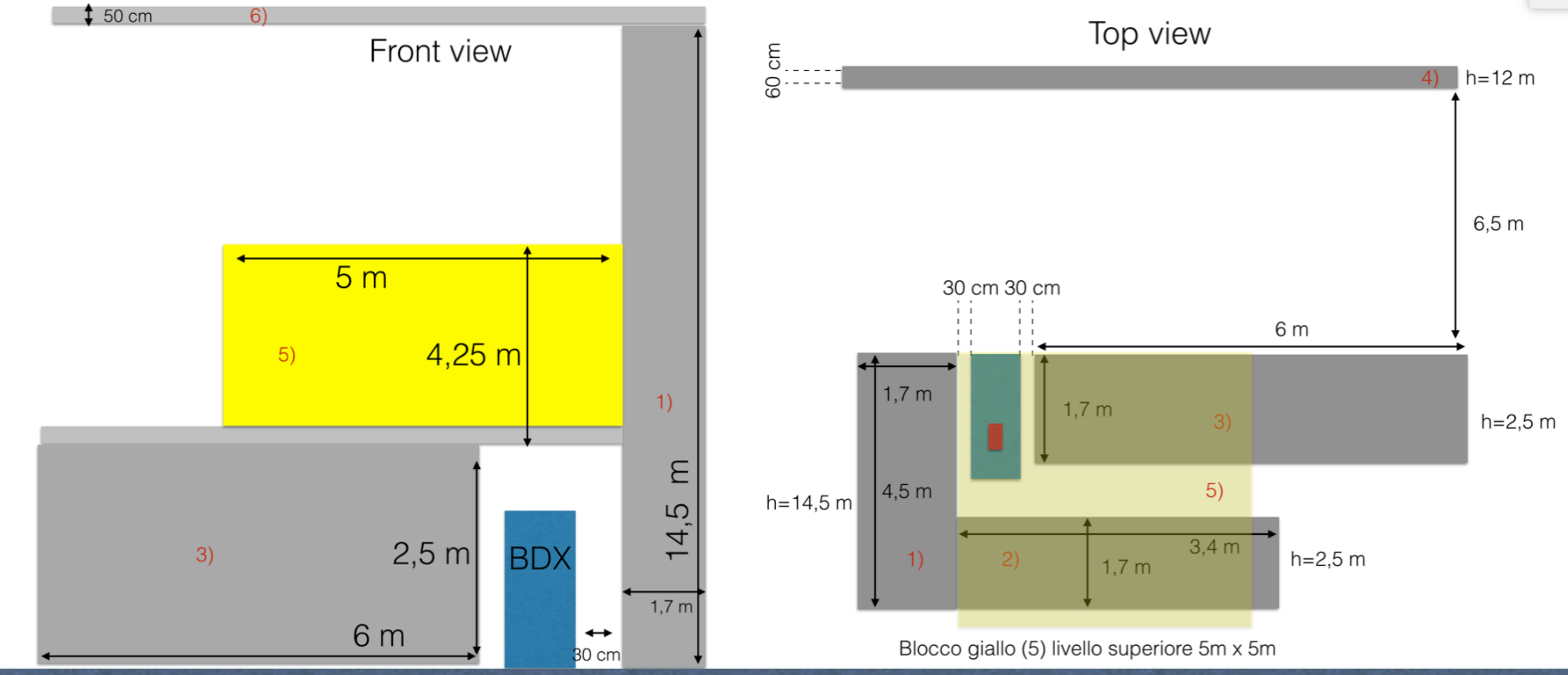}
 \caption{Front (left) and top (right) view of the LNS set up. The precise BDX prototype position is shown in the two drawings.}\label{fig:LNS-conf}
\end{figure}

As mentioned at the beginning of this Section, we exposed the prototype to cosmic rays in two different experimental set up. The first set of data was taken with the prototype placed in a room on the last floor of the INFN-CT. The concrete roof was the only shielding present.  We use this configuration as a benchmark to validate the model of the cosmic muons and cosmic neutrons implemented in the simulations.
The second, and current, set-up positioned the prototype in a bunker at LNS shielded by 470 cm of concrete walls. This configuration is very similar to the JLab overburden proposed for this experiment. Figure~\ref{fig:LNS-conf} shows the front and the top view of LNS set up. The two configurations were implemented in a simplified way into GEMC considering the detector located within a cube made of concrete 15 cm (470 cm) thick for the INFN-CT (LNS) configurations, respectively. 

\subsection{Cosmic muons and cosmic neutrons}
\begin{figure}[t!] 
\center
\includegraphics[width=16cm]{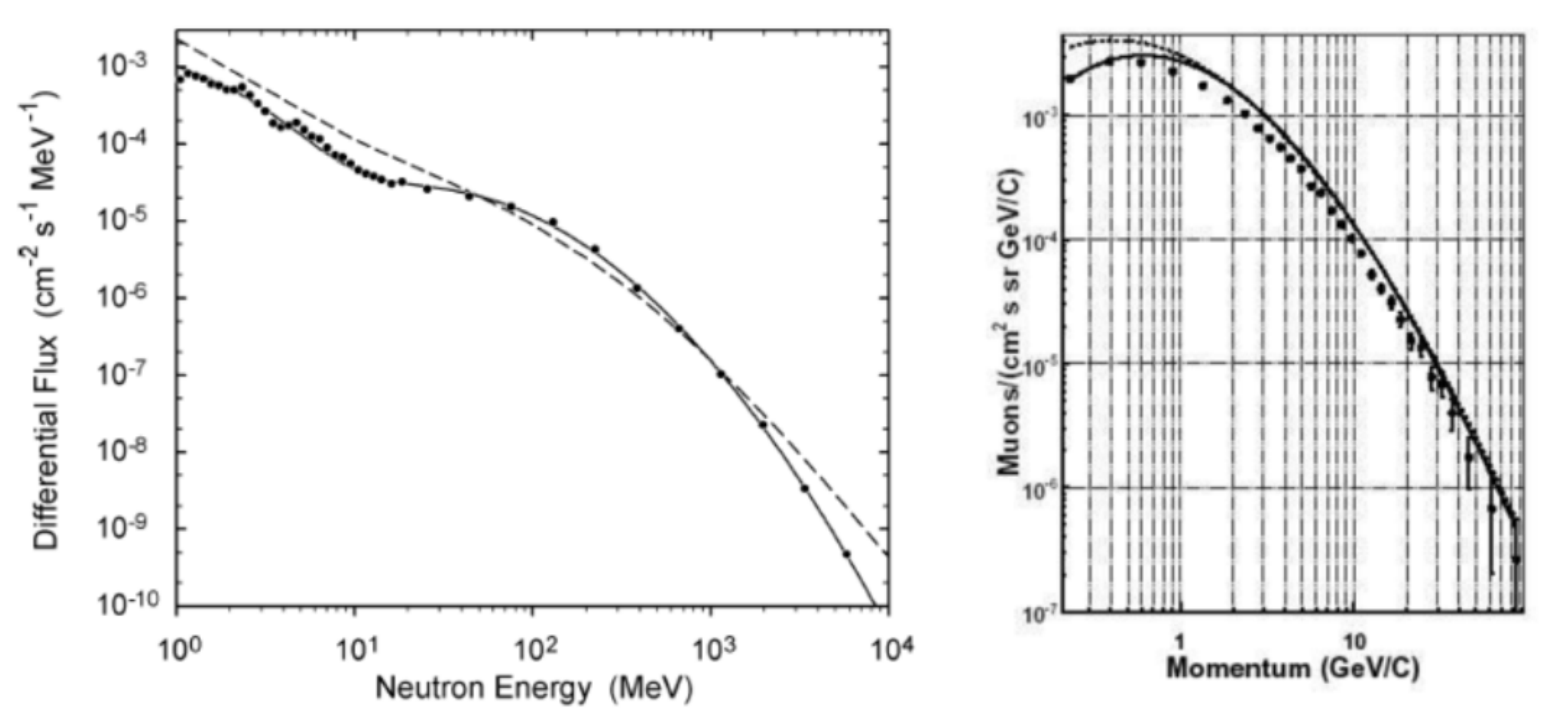} 
 \caption{Flux of cosmic neutron (left) and muons (right). }
 \label{fig:cosmic} 
\end{figure}

The cosmic muon and cosmic neutron energy spectra have been implemented directly in GEMC.  For muons, we used the energy and angular distribution reported in Ref.~\cite{Appleton:1971kz}, for neutrons the parametrization from Ref.~\cite{1369506,1589260}. Fig.~\ref{fig:cosmic} shows the two energy spectra. Cosmic particles were generated in a fiducial volume big enough to  contain the detector and a careful normalization has been performed to correctly take into account the crossing on the lateral sides (the correction with respect to a flat top surface is on the order, at most, of 20\%, depending on the generation volume). Particles found to cross the fiducial volume where then projected far away and the production vertex extracted outside the shielding. This procedure is not completely correct for neutrons and low energy muons that may undergo a significant multiple scattering effect or produce other particles hitting the shielding and, indeed, not crossing any more the detector.
This effect was corrected keeping the generation volume big enough to account for  deviation in the trajectory and include contributions from secondary produced in the surrounding materials. The generation has been limited to the upper half of the solid angle, considering a null direct  flux from the bottom (this does not prevent secondaries to bounce on the floor and enter in the fiducial volume).
\subsection{Results of cosmogenic background}
The absolute rates of  muons crossing the crystal considering  different conditions on veto counters (no conditions, OV anti-coincidence, IV anti-coincidence) compare well (at level of 10\%)  to the experimental rates  providing a validation of the simulations (for both the used cosmic spectra and the detector response). The remaining  rate ($\sim 1.8 10^{-3}$ Hz for the LNS configuration, with 10 MeV threshold) measured by the crystal  when  both OV and IV are in anti-coincidence shows a significant difference indicating that a neutral component is present in the data and not fully understood\footnote{The experimental rate of the crystal when both the OV and IV are put in anti-coincidence, does not scale as the product of the two separate OV and IV ineffiency as indicated by the simulation. We believe that a contribution from neutron or gamma, not correctly implemented or reproduced by the simulations, is responsible for this tiny but important contribution.}. Nevertheless the good agreement obtained between simulations and data for crossing muons in different experimental conditions (INFN-CT and LNS) and the good agreement between simulated LNS and Jefferson Lab set up proof that experimental results obtained with the prototype at LNS can be safely extrapolated to the full detector in the JLab setup.

\subsubsection{Results from INFN-CT data}
The first measurement campaign with the BDX-prototype has been conducted at INFN-CT, with the primary goal of measuring the absolute rates of muons crossing the crystal to compare it with simulations. Also, the simplified setup permitted to commission the detector before moving it to the final position at LNS, as well as to implement and validate the data-reconstruction framework. Data was measured in February - March 2016, for about 1 month of data-taking.

The following data-reconstruction procedure was adopted. For the CsI crystal, both waveforms were numerically integrated within a 1 $\mu$s time windows to obtain the corresponding charge, that was then converted in MeV units by using calibration constants deduced from the cosmic-rays Landau distribution most probable value. In doing so, the following effect was observed: due to the long CsI scintillation decay time, events with low energy deposition (less than  10 MeV) could result in a signal with a rising edge not monotonic, but dominated by single phe signals. The trigger crossing time could thus be artificially delayed, due to the signal reaching the analog threshold later in time. This results in a time disalignement between the FADC sampling window and the veto signals, that are not recorded for these events. Therefore, particular care was taken in order to reject these cases from the analysis, by evaluating the CsI SiPM waveforms pedestal average and standard deviation, event by event, and comparing these with reference values. The fraction of rejected events is $\simeq 55\%$ for $10$ MeV deposited energy, and flattens to $\simeq 70\%$ at higher energies. All the event rates have been corrected for this effect.


For the inner-veto, after numerical integration, a tight coincidence cut (100 ns) was applied for SiPM signals referring to the same plastic scintillator counter, requiring a multiplicity equal or greater than two SiPMs. This permitted to use a very-low threshold (2.5 phe), while maintaining the contribution of thermal-induced signals negligible (the latter was evaluated by a random pulser trigger, and found to be less than 5$\%$). For counters with single-SiPM readout, an higher threshold was applied (12.5 phe), resulting to the same thermal-induced signals contribution. For the outer-veto and for the two small inner paddles, each PMT signal was numerically integrated, and after this a 100 keV energy threshold was applied. 

\begin{figure}[t]
\centering
\includegraphics[width=.8\textwidth]{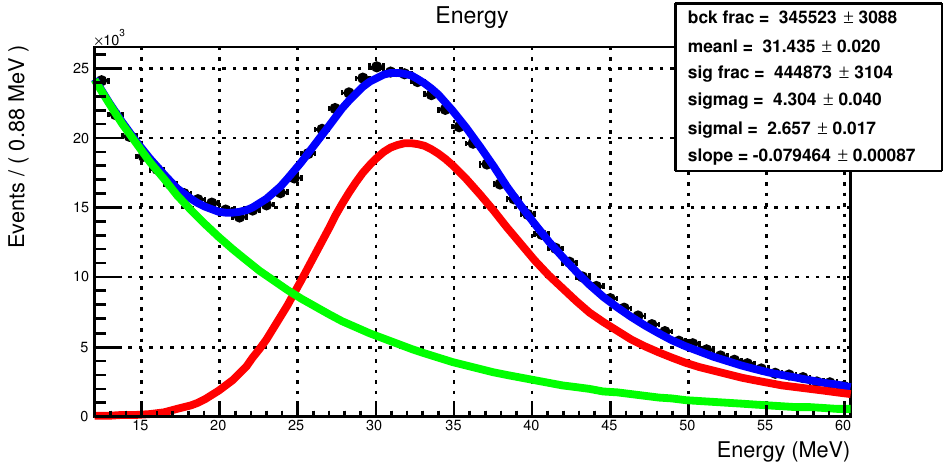}
\caption{\label{fig:INFN-CT-Cosmics} Example of a Maximum Likelihood fit to the CsI energy spectrum to measure the rate of cosmic muons crossing the crystal. The fit has been performed modeling the background (green) with an exponential function. The event yield has been obtained by integrating the signal (red) in the full energy range.}
\end{figure}

To evaluate the rate of cosmic muons crossing the crystal, the following events topologies were considered:
\begin{itemize}
\item{All events}
\item{Events with the Top and Bottom inner-veto counters in coincidence}
\item{Events with the Top and Bottom outer-veto counters in coincidence}
\item{Events with both inner-veto and outer-veto Top and Bottom counters in coincidence}
\end{itemize}
For each topology, the CsI energy distribution of selected events was fitted with a Landau function convoluted with a Gaussian response, plus a proper background contribution. For the latter, different functional forms were employed for each topology (exponential, polynomial, \ldots). An example of fit is reported in Fig.~\ref{fig:INFN-CT-Cosmics}. The expected rate was obtained by integrating the signal component, and then dividing by the corresponding measurement time. The same procedure was adopted for MC simulations. The most important uncertainty contribution to the results, reported in Table~\ref{tab:protoCosmics}, is the systematic error in the background parametrization and in the fitting range. This has been evaluated to be of the order of $10-20\%$, depending on the considered topology. 

\begin{table}[htpb]
\centering
\begin{tabular}{|c||c|c|c||c|c|c|}
\hline
\multirow{2}{*}{\textbf{Event Topology}} & \multicolumn{3}{|c||}{\textbf{INFN-CT}} & \multicolumn{3}{|c|}{\textbf{LNS}} \\
\cline{2-7}
& Data rate & MC rate  & Ratio  & Data rate & MC rate& Ratio \\
\hline
All events & 1.40 Hz & 1.32 Hz & 1.06 & 0.46 Hz &0.31 Hz &1.48 \\
\hline
Inner-veto coincidence & 1.11 Hz & 1.38 Hz & 0.80 & 0.36 Hz & 0.37 Hz & 0.97 \\
\hline
Outer-veto coincidence & 0.99 Hz & 1.25 Hz & 0.79 & 0.33 Hz & 0.36 Hz & 0.92 \\
\hline
Both veto systems coincidence & 0.95 Hz & 1.24 Hz & 0.77 & 0.32 Hz & 0.36 Hz & 0.89 \\
\hline
\end{tabular}
\caption{\label{tab:protoCosmics} Comparison between the measured rate of cosmic muons crossing the CsI crystal, for different selection criteria, and the predictions from MC simulations, for both the INFN-CT and LNS measurement campaigns. The statystical error on the above results is negligible, compared to the systematic contribution, of the order of $10-20\%$, depending on the event topology.}
\end{table}

\subsubsection{Results from LNS data}

After commissioning the detector and validating MC results, the BDX-prototype was moved to the final position at LNS. Here, after repeating the cosmic ray measurement rate and comparing it again with simulations, absolute rates for different anti-coincidence configurations were measured, as a function of the deposited energy in the crystal. The results were then extrapolated to the full BDX detector configuration, as described in Sec.~\ref{sec:beam_unrelated_bck}. The measurement campaign started in April 2016: the results here presented corresponds to about 1 month of data-taking. 

\begin{figure}[t]
\centering
\includegraphics[width=\textwidth]{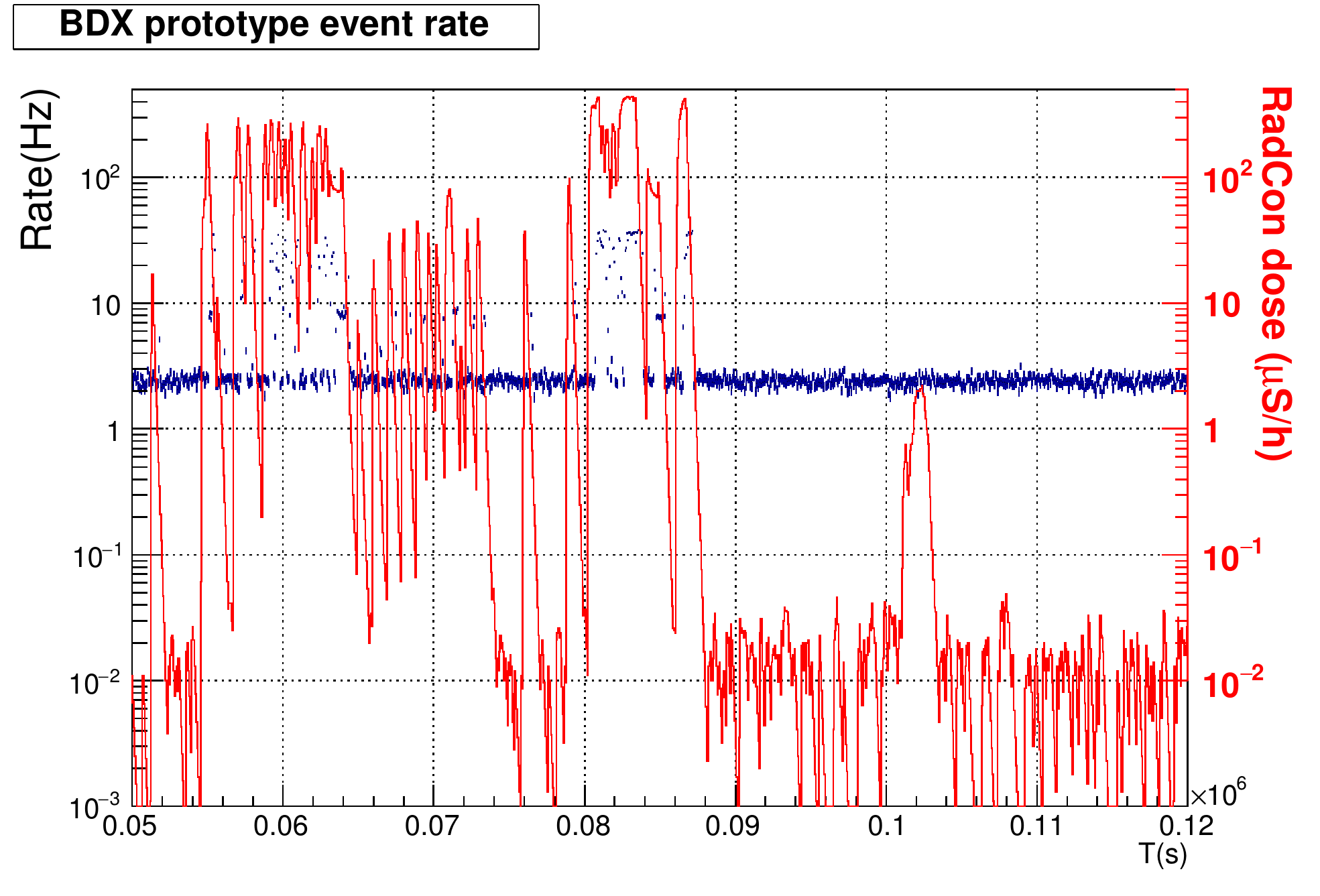}
\caption{\label{fig:INFN-LNSratedose} Comparison between the absolute event rate measured with the BDX prototype in the LNS configuration and the instantaneous dose measured close to the nearby accelerator, reported by the LNS RadCon, for a selected data-taking period. A clear correlation is visible. We also observed periods of significant accelerator activity with no corresponding increase in the overall event rate (for example, close to $T=0.1\cdot 10^{6}$s).}
\end{figure}

The same event reconstruction procedure described in the previous section was adopted. We also excluded from the analysis events corresponding to periods with activity in one of the nearby LNS accelerators, as reported by the LNS RadCon service. A comparison of the measured rate in the BDX-detector with RadCon data is reported in Fig.~\ref{fig:INFN-LNSratedose}: a clear rate enhancement correlated with accelerator activities is visible. Although one would expect that this events would result in a low ($<$ 1 MeV) energy deposition in the crystal, we saw contributions up to $10$ MeV, hence the decision of excluding these periods from the data-analysis.

The following anti-coincidence configurations were considered. By ``anti-coincidence'' we mean that we selected events with no activity in any of the veto counters.
\begin{itemize}
\item{Anti-coincidence with inner veto}
\item{Anti-coincidence with outer veto}
\item{Anti-coincidence with both veto systems}
\end{itemize}
Preliminary studies with anti-coincidence with one of the veto systems, but a signal in the other system were also performed, in order to study the possible inefficiencies.  

\begin{figure}[htpb]
\centering
\includegraphics[width=.48\textwidth]{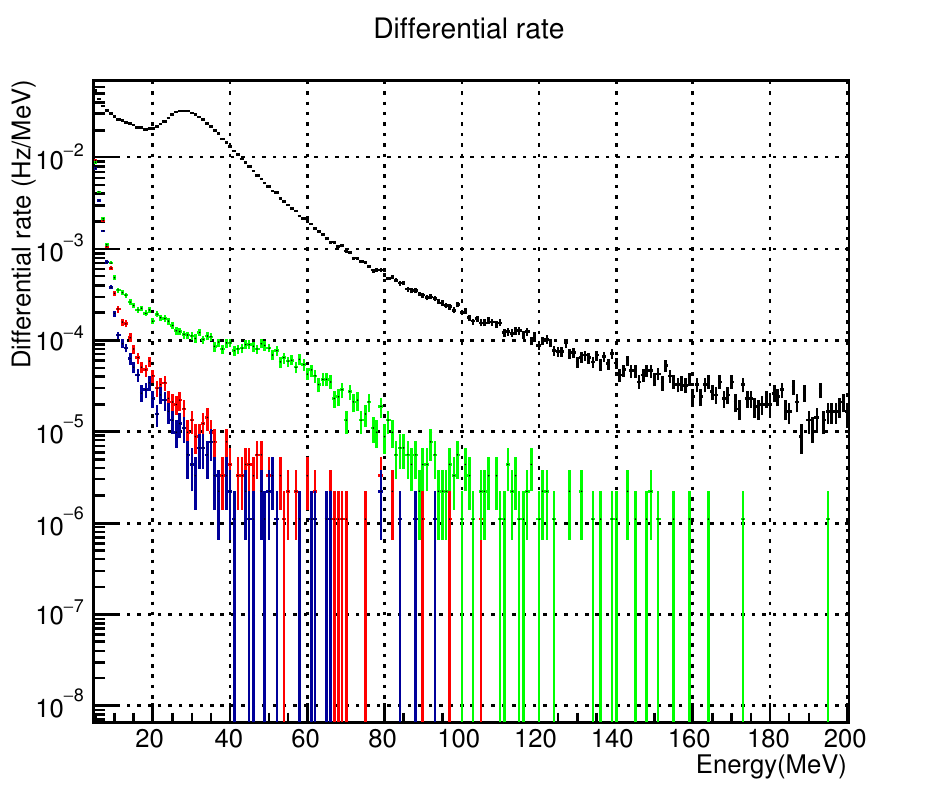}
\quad
\includegraphics[width=.48\textwidth]{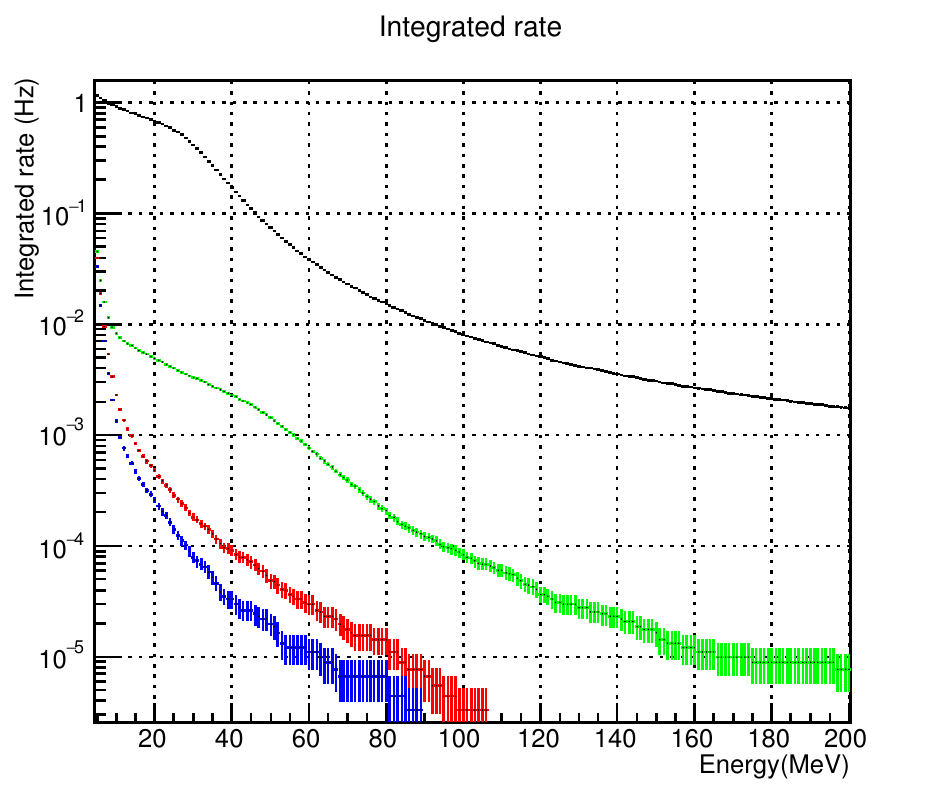}
\caption{\label{fig:INFN-LNSlowenergy} Left plot: the measured event rate as a function of deposited energy in the crystal. Different colors refer to the different anti-coincidence selections: black - all events, red - anti-coincidence with inner veto, green - anti-coincidence with outer veto, blue - anti-coincidence with both veto systems. Right plot: the integrated event rate as a function of the energy threshold (same color scheme).}
\end{figure}

Results corresponding to the low-energy part of the CsI energy spectrum are shown in Fig.~\ref{fig:INFN-LNSlowenergy}. The integrated rate, requiring the anti-coincidence with both veto systems, for a 10 Mev (20 MeV) threshold, is 1.7 mHz (0.37 mHz) - after applying all the necessary corrections. Results are summarized in Table~\ref{tab:INFN-LNSlowenergy}. 

\begin{table}[tpb]
\centering
\begin{tabular}{|c|c|}
\hline
Energy threshold & Measured rate \\
\hline
5 MeV & $(4.63\pm 0.03)\cdot10^{-2}$ Hz \\
\hline
10 MeV & $(1.88\pm 0.05)\cdot10^{-3}$ Hz \\
\hline
15 MeV & $(6.58\pm 0.03)\cdot10^{-4}$ Hz \\
\hline
20 MeV & $(3.67\pm0.02)\cdot10^{-4}$ Hz \\
\hline
\end{tabular}
\caption{\label{tab:INFN-LNSlowenergy} The inner-veto and outer-veto anti-coincidence rate for the BDX-prototype, measured in the LNS configuration.}
\end{table}

The analysis of the high-energy part of the spectrum (E$>$100 MeV) is, instead, more critical, due to the low statistics. In particular, for the accumulated statistics, corresponding to $1.9\cdot 10^6$ s of measurement, the following events were observed:
\begin{itemize}
\item Two events with $E\simeq 500$ MeV, in anti-coincidence with the outer veto, but with an intense activity in the inner veto. One of these two events was measured in a period with a non-zero activity of the nearby accelerators.
\item One event with $E=260$ MeV, with no activity in any veto. A deep scrutiny of this event (looking at the acquired waveforms) shows the following criticities: the signal decay time is a factor 2 shorter than the expected CsI value (500 ns vs 1$\mu$s), and the signal measured by the 50 $\mu$m SiPM is $98\%$ lower - although the corresponding PDE is about twice. Therefore, we decided to exclude this event from the analysis.
\end{itemize}

\begin{figure}[htpb]
\centering
\includegraphics{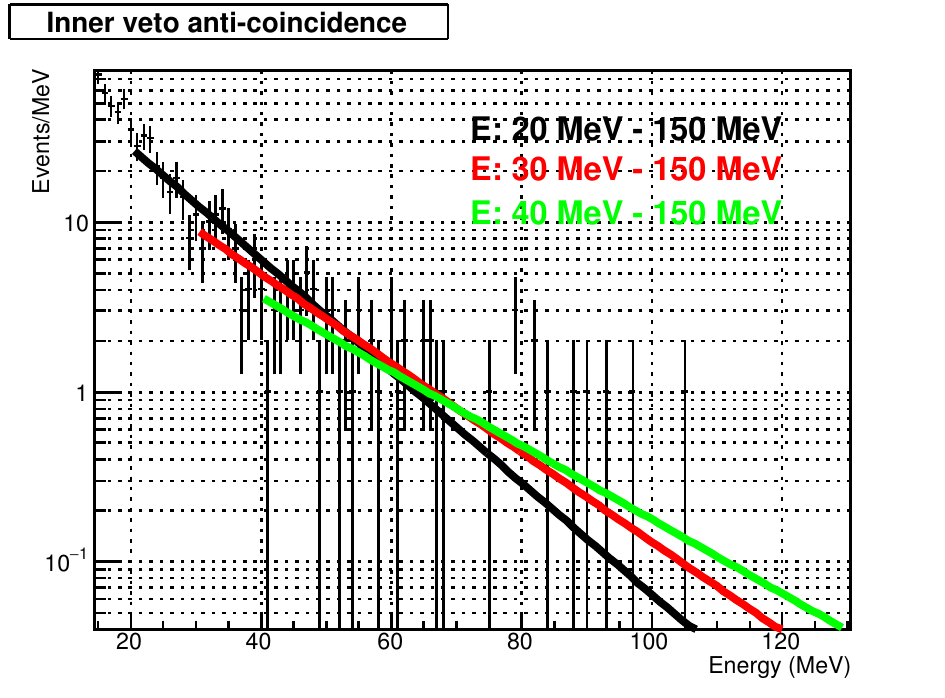}
\caption{\label{fig:INFN-LNShighenergy} Extrapolation of the measured CsI energy spectrum in anti-coincidence with the inner veto from the low-energy region, where a non-zero event rate was measured, to the high energy region, where no events were detected. The three curves refer to Maximum Likelihood fits performed in different energy ranges.}
\end{figure}

For the outer-veto anti-coincidence selection, the extrapolation of the single high-energy event (with $E\simeq 500$ MeV, no activity in the outer veto, intense activity in the inner veto) to a final result is clearly problematic, and for the inner-veto anti-coincidence, with no measured events, the procedure is even more critical. Therefore, we decided to proceed as follows. We considered the lower part of the spectrum - where a significant statistics has been measured - and performed an extrapolation to the higher energy regime trough a Maximum Likelihood fit, using an exponential function. The systematics associated with this procedure has been evaluated by changing the fit range and comparing the obtained results (see Fig.~\ref{fig:INFN-LNShighenergy}). Results are reported in Tab.~\ref{tab:INFN-LNShighenergy} for the inner-veto anti-coincidence selection and for different energy thresholds. In a conservative approach, the reported values refer to fit performed in the range 40 MeV - 150 MeV, resulting in the largest projection at high energy, and the quoted error is the RMS of the three results from fits performed starting from 20, 30, and 40 MeV. No higher ranges were considered for fitting, due to the lack of statistics.

\begin{table}[tpb]
\centering
\begin{tabular}{|c|c|}
\hline
Energy threshold & Extrapolated rate \\
\hline
200 MeV & $(3.6\pm1.5)\cdot10^{-8}$ Hz \\
\hline
250 MeV & $(2.9\pm1.3)\cdot10^{-9}$ Hz \\
\hline
300 MeV & $(2.4\pm1.1)\cdot10^{-10}$ Hz \\
\hline
350 MeV & $(1.9\pm0.9)\cdot10^{-12}$ Hz \\
\hline
\end{tabular}
\caption{\label{tab:INFN-LNShighenergy} The inner-veto anti-coincidence rate for the BDX-prototype, measured in the LNS configuration, obtained by extrapolating from the low energy part of the spectrum.}
\end{table}
\subsection{Results of on-beam measurements}\label{sec:onbeam}

\begin{figure}[htbp]
   \centering
   \includegraphics[height=6cm,clip=true] {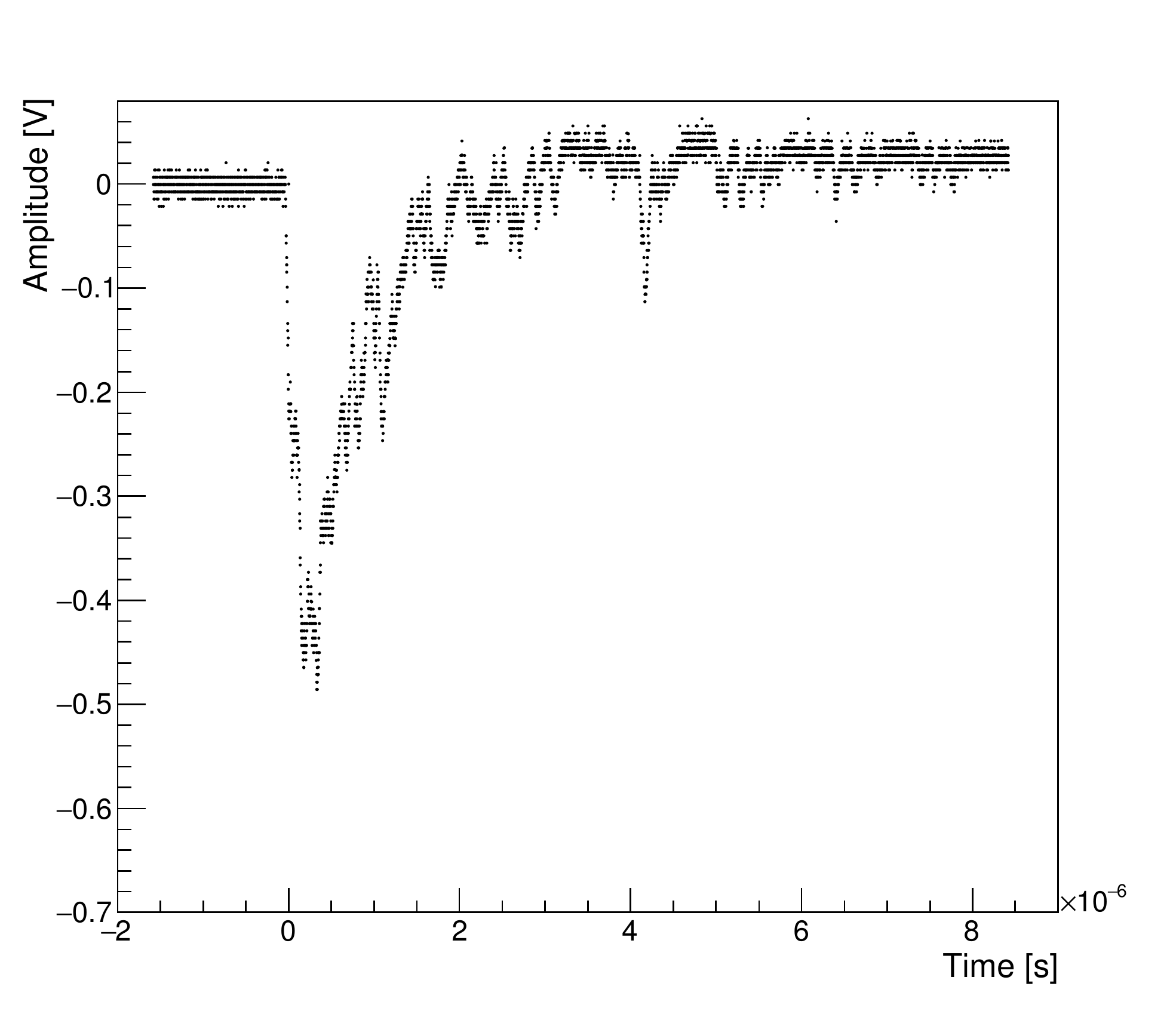}
   \includegraphics[height=6cm,clip=true] {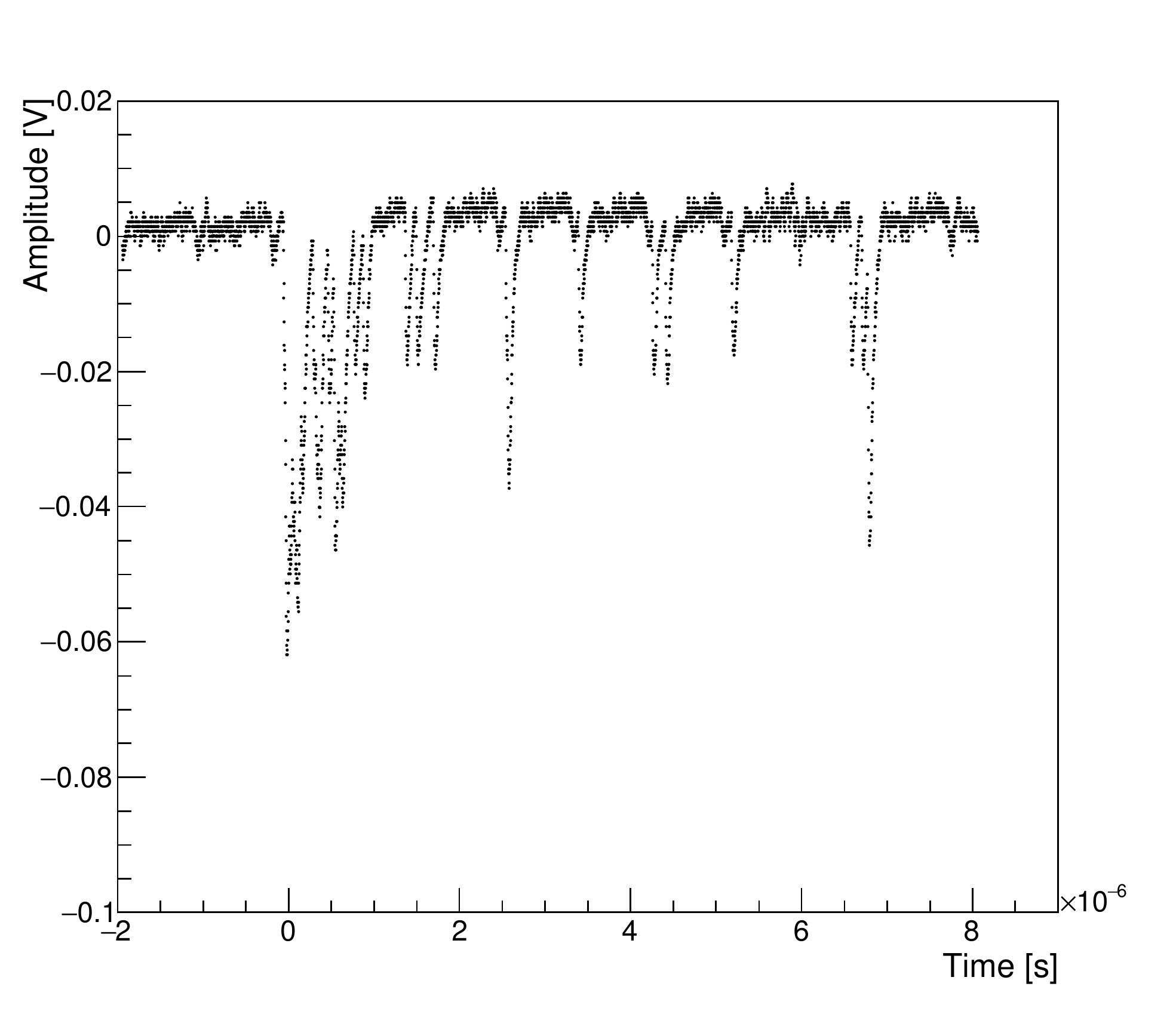}
   \caption{Signals generated by protons of 24 MeV (left) and 2.5 MeV (right) for the SiPM with a pixel size of 25 $\mu$m.}
   \label{fig:signals}
\end{figure}

The CsI(Tl) crystal was irradiated with mono-energetic protons of T$_p$ 24, 20, 18, 16, 14, 12, 10, 9, 8, 7, 6, 5, 4, 3 and 2.5 MeV. For each incident energy the beam was focused on 4 positions along the long side of the crystal. Each point was located at  a different distance from the face hosting the  two SiPM. For this purpose the long side of the crystal facing the beam was covered with a brass mask provided with four holes of 5mm of diameter located at 5, 12, 18.9 and 25.8 cm from the SiPMs. The 2 mm mask thickness prevents protons to hit the crystal out of the holes. For avoiding any possible energy loss, the measurement was performed inside a vacuum chamber
. The beam was collimated by a 5 mm collimator, 60 cm long. The detector was placed on a moving plate which allowed to remotely center the beam on each hole. Two thermocouples have been used to monitor the SiPM temperature during the test. The signal waveforms were digitized at 500 Msamples by a Lecroy WwaveRunner 620Zi oscilloscope. Fig.~\ref{fig:signals}  shows two typical signals observed for protons of 24 and 2.5 MeV. 

\newpage

\begin{figure}[htbp]
   \centering
   \includegraphics[width=12cm,clip=true] {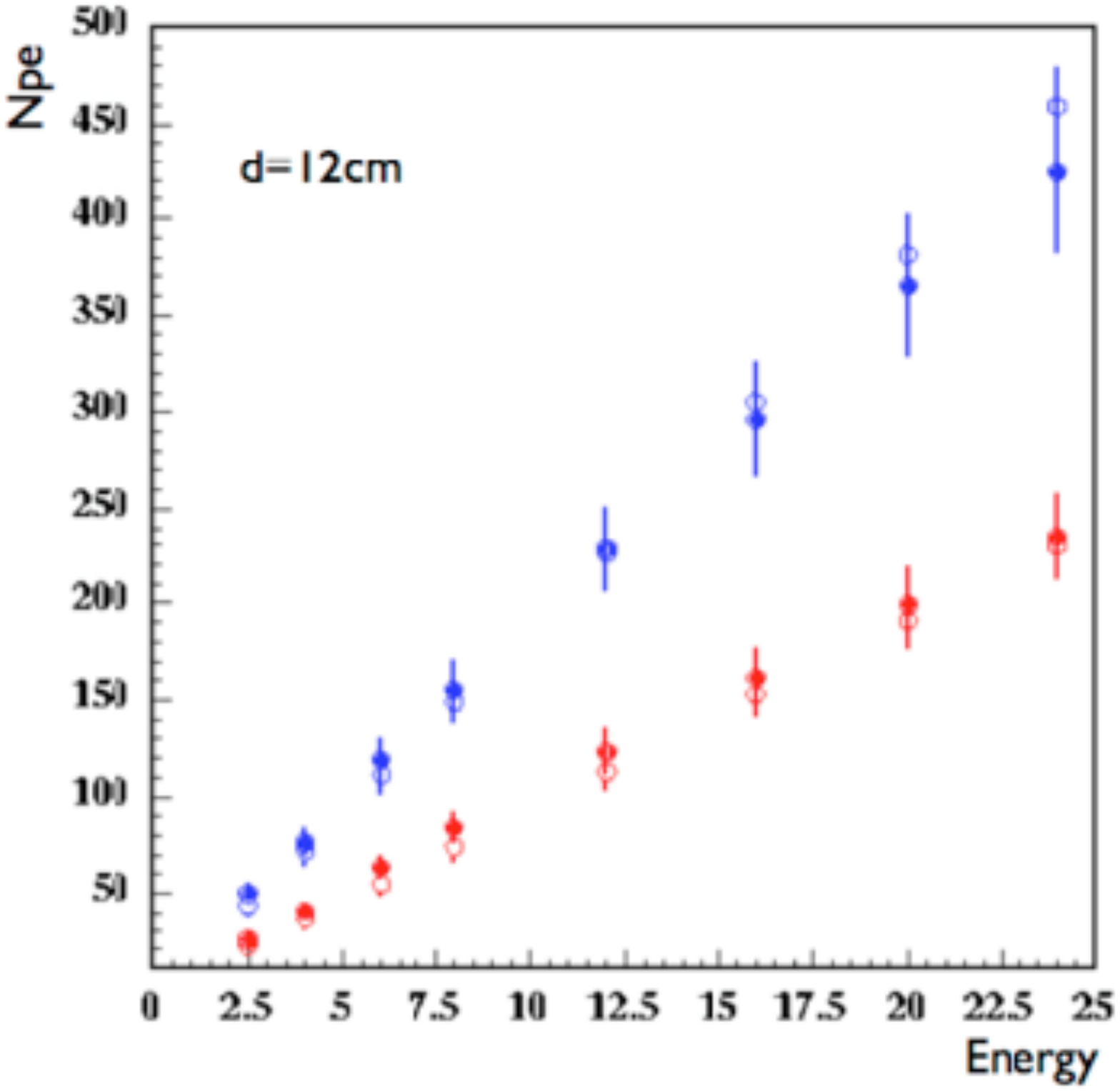}
   \caption{Number of detected p.e. as a function of the proton incident energy for the two SiPMs (full dots, blue for the 50 $\mu$m and red for the 25 $\mu$m pixel size). The beamÐSiPMs distance in this case was 12 cm. Empty dots represent the results of Monte Carlo simulations.  }
   \label{fig:pe_vs_energy}
\end{figure}

Time jitter, calculated as the signal rise-time over the signal to noise ratio, is of the order of 1 ns for protons of 24 MeV, for both SiPMs. This result suggests that, despite the long scintillation time of the crystal (about 2-3 $\mu$s), a few ns time coincidence is possible. Fig.~\ref{fig:pe_vs_energy} shows the number of p.e. collected by the two SiPMs as a function of the proton energy. The light quenching is correctly described by a Birk constant in the range of 3.2e-3 g/(MeV cm$^2$). The number of collected p.e. decreases linearly from 24 down to 2.5 MeV. It is worth noticing that protons of 2.5 MeV produce about 20 p.e. 25 $\mu$m SiPM.

\newpage

\begin{figure}[htbp]
   \centering
   \includegraphics[width=8cm,clip=true] {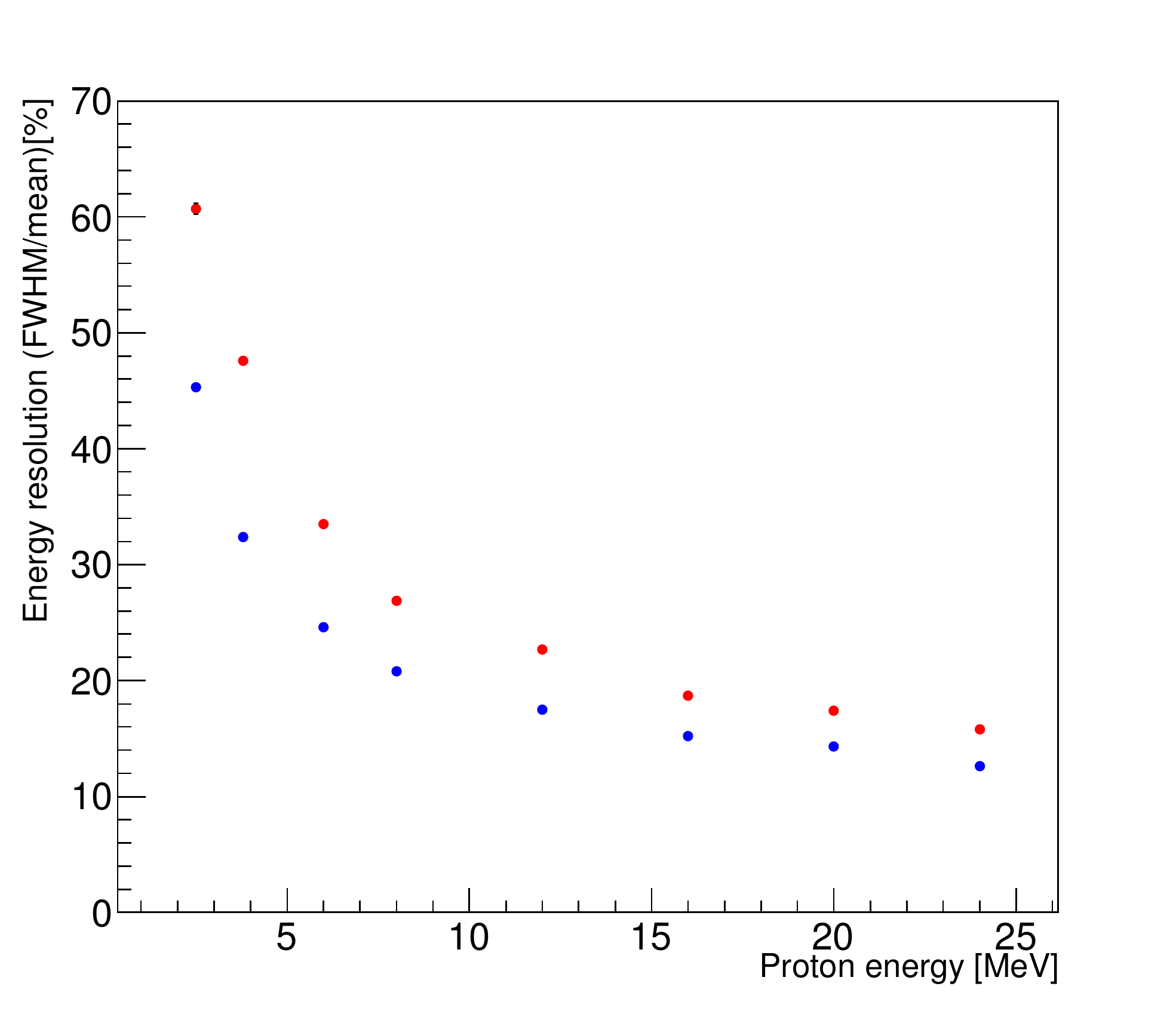}
   \caption{Relative energy resolution as a function of the proton incident energy for protons hitting the crystal at 12 cm from the SiPM. Blue dots refer to the 50 $\mu$m SiPM and red dots to the 25 $\mu$m one.  }
   \label{fig:ene_res}
\end{figure}

Fig.~\ref{fig:ene_res} shows the dependence of the relative energy resolution of the two SiPMs, defined as the FWHM over the mean value of the detected charge, as a function of the proton energy. As expected, resolution increases at higher energies reaching a value of $\sim$15\%. 

\newpage

\begin{figure}[htbp]
   \centering
   \includegraphics[width=12cm,clip=true] {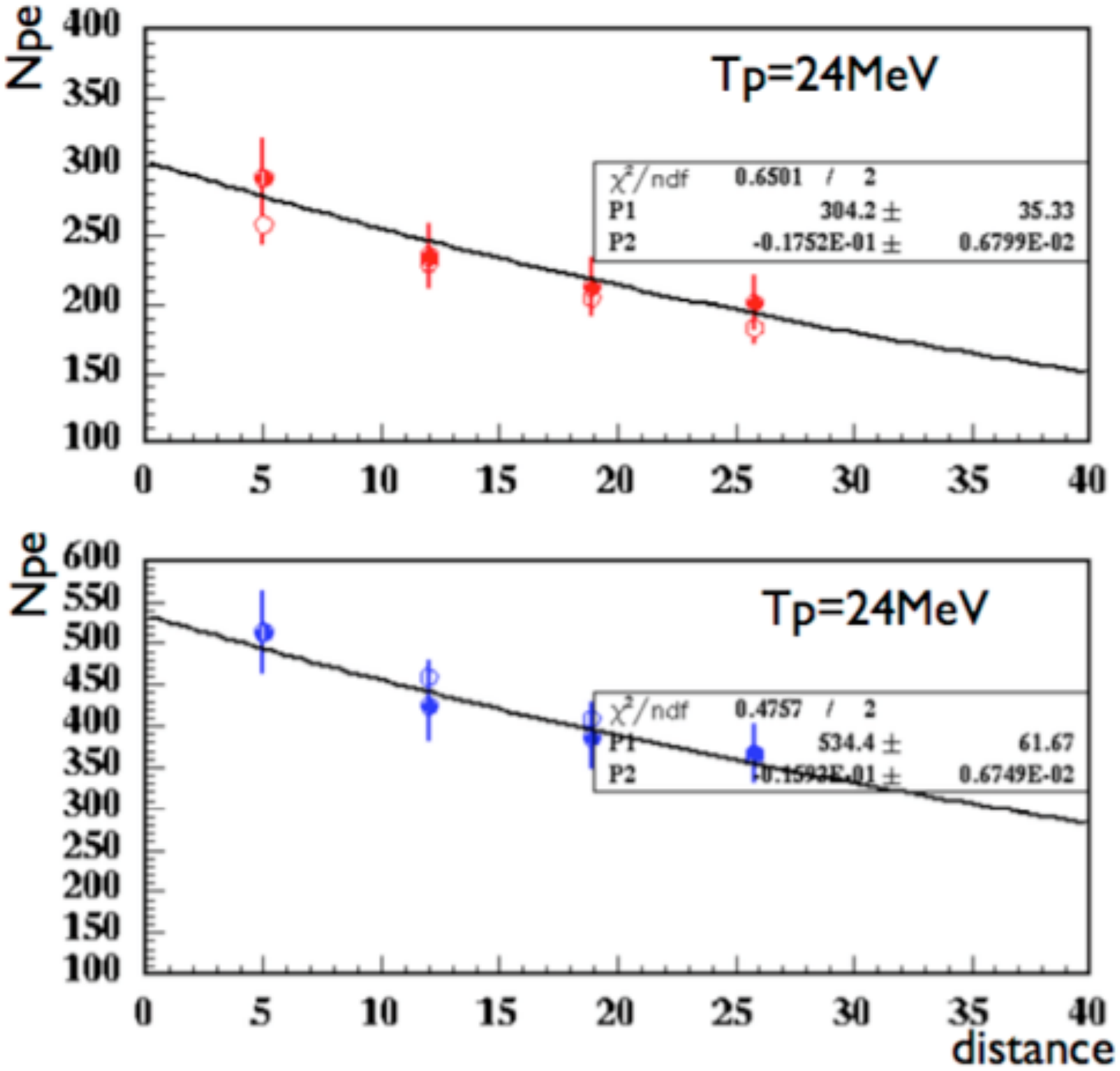}
   \caption{Number of detected photons as a function of the distance between the SiPMs and the point where the protons at 24 MeV hit the crystal, for the 50 $\mu$m SiPM  (top) and 25 $\mu$m one (bottom). Empty dots represent the results of Monte Carlo simulations. The black line, for each SiPM, is the exponential function that best fit the data.}
   \label{fig:pe_vs_position}
\end{figure}

Fig.~\ref{fig:pe_vs_position} shows, for protons of 24 MeV,  the number of collected p.e.  as a function of the distance between the SiPMs and the hit position. The behavior  is well described  by an exponential function with nearly the same attenuation length  of about 1/.016 cm$^{-1}$ ($\sim$ 60 cm) for both SiPMs.

\newpage

Notice that again a significant amount of p.e. is detected for protons of 2.5 MeV, even for those hitting the crystal at 25 cm from the SiPMs. These results prove the possibility of detecting low energy protons by using a large volume CsI(Tl) crystal coupled to the new readout based on small area SiPMs sensors. Figs.~\ref{fig:pe_vs_energy} and ~\ref{fig:pe_vs_position}  show how data  compare to GEANT4 Monte Carlo simulations (see Sec.~\ref{sec:crys-response} for details).  As it can be seen, the simulation well reproduces the number of p.e. as a function of both the proton energy and the incident position.

\clearpage

\clearpage
\newpage
\section{DRIFT-BDX}\label{sec:drift}

One of the advantages of a beam dump experiment is multiplicity.  Having invested in shielding sufficient
to block charged components and neutral hadronic components from entering the experimental hall behind the beam dump,
multiple experiments can then look for a dark matter beam with high sensitivity.  Because of the rare
interaction rate these experiments can be stacked one behind the other.  With the addition of the
DRIFT-BDX detector, described below, the collaboration intends to take advantage of this feature and at
the same time provide complementarity needed for a robust detection of dark sector dark matter.
Specifically the addition of DRIFT-BDX will provide powerful cross-checks on backgrounds in the beam
dump lab, utilizes a different physics channel for detecting dark sector dark matter and offers a
powerful directional signature of dark matter recoils.  Thus for marginal extra cost the DRIFT-BDX
experiment adds much to the physics potential of this proposal.

\begin{figure}[bp]
\begin{center}
\includegraphics[height=10cm,clip=true]{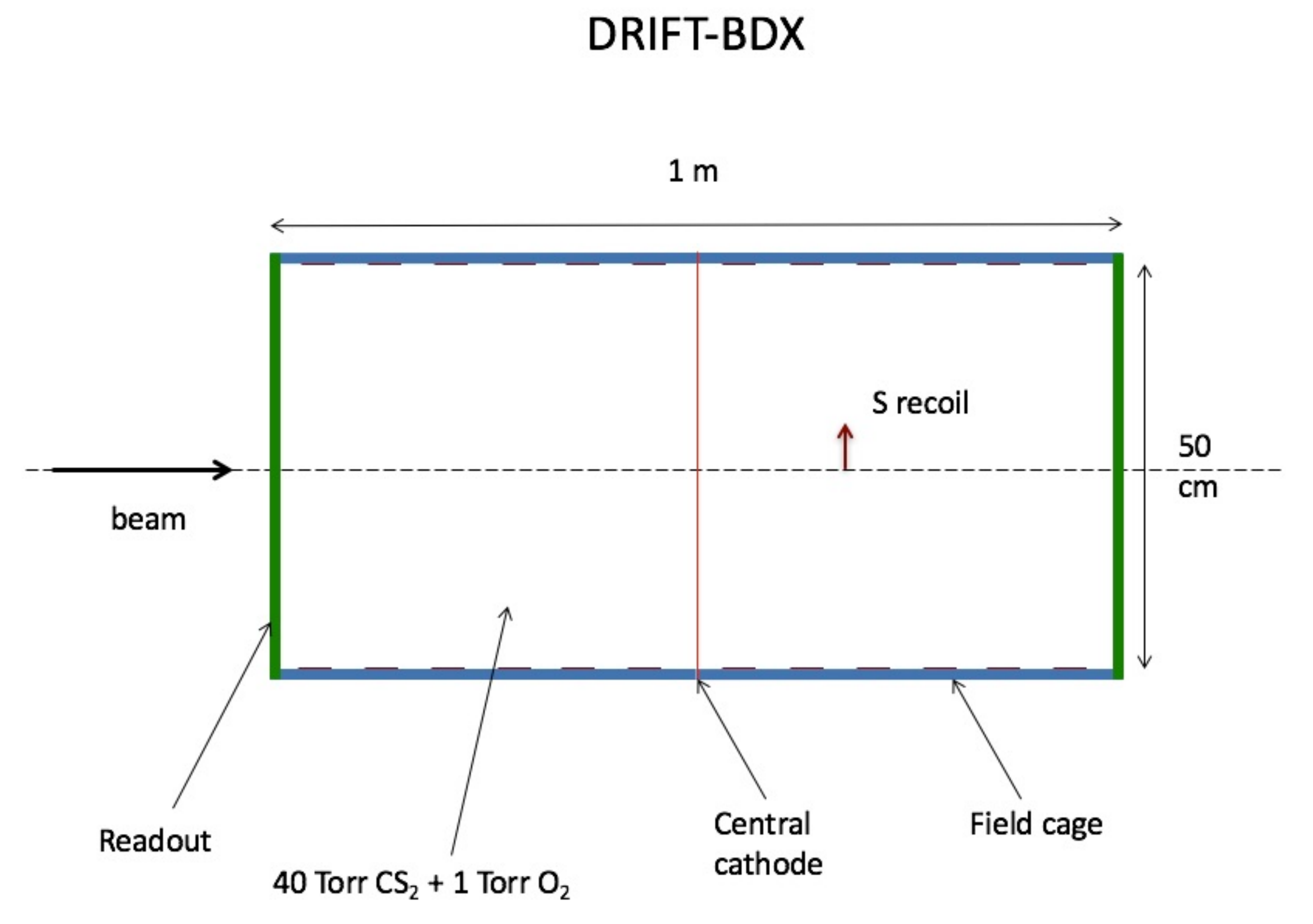}
\caption{A schematic for the DRIFT-BDX experiment.  Because of the prevalence of sulfur in the gas and the Z$^2$
dependence for elastic, low-energy, coherent scattering, the recoils would be predominantly S.
\label{fig:Drift_Schematic}}
\end{center}
\end{figure}

\subsection{Capabilities}
The Directional Recoil Identification From Tracks (DRIFT) project, with strong support from the NSF,
has been in continuous development since 1998.  The goal of the DRIFT collaboration is the detection of a
directional signal from WIMP dark matter in our galaxy \cite{DRIFTPRL}.  In order to accomplish this goal a
unique, low-pressure, Negative Ion Time Projection Chamber (NITPC) technology has been developed \cite{DRIFT-mobility}.
The negative-ion drift allows DRIFT to have the lowest energy threshold and best inherent directionality of
any limit-setting, directional, dark matter detector \cite{Burgos2009417,Burgos2009261}.  In 2013 another feature of negative-ion
drift was discovered paving the way for zero-background running of DRIFT \cite{Drift-MinorityCarriers}.  As a
consequence, DRIFT’s sensitivity to dark matter, utilizing the current DRIFT-IId detector in the Boulby
Mine, is almost 1,000$\times$ better than the competition’s \cite{Battat20151}.  With its unique directional
and background rejection capabilities, the DRIFT NITPC technology is ideally suited to search for light
dark matter at accelerators (LDMA).  We propose to search for directional low-energy LDMA-induced, recoils
utilizing the low-background NITPC technology developed for DRIFT.  There is no other current dark sector
experiment looking for dark matter via this coherent ($Z^2$) channel.

\subsection{Detector}
The design for the DRIFT-BDX detector is shown in Fig.\,\ref{fig:Drift_Schematic}.
This design was developed based on our experience operating DRIFT; we know that such a design is feasible.
The detector prototype would be 1 m long and 50 cm on a side filled with a mixture of 40 Torr
CS$_2$ and 1 Torr O$_2$ and placed in the beamline, as shown.  In the event that this prototype can successfully 
be operated
a larger, 10 m long detector could be made by replicating the 1 m long prototype 10 times.

\begin{figure}[tp]
\begin{center}
\includegraphics[height=10cm,clip=true]{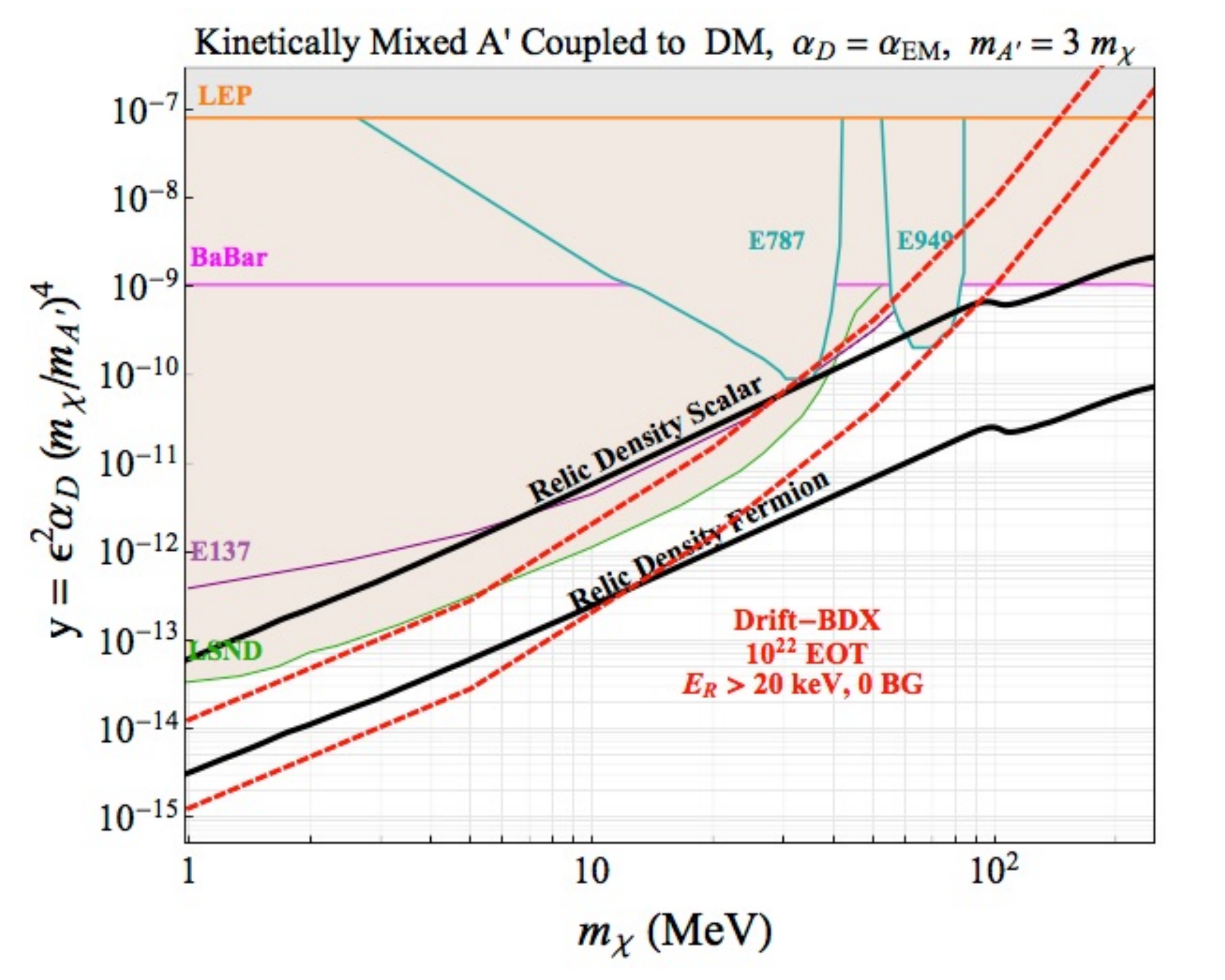}
\caption{90\% C.L. exclusion bands for DRIFT-BDX shown in dashed red for a 1 m long DRIFT-BDX prototype detector, upper curve, and 
for a 10 m long detector, lower curve, in comparison with other limits.
\label{fig:Drift_LDMA_Limits}}
\end{center}
\end{figure}

The benefit of being able to detect 10-100 keV
recoils is enormously enhanced sensitivity governed by the differential scattering cross section for coherent
LDMA detection as shown in Eq.\,\ref{eq:driftscat}.
\begin{eqnarray}
{d\sigma \over dT} & \approx & {8 \pi \alpha \alpha_D \epsilon^2 Z^2 M \over (m_{A'}^2 + 2 MT)^2 }
      \label{eq:driftscat}
\end{eqnarray}
where $\alpha$ is the fine structure constant, $\alpha_D$ is the dark sector fine structure constant,
$\epsilon$ is the coupling
to the dark sector, $Z$ is the charge of the nucleus, $M$ is the mass of the nucleus, $m_{A'}$ is the mass of the
mediator and $T$ is the kinetic energy of the recoil.  Even for small thresholds the second term in the
denominator tends to dominate so the full, coherent, scattering cross-section goes as $\sim (1/T_{thresh})^3$.
With an order of magnitude, or more, lower threshold than other experiments this, alone, confers to
DRIFT-BDX a huge increase in sensitivity.  In addition because the momentum transfer is so small
DRIFT-BDX can take full advantage of the $Z^2$, coherent, term in the numerator.  These factors largely
negate the density penalty for using a low-pressure gas detector. The limits achievable from DRIFT-BDX
running in parallel with the BDX scintillator experiment are shown in Fig.\,\ref{fig:Drift_LDMA_Limits}. As can be 
seen the sensitivity of DRIFT-BDX is comparable to BDX.

\subsection{Backgrounds}
Backgrounds in LDMA experiments are beam-related or beam-unrelated.  The goal, as with the DRIFT experiment,
is zero accepted background events.  In the event that is not possible DRIFT-BDX has powerful signatures,
discussed below, to pull signals out of the noise.
\begin{description}
\item[Beam-related $\nu$s] Generically electron beam-dump experiments generate far fewer $\nu$s than proton beam-dump
experiments \cite{Izaguirre:2013uxa}.  The collaboration estimates $\sim$10$^{17}$ will be produced by 10$^{22}$ EOT.  With a typical coherent, elastic scattering cross-section of 10$^{-39}$ cm$^2$ \cite{Drift-ScholbergNeutrino} a
back-of-the-envelope calculation gives a background for DRIFT-BDX of $\sim$0.03 events.  DRIFT-BDX will not observe
backgrounds from coherent elastic neutrino-nucleus scattering in this experiment.

\item[Beam-related n’s] It is difficult to simulate neutron backgrounds from 10$^{22}$ EOT.  In the biggest GEANT
simulation to date 1.4$\times 10^9$ EOT produced no neutrons in the lab.  Fortunately beam related neutrons were
measured in the Stanford Linear Accelerator Center (SLAC) milliQ experiment
\cite{Drift-milliQThesis}.
In that experiment 29.5 GeV electrons produced muons in the beam dump which then came to rest in sandstone
shielding, 5,000 g/cm$^2$ upstream of the detector.  Neutron recoils were measured in the milliQ scintillators
at a rate of 1 neutron recoil for every 10$^{12}$ EOT above a threshold of ~20 keV proton recoil energy \cite{SLACLDMA}.

The current, preliminary design for the shielding has the muons ranging out 3300\,g/cm$^2$ upstream of the detector though
there is ample room to add the remaining 1700 g/cm$^2$ shielding.  This calculation will assume the additional shielding.
Scaling conservatively DRIFT-BDX could expect 400,000 neutron recoils from 10$^{22}$ EOT.  GEANT-based simulations
of neutron shielding for the DRIFT experiment we calculate that only 100 g/cm$^2$ of plastic shielding will be required to shield
the DRIFT-BDX detector from beam-related neutron backgrounds.

This estimate is very conservative.  The 29.5 GeV muons in the milliQ experiment would produce many more neutrons than the
11 GeV muons in the BDX JLab experiment.  The beam overburden in the JLab experiment is significantly greater
($\sim$1700 g/cm$^2$ vs 1160 g/cm$^2$) than for the SLAC milliQ experiment reducing neutron scatters in the atmosphere.  And the milliQ
detector had no overburden whereas DRIFT-BDX will have 1,000 g/cm$^2$ overburden.  Finally the $\sim$20 keV proton-recoil threshold 
for the milliQ experiment corresponds to a $\sim$5 keV S recoil threshold for the DRIFT-BDX experiment after
accounting for mass and quenching effects, far below the effective 20 keV S recoil threshold of DRIFT.

\item[Off-axis counting] Because $\chi \overline{\chi}$ pairs are forward peaked (because of their small mass) and the proximity of the detector to the beam-dump, the event profile is expected to fall off rapidly from the beamline, especially at low masses.  Backgrounds, however, are likely to be uniformly distributed.  This will allow backgrounds to be measured and subtracted even for beam-related backgrounds.

\item[Beam-unrelated detector backgrounds] The underground operation of the DRIFT-IId detector suggests that backgrounds from the detector
are under control.  In the strongest (unpublished) limit no events were observed in 55 live days of running at low threshold.
Given the size difference between DRIFT-IId and DRIFT-BDX this is equivalent to running DRIFT-BDX for 175 days background-free.  
This limit is expected to improve.

\item[Beam-unrelated neutrons] A DRIFT-BDX experiment run on the surface would be dominated by cosmic-ray neutron recoils.  A
GEANT-based simulation suggests 435 events per day above 20 keV.  However DRIFT-BDX will be placed underground.  An estimate made
based on measurements done at almost identical depth \cite{DaSilva1995553} suggests that the flux of neutrons at the detector would be approximately
10$\times$ less than the flux coming from the walls of the cavern.  GEANT-based results give similar results.  DRIFT has experience shielding
neutrons coming from the walls.  With only 50 g/cm$^2$ plastic shielding the 55 live day results suggest, for the same reasons as above,
that DRIFT-BDX can run free from wall-neutron interactions in the fiducial volume of the detector for the requisite time.  In this
proposed experiment 2$\times$ this amount of shielding will be used to reduce beam-related neutrons to acceptable levels.  Thus DRIFT-BDX
will not be limited by cosmic-ray neutron recoils.

\item[Beam-unrelated muons] No reasonable amount of overhead shielding will reduce the flux of cosmic ray muons through the detector.
However, the ionization density (due to the 1/20 atmosphere gas) of relativistic muons is 350$\times$ lower than a typical nuclear recoil.
Thus unlike solid state detectors DRIFT-BDX can easily distinguish nuclear recoils from relativistic muons.  In the event that muons
are found to induce nuclear recoils, by muon induced neutrons inside the detector for instance, a muon veto will be considered.

\item[Beam-unrelated gammas] Gammas from the environment can Compton scatter in the gas of DRIFT-BDX and deposit ionization in
magnitudes similar to nuclear recoils.  However in a recent Co-60 exposure of DRIFT-IId the equivalent of 24 live days of
exposure of gammas from the walls of the underground lab was done and no events made it through the standard analysis.  Gamma
fluxes from shallow sites are thought to be equivalent to deep sites \cite{Gammas}.  The size difference between DRIFT-IId
and DRIFT-BDX and the addition of shielding relative to the DRIFT-IId experiment imply that beam-unrelated gammas will not affect 
DRIFT-BDX for the proposed experiment.
\end{description}

\begin{figure}[htp]
\begin{center}
\includegraphics[height=5cm,clip=true]{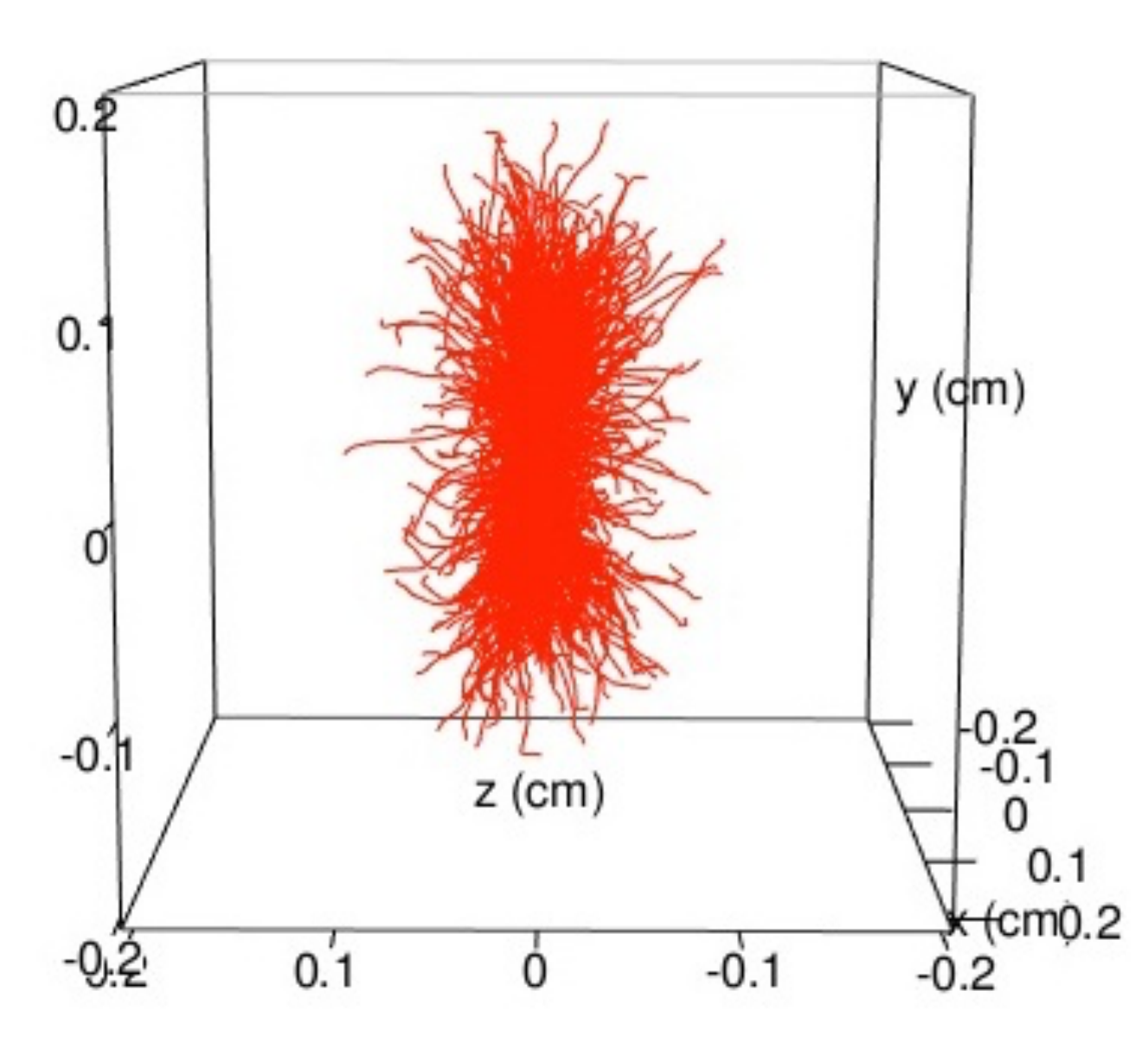} \hspace{2cm}
\includegraphics[height=5cm,clip=true]{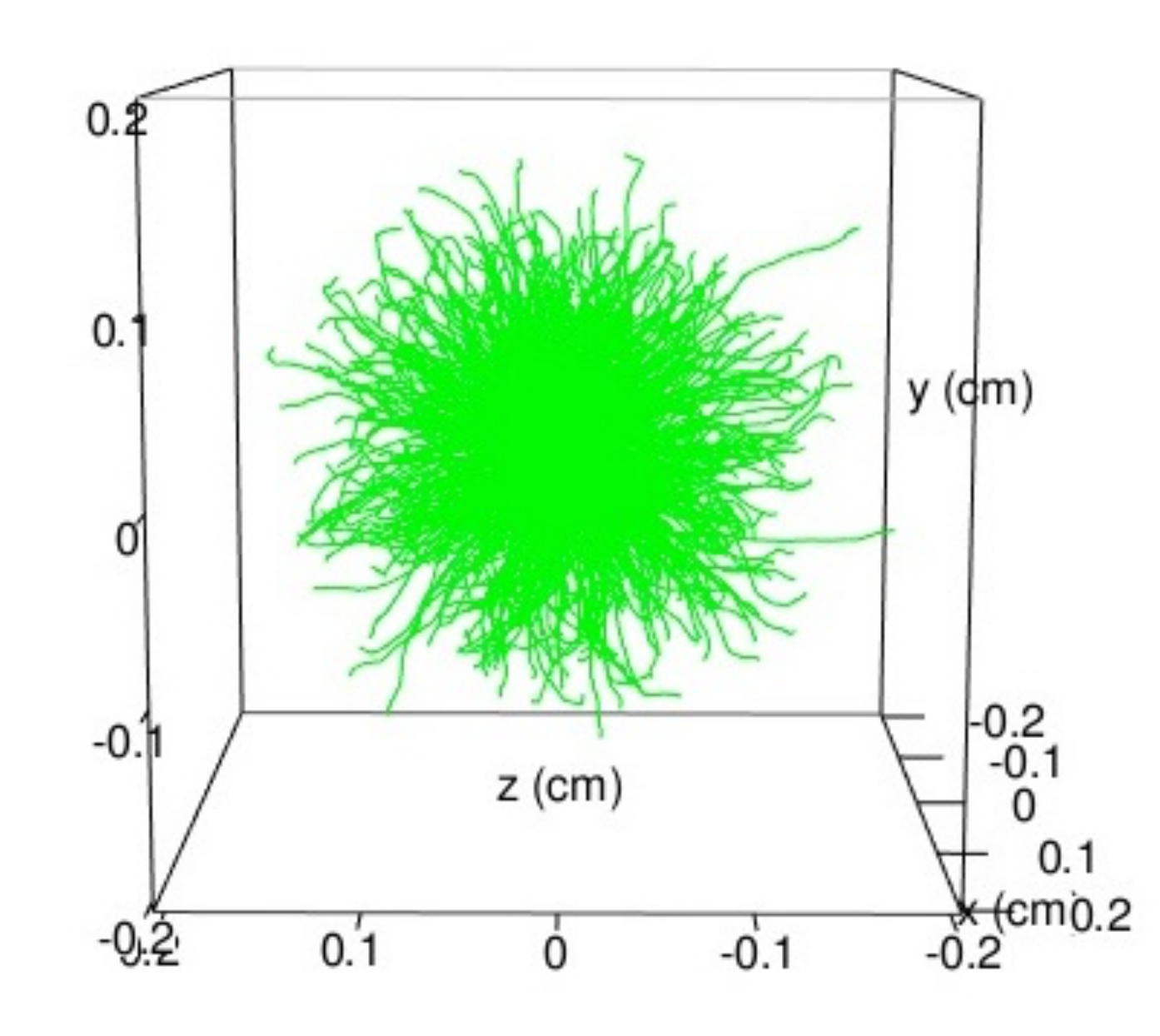}
\caption{a) This figure shows the tracks produced by 1,000 50 keV S recoils originally oriented perpendicular to the beam or
z axis according to an SRIM \cite{SRIM} simulation. b) This figure shows 1,000 50 keV S recoils oriented randomly as a comparison background.
\label{fig:Drift_PrettyWorms_S_50}}
\end{center}
\end{figure}

\subsection{Signature}
\begin{figure}[tbh]
\begin{center}
\includegraphics[height=10cm,clip=true]{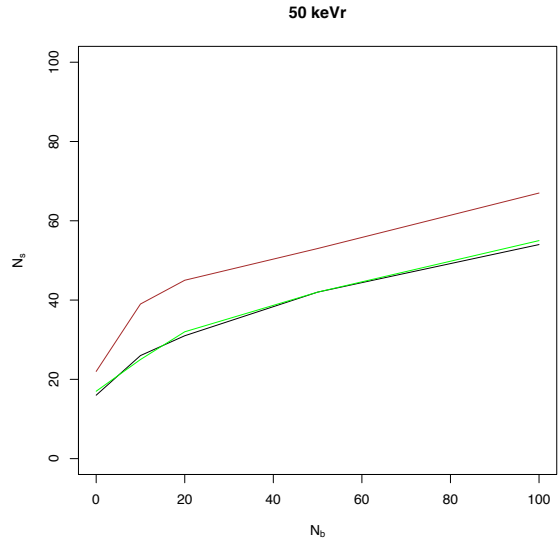}
\caption{The figures above show the number of signal events, $N_s$, on the vertical axis required for a 90\% C.L. detection in the
presence of, $N_b$, background events for 50 keV S recoils.  The black curve is for perfect detector
residual resolution, see text.  The green curve is are for a residual resolution of 0.02 cm.  And the brown curve is for
a residual resolution of 0.05 cm.  For reference the range of a 50 keV S recoil in 40-1 Torr CS$_2$ – O$_2$ is 0.086 cm.
\label{fig:Drift_Ns_vs_Nb_summary_50_keV}}
\end{center}
\end{figure}

Because of the prevalence of sulfur, S, in the gas and the $Z^2$ dependence for elastic, low-energy, coherent scattering, the recoils
would be predominantly S.  The 10-100 keV S recoils produced by light dark matter would be scattered within a few degrees of perpendicular
to the beam line due to extremely low momentum transfer scattering kinematics.  The signature of light dark matter interactions,
therefore, would be a population of S recoils with ionization parallel to the detector readout planes and uniformly oriented azimuthally.
Straggling of recoils at these low energies is significant.  Figure 48a shows the results of an SRIM \cite{SRIM} simulation of 1,000 50
keV S recoils oriented, originally, perpendicular to the beam, or $z$, or horizontal direction.  As can be seen the signature, small
dispersion in $z$, is degraded by straggling.  In order to quantify the effect an assumption must be made about backgrounds.
The simplest model for backgrounds is a uniform background.  Fig.\,\ref{fig:Drift_PrettyWorms_S_50} shows an SRIM simulation of 1,000 50 
keV S recoils oriented randomly.  For each event, signal or background, the dispersion of the ionization of the track in $z$, $\sigma_z$, was calculated.
A simple Kolmogorov-Smirnov-based test then determined the number of signal events, $N_s$, required in the presence of a number, $N_b$, of
background events for a 90\% C.L. rejection of an isotropic background only.  The results
are shown in Fig.\,\ref{fig:Drift_Ns_vs_Nb_summary_50_keV}.  The black curve in Fig.\,\ref{fig:Drift_Ns_vs_Nb_summary_50_keV}b shows 
$N_s$ signal events required for a 90\% C.L. detection in the presence of
$N_b$ background events for 50 keV S recoils.  For zero-background 16 events would be required.

Thermal diffusion and various detector effects will contribute to the measured dispersion in $z$ as discussed in Ref.\,\cite{DRIFT-mobility}.  The
largest of these is thermal diffusion from a track 50 cm from the detector plane.  Fortunately because the absolute position of
the event, $z$, can be measured \cite{Drift-MinorityCarriers} this contribution to the measured dispersion can be subtracted in quadrature.
We believe that various detector contributions can also be removed based on though the residual resolution, after
subtraction, has yet to be fully characterized.  As a result Figure 5 shows curves for 0 (black), 0.02 (green), and 0.05 (brown) cm
residual resolution for 50 keV S recoils.

\subsection{Conclusion}
The inclusion of the gas based DRIFT-BDX detector in the beamline with the scintillator based BDX experiment at JLab will allow for
significantly enhanced capabilities.  First both detectors are susceptible to different backgrounds with different thresholds allowing
for a broader understanding of facility backgrounds.  Second the sensitivity reach of the combined experiments will be significantly
enhanced.  And third, in the event of a detection DRIFT-BDX allows a powerful directional signature for verification.

\clearpage
\section{Required equipment and future plans}\label{sec:costs}
In this section we report the list of the equipment required to run the BDX experiment, the future plans and possible improvement of the detector and data analysis. 

\subsection{The BDX Collaboration}
The BDX Collaboration is composed by more than hundred physicists from over seven countries in the World. It merges competences in  scintillating and gaseous detector technology, high intensity, medium energy electron beam experiments and low energy proton/neutron detection. The BDX experimental program is  supported by a strong theory group, whose leadership in the Light Dark Matter field is widely recognized.  The BDX Collaboration is constantly in touch with the other Collaborations that are running or planning experiments to search for  light dark matter with electron and  proton beams: HPS, APEX, DarkLight, PADME, MMAPS, LDMX, NA64, MiniBoone.  It is an active part and significant contributor of the rapidly growing  Light Dark Matter search at Accelerator working group, organizing and participating to workshop conferences and common projects. The BDX  $R\&D$ activity performed so far  has been funded and supported by H2020-EU,  INFN-Italy, DOE-US and NSF-US. Requests for grants have already been submitted to these and other (MIUR-Italy) funding agencies. If  approved, the BDX Collaboration will be able to provide the necessary man power and seek for the necessary funds to  target the experimental program described in this proposal.

\subsection{Required equipment} 
The most part of the equipment necessary to run the BDX experiment will be provided by the BDX Collaboration. The detector as well as the readout electronic will be assembled, tested and deployed at JLab by the BDX Collaboration requiring a  minimal support form the lab. The Collaboration will be also responsible for detector  maintenance and operation and  data taking shifts. 
We rely on Jefferson Lab to build  the new underground facility  and  instrument it with the necessary services (power, A/C, networking). JLab is also expected to be in charge of networking, data processing and storage as for the other experiments running in Hall A, B, C and D. Here below a summary of the different component of the BDX experiments.\\

{\bf Detector: Calorimeter} 
\begin{itemize}
{\item CsI(Tl) crystals: provided by SLAC, refurbished and reassembled by the BDX Collaboration;}
{\item Crystal readout: SIPM, 6x6 mm$^2$ area , 25$\mu$m pixel size, to be procured by the BDX Collaboration;}
{\item FE electronics: trans-impedence preamplifiers, to be procured by the BDX Collaboration;}
{\item Services: LV, HV distribution, slow controls for SIPM to be procured by the BDX Collaboration;}
{\item Mechanics:  including crystal alveolus, supports, ..., to be  provided by the BDX Collaboration.}
\end{itemize}

{\bf Detector: Vetos} 
\begin{itemize}
{\item Plastic scintillator: BC412 paddles 1 and 2 cm thick in various sizes, to be procured by the  BDX Collaboration;}
{\item Light guides: PMMA, tapered, to be procured and polished  by the  BDX Collaboration;}
{\item WLS: Y-11, 1mm diameter, double cladding procured  by the  BDX Collaboration;}
{\item Photomultipliers: 1 inch area , bialkali photocathode, to be procured by the BDX Collaboration;}
{\item Photosensors: SIPM, 3x3 mm$^2$ area , 100$\mu$m pixel size, to be procured by the BDX Collaboration ;}
{\item FE electronics: trans-impedence preamplifiers, to be procured by the BDX Collaboration;}
{\item services: LV/HV, slow controls for SIPM and PMTs to be procured by the BDX Collaboration;}
{\item Mechanics:  including support for PMTs and SIPM, walls, legs, ..., to be  provided by the BDX Collaboration.}
\end{itemize}

{\bf Detector: Lead vault} 
\begin{itemize}
{\item lead bricks: 20x10x5 cm$^3$ to be procured by the  BDX Collaboration;}
\end{itemize}

{\bf Detector: DAQ} 
\begin{itemize}
{\item fADC:  250Mhz, 12 bit, 16 ch to be procured by the  BDX Collaboration;}
{\item Trigger:  VTP boards to be procured by the  BDX Collaboration;}
{\item Other: VME-VXS crates, on-board CPU, workstations  to be procured by the  BDX Collaboration.}
\end{itemize}

{\bf Data storage and data analysis Lab} 
\begin{itemize}
{\item Networking: fast connection between the Computer Center and the new facility to be provided by JLab;}
{\item Data processing:  $8\cdot 10^6$ CPU hours (single core) required to filter and process raw data, and run Monte Carlo simulations, to be provided by JLab;}
{\item Data storage: 600 TB of permanent storage (tape) and 100 TB of disk space for raw, reconstructed and simulated data.}
\end{itemize}

{\bf Beam dump facility} 
\begin{itemize}
{\item New underground beam-dump lab: including all necessary services to be provided by JLab;}
{\item Shielding: iron, concrete and dirt to be provided by JLab.}
\end{itemize}


\subsection{Further improvements and tests} 
Results show that, running at JLab high intensity Hall A beam-dump, it is possible to explore  a wide range of  model parameters.
We believe that there is still room for improvement in the detector concept,  active veto and  shielding design. Moreover, the expected low counting rate coupled with the signal digitization of  fADCs will allow us to optimize the off-line reconstruction algorithms, further  enhancing the rejection capability.\\
In deriving the sensitivity curves we always applied a conservative approach. Here below is a list of items we are currently investigating or we are planning to implement in an upgraded version of the prototype and MC simulations: \\ 

{\bf Cosmogenic background} 
\begin{itemize}
{\item the cosmogenic background was evaluated scaling the single-crystal rate measured in the  LNS set-up to a full detector: we are assembling  a matrix of 4x5 BaBar crystals CsI(Tl)  to obtain more accurate and realistic information about the possible correlations of inter-crystals rates; }
{\item we are planning to continue the LNS prototype measurement campaign with the crystal matrix. A long statistics run ($\sim$6 months) will provide insight on high energy events (E$_{seed}>$300 MeV) for a safer extrapolation to JLab set up; }
{\item the internal lead shielding could be further optimized changing thickness and number of layers, if necessary, and  the effect of adding  a further layer of an active inner veto is under investigation with both MC simulations and prototyping; }
{\item the JLab overburden described in the proposal  was chosen to mimic LNS conditions and provide a validated upper limit on cosmogenic rates in the detector: there is still some room  for further optimization (e.g. replacing the dirt with concrete or heavier material or accumulating more dirt on top of the beam-dump facility); a more effective shielding will be studied by MC simulation and validated with measurement locating the prototype in a similar configuration at LNS; }
{\item we are investigating the possibility of running the CEBAF machine with a beam macro-structure of 1$\mu$s and keeping the bunch micro-structure of 250 MHz; if a charge per bunch of 1.3 pC (the same configuration of G0 experiment) would be possible, we could achieve a factor of 5 of cosmic background rejection;  this would imply to run in  dedicated-mode, whose compatibility with the Hall-A current physics program should be checked;.}
\end{itemize}
  
{\bf Beam-related  background} 
\begin{itemize}
{\item a thorough scrutiny of how hadronic processes contributing to the predicted  neutrino flux are implemented in GEANT4 is underway;}
{\item a detailed implementation of the $\nu_e$-nucleon scattering in the detector is  being  implemented for a precise  evaluation of  the electromagnetic shower  induced by the $\nu$-N CC interaction;}
{\item the use of other simulation tools, such as MCNP or FLUKA, may provide a more reliable treatment of the predicted low-energy background (gamma and neutrons); in addition, these tools, providing a statistical description rather than a particle-by-particle tracking,  are better suited to deal with the large number of the EOT expected in BDX and, indeed, requiring a more limited extrapolation.}
\end{itemize}

{\bf Signal optimization and data analysis}
\begin{itemize}
{\item no directionality or other signal/background discriminating cuts, such as time correlation between veto and calorimeter or crystal multiplicity in the calorimeter, have been applied yet in the data analysis;
we expect that the information collected by the different sub-components of the BDX detectors will provide further discriminating power between signal and background;}
{\item the effect of limiting the signal detection to a inner   fiducial volume   by exploiting the calorimeter  segmentation and using the last layer of crystals as an additional veto is currently being studied by MC simulations;}
{\item keeping the number of crystals compatible with the BaBar ECal end cap supply, changing the geometrical arrangement of the crystal array is under study; }
\end{itemize}

The proposed BDX experiment, tacking advantage of the high intensity, high energy electron beam available at JLab has the unique capability of extending the possible reach by an order of magnitude with respect to the previous (un-optimised) measurements getting close to the unreducible background due to the neutrinos produced in the beam-dump interaction.  The BDX experiment at Jlab may represent the ultimate beam-dump experiment with an intense electron beam challenging a wide category of light DM models.
\clearpage
\section{Cover letter for BDX proposal submission to PAC44}
This Proposal follows the  Letter of Intent LOI-12-14-006 {\it Dark matter search in a Beam-Dump eXperiment (BDX) at Jefferson Lab} presented to PAC42 in 2014.\\
The  recommendations included in the PAC42 Final Report document read as follow:\\

\textbf{\textit{``Summary and Recommendation:}}{\it BDX could become the definitive beam dump experiment at electron accelerators. Sited at Jefferson Lab, it would use the CEBAF high intensity beam and modern technologies for detector design, trigger, and data acquisition, to achieve the most stringent limits (or to make the first discovery) of a class of dark matter particles.
The collaboration is encouraged to proceed with a full proposal to the laboratory, but the PAC emphasizes that the collaboration needs to meet a high standard in order to be eventually approved. Experimentally, a fully fleshed-out detector design needs to be presented, including both simulations and measurements (with CORMORINO or otherwise) that demonstrate its sensitivity to both detection channels as well as its ability to reject cosmic ray backgrounds with whatever necessary overburden. Theoretically, it must be made clear what models and attendant assumptions motivate this particular measurement, as well as the extent to which these models are (or are not) addressed in other experiments at other laboratories. Finally, the PAC realizes that the infrastructure costs to build and instrument a pit that would house this experiment will be extensive, and recommends that the laboratory require an approved proposal before scheduling onsite tests with beam as part of the design process.\\
Finally, we comment that BDX would obviously benefit from a low duty factor beam, as opposed to that provided by CEBAF, if a suitable high energy, high intensity accelerator could be identified.''}\\
 
The BDX Collaboration believes that this proposal addressed all the concerns following the  recommendations expressed by the PAC42:
\begin{itemize}
\item{the theoretical motivation have been revised based on two-years of rapid evolution in the field: the thermal target is clearly indicated as a reasonable, well
motivated, limit that BDX will be able to reach (at least for the {\it Relic density Scalar} scenario); }
\item{we investigated options for beam dump experiments at other electron-beam facilities, including DESY, Frascatti, Mainz and SLAC, and only JLab has the
energy, high current, practical access and beam availability for this program;}
\item{the growing  activity at proton machines (FNAL and CERN) as well as other experiments planned at electron/positron accelerator facilities (SLAC, Cornell and LNF)
have been mapped and demonstrate that, due to the unique combination of high energy and intensity, BDX at JLab represents one of  the best options
for dark matter (DM) searches in beam-dump experiments; }
\item{a full model of DM production in the beam dump, that realistically includes the effects of the electromagnetic shower formation, has been developed based on
the state-of-the art tools (MADGRAPH and GEANT4);}
\item{the BDX detector has been fully fleshed out: the core of the detector is an electromagnetic calorimeter that  reuses the scintillating CsI(Tl) crystals formerly
used in BaBar Ecal with a modern readout based on SIPMs and fADCs; active and passive vetos complement  the calorimeter by reducing and/or vetoing  cosmic background;}
\item{a detector prototype has been built and has been running for several months under experimental conditions similar to those expected at JLab; the measured cosmic
rates were extrapolated to the JLab configuration providing a solid, although conservative, basis for  the expected beam unrelated backgrounds in the detector;}
\item{full GEANT4 simulations of the detector, the beam dump,  the new underground facility and the shielding have been validated with real data down to a detected energy of  few MeV,  and used to predict the beam-related background and the expected signal produced by DM interactions; }
\item{the reach of the proposed experiment, for some of the predicted DM interaction processes, is only limited by the irreducible backgound (Charged Current $\nu_e$ interaction) showing that BDX at Jefferson Lab represents the ultimate beam-dump experiment with intense electron beams; }
\item{the concern related to the extensive infrastructure costs to build and instrument a new undeground facility for this and other beam-dump experiments, has been addressed by providing a realistic and detailed quote that includes not only the construction but also instrumentation and  services. }
\end{itemize}

\clearpage

\clearpage
\newpage

\bibliographystyle{unsrt}                                                                              
\bibliography{BDX-proposal_main}

\end{document}